## What is the Observer generated information process?

Vladimir S. Lerner, USA

*What is scientific path to fact of realty through its observed information? Is reality continuous or discrete as* ***0****s,* ***1****s information?*
*However, up to now both information and its connection to reality have not scientifically conclusive definitions, neither implicit origin.*
*What is information being observed? And how is the observed information connected with reality of observation.*
*The scientific path emerge in observing a multiple impulses inter-active yes-no actions modeling information Bits. Multiple interactions build Universe independently of their origin, and reality is only the emerging interactions as universal actions. Or information is phenomenon of interactions bringing certain reality with information of observing bits encoding an image. Multiple interactions are random composing random process which covers both the interacting bits and the process certainty. Uncovering the bits and information process, carrying certainty through observation of the random process, runs an Observer. Otherwise, the information carries reality of its observation and creates certain information process from which emerges the Information Observer. The observed information creating its Observer connects reality, information, and Observer. Solving these problems requires formal probabilistic approach where the notions of object, particle, and image are replaced with the probability of their events observation. Observing these events' probabilities models Markov process whose sequence of interacting yes-no impulses carry both bits' hidden information, covered by entropy of the process correlations and certainty, revealed by maximal observing probability. The impulse of the interactive No-action cuts the maximum entropy, while its Yes-action transfers a cutting minimum to the next impulse, thus creating the maxmin-minimax principle, decreasing the uncertainty along the observing process.*
*The hidden entropy releases an attraction of the correlation being cut. The impulse minimax probabilistic' a priori and a posteriori observing actions convey the process probabilistic causality in the entropy logic along the observations. When observation becomes more frequent, its probability grows, the observing impulses can merge. Merging action and reaction of the interactive impulses generate a microprocess within a bordered impulse bringing together probabilistic a priori and a posteriori actions on edge of predictability.*
*The emerging microprocess runs superposition and entanglement of the impulse yes-no conjugated entropy fractions. The fractions entangle during the time interval before the space is formed, composing two qubits and/or Bit which set up the information process, beginning with reversible logic within this impulse and creating geometry. When the impulse interacting actions curve the impulse space geometry, such interaction logically erases previous form of the entangled space entropy creating asymmetrical logic Bit as a logical Maxwell demon. With approaching probability one, the impulse' attracting interaction captures energy of such a real-certain interactive action physically erasing the entropy of bit logic. The erasure costs compensation of high–quality (quantum) energy of the physical process for entropy of lesser quality carrying the initial Markov diffusion. That removes the causal entropy of the asymmetrical logic, bringing asymmetrical Information logical Bit as certain impulse Bit. Such a Bit is naturally extracted at minimal energy measure equivalent to entropy ln2. This virtual probabilistic (entropy) hidden in interactions becomes real physical Bit. Bit is memorized at cost of Landauer's energy, working as Maxwell Demon.*
*The memorized impulse includes information Bit and free information, enclosing hidden information cutting correlation of the Markov process impulses, which measures interaction. Statistical micro-thermodynamics describe the interacting energy of the microprocess ensemble.*
*Multiple Bits self-organize information macroprocess, performing functions of Weller's Bit-participator. Along the multi-dimensional observing process, each memorized information binds the reversible microprocess with irreversible information macroprocess. The cutting entropy is automatically converting to information conveying the process logical information causality, certain logic, and complexity.*
*The process free information self-cooperate the Bits in triple information units. The triplet units assemble information network (IN) encoding the units in information three-geometrical structures enclosing the triplets' code. The IN triplets request the needed information generating a logic of probing impulses, sequentially cutting the observing process entropy measure and encoding new information units in the IN. The memorized and encoding information frizzed the interactive reality units. Multiple INs bind their ending triplets, enclosing information, cognition, and intelligence of the multilevel' interactive reality units. The Observer cognition assembles common units through multiple attractions in resonances loops at forming IN triplet hierarchy. The distributed cognitive logic self-controls encoding of the intelligence in a double helix coding structure (DSS). The Observer clock time intervals open access to external energy at each specific level of the IN multiple hierarchy, enabling memorization the hierarchy of the Bits encoding an explicit image. Information Bits become hierarchical measure of certain reality on the path from uncertainty. Multilevel interactions bring the multilevels reality and its information hierarchical measure. The intelligent Observer, self-reflective to the DSS logic, enables reading and understanding the message meaning. The Information Path Functional integrates the multiple interactive dynamics in the finite bits, which observe and measure reality during the time integrating path to reality.*
*The time and space of reality exists only as the discrete units of information. That brings the information measure to Wheeler's fundamental constant of Nature .All these uncover information-physical structure of interactions. The basic physical laws describe continuous reality.*
*The Interactive Integrated Information Dynamics describe the observing interactive dynamics of doublet-triplet logic, the IN formation, integration the logic, memory, cognition, encoding the intelligent observer in triplet code, which further encodes physical information, cognition, intelligence including consciousness, motor function, enable creating AI intellectual brain in the discrete reality.*

## I.INTRODUCTORY

**Introducing notions: reality, observation, uncertainty, certainty, information, information observer**

In book [1], the eminent scientists discussed various aspects of information, from quantum information to biological and digital information, in order to understand how Nature works.
Most discussions focus on unproven physical phenomena (like multiple Universe, "everywhere is conscience", complexity emerging through cosmological and biological evolution, others whose origin is still unknown. However a concept "that all reality is information and all information is reality, and the universe is composed of bits of information" is close to ours.
But how the bits of information emerge from reality is not discussed.
According to concept [2]:"It from bit symbolizes the idea that every item of the physical world has at bottom—at a very deep bottom, in most instances—an immaterial source and explanation; that what we call reality arises in the last analysis from the posing of yes-no questions".
*This concept is most closely related to ours.* We add that "information bits models multiple yes-no *actions*. Inter-actions unify all yes–no actions in physics and human being behavior. That why reality is not inactions–statists but a *dynamic phenomenon of interactions*, which create information".
Other concept [3] is that "information makes no sense in the absence of something to be informed—that is, a conscious observer capable of choice, or free will."
Moreover, recent study shows how brain gains knowledge through observation [4].
We agree that information should be observed, adding concept observation and its observer to ours.
For example, searching information on Web, a potential observer of this information sends probing impulses interacting with Web observing events. That creates multiple probabilistic events for the observer' choice activating its brain neurons impulses until actual information fact-events appears for the observer.
*This observer becomes observer of this information or the Information Observer.*
Similar examples are in scientific research, searching certain facts-information by multiple experiments-probes, or observing unknown particles, planets in a yet unknown Galaxy, tracking their probable or real interaction.
Like an astronomer traces unobserving planet measuring image of its probabilistic trajectory until it become most probable and informative; or a physicist traces a trajectory interactive particles in an Accelerator.
Observers are everywhere from communicating people, animals, different species up to any *interacting* subjects, accepting, transforming and exchanging information. This brings concept of *Observer interacting in Reality.*
The questions are: What they actually observe? Do they observe Reality? What is Information they observe?
And how is the observed information connected with reality of observation? What is the time of observing reality?
And how reality, information, observation, and the time, integrating a path to reality, are connected?
All these questions are still unanswered.
The known publications *do not explain* the connection *Reality-Information-Information Observer.*
Interaction is fundamental reality building structure of Universe from deep levels of physics: four fundamental interactions (gravitation, electromagnetic, weak, and strong) toward different chemical, thermodynamic, biological, human observations, up to brain neurons' inter-active actions, cognitive, intellectual, communication, all forms of life. They evolve in various substances of our world. *Inter-action includes act (act-ual) which causes real.*
What is scientific path to uncover the fact of realty through its observed information?
However, up to now both information and its connection to reality have not scientifically conclusive definitions, neither implicit origin, and *interactions were not considered as phenomena of reality.*
The formal recursive yes-no inter-action is an impulse modeling elementary information unit-bit (or 1|0,0|1 qubit) independently on their physical nature-from elementary particles up to any multiple interacting objects.
Since interactions are fundamental phenomena building various structure of Universe, they bring our concept: "*information is natural phenomenon of interactions, and observing information measures a certain reality".*

Reality as a natural physical existence is abstract entity only as universal interaction. But each observation creates specific certain reality of the observed interaction as the bits of discrete impulses encoding explicit image.
Hence, to observe reality, as inter-actions, the observation, carrying reality, should convey impulses.
The sequential observing interacting impulses convey information process that integrates the discrete observations. *Where interaction is, there is a reality, and information transferring the interaction.*
Each observation is act changing an observing process like its measurement. That generates an interactive observation. Exchanges between the interactive actions contribute connecting and binding these actions.
Multiple interactive impulses compose random process which conceals (hides) the interacting information bits and random process' reality. Such *observation* is a *random process of interactive impulses.*
The observation is a form of measurement: classical or virtual when, for example, Bayes probabilistic observing interactive impulse is a primary indicator of a potential probabilistic certain reality.
While before the measurement, at beginning of this process there are no facts about reality, the beginning is uncertain or imaginary, *virtual* regarding the facts certainty.
How uncover a path from uncertainty to certainty-as the fact of realty, focusing not on physics of observing process but on its information-theoretical essence?
*In observing interactive processes, the subsequent yes-no interactive actions, converting the observed uncertainty to observing certainty, creates information Bits carrying certain reality.*
Uncovering the bits and hidden reality through observation of the random process reveals the information process that *creates Information Observer*. What *runs* binding the Bits in *information process* during the observation of the multiple interactive impulses? How to find the information process uniting multiple certain bits that reveals multiple facts? The other question are: How to restore the information impulses carrying certain Bits, hidden in random observations, using the observing probes, which primarily interacting, able generating the randomness?
What is fundamental source of random interacting impulses, which potentially cover the information bit and their sequence in information process? How to measure the rapprochement of the observing process uncertainty to certainty and extract a certain physical bit? How to rebuild the information process from multiple sequence of observing bits hidden in interactive random process? How does the observing information create the Information Observer? These are fundamental questions, essential for Science and Applications, rising at observing multiple events and processes in physics, biology, cognition, cosmos, economy, sociology, experimental science, learning, reading, acquiring knowledge, examining investigations, playing games, dealing with many human activities; for example, in human interactive communications, discussions searching facts (truths).
Consequently, "what is the common in all of these" and how to use that common ground for answering the essential questions above? *The common is reality which interaction brings to the discrete interactive reality as a source of information in the interactive observations.* That primary includes the interactive processes at different levels: from quantum microprocess to multiple physical macroprocesses and up to human interactions.
That creates different levels of information quantity and quality evolving in an intellect and conscience.
Multilevel interactions bring the *multilevels' reality and its information measure. The information, measuring each level, allows the level's quantitative comparison for classifying their difference by minimal information lost.*
Understanding these processes and their regularities requires information laws describing development of the Information Observer in a discrete observing reality. Since basic physical laws describe continuous reality, the sought information laws should describe interactive processes resulting from integration the impulse observation in discrete reality. Such study requires more general and formal approach aimed on artificial design of human thoughts.
Here we focus on the joint notion of information and reality, the Observer's generating multilevel micro-macro processes, emerging cognition, intelligence, and their information-physical regularities.

## 1.Foundation of the Approach

1.1. Axiom

Multiple natural interactions build Universe.
Emergent interactions form Reality.
Inter-action is interconnection that includes act of probing or observing reality.
The observed interaction creates a specific certain reality.

1.2. Corollaries

1. Natural interactions unify a sequence of interactive impulses Yes-No (or No-Yes) actions. Each real (certain) inter-action is opposite yes-no action modeling elementary Bit of information, or a discrete impulse of 1|0, 0|1 qubit. Thus, information is a universal phenomenon of interactions which also 'an engine' to certain Reality.
2. The interaction of multiple impulses is source of randomness. Multiple interactive impulses compose random processes in surrounding random field [5:26]. The field connects sets of possible and actual events with their probabilities. This connection forms a triad. The actual events store energy.
Each triad of interacting events emanates random process. The interacting events hide Bits which covers the random process.
3. A manifold of observing actions identifies particular observer which observes the random process of specific triad. Each observation changes an observing process, generating an interactive observation. Disclosing the hidden Bit discrete impulses requires interactive cuts of the random process.The observation starts with observing the random process by probabilistic impulse virtual observations, which could uncover a vitrtual observer.
4. Reaching the observing certain reality runs the interactive process probability approaching one when the real cut delivers the actual events' energy. That exposes Bit emerging from the observing specific impulse interactions as unit of information in information process, carrying explicit image and certain reality. The information process dynamics assemble hierarchical Nets with multilevel interactive information units. The integrated units dynamics compose Information Observer enables ingoing observations.
5. Information path to reality runs integration of observation which emerges as information measure of the reality where searching fact is.

*Information emerges as a phenomenon of interactions and an observed measure of the reality.*

### 1.2.Essence of mathematical modeling methodology implementing the Axiom and Corollaries

The modeling uses example with an astronomer (an observer) that traces unobserving yet planet image by its probabilistic trajectory.
All potential observed probabilistic images, considering as abstract entity, we model by axiomatic probabilities of a random field[5].
An observing planet image models a specific triad of the field, while its probabilistic trajectory models Markov chain [5:25]. The observer sends discrete Yes-No ↓↑ probing impulses acting on his Markov chain model, imitating the image trajectory. These actions carry the impulses changing probabilities of the events in the specific triad where all set of trajectories images defines the observing events, the probabilities of events define the set of possible events, and the events' most probable copy identifies set of actual events.
The most probable copy, as an actual event, is memorized–registered using the actual event energy.
These impulse probabilities virtually observe the running Markov chain by multiple interactive acts.
A probabilistic observation of at least two discrete impulses of the chain connects a minimum of three its random events in mutual correlation during the observing random time.
The Markov chain under such observation transforms into Markov diffusion process [6].
The process transitive probabilities define Bayes probabilities [5:77], whose sequential a priori-a posteriori probabilities bring probabilistic impulse observation of the Markov diffusion process.

Each observing No action $\downarrow$, in a sequence of these probabilities, virtually cuts entropy of the impulse correlation from the observing Markov diffusion process and transfers Yes action $\uparrow$ to the Bayes observing-process impulses. During the transition switching time, each impulse accumulates entropy spent on the interactive transition, sequentially increasing the entropy of the observing impulses. The entropy being cut from the Markov process decreases, while entropy enclosed within each impulse interactive action $\downarrow\uparrow$ increases. Each following interactive impulse of the Bayes probabilities, observing the Markov process, maximizes the entropy of the Bayes process impulses.
The multiple Bayes impulses generate the secondary Markov diffusion whose entropy increases comparing with that in the observing Markov process. That brings max-min principle for each interacting impulse.
Under such observations, the Markov diffusion evolves to a Lévy process [7].
The Lévy process independent random increments model multiple observing impulses of the random bits.
The Lévy-Khintchine decomposition [7:113-117] characterizes any Lévy process in terms of three components (the Lévy triplet) one of which is the Lévy measure.
This measure describes the distribution of the Lévy process potential jumps.
Until any two time intervals of the process' independent random increments do not overlap, the jump absent and Lévy measure is zero.
The overlap is a merge of the yes-no impulses, when Lévy measure is nonzero, finite.
The merge of the yes-no impulses indicates zero random bit.
When Levy measure becomes finite, the jumps have started.
The jumps initiate a microprocess within the observing Levy process, which releases the entropy-uncertainty hidden in the cutting correlation, finally generating a non-random real bit.
At the merge, the probability of the impulse carrying a bit is zero.
It means there is no reality, defined by the interaction, when the microprocess starts.
With zero probability of the microprocess start, it is unpredictable.
The information process starts when the microprocess generates real qubit and/or bit.
Each specific triad through an observing microprocess is virtually forming related physical bits.
The considered path from emerging probabilistic impulse observation up to zero impulse identifies virtual (uncertain) process covering hidden information (as uncertain reality), which precedes the starting information process (beginning a certain reality).

**1.2.1.A Mathematical model of raising the jumps**

Consider a one dimensional Levy process [8] $\{X(t), t \in [0,\infty)\}$ defined by a Brownian motion $W(t)$ and a jump process $\varsigma(t)$, where $X(t) = \alpha t + \sigma W(t)$ is Markov process with continuous part and $\varsigma(t) = \int_R zN(t,dz)$ is the jump process with independed components. It is peace-wise constant with right–continuous and left limits, the jump height at time $t$ as $\Delta\varsigma(t) : \varsigma(t) - \varsigma(-t)$.
For any Lebesque measurable set $D \notin 0$, the jump measure of $\varsigma(t)$ defines
$N(t,D) := \sum_{s:0<s\leq t} I(D \mid \Delta\varsigma(s))$, where $I(D \mid x) = 1$ if and only if $x \in D$ is the identity function.
Here $N(t,D)$ is number of jumps of size $\Delta\varsigma(t) \in D$ which occure before or at time $t$, which is a random number depending on the sample path of $\{\varsigma(t), t \in [0,\infty)\}$.

The average number of jumps of size $\Delta\varsigma(t) \in D$ in a unit of time, as the jump rate $\nu$, is called Levy measure:

$$\nu(D) := E[N(1,D)] = E[\sum_{s:0<s\leq t} I(D \mid \Delta\varsigma(s)] . \qquad (1)$$

The Levy measure can be extended on the n-dimensional Levy process [7].
It follows, if Lévy measure is zero, the jump absent.

**1.2.2. About the experimental frequencies forming Markov chain**

Probability of random events (inter-actions) defines ratio of a favorable event numbers to total number of events. This ratio measures frequency of the multiple events. When this ratio grows up to satisfying symmetry, or equal possibility for the experimental frequencies, and independence of the events, with Big Number Law, defining frequency stability; *then* formulas below counts probability of the random events for that experimental frequency.

Specifically, with occurrence in multiple tests with *m*-events frequencies $f_m^i$ in *n*-sequence they approach experimental probability:

$$E[f_m^i]_{n\to\infty} = P_m^i , \qquad (2)$$

where mathematical expectation for these frequencies theoretically satisfies to Big Number Law.
Hence the multiple interactive frequencies produce the random events' probabilities, while a sequence of the independent random events, with conditional probabilities depending only on previous event, forms a random process-*Markov chain* [5:25].
That also leads to the Markov chain our initial random model of observation where increasing frequency grows probabilities.

**1.3.The integral measures of the virtual and information processes**

The *integral measure* of the observing *process* trajectories formalizes an *Entropy Functional* (EF)[9,10] which is expressed through functions drift $a^u(t,\tilde{x}_t)$ and diffusion $b(t,\tilde{x}_t) = 1/2\sigma(t,\tilde{x}_t)\sigma(t,\tilde{x}_t)^T$ of Markov diffusion process $\tilde{x}_t$: $\Delta S[\tilde{x}_t]|_s^T = 1/2E_{s,x}\{\int_s^T a^u(t,\tilde{x}_t)^T(2b(t,\tilde{x}_t))^{-1}a^u(t,\tilde{x}_t)dt\} = \int_{\tilde{x}(t)\in B} -\ln[p(\omega)]P_{s,x}(d\omega) = -E_{s,x}[\ln p(\omega)]$, (3) and the probability density measure $p(\omega)$ on the process trajectories.

The process entropy EF specific is its connection with time of observation $(T-s) = T_o$.

Cutting the EF by the impulse Kronicker function determines the discrete increments of information $\Delta I_k[\tau_k]$ for each impulse during its time interval $\tau_k$.

(The EF cut off [10] models the observing discrete interacting impulse).

*Information path functional* (IPF) [11] integrates the cutting information increments during time interval $T_I$ of observing information, where $T_I = \sum_{k=1}^{m}\tau_k$.

The IPF integrated impulse information measures a certain reality. The sum of the discrete increments along the path of the cutting process intervals approximate the IPF $I_p$ in a limit

$$I_p = \lim_{m\to\infty}\sum_{k=1}^{m}\Delta I_k \times [\tau_k] \qquad (4)$$

Each EF $\Delta S_k[x(\tau_k)]$, cutting on invariant impulse measure $[x(\tau_k)]$, is proportional to the impulse time interval $[x(\tau_k)] \sim [\tau_k]$, it satisfies $\Delta S_k[x(\tau_k)] \to \Delta I_k[\tau_k]$.

The minimax variation principle (VP) [12] for the EF-IPF determines the invariant information measure $\Delta I_k[\tau_k]$ decrease proportional to size of the cutting impulse time intervals $[\tau_k]$.

Along the VP extreme trajectory, these invariants satisfy the following relations: $\Delta I_k = \Delta I_{k=1} Q^{m-1}$ and $\tau_k = \tau_{k=1} Q^{-(m-1)}$, at coefficient of decreasing $[\tau_k]: Q = 3$.
That leads to the IPF in form

$$I_p = \lim_{m\to\infty} \sum_{k=1}^{m} \Delta I_{k=1} 3^{m-1} \times \tau_{k=1} 3^{-(m-1)} = \lim_{m\to\infty} \sum_{k=1}^{m} [\Delta I_{k=1} \times \tau_{k=1}], \quad (5)$$

which, for invariant impulse information measure $\Delta I_{k=1} = |1|_k = 1Nat$, brings

$$I_p = \lim_{m\to\infty} \sum_{k=1}^{m} [|1|_{k=1} \times \tau_{k=1}] = \lim_{m\to\infty} m[|1|_{k=1} \times \tau_{k=1}]. \quad (6)$$

Here $\tau_{k=1}$ is time interval of the impulse starting information process; thus, the starting $[|1|_{k=1} \times \tau_{k=1}]$ is fixed.
The observing time $T_I = \sum_{k=1}^{m} \tau_k$ determines the sum of geometrical series sequence of $\tau_k = \tau_{k=1} 3^{-(m-1)}$:
$T_I = (1 - 3^{-(m-1)}) / 3^{-1)}$ which at the limit $T_{\mathrm{Im}} = \lim_{m\to\infty} T_I(m) = 3/2$ identifies the finite time on the observing extreme trajectory. During this time, the number $m$ of the observing the IPF contributions is limited:

$$T_{\mathrm{Im}} = \tau_{k=1} \sum_{m} 3^{-(m-1)} = \tau_{k=1} 3^{-1} m, m = 9/2\tau_{k=1} \text{ at limited } \tau_{k=1}.$$

That also limits value of the IPF observing during $T_{\mathrm{Im}}$: $I_p = m[|1|_{k=1} \times \tau_{k=1}] = |1|_{k=1} 9/2$.
Generally, at any $Q > 0$, upper time $T_I$ in the IPF functional is limited by $T_I = \lim_{m\to\infty} \tau_{k=1} Q^{-(m-1)} \to 0$, (7)
as well as $\min I_p = \min[\lim_{m\to\infty} \Delta I_{m-1} x(\tau_{m-1})] = \min[\lim_{m\to\infty} \Delta I_{m-1}[\tau_{m-1}] \to 0$,
which limits the above sum of $I_p$ at $m \to \infty$.
Therefore, the cutting *Entropy Integral converges in the Path Functional*, and both of them are restricted at the unlimited dimension number. (The sum $I_p$ at $m \to \infty$ does not diverge since the limit of the min $I_p$ is zero).
Moreover, for the invariant impulse, its geometrical measure $M_{gk} = |1|_{gk} = \pi$ is proportional to the time measure $[\tau_k]: [\tau_k] = \pi p, p = 1/\sqrt{2}: M_{gk} = p \times [\tau_k]$ at constant $p = 1/\sqrt{2}$, while the information and geometrical measures hold invariant [10]. That leads to limited

$$I_p = \lim_{m\to\infty} \sum_{k=1}^{m} [\Delta I_k \times [\tau_k]] \to \sum_{k=1}^{m} p \times [\tau_k]^2 \to 0 \text{ at } \tau_k = \tau_{k=1} Q^{-(m-1)}. \quad (8)$$

The density of information for each following impulse
$D_k[\Delta I_k] = \Delta I_k / \tau_k = \Delta I_{k=1} Q^{m-1} / \tau_{k=1} Q^{-(m-1)} = (\Delta I_{k=1} / \tau_{k=1}) Q^{2(m-1)}$
is substantially growing along the observing extreme trajectory. In the limit, at $Q > 0$:

$$D_m = \lim_{m\to\infty} \sum_{k=1}^{m} [(\Delta I_{k=1} / \tau_{k=1}) Q^{2(m-1)}] \to \infty. \quad (9)$$

Since the time transition to each following IPF' Nat decreases, each such Nat integrates all preceding Nats concentrating the integral information in the final IPF Nat.
Such a Nat of infinite density cannot exist as real.
Or such reality measure is unobservable even during finite observation time (7). It measures *absolute reality*.
The time, starting the information process of the discrete information impulses, begins its IPF integral information measure in Information Macrodynamics [13,14].

This integral information measure on the extreme trajectory of information impulses is limited measuring certain reality; while the impulse information measure density growing to infinity reaches an absolute and unobservable reality.
The IMD formalism evolves to Interactive Integrated Information Dynamics (IIID).
The minimax variation principle for the EF-IPF determines equations of IIID which formalize the information regularities of observing random process, and information measures of its interactive reality.
The IIID integrates the multiple interactive dynamics in finite bits, which observe and measure reality during the time integrating path to reality. All these uncover information-physical structure of interactions.

### 1.4.Essence of Information Process and Intelligent Observer

The Information process emerges within observing random Levy process, starting with unpredictable (probability zero) merge-overlap of the yes-no impulses creating random jump with conjugated entropy fractions beginning a microprocess.
The emerging microprocess runs superposition and entanglement of the impulse yes-no conjugated entropy fractions. The fractions entangle during the time interval before its space is formed, composing two qubits and/or Bit which set up a reversible logic within this impulse created geometry.
When the impulse interacting actions curve the impulse space geometry, such interaction logically erases previous form of the entangled space entropy creating asymmetrical logic Bit as a logical Maxwell demon.
With approaching probability one, the impulse‘ attracting interaction captures energy of such a real-certain interactive action physically erasing the entropy of bit logic.
The erasure costs compensation of a high–quality energy of the physical (quantum) process for entropy of lesser quality carrying the initial Markov diffusion. That removes the causal entropy of the asymmetrical logic, bringing asymmetrical Information logical Bit as certain impulse Bit.*’Entropy is a measure of lost information*’[2]. Such a Bit is naturally extracted at minimal quality energy measure equivalent to entropy ln2.
This virtual probabilistic (entropy) hidden in interactions becomes real physical Bit.
The Bit is memorized at cost of Landauer’s energy, working as Maxwell Demon.
The memorized impulse includes information Bit and free information, enclosing hidden information of cutting correlation of the Markov process impulses.
The initial probabilistic events through the free information virtually form related physical bits.
Statistical micro-thermodynamics describe the interacting energy of the microprocess ensemble.
Multiple Bits self-organize information macroprocess, performing functions of Weller‘s Bit-participator [2].
Along the multi-dimensional observation process, each memorized information binds the reversible microprocess with irreversible information macroprocess.
The cutting entropy is automatically converting to information conveying the process logical information causality, certain logic, and complexity.
The process free information self-cooperate the Bits in triple information units. The triplet units assemble information network (IN) encoding the units in information geometrical structures enclosing the triplets’ code.
The IN triplets request the needed information generating a logic of probing impulses, sequentially cutting the observing process entropy measure and encoding new information units in the hierarchical IN.
Multiple INs bind their ending triplets, enclosing Observer Information, hierarchical reality, cognition, and intelligence [16].

The Observer cognition assembles common units through multiple attractions in resonances loops at the forming IN triplet hierarchy.
The distributed cognitive logic self-controls encoding of the intelligence in a double helix coding structure (DSS).
The Observer clock time intervals open access to external energy at each specific level of the IN multiple hierarchy, enabling the memorization and encoding the hierarchy of these Bits.
These Bits become hierarchical measure of certain reality on the path from uncertainty[17].
The Intelligent Observer, self-reflective to the DSS logic, enables reading and understanding the message meaning.
The IIID describes the emerging doublet-triplet logic, the evolving hierarchy of the information process with IN formation, integration the logic, memory, cognition, encoding the intelligent observer in triplet code, which further encodes physical information, cognition, intelligence including consciousness, motor function, enable create AI with intellectual brain.

**References to Introductory**

[1]. P.Davies, N.H. Gregersen, *Information and the Nature of Reality: From Physics to Metaphysics*, 2011, Cambridge Univ. Press.
[2]. J. A. Wheeler, Hermann Weyl and Unity of Knowledge, *American Scientist*, **74**(4): 366-375, 1986.
[3]. J.Horgan, Why information can't be the basis of reality, *Scientific American*, 127, March 7, 2011.
[4] C C. Charpentier et al. Neuro-computational account of arbitration between imitation and emulation during human observational learning, Neuron ,2019, DOI: 10.1101/828723.
[5]. A.N. Kolmogorov, *Foundations of the Theory of Probability*, Chelsea, NY, 1956.
[6]. E.B.Dynkin, *Theory of Markov Processes,* Prentice-Hall. 1961.
[7]. P.P. Levy, *Stochasic Processes and Brownian movement,* Deuxieme Edition, Paris, 1965.
[8]. Xi-Ren Cao, A new model of continuous-time Markov process and impulse stochastic control, IEEE 978-1-4244-3872-3/09, 2009.
[9]. V.S. Lerner, The boundary value problem and the Jensen inequality for an entropy functional of a Markov diffusion process, *Journal of Mathematical Analysis and Applications*, **353** (1): 154-160, 2009.
[10]. V. S. Lerner, The impulse cutoff an entropy functional measure on trajectories of Markov diffusion process integrating in information path functional, *arXiv*: 1204.5513, 2016.
[11] V.S. Lerner, Information Path from Randomness and Uncertainty to Information, Thermodynamics, and Intelligence of Observer, *arXiv:*1401.7041.
[12]. V.S. Lerner**,** Solution to the variation problem for information path functional of a controlled random process functional, *Journal of Mathematical Analysis and Applications*, **334**:441-466, 2007.
[13]. V.S. Lerner, Information Systems Theory and Informational Macrodynamics: Review of the Main Results, *IEEE Transactions on systems, man, and cybernetics*—Part C: Applications and reviews, **37** (6):1050-1066, 2007.
[14]. V.S. Lerner, *Information Path Functional and Informational Macrodynamics,* Nova Sc.Publ., NY, 2010.
[15].V.S.Lerner, *How Information creates its Observer. The Emergence of the Information Observer with Regularities*, Nova Sc. Publ, NY, 2019.
[16]. V.S. Lerner, Reality, Information, and Intelligent Observer, *European Journal of Computer Science and Information Technology,* **8**(1): 39-175, 2020.
[17]. V.S. Lerner, Reality, Information, and Information Observer, *Proceedings.* Presented at the Conference Theoretical Information Studies (TIS), Berkeley, CA, USA, 2–6 June 2019, mdpi.com/journal/proceedings 2020, 47 (10), doi 10.390.

## II. Emergence of the Discrete Time-Space, Information, and Constrains

### 2.0.Transition to Markov disffusion process. Specific of Rising the Entropy Functional

**A.** Let us substantiate how probabilistic observation of the Markov chain starts Markov division process.
A continuous random time Markov chain can converge to diffusion [1] if observation is a three time-scale motion transiting an average sequence of each three time states of the chain along the Markov chain.
Since the Markov chain is a sequence of possible events in which the probability of each event depends only on the state attained in the previous event, the three-time observation can start at time states of the previous event $i-1$, continue at time state of a current event $i$ and ends at the time state of a following event $i=2$.
Suppose random states $\xi_m = \xi_m(t_m), \xi_n = \xi_n(t_n)$ at a fixed random time $\mathrm{t}_m, \mathrm{t}_n$ are observed within three moments $t-t_o, u-t, t_1-u$ at $t_o < t < u < t_1$, where observation begins at time $\mathrm{t}_m$ and continued up to time $\mathrm{t}_n$, which covers above three time intervals. Then, between random $\xi_m, \xi_n$ establishes a coefficient of correlation [2:23]:

$\mathrm{r}_{mn} = \sqrt{\frac{t_m}{\mathrm{t}_n}}$ which exists if $|\sqrt{\frac{t_m}{\mathrm{t}_n}}| > 0$ .

The impulse probabilities of the running chain may virtually observe two states $\xi_m = \xi_m(t_m), \xi_n = \xi_n(t_n)$ at a fixed random time $\mathrm{t}_m, \mathrm{t}_n$ within three moments $t-t_o, u-t, t_1-u$ at $t_o < t < u < t_1$, where the last holds the two current impulses. This three time-scale motion, transiting two impulses, correlates the beginning $\xi_m$ and ending state $\xi_n$ of the observing impulses. Those cooperates-binds the three times of Markov chain in Markov diffusion.
Hence, when at least two discrete impulses of the Markov chain connect a minimum of three its random events, their transitional probabilities transform to Markov diffusion process.

**B.**The Markov diffusion process trajectories are defined on $n$-dimensional probability distribution $P_n = P_n[X(\omega)]$ with transitional probabilities $P(s,\tilde{x},t,B)$, $\sigma$-algebra $F(s,t)$ created by events $\{\tilde{x}(\tau)\in B\}$ at $s \le \tau \le t$, and conditional probability distributions $P_{s,x}$ on $F(s,t)$, where transitional probability is equivalent to the conditional.
Transformation of this probability measures relation [3]:

$$\tilde{P}_{s,x}(d\omega) = p(\omega)P_{s,x}(d\omega) \qquad (1.0)$$

on trajectories of Markov process $(\tilde{x}_t, P_{s,x})$ which holds distributions $\tilde{P}_{s,x} = \tilde{P}_{s,x}(A)$ on extensive $\sigma$-algebra $F(s,\infty)$ with density measure:

$$p(\omega) = \frac{\tilde{P}_{s,x}(d\omega)}{P_{s,x}(d\omega)} \qquad (1.1)$$

Applying the definition of conditional entropy to mathematical expectation of logarithmic probability *functional density measure* (0.1), we introduce *entropy functional on trajectories of Markov diffusion process* [4]:

$$S = E\{-\ln[p(\omega)]\} = \int_{\tilde{x}(t)\in B} -\ln[p(\omega)]P_{s,x}(d\omega), \qquad (1.2)$$

where $E = E_{x,s,\tilde{x}_t}$ is conditional mathematical expectation, taken along the process trajectories $\tilde{x}_t$ at a varied $(\tilde{x}, s)$ (by analogy with M.Kac [5]).
Markov diffusion process describes drift function $a = a(t,x)$ and function of diffusion $\sigma = \sigma(t,x)$ which define additive functional [6-9]:

$$\varphi_s^T = 1/2\int_s^T a(t,\tilde{x})^T (2b(t,\tilde{x}))^{-1} a(t,\tilde{x}_t)dt + \int_s^T \sigma(t,\tilde{x})^{-1} a(t,\tilde{x}) d\xi(t), 2b(t,\tilde{x}) = \sigma(t,\tilde{x})\sigma^T(t,\tilde{x}) > 0 \ . \quad (1.3)$$

Functional (1.3) describes transformation of the Markov processes' random *time* traversing the process trajectory. Additive functional (1.3) measures probability density (1.1) which also can define $\varphi_s^t(\omega)$ :

$$p(\omega) = \exp\{-\varphi_s^t(\omega)\} \text{(a), or } \varphi_s^t(\omega) = -\ln p(\omega) \text{(b)} \quad (1.4)$$

Transitional probabilities for a diffusion process $\varsigma_t$, transforming density (1.4) with aid of additive functional (1.3), satisfy relation (1.0) in form

$$\tilde{P}(s,\varsigma,t,B) = \int_{\tilde{x}(t)\in B} \exp\{-\varphi_s^t(\omega)\} P_{s,x}(d\omega). \quad (1.5)$$

Applying definition of entropy *functional* (1.2) to process $\tilde{x}_t$ conditional to process $\varsigma_t$, we get the entropy functional measure expressed via additive functional $\varphi_s^t(\omega)$ on the trajectories of the diffusion processes:

$$S[\tilde{x}_t / \varsigma_t] = E[\varphi_s^t(\omega)]. \quad (1.6)$$

Minimum of this functional, depending on the additive functional, measures closeness the above distributions:

$$\min_{\varphi_s^t} S[\tilde{x}_t / \varsigma_t] = S^o \ . \quad (1.7)$$

Let the transformed process be

$$\varsigma_t = \int_s^t \sigma(v, \xi_v) d\xi_v \quad (1.8)$$

having the same diffusion as initial process $\tilde{x}_t$, but the zero drift, then $\varsigma_t$ models a standard perturbations -"white noise". Process $\varsigma_t$ is a transformed version of process $\tilde{x}_t$ whose transition probability satisfies (1.1). Transformed probability $\tilde{P}_{s,x}$ for this process evaluates the Feller kernel measure [10, 11].

Since transformed process $\varsigma_t$ (1.8) has the same diffusion matrix but zero drift, the right part of additive functional in (1.3) satisfies

$$E[\int_s^T (\sigma(t,\tilde{x})^{-1} a(t,\tilde{x}) d\xi(t)] = 0. \quad (1.9)$$

It brings the integral measure of the entropy functional expressed via parameters of Ito stochastic equation [4,12]:

$$\Delta S[\tilde{x}_t]|_s^T = 1/2 E_{s,x}\{\int_s^T a^u(t,\tilde{x}_t)^T (2b(t,\tilde{x}_t))^{-1} a^u(t,\tilde{x}_t) dt\} = \int_{\tilde{x}(t)\in B} -\ln[p(\omega)] P_{s,x}(d\omega) = -E_{s,x}[\ln p(\omega)]. \quad (1.10)$$

Formulas (1.2-1.5), (1.6), (1.7), (1.8) and (1.10) are in [13,14] with related citations and references.

The entropy functional (EF) in forms (1.6, 1.10) is an *information indicator* of distinction between the probability measures of the processes $\tilde{x}_t$ and $\varsigma_t$; it measures a *quantity of information* of process $\tilde{x}_t$ conditional to process $\varsigma_t$.

The right side of (1.10) is the EF equivalent formula, expressing it via probability density $p(\omega)$ of random events $\omega$, integrated with the probability measure $P_{s,x}(d\omega)$ along the process trajectories $\tilde{x}(t) \in B$, defined at the set $B$.

*Formula (1.10) directly connects the probabilities, defining the EF, and the Markov process functions drift and diffusion without necessary to the implement probabilities measurement for the given process.*

The process is *irreversible* if its entropy (1.10) is positive, or

$$-E_{s,x}[\ln p(\omega)] > 0 \text{ at } 0 < p(\omega) \le 1. \quad (1.11)$$

The observing probability of triple cooperative events above leads to their forth probability more predictable with growing the density and satisfaction of (1.11). Observing more cooperating events increses the density which is limited by $p(\omega)=1$. That leads to zero entropy in (1.2).

It requires zero additive functional $\varphi_s^t(\omega)=0$, or its negative diffusion part in Eq. (1.3).

This is possible at negative drift, since $\sigma=\sigma(t,x)>0$ and the quadratic form in first part of (1.3) is positive.

The negative drift decreases speed of the Markov diffusion following from the increasing cooperation.
The decreased speed compensates for the diffusion part at the same $\sigma=\sigma(t,x)>0$ in Eq.

$$1/2\int_s^T a(t,\tilde{x})^T(2b(t,\tilde{x}))^{-1}a(t,\tilde{x}_t)dt=\int_s^T \sigma(t,\tilde{x})^{-1}a(t,\tilde{x})d\xi(t) \quad (1.3a)$$

before math expectation in (1.6) is taken. Eq. (1.3a) defines zero EF at satisfaction (1.9), while diffusion part of (1.3a) limits its minimal drift of the Markov process.
The limit holds start of the process dynamics running only its drifts. Since the time, following from the zero EF, is also equal to zero, the observation of the process dynamics has *not* begun, being indicator of the starting observation.

According to [9:358], if after some moment $t>\tau$, additive functional satisfies $\varphi_{t>\tau}^t(\omega)=0(a)$, while at previous moment $t\le\tau$, it holds $\varphi_{t\le\tau}^t(\omega)=1$(b); such situation calls the Markov process' break off. If at the previous moment approaching $t<\tau$, $p_{t\to\tau}(\omega)\to 0.3679$, then following 1(b): $\varphi_{t\to\tau}^t(\omega)=-\ln p_{t\to\tau}(\omega)\to 1$. Therefore, at the moment approaching the break off , the additive functional satisfies 1(b), and at the following moment holds 1(a) that satisfies $p_{t=\tau}(\omega)=1$ and limits the cooperation. Hence, the cooperation may start with $p_{t\to\tau}(\omega)$ and end with the process break off. Then, at the moment following beak off, after $p_{t=\tau}(\omega)|_{\tau_o}=1$ at $t>\tau_o$ and $p_{t\to t>\tau_0}(\omega)\to 0.3679$ the cooperation may restart. It involves altering 0-1 values for $\varphi_s^t(\omega)$ (which ref.[1] formulates as a law).

Zero entropy limits the observed cooperation, but it also lowers limit of minimal entropy (1.10). The minimal entropy imposes extreme principle on the cooperating process probabilities, which leads to invariant maximal $p(\omega)\to 1$. Thus, the maximal predicted events may follow the beak off unpredictability (in virtual reality).

The extreme EF determines equations of Information MacroDynamics [15] as part of the IIID [16].
Here, the minimal entropy principle limits the admissible cooperation, at decreasing speed of the cooperative Markov process. Such Markov diffusion imposes a limit of its *irreversibility*. The IMD starts on this limit.
When the Markov diffusion continues approaching the break off, its probability density $p_{t\to\tau}(\omega)\to 0.3679$ binds the triple connected events' probabilities, starting on step $i=1$ with $p_{t<\tau}(\omega)|_i=\sqrt[3]{p_{t\to\tau}(\omega)}\to\sqrt[3]{0.3679}\approx 0.7788$. •

### 2.1. The impulse cutting action on the entropy integral during observation

Each observation is act changing the observing process that generates the interactive impulse observation.
During the observation, the impulses sequentially cut the observing process. The impulse probabilities bring virtual cuts, which provide Markov transitional probabilities. With probability approaching one, the cut provides the real no-yes impulse actions bringing physical cuts. The transitional probabilities, transferring along the Markov process trajectories, defines the Entropy Functional (EF) on these trajectories.
When a certain probability one brings real impulse, physically cutting-erasing its entropy, the impulse entropy converges to equivalent information. Then, the cutting EF converges to related Information Functional (IPF) defined along the cutting path on trajectories of an information process.
These operations need detail analysis and mathematical substantiation.

For the formal analysis of the cutting action of EF and the produced IPF, we introduce an abstract impulse control below. Since such controls actually belong to the Markov process impulse, we identify a class of the discrete controls satisfying the cutting action of the Markov process in Sec.2.1.1.
Using these discrete controls properties, we find specific their cutting fraction of the EF.(Secs.2.1.3, 2.2).

**2.1.1 Evaluation of the cutting process EF fractions by an impulse control**
Let us define control on the space $KC(\Delta, U)$ of a piece-wise continuous step functions $u_t$ at $t \in \Delta$:

$$u_- \overset{def}{=} \lim_{t \to \tau_k - o} u(t, \tilde{x}_{\tau_k}),\; u_+ \overset{def}{=} \lim_{t \to \tau_k + o} u(t, \tilde{x}_{\tau_k}) \tag{2.1}$$

which are differentiable on the set

$$\Delta^o = \Delta \setminus \{\tau_k\}_{k=1}^m,\; k = 1,...,m, \tag{2.1a}$$

and applied on diffusion process $\tilde{x}_t$ from moment $\tau_{k-o}$ to $\tau_k$, and then from moment $\tau_k$ to $\tau_{k+o}$, implementing the process' transformations $\tilde{x}_t(\tau_{k-o}) \to \varsigma_t(\tau_k) \to \tilde{x}_t(\tau_{k+o})$; $n$-dimensional process holds $m$ such transformations.
At a vicinity of moment $\tau_k$, between the jump of control $u_-$ and the jump of control $u_+$, we consider a control *impulse*

$$\delta u_{\pm}(\tau_k) = u_-(\tau_{k-o}) + u_+(\tau_{k+o}). \tag{2.2}$$

The following statement evaluates the EF information contributions at such transformations.

<u>Proposition 2.1.</u>
Entropy functional (1.10) at the switching moments $t = \tau_k$ of control (2.2) takes values

$$\Delta S[\tilde{x}_t(\delta u_{\pm}(\tau_k)] = 1/2, \tag{2.3}$$

and at locality of $t = \tau_k$: at $\tau_{k-o} \to \tau_k$ and $\tau_k \to \tau_{k+o}$, produced by each of the impulse control's step functions in (2.1), is estimated by

$$\Delta S[\tilde{x}_t(u_-(\tau_k)] = 1/4,\; u_- = u_-(\tau_k),\; \tau_{k-o} \to \tau_k \tag{2.3a}$$

and

$$\Delta S[\tilde{x}_t(u_+(\tau_k)] = 1/4, u_+ = u_+(\tau_k), \tau_k \to \tau_{k+o}. \tag{2.3b}$$

*Proof.* The jump of control function $u_-$ in (2.1) from moment $\tau_{k-o}$ to $\tau_k$, acting on the diffusion process, might cut off this process after moment $\tau_{k-o} \to \tau_k$.
The cutoff diffusion process has the same drift vector and the diffusion matrix as the initial diffusion process.
The additive functional for this cut off has the form [6-9]:

$$\varphi_s^{t-} = \begin{cases} 0, t \le \tau_{k-o} \\ \infty, t > \tau_k \end{cases}. \tag{2.4}$$

The jump of the control function $u_+$ (2.1) from $\tau_k$ to $\tau_{k+o}$ might cut off the diffusion process *after* moment $\tau_k \to \tau_{k+o}$ with the related additive functional

$$\varphi_s^{t+} = \begin{cases} \infty, t > \tau_k \\ 0, t \le \tau_{k+o} \end{cases}. \tag{2.5}$$

For the control impulse (2.2), the additive functional at a vicinity of $t = \tau_k$ acquires the form of an *impulse function*

$$\varphi_s^{t-} + \varphi_s^{t+} = \delta \varphi_s^{\mp}, \tag{2.6}$$

which summarizes (2.3) and (2.4).
Entropy functional (1.10) following from (2.4-2.5) takes values

$$\Delta S[\tilde{x}_t(u_-(t \le \tau_{k-o}; t > \tau_k))] = E[\varphi_s^{t-}] = \begin{cases} 0, t \le \tau_{k-o} \\ \infty, t > \tau_k \end{cases}, \quad (2.7a)$$

$$\Delta S[\tilde{x}_t(u_+(t > \tau_k; t \le \tau_{k+o}))] = E[\varphi_s^{t+}] = \begin{cases} \infty, t > \tau_k \\ 0, t \le \tau_{k+o} \end{cases}, \quad (2.7b)$$

from 0 to $\infty$ acquiring *absolute maximum* at $t > \tau_k$ ,and back from $\infty$ to 0, acquiring *absolute minimum* at $\tau_{k-o}$ and $\tau_{k+o}$.

The multiplicative functional [9], related to (2.4-2.5), are:

$$p_s^{t-} = \begin{cases} 0, t \le \tau_{k-o} \\ 1, t > \tau_k \end{cases}, \ p_s^{t+} = \begin{cases} 1, t > \tau_k \\ 0, t \le \tau_{k+o} \end{cases}. \quad (2.8)$$

Control impulse (2.2) provides an impulse probability density in form of multiplicative functional

$$\delta p_s^{\mp} = p_s^{t-} p_s^{t+}, \quad (2.9)$$

where $\delta p_s^{\mp}$ holds $\delta[\tau_k]$-function, which determines probabilities

$$\tilde{P}_{s,x}(d\omega) = 0 \ \text{ at } t \le \tau_{k-o}, t \le \tau_{k+o}, \text{ and } \tilde{P}_{s,x}(d\omega) = P_{s,x}(d\omega) \ \text{ at } t > \tau_k. \quad (2.9a)$$

For the cutoff diffusion process, transitional probability (at $t \le \tau_{k-o}$ and $t \le \tau_{k+o}$) turns to zero, and states $\tilde{x}(\tau_k - o), \tilde{x}(\tau_k + o)$ become independent, while their mutual time correlations *are dissolved*:

$$r_{\tau_{k-o},\tau_{k+o}} = E[\tilde{x}(\tau_k - o), \tilde{x}(\tau_k + o)] \to 0. \quad (2.10)$$

Entropy increment $\Delta S[\tilde{x}_t(\delta u_{\pm}(\tau_k))]$ of additive functional $\delta\varphi_s^{\mp}$ (2.5), produced within or at a border of control impulse (2.2), defines equality

$$E[\varphi_s^{t-} + \varphi_s^{t+}] = E[\delta\varphi_s^{\mp}] = \int_{\tau_{k-o}}^{\tau_{k+o}} \delta\varphi_s^{\mp}(\omega) P_{\delta}(d\omega), \quad (2.11)$$

where $P_{\delta}(d\omega)$ is a probability evaluation of impulse $\delta\varphi_s^{\mp}$.

Integral of symmetric $\delta$ -function $\delta\varphi_s^{\mp}$ between the above time intervals on the border is

$$E[\delta\varphi_s^{\mp}] = 1/2 P_{\delta}(\tau_k) \text{ at } \tau_k = \tau_{k-o}, \text{ or } \tau_k = \tau_{k+o}. \quad (2.12)$$

The impulse, produced by deterministic controls (2.2) for each process dimension, is random with probability

$$P_{\delta c}(\tau_k) = 1, k = 1,...,m \quad (2.13)$$

at each $\tau_k$ -locality.

This probability holds a jump-diffusion transition probability in (2.12) (according to [17]), which is conserved during the jump. For each jump, condition (2.4) leads to $a^u(t, \tilde{x}_t) \to \infty$..

*Therefore, each jump increases Markov speed up to infinitely within a finite impulse with fixed* $\tau_{k-o}$ *and* $\tau_{k+o}$.

From (2.11)-(2.13) it follows estimation of the EF increment under impulse control (2.2) applying at $t = \tau_k$ in form

$$\Delta S[\tilde{x}_t(\delta u_{\pm}(\tau_k))] = E[\delta\varphi_s^{\mp}] = 1/2 \quad (2.14)$$

which proves (2.3), while delta impulse at $\delta\varphi_s^{\mp} \to \infty$ brings absolute maximum to (2.14) within each $k$ cutoff impulse.

Symmetrical entropy contributions (2.6) at a vicinity of $t = \tau_k$ :

$$E[\varphi_s^{t-}] = \Delta S[\tilde{x}_t(u_-(t \le \tau_{k-o}; t > \tau_k))] \quad (2.15a)$$

$$E[\varphi_s^{t+}] = \Delta S[\tilde{x}_t(u_+(t > \tau_k; t \le \tau_{k+o}))] \quad (2.15b)$$

estimate relations

$$\Delta S[\tilde{x}_t(u_-(t \le \tau_{k-o}; t > \tau_k))] = 1/4, u_- = u_-(\tau_k), \tau_k \to \tau_{k-o}, \quad (2.16a)$$

$$\Delta S[\tilde{x}_t(u_+(t > \tau_k; t \le \tau_{k+o}))] = 1/4, u_+ = u_+(t > \tau_k), \tau_k \to \tau_{k+o}, \quad (2.16b)$$

which prove (2.3a,b).
Entropy functional (1.10), defined through Radon-Nikodym's probability density measure (1.3), holds all properties of the considered cutoff controllable process, where both $P_{s,x}$ and $\tilde{P}_{s,x}$ are defined.

Thus, cutting correlations (2.10) extracts entropy of hidden process information which directly measures each $\delta-$ cutoff:

$$\Delta I_k[\tilde{x}_t(\delta u(\tau_k))] = \Delta S[\tilde{x}_t(\delta u_{\pm}(\tau_k))] = 1/2 \ . \quad (2.17)$$

The known information measures do not provide such direct measuring.
According the definition of entropy functional (1.2), it is measured in natural $\ln$ where each its Nat equals $\log_2 e \cong 1.44 bits$; therefore, the EF does not using Shannon entropy measure. •

### 2.1.2. The Proposition Corollaries

From Proposition 1.2 it follows that:

**(a)**-Stepwise control function $u_- = u_-(\tau_k)$, implementing transformation $\tilde{x}_t(\tau_{k-o}) \to \varsigma_t(\tau_k)$, converts the EF from its minimum at $t \le \tau_{k-o}$ (2.16a) to maximum at $\tau_{k-o} \to \tau_k$ (2.17);

**(b)**-Stepwise control function $u_+ = u_+(\tau_k)$, implementing transformation $\varsigma_t(\tau_k) \to \tilde{x}_t(\tau_{k+o})$, converts the EF from its maximum at $\tau_{k-o} \to \tau_k$ (2.17) to minimum at $\tau_k \to \tau_{k+o}$ (2.16b);

**(c)**-Impulse control function $\delta u^{\mp}_{\tau_k}$, implementing transformations $\tilde{x}_t(\tau_{k-o}) \to \varsigma_t(\tau_k) \to \tilde{x}_t(\tau_{k+o})$, switches the EF from its minimum to maximum and back from maximum to minimum, while the maximum of the entropy functional at a vicinity of $t = \tau_k$ allows the impulse control to deliver *maximal amount* of information (2.17) from these transformations;

(That maximum relates to $p_{t=\tau}(\omega) = 1$ maximum following break off that limits the speed of correlated cooperation).

**(d)**-The dissolving (cutting) correlation between the process' cutoff points (2.10) cuts *functional connections* at these discrete points which border the Feller kernel measure [11];

**(e)**-The relation of that measure to additive functional in form (1.3) allows evaluating the *kernel's information* by the EF (1.5). The jump action (2.2) on Markov process, associated with "killing its drift", selects the Feller measure of the kernel [18, 19], while the cutoff of EF *provides information measure* of the Feller kernel (2.17);

**(f)**-Stepwise control $u_- = u_-(\tau_k)$, transferring the EF from $\tau_{k-o} \to \tau_k$, maximizes (by moment $\tau_k$) the minimal information increment (brought at $t \to \tau_{k-o}$), implementing condition

$$\max_{\tau_k} \min_{\tau_{k-o}} \Delta I_k[\tilde{x}_t(\delta u(\tau_k))]; \quad (2.17a)$$

**(g)**-Stepwise control $u_+ = u_+(\tau_k)$, transferring the EF from $\tau_k \to \tau_{k+o}$, kills the additive functional at stopping moment $\tau_{k+o}$ minimizing the maximal information increment by the end of this transformation, implementing condition

$$\min_{\tau_{k+o}} \max_{\tau_k} \Delta I_k[\tilde{x}_t(\delta u(\tau_k))]. \quad (2.17b)$$

Such transformation associates with killing the Markov process at the rate of increment of related additive functional $d\varphi_s^{ti} / \varphi_s^{ti}$ for each single dimension *i* [20].

Control $u_+ = u_+(\tau_k)$ transfers the rate of killed Markov process to a process with probability (2.13), which is conserved during the jump, and starts a maximal probable (non-random) process with the eigenvalue of diffusion operator [21].That process balances the killing at the same rate [22].
The step-wise controls, acting on the multi-dimensional diffusion process dimensions, sequentially stops and starts the process, evaluating information of multiple functional, and changes probabilities 0-1 of each impulse.
That impulse observation (Sec.2.0(A)) cooperates the random events starting the minimal entropy principle at random observations. The dissolved element of the correlation matrix at these moments provides independence of the cutting off fractions, leading to orthogonality of their correlation matrix. •

*The control action on the Markov additive functional with drift function* $a^u = a(x,t,u)$ *is an equivalent of changing* the EF *measure under the No-Yes actions. The above transformation carries the jump of Markov probabilities densities* $p(\omega)$ *in Sec.2.1.1, which in the Proposition 2.1 runs* $\delta p_s^{\mp}$ *under the controls.*

*Each observing impulse, acting on the Markov probabilities, virtually cuts the EF until real control impulse converts the cutting EF entropy to information unit. Mover, as it follows from that Proposition, each jump increases Markov speed allowing the arising maximin-minimax impulses running faster than the initial Markov movement. That speeds up getting information unit from the impulse observation compare to any of non-impulse observations.*

### 2.1.3.The entropy integral functional for a single dimensional Markov process

The EF integrant in (1.10) is *partially observable* via measuring only covariation function on the process' trajectories.
For a single dimensional Markov process with drift function $a = c\tilde{x}(t)$ at given nonrandom function $c = c(t)$ and diffusion $\sigma = \sigma(t)$ entropy functional (1.10) acquires form

$$S[\tilde{x}_t / \varsigma_t] = 1/2\int_s^T E[c^2(t)\tilde{x}^2(t)\sigma^{-2}(t)]dt\,, \qquad (3.1.1)$$

from which, at function $\sigma(t)$ and nonrandom function $c(t)$, we get

$$S[\tilde{x}_t / \varsigma_t] = 1/2\int_s^T [c^2(t)\sigma^{-2}(t)E_{s,x}[x^2(t)]dt = 1/2\int_s^T c^2[2b(t)]^{-1}r_s dt\ . \qquad (3.1.2)$$

For the Markov diffusion process, the following relations hold true:

$$2b(t) = \sigma(t)^2 = dr/dt = \dot{r}_t\,, E_{s,x}[x^2(t)] = r_s\,. \qquad (3.1.3)$$

This allows *identifying* the EF on observed Markov process $\tilde{x}_t = \tilde{x}(t)$ by measuring above correlation functions, applying positive function $u(t) = c^2(t)$, and *representing* functional (1.10) through a regular integral

$$S[\tilde{x}_t / \varsigma_t] = 1/2\int_s^T u(t)A(t,s)dt \qquad (3.1.4)$$

with non-random functions

$$A(s,t) = [2b(t)]^{-1}r_s = \dot{r}_t^{-1}r_s \quad (3.1.4a) \quad \text{and} \quad u(t) = c^2(t) \qquad (3.1.4b)$$

The *n*-dimensional form of functional integrant (3.1.4a,b) follows directly from the related *n*-dimensional covariations (3.1.3), dispersion matrix, and applying *n*-dimensional function $u(t)$.
At given nonrandom function $u(t)$, regular integral (3.1.4) measures the entropy functional of Markov process at the probability transformation (1.2) with additive functional (1.3), where the integrant averages the additive functional.

Proposition 3.1.

Integral (3.1.4), satisfying variation condition (1.15) at linear function $c^2(t) = u \bullet t = c^2 \bullet t$, forms

$$S[\tilde{x}_t / \varsigma_t] = 1/2\int_s^T u(t)o(t)dt\,, \quad (3.1.5)$$

where the extreme of function (3.1.4a) holds minimum

$$A(t, s_k^{+o}) = o(s)b_k(s_k^{+o}) / b_k(t) = o(t)\,, \quad (3.1.5a)$$

which decreases with growing time $t = s_k^{+o} + o(t)$ at $t \to T$ and fixed both $b_k(s_k^{+o})$ and the starting interval

$$o(s) = A(s,s)\,. \quad (3.1.5b)$$

Since satisfaction of this variation condition includes transitive transformation of a current distribution to that of Feller kernel, the $b_k(t)$ is the transition dispersion at this transformation, which is growing with the time of the transformation.

Proposition 3.2.

Entropy integral (3.1.5) under impulse control $c^2(t,\tau_k) = \delta u_t(t-\tau_k)$ takes the following information values:

(a)-at a switching impulse middle locality $t = \tau_k$ :

$$S[\tilde{x}_t / \varsigma_t]_{t=\tau_k} = 1/2 \text{ Nats}, \quad (3.1.6)$$

(b)- at switching impulse left locality $t = \tau_k^{-o}$ :

$$S[\tilde{x}_t / \varsigma_t]_{t=\tau_k^{-o}} = 1/4 Nats, \quad (3.1.6a)$$

(c)-and at switching impulse right locality $t = \tau_k^{+o}$ :

$$S[\tilde{x}_t / \varsigma_t]_{t=\tau_k^{+}} = 1/4 Nats. \quad (3.1.6b)$$

*Proof.* Applying delta-function $c^2(t,\tau_k) = \delta u_t(t-\tau_k)$ to integral

$$\Delta S[\tilde{x}_t / \varsigma_t]|_{\tau_k^{-o}}^{\tau_k^{+o}} = 1/2\int_{\tau_k^-}^{\tau_k^+} \delta u_t(t-\tau_k)o(t)dt, \tau_k^{-o} < \tau_k < \tau_k^{+o}\,, \quad (3.1.7)$$

determines functions [23:678-681]:

$$\Delta S[\tilde{x}_t / \varsigma_t]|_{t=\tau_k^{-o}}^{t=\tau_k^{+o}} = \begin{cases} 0, t < \tau_k^{-o} \\ 1/4o(\tau_k^{-o}), t = \tau_k^{-o} \\ 1/4o(\tau_k^{+o}), t = \tau_k^{+o} \\ 1/2o(\tau_k), t = \tau_k, \tau_k^{-o} < \tau_k < \tau_k^{+o} \end{cases} \quad (3.1.8)$$

Or such cutoff impulse $o(\tau_k)$ brings amount of the EF on its middle locality $\Delta S[\tilde{x}_t / \varsigma_t]_{t=\tau_k} = 1/2o(\tau_k) = 1/2$ Nats, and on borders of interval $o(\tau_k)$ the EF amounts are $\Delta S[\tilde{x}_t / \varsigma_t]_{t=\tau_k^{-o}} = 1/4o(\tau_k^{-o})Nats$ and $\Delta S[\tilde{x}_t / \varsigma_t]_{t=\tau_k^{+o}} = 1/4o(\tau_k^{+o})Nats$ accordingly.•

The results concur with (2.2.3, 2.2.3a,b).

The cutoff delivers entropy $3/4Nats$ transferring $1/4Nats$ to the right border of the impulse.

Sum of the impulse localities of cutting interval $o(\tau_k)$ :

$$\sum_{t=\tau_k^{-o}}^{t=\tau_k^{+o}} \Delta S[\tilde{x}_t / \varsigma_t]_t = 1/4o(\tau_k^{-o}) + 1/2o(\tau_k) + 1/4o(\tau_k^{+o}) = o_k \quad (3.1.9)$$

evaluates constant-the invariant $1Nat$ fraction of the EF cutoff on this interval, which the interval encloses.

The invariant cutting fractions follow from variation condition (1.1.7) imposed on (3.1.4), which leads to (3.1.9). The impulse 0-1 probability, virtually cutting minimax entropy measure: $\Delta S_{P=0} \to -\infty, \Delta S_{P=1} = 0$ also applies to a virtual observer with virtual–probabilistic density $\overline{p}(\omega)$ and entropy functional measure $S_B = E_{s,x}[-\ln \overline{p}(\omega)]$.

**2.1.4.The Information Path Functional**

*Information path functional* (IPF) defines the distributed actions of multi-dimensional delta-function on entropy functional (1.10) through the additive functional for all dimensions:

$$I_{pm} = \delta_m\{S[\tilde{x}_t / \varsigma_t]|\} = 1/2E\{\int_s^T \delta_m[a(t,\tilde{x})^T (2b(t,\tilde{x}))^{-1} a(t,\tilde{x})dt)]\} \ . \qquad (4.4.1)$$

Relation (4.4.1) is summing discrete information measures $\Delta I_k[\tilde{x}_t(\delta u(\tau_k))]$ along the path on cutting process intervals (3.1.9). In a limit it leads to

$$I_p = \lim_{m\to\infty} \sum_{k=1}^{m} \Delta I_k[\tilde{x}_t(\delta u(\tau_k))]. \qquad (4.4.2)$$

Formal definition (4.4.1) allows the IPF representation by Furies integral [23] leading to the frequency analysis with Furies series.

.The IPF is the sum of *extracted* information which approaches theoretical measure (1.10):

$$I_p = \lim_{m\to\infty} I_{mo} |_s^T = \lim_{m\to\infty} S_{mo} |_s^T \to S[\tilde{x}_t / \varsigma_t]_s^T, \qquad (4.4.3)$$

if all finite time intervals $t_1 - s = o_1, t_2 - t_1 = o_2, ..., t_{k-1} - t_k = o_k, ..., t_m - t_{m-1} = o_m$, at $t_m = T$ satisfy condition

$$(T - s) = \lim_{m\to\infty} \sum_{t=s,m}^{t=T} o_m(t). \qquad (4.4.4)$$

According to (3.1.9) each cutting interval $o(\tau_k)$ encloses invariant entropy measure converting to equivalent information measure $\Delta I_k[\tilde{x}_t(\delta u(\tau_k))]$. The $I_p$ (4.4.3) limits the initially undefined upper time $T$ of the EF integral (1.10), which brings direct connection $I_p$ and $T - s$, where each $o(\tau_k)$ limits invariant discreet measure (3.1.9). Therefore, at infinite sequence of the integrated time intervals, this sequence limits only zero discreet measure:

$$\lim_{m\to\infty} o(t_m) \to 0, \qquad (4.4.5)$$

and sum of such sequence is finite [23:130, 4.8-1].

A *random ensemble of the sequence, moving with time (4.4.5), would have infinite velocity at any finite size of its space distribution.*

Additive functional (1.3), EF (3.14), and relation (4.4.4) *directly connect (1.10) with the process' time* $T$ **.**

Realization (4.4.1), (4.4.4), (4.4.5) requires applying the impulse controls at each instant $(\tilde{x}, s), (\tilde{x}, s + o(s)), ...$ along the process trajectories with conditional math expectation (1.10).

However for any *finite* number $m$ of these instants, the *integral* process information, composed from the discreet information measured for the process' fractions, is not complete.

*The* $I_p$ *properties*:

1. The IPF measures information of the cutting process's interstate connections hidden by the states correlations, which are not covered by traditional Shannon information measure.

2. Since each cutting value $\Delta S_k[\tilde{x}_t(\delta u(\tau_k))] = \Delta I_k[\tilde{x}_t(\delta u(\tau_k))]$ maximizes the cutting information on the path intervals, $I_p$ measures total (integral) maximal information on this path.

The cutting control provides the equal maxmin-minimax information contributions

$$\max_{\tau_k} \min_{\tau_{k-o}} \Delta I_k[\tilde{x}_t(\delta u(\tau_k))] = \min_{\tau_{k+1}} \max_{\tau_k} \Delta I_k[\tilde{x}_t(\delta u(\tau_k))] \qquad (4.4.6)$$

on each path $t_{k-1} \to (\tau_{k-o} \to \tau_k \to \tau_{k+o}) \to t_k$ from cutting $t_{k-1}$ to following cut $t_k$ (Corollaries a-c).

3. If each $k$ -cutoff "kills" the $m$ dimensional process' correlation at moment $\tau_{k+o}$, then at $m=n$ relations (4.4.1-4.4.6) require infinite process dimensions.
4. At $m=n\to\infty, o_k=t_k-t_{k-1}\to\tau_k$, the process time

$$(T-s)=\lim_{n\to\infty}\sum_{t=s,k=1}^{t=T,k=n}\tau_k \qquad (4.4.7)$$

approaches the summary of the discrete time intervals cutting all correlations on the path.
5. Sequential cuts transform the IPF discrete information contributions from each maximum through minimum to next maximal information contributions

$$\max_{\tau_k}\Delta I_k[\tilde{x}_t(\delta u(\tau_k))]\to\min_{\tau_{k+o}}\Delta I_k[\tilde{x}_t(\delta u(\tau_k))]\to\max_{\tau_k+1}\Delta I_k[\tilde{x}_t(\delta u(\tau_{k+1}))], \qquad (4.4.8)$$

where each next maximum decreases at the cutoff moments

$$\max_{\tau_k+1}\Delta I_k[\tilde{x}_t(\delta u(\tau_{k+1}))]<\max_{\tau_k}\Delta I_k[\tilde{x}_t(\delta u(\tau_k))]. \qquad (4.4.9)$$

Each Dirac delta-function preserves its cutting information (3.1.8).
The information contribution by final interval $\tau_m=o_m$ at its inner ending moment $\tau_{m+o}$, according to (4.4.2-4.4.4), satisfies

$$\min_{\tau_{m+o}}\Delta I_m[\tilde{x}_t(\delta u(\tau_m))]\to 0, \qquad (4.4.10)$$

which limits sum (4.4.3) at $m=n\to\infty$.
6. Since EF functional $S[\tilde{x}_t/\varsigma_t]_s^T$ limits the growth of $S_{mo}|_s^T$ in (4.4.3), it limits the IPF in (4.4.8), (4.4.9). Hence, the IPF approaches the EF functional during time (4.4.7) at the unlimited increase of the process dimensions.
7. Because upper time $T$ in both the EF integral and IPF functional is limited by (4.4.3),(4.4.4) and (4.4.10), the cutting entropy integral converges in the path functional, and both of them are restricted at the unlimited dimension number. (See also Sec.1.3.3 in Introductory).
8. For any of these limitations, EF measure, taken along the process trajectories for time $(T-s)$, limits maximum of total process information, while IPF extracts maximum of the process hidden information during the same time and brings more information than Shannon traditional information measure for multiple states of the process.
9. Information density of cutting impulse

$$I_{ko_k}=I_k[\tilde{x}_t(\delta u(\tau_k))]/\tau_k \qquad (4.4.11)$$

grows to absolute maximum at $\tau_k\to 0$:

$$I_{ko_n}\to\infty. \qquad (4.4.12)$$

The time transition to each following EF' Nat decreases. Each such Nat integrates all preceding Nats concentrating the integral information in the final IPF Nat, which Feller kernel absorbs.
The final finite integral information impulse approaches the Kronicker 0-1 impulse generated during finite $\tau_{k=n}$, which is preserving measure (3.3.6).
The invariant probability measure applies to the impulse probes on an observable random process, which hold opposite Yes-No probabilities–as the unit of probability impulse' step-function [14] preserving the max-min.
Comments. Number $M$ of equal probable possibilities determines Hartley's quantity of information $H=\ln M$ measured on Nats, which for the impulse $M=2$ holds $H=\ln 2Nat$. The impulse information measured in bits is $I=1/\ln 2\ln M=\lg Mbit=1bit$.
The correlation cutting by the impulse brings information $0.75Nat$ from which $\delta S_u\cong 0.0568Nat$ delivers the impulse controls. Since each cutoff brings invariant $1Nat$ (3.3.6), the difference $(1-0.7)Nat\cong 0.3Nat$ presents "free information" for each cutting impulse. •

The IPF final impulse evaluates maximal speed of enclosing information (4.4.11) in the finite impulse time interval. That impulse time of cutting correlation evaluates hidden information equals to Bit $= \ln 2$.
All integrated information encloses the Feller kernel whose time and energy evaluate results [18].
Minimal physical time interval limits the light time interval $\delta t_\tau \cong 1.33 \times 10^{-15}$ sec defined by the light wavelength $\delta l_m \cong 4 \times 10^{-7} m$. That allows us to estimate maximal information density (4.4.14) for 1 bit:

$$I_{ko_k} \cong \ln 2 / 1.33 \times 10^{-15} \cong 5.2116 \times 10^{+15} Nat / s. \quad (4.4.13)$$

Or for each invariant impulse $1Nat$, the maximal density estimates

$$I_{ko_{k1}} \cong 1 / 1.33 \times 10^{-15} \cong 7.5188 \times 10^{+15} Nat / s. \quad (4.4.14)$$

Variety of the impulse $\downarrow\uparrow$ physical interactions unites the impulse information model, which the EF-IPF integrate.

## 2.2. The evaluation of impulses in the interactive observing process

### 2.2.1. Discrete control action on the entropy functional

Let us find a class of step-down $u_-^t = u_-(\tau_k^{-o})$ and step-up $u_+^t = u_+(\tau_k^{+o})$ functions acting on discrete interval $o(\tau_k) = \tau_k^{+o} - \tau_k^{-o}$,which will preserve the Markov diffusion process' additive and multiplicative functions within the cutting process of each impulse.

(These functions are controls implementing the interactive cutting actions required to reveal the Bit discrete impulse.)

Lemma 1.1.

1. Opposite discrete functions $u_-^t$ and $u_+^t$ in form

$$u_-(\tau_k^{-o}) = \downarrow_{\tau_k^{-o}} \bar{u}_-, u_+(\tau_k^{+o}) = \uparrow_{\tau_k^{+o}} \bar{u}_+ \quad (1.1)$$

satisfy conditions of additivity

$$[u_+^t - u_-^t] = U_a \quad (1.1A)$$

and multiplicativity

$$[u_+^t \times u_-^t] = U_m \quad (1.1B)$$

at real

$$\mathrm{U}_a = U_m = c^2, \quad (1.1C)$$

where $U_a$ is measurable function of additive events, $U_m$ is measurable function of multiplicative events.

Example 1. The orthogonal functions $u_-^t = \bar{u}_- \times \xrightarrow{\tau_k^{-o}}$ and $\uparrow_{\tau_k^{+o}} \bar{u}_+$, where instance-jump $\downarrow_{\tau_k^{-o}}$ has time interval $\bar{u}_-$ and instance jump $\uparrow_{\tau_k^{+o}} \bar{u}_+$ has high $\bar{u}_+$, represent complex functions

$$u_t(u_\pm^{t1}), u_\pm^{t1} = [u_+ = (j-1), u_- = (j+1)]\ , j = \sqrt{-1} \quad (1.2a)$$

which satisfy conditions (1.1A), (1.1B) in forms

$$u_+ - u_- = (j-1) - (j+1) = -2,\ u_+ \times u_- = (j-1) \times (j+1) = (j^2 - 1) = -2,$$

and preserve negative real $c^2 < 0$ in (1.1C).

Example 2. The opposite complex functions

$$u_t(-u_\pm^{t1}) = u_t(u_\pm^{t2}), u_\pm^{t2} = [u_+ = (j+1), u_- = (j-1)], \quad (1.2b)$$

represent orthogonal functions $u_+^t = \bar{u}_+ \times \xrightarrow{\tau_k^{-o}}$ and $u_- = \uparrow_{\tau_k^{+o}} \bar{u}_-$ where instance jump $\downarrow_{\tau_k^{-o}}$ has time interval $\bar{u}_+$ and instance jump $\uparrow_{\tau_k^{+o}}$ has high $\bar{u}_-$ .

The complex functions (1.2b) satisfy (1.1A), (1.1B) in form

$u_+ - u_- = (j+1)-(j-1) = 2, u_+ \times u_- = (j+1)\times(j-1) = (j^2-1) = -2$

changing positive $c^2$ on negative one, and therefore not satisfying (1.1C), violating this Markov property.

Example 3. If function (1.2a), (1.2b) apply on borders of impulse interval $o(\tau_k) = (\tau_k^{-o}, \tau_k^{+o})$ in forms

$$u_-^{t1} = u_-(\tau_k^{-o}), u_+^{t1} = u_+(\tau_k^{-o}) \text{ and } u_-^{t2} = u_-(\tau_k^{+o}), u_+^{t2} = u_+(\tau_k^{+o}), \quad (1.2d)$$

at $u_-^{t1}u_+^{t1} = c^2(\tau_k^{-o}),\ u_-^{t2}u_+^{t2} = c^2(\tau_k^{+o}),\ t = \tau_k^{+o}$, (1.2e)

then $c^2(\tau_k^{-o}), c^2(\tau_k^{+o})$ are discrete functions (1.1a), (1.2e) switching on interval $\Delta_\tau = \tau_k^{+o} - \tau_k^{-o}$ in form $c^2(\tau_k^{-o}) = \downarrow_{\tau_k^{-o}} c^2, c^2(\tau_k^{+o}) = \uparrow_{\tau_k^{+o}} \pm c^2$.

It follows that only at beginning of this time interval at $t = \tau_k^{-o}$, all Markov properties (1.1A,B,C) satisfy, while by the end of this time interval at at $t = \tau_k^{+o}$ the process satisfies only (1.1A,B). •

Example 4. At equal absolute values of actions $|u_+^t| = |u_-^t|$, imaginary functions

$$u_+^t = j\sqrt{2}, u_-^t = -j\sqrt{2} \quad (1.2c)$$

satisfy only multiplicative part $U_m = u_+^t \times u_-^t = -j^2 2 = 2$ at $c^2 > 0$ while the impulse additive measure holds $U_a = 0$. •

<u>Corollary 1.1</u>.

1. The step-functions $u_-(\tau_k^{-o}) = -1_{\tau_k^{-o}} \bar{u}_-, u_+(\tau_k^{+o}) = +1_{\tau_k^{+o}} \bar{u}_+$ forming impulse $\delta u_t = [u_-(\tau_k^{-o}) - u_+(\tau_k^{+o})]$ on interval $\Delta_\tau = \tau_k^{+o} - \tau_k^{-o}$ holds discrete function

$$\delta^o u_{t=\Delta_\tau} = [u_-(\tau_k^{-o}) - u_+(\tau_k^{+o})]/(\tau_k^{+o} - \tau_k^{-o}) \quad (1.3a)$$

2. Discrete functions on $\Delta = (s_k^{+o}, \tau_k^{-o})$ :

$$u_+(s_k^{+o}) = +1_{s_k^{+o}} \bar{u}_+, u_-(\tau_k^{-o}) = -1_{\tau k}^{+o} \bar{u}_- \quad (1.3b)$$

are multiplicative:

$(u_-(\tau_k^{-o}) - u_+(s_k^{+o})) \times (u_-(\tau_k^{-o}) - u_+(s_k^{+o})) = [u_-(\tau_k^{-o}) - u_+(s_k^{+o})]^2$,

which for functions $\bar{u}_+ = j\bar{u}, \bar{u}_- = -j\bar{u}, \bar{u} \neq 0$ holds

$$[u_-(\tau_k^{-o}) - u_+(s_k^{+o})]^2 = -(j\bar{u})^2[-1_{\tau_k^{-o}} - 1_{s_k^{+o}}]^2 > 0. \bullet \quad (1.3d$$

Let us find discrete analog of the integral increments under *discrete* function (1.3a) in a form of delta-function on $\delta_o$:

$$\delta[u_-(\tau_k^{-o}), u_+(\tau_k^{+o})], u_-(\tau_k^{-o}) = -1_{\tau_k^{-o}} \bar{u}_-, u_+(\tau_k^{+o}) = +1_{\tau_k^{+o}} \bar{u}_+ . \quad (1.3e)$$

<u>Proposition 1.2.</u>

1. Applying discrete delta-function (1.3e) to the EF integral in form (Sec.1.3.3) leads to

$$\Delta S[\tilde{x}_t / \varsigma_t]|_{t=\tau_k^{-o}}^{t=\tau_k^{+o}} = \begin{cases} 0, t < \tau_k^{-o} \\ 1/4u_-(\tau_k^{-o})o(\tau_k^{-o})/\tau_k^{-o}, t = \tau_k^{-o}, 1/4 \downarrow 1_{\tau_k^{-o}} \bar{u}_{ko} \\ 1/2(u_-(\tau_k^{-o}) - u_+(\tau_k^{+o}))o(\tau_k)/(\tau_k^{+o} - \tau_k^{-o}), t = \tau_k, \tau_k^{-o} < \tau_k < \tau_k^{+o}, 1/2(\downarrow 1_{\tau_k^{-o}} - \uparrow 1_{\tau_k^{+o}})\bar{u}_{km} \\ 1/4u_+(\tau_k^{+o})o(\tau_k^{+o})/\tau_k^{+o}, t = \tau_k^{+o}, 1/4 \uparrow 1_{\tau_k^{+o}} \bar{u}_{k1} \end{cases} \quad (1.4)$$

where

$$\bar{u}_{ko} = \bar{u}_- \times o(\tau_k^{-o})/\tau_k^{-o}, \bar{u}_{km} = (\bar{u}_+ - \bar{u}_-) \times o(\tau_k)/(\tau_k^{+o} - \tau_k^{-o}), \bar{u}_{k1} = \bar{u}_+ \times o(\tau_k^{+o})/\tau_k^{+o},$$
$$\bar{u}_{km} = 1/2(\bar{u}_+ - \bar{u}_-) = 0.75, o(\tau_k) = \tau_k^{+o} - \tau_k^{-o}, o(\tau_k^{-o})/\tau_k^{-o} = 0.5, o(\tau_k^{+o})/\tau_k^{+o} = 0.1875 \quad , \quad (1.5)$$

At $\bar{u}_+ = 2, \bar{u}_- = 0.5,$ and $|\bar{u}_- \times \bar{u}_+| = |1/2 \times 2| = |\bar{u}_k| = |1|_k$ is multiplicative measure of the EF impulse $(\downarrow 1_{\tau_k^{-o}} - \uparrow 1_{\tau_k^{+o}})\bar{u}_k$.

Measuring the EF middle interval in (1.4), (1.5) by single impulse information unit $\bar{u}_k = |1|_k$, determines finite size of the EF impulse unit parameters $\bar{u}_{ko}, \bar{u}_k, \bar{u}_{k1}$ in (1.5), which estimate value $\bar{u}_{km}$ on the unit border:

$$\bar{u}_{ko} = 0.25 = 1/3\bar{u}_{km}, \bar{u}_{k1} = 2 \times 0.1875 = 0.375 = 0.5\bar{u}_{km} \ . \bullet \quad (1.6)$$

Proofs follow from Proposition 1.3 below. •

Let us introduce an entropy unit impulse $\bar{u}_s = |1|_s$ with moments $(s_k^{-o}, s_k^{o}, s_k^{+o})$ prior to impulse $\bar{u}_k = |1|_k$, which measures interval of impulse entropy $\bar{u}_{sm}$.

Then we will find increment of entropy $\Delta S[\tilde{x}_t / \varsigma_t]|_{s_k^+}^{\tau_k^{-o}}$ on a border of impulse $\bar{u}_k$ at prior $\Delta_{\tau s+} = \delta_{sk\pm} = (s_k^{+o} - \delta_k^{\tau-})$ and posterior $\Delta_{\tau s-} = \delta_{sk\mp} = (\delta_k^{\tau-} - \delta_k^{\tau+})$ moments under the impulse functions with unit $\bar{u}_s = |1|_s$:

$$\delta^o u_{\tau=(s_k^{+o} - \delta_k^{\tau-})} = (u_+(s_k^{+o}) - u_-(\delta_k^{\tau-}))(s_k^{+o} - \delta_k^{\tau-})^{-1} = \uparrow 1_{s_k^{+o}} \bar{u} - \downarrow 1_{\delta_k^{\tau-}} \bar{u} = [\uparrow 1_{s_k^{+o}} - \downarrow 1_{\delta_k^{\tau-}}]\bar{u}, \quad (1.7)$$

$$\delta^o u_{\tau=(\delta_k^{\tau-} - \delta_k^{\tau+})} = (u_-(\delta_k^{\tau-}) - u_+(\delta_k^{\tau+}))(\delta_k^{\tau-} - \delta_k^{\tau+})^{-1} = \downarrow 1_{\delta_k^{\tau-}} \bar{u} - \uparrow 1_{\delta_k^{\tau+}} \bar{u} = [\downarrow 1_{\delta_k^{\tau-}} - \uparrow 1_{\delta_k^{\tau+}}]\bar{u}, \quad (1.8)$$

$$\delta^o u_{\tau=(\delta_k^{\tau+} - \tau_k^{-o})} = (u_+(\delta_k^{\tau+}) - u_-(\tau_k^{-o}))(\delta_k^{\tau+} - \tau_k^{-o})^{-1} = \uparrow 1_{\delta_k^{\tau+}} \bar{u} - \downarrow 1_{\tau_k^{-o}} \bar{u} = [\uparrow 1_{\delta_k^{\tau+}} - \downarrow 1_{\tau_k^{-o}}]\bar{u}. \quad (1.9)$$

Here $\bar{u}$ evaluates each impulse interval, which according to the optimal principle is an invariant.
Since the EF is additive functional, applying functions (1.7)-(1.9) leads to additive discrete sum of its' increments:

$$\Delta S[\tilde{x}_t / \varsigma_t]|_{s_k^+}^{\tau_k^{-o}} = \Delta S[\tilde{x}_t / \varsigma_t]|_{s_k^+}^{\delta_k^{\tau-}} + \Delta S[\tilde{x}_t / \varsigma_t]|_{\delta_k^{\tau-}}^{\delta_k^{\tau+}} + \Delta S[\tilde{x}_t / \varsigma_t]|_{\delta_k^{\tau+}}^{\tau_k^{-o}} \quad (1.10)$$

along the time interval

$$\Delta_{\tau sk\pm} = s_k^{+o} - \delta_k^{\tau-} + \delta_k^{\tau-} - \delta_k^{\tau+} + \delta_k^{\tau+} - \tau_k^{-o} = s_k^{+o} - \tau_k^{-o} = \Delta_{\tau s}. \quad (1.10a)$$

Proposition 1.3.

**A**. The increments of the entropy functional (1.10) collected on intervals (1.10a), under functions (1.7)-(1.9), bring the following entropy contributions:

$$\Delta S[\tilde{x}_t / \varsigma_t]|_{s_k^+}^{\delta_k^{\tau-}} = 1/2(u_+(s_k^{+o}) - u_-(\delta_k^{\tau-}))o(s_k^{+o} - \delta_k^{\tau-}))(s_k^{+o} - \delta_k^{\tau-})^{-1} = 1/2[\uparrow 1_{s_k^{+o}} - \downarrow 1_{\delta_k^{\tau-}}]\bar{u}_{ks} \quad (1.11)$$

at interval

$$\bar{u}_{ks} = \bar{u}(o(s_k^{+o} - \delta_k^{\tau-})(s_k^{+o} - \delta_k^{\tau-})^{-1}; \quad (1.11a)$$

$$\Delta S[\tilde{x}_t / \varsigma_t]|_{\delta_k^{\tau-}}^{\delta_k^{\tau+}} = 1/2(u_-(\delta_k^{\tau-}) - u_+(\delta_k^{\tau+}))o(\delta_k^{\tau-} - \delta_k^{\tau+}))(\delta_k^{\tau-} - \delta_k^{\tau+})^{-1} = 1/2[\downarrow 1_{\delta_k^{\tau-}} - \uparrow 1_{\delta_k^{\tau+}}]\bar{u}_{k\delta s}, \quad (1.12)$$

at interval

$$\bar{u}_{k\delta s} = \bar{u} \times (o(\delta_k^{\tau-} - \delta_k^{\tau+}))(\delta_k^{\tau-} - \delta_k^{\tau+})^{-1}, \quad (1.12a)$$

and

$$\Delta S[\tilde{x}_t / \varsigma_t]|_{\delta_k^{\tau+}}^{\tau_k^{-o}} = 1/2(u_+(\delta_k^{\tau+}) - u_-(\tau_k^{-o}))o(\delta_k^{\tau+} - \tau_k^{-o})(\delta_k^{\tau+} - \tau_k^{-o})^{-1} = 1/2[\uparrow 1_{\delta_k^{\tau+}} - \downarrow 1_{\tau_k^{-o}}]\bar{u}_{k\delta}, \quad (1.13)$$

under function

$$[\uparrow 1_{\delta_k^{\tau+}} - \downarrow 1_{\tau_k^{-o}}]\bar{u}_{k\delta} = [\uparrow 1_{\delta_k^{\tau+}} + \uparrow 1_{\tau_k^{-o}}]\bar{u}_{k\delta}\ . \quad (1.13a)$$

Here each impulse interval acquires a specific entropy measure at known $\bar{u}_{k\delta}, \bar{u}_{k\delta s}, \bar{u}_{ks}$.
Let us find them.
The $\bar{u}_{k\delta}$ at on the impulse invariant interval $\bar{u}$ acquires form

$$\bar{u}_{k\delta} = \bar{u} \times o(\delta_k^{\tau+} - \tau_k^{-o})(\delta_k^{\tau+} - \tau_k^{-o})^{-1} = \bar{u} \times o(\delta_k^{\tau+})(\delta_k^{\tau+} - \tau_k^{-o})^{-1} + \bar{u} \times o(\tau_k^{-o})(\tau_k^{-o})^{-1}(\delta_k^{\tau+} - \tau_k^{-o})^{-1}\tau_k^{-o} \quad (1.14)$$

Relation (1.14) leads to the impulse interval

$$\bar{u}_{k\delta} = \bar{u}_{k\delta o} + \bar{u}_{k\delta 1} \quad (1.14a)$$

with its parts

$$\bar{u}_{k\delta o} = \bar{u} \times (o(\delta_k^{\tau+}))(\delta_k^{\tau+} - \tau_k^{-o})^{-1}),\ \bar{u}_{k\delta 1} = \bar{u}_{ko1} \times \bar{u}_{ko2}, \quad (1.14b)$$

$$\bar{u}_{ko1} = \bar{u} \times (o(\tau_k^{-o}))(\tau_k^{-o})^{-1},\ \bar{u}_{ko2} = \bar{u}^{-1} \times \tau_k^{-o}(\delta_k^{\tau+} - \tau_k^{-o})^{-1}\ . \quad (1.14c)$$

**B**. Intervals $\bar{u}_{ko1}$ and $\bar{u}_{ko2}$ are multiplicative parts of impulse step-up interval $\bar{u}_{k\delta 1}$, which satisfies relations

$$\bar{u}_{k\delta o} = \bar{u}_{k\delta 1} = 1/2\bar{u}_{k\delta},\ \bar{u}_{k\delta 1} = \bar{u}_{ko1}\ , \quad (1.15)$$

where invariant impulse $|\bar{u}_{k\delta}| = |1|_s$, acting on time interval $\delta_k^{\tau+} = 2\tau_k^{-o}$, measures

$$\bar{u}_{k\delta} = \bar{u}_{ks} \text{ at} |\bar{u}_{k\delta 1}| = 1/2\bar{u}_{sm}, \quad (1.15a)$$

and relative time intervals of $\bar{u}_{ko}$ and $\bar{u}_{k1}$ accordingly are

$$o(\tau_k^{-o})(\tau_k^{-o})^{-1} = 0.5,\ 1/4\bar{u}_{km} = o(\tau_k^{+o})(\tau_k^{+o})^{-1} = 0.1875. \quad (1.15b)$$

Step-controls of impulse $\bar{u}_{k\delta}$ apply on two equal time intervals:

$$(\delta_k^{\tau+} - \tau_k^{-o}) = \delta_k^{\tau+}/2 \ (1.16a) \qquad \text{and} \qquad \tau_k^{-o} = \delta_k^{\tau+}/2. \quad (1.16b)$$

On first (1.16a), its step-up part $[\uparrow 1_{\delta_k^{\tau+}}]$ captures entropy increment

$$\Delta S[\tilde{x}_t / \varsigma_t]|_{\delta_k^{\tau+}}^{\tau_k^{-o}} = 1/2[\uparrow 1_{\delta_k^{\tau+}}]\bar{u}_- = 1/8[\uparrow 1_{\delta_k^{\tau+}}], \qquad (1.16)$$

on second (1.16b), its step-down multiplicative part in (1.14b) at $\bar{u}_{ko2} = \bar{u}^{-1}$ transfers entropy (1.16) to starting impulse action $[\downarrow 1_{\tau_k^{-o}}]$ which cuts it within impulse (1.4) at $\bar{u}_{ko1} = 1/2\bar{u}_{ko}$;

where $\bar{u}_{k\delta 1}$ in (1.14b) multiplies parts

$$\bar{u}[\uparrow 1_{\delta_k^{\tau+}}]\delta_k^{\tau+} / 2 \times \bar{u}^{-1}[\downarrow 1_{\tau_k^{-o}}]\tau_k^{-o}. \qquad (1.16c)$$

Both equal time intervals in (1.16b) are on the impulse border where the opposite inverse entropy increments are orthogonal.

**C**. The applied *extreme* solutions (1.11, 1.13), decreasing the time impulse intervals, bring minimal increment (1.10a) and

(a)-persistence continuation a sequence of the process impulses;

(b)- balance condition for the entropy contributions;

(c)-each impulse invariant unit $\bar{u}_k = |1|_k$, supplied by entropy unit $\bar{u}_s = |1|_s$, *triples* information *increasing information density* in each following information unit.•

Proofs.

The additive sum of entropy increments under invariant impulses (1.7-1.9) satisfies balance condition:

$$\Delta S[\tilde{x}_t / \varsigma_t]|_{s_k^+}^{\tau_k^{-o}} = \Delta S[\tilde{x}_t / \varsigma_t]|_{s_k^+}^{\delta_k^{\tau-}} + \Delta S[\tilde{x}_t / \varsigma_t]|_{\delta_k^{\tau-}}^{\delta_k^{\tau+}} + \Delta S[\tilde{x}_t / \varsigma_t]|_{\delta_k^{\tau+}}^{\tau_k^{-o}} =$$

$$1/2[\uparrow 1_{s_k^{+o}} - \downarrow 1_{\delta_k^{\tau-}}]\bar{u}_{ks} + 1/2[\uparrow 1_{\delta_k^{\tau-}} - \uparrow 1_{\delta_k^{\tau+}}]\bar{u}_{k\delta s} + 1/2\uparrow 1_{\delta_k^{\tau+}}\bar{u}_{k\delta o} - 1/2\downarrow 1_{|\tau|_k^{-o}}\bar{u}_{k\delta 1} = 0, \qquad (1.17)$$

$$1/2\downarrow 1_{|\tau|_k^{-o}}\bar{u}_{k\delta 1} \Rightarrow 1/4\downarrow 1_{|\tau|_k^{-o}}\bar{u}_{ko} \qquad (1.17a)$$

where action transfers entropy increment $\Delta S[\tilde{x}_t / \varsigma_t](\tau_k^{-o}) = 1/4\downarrow 1_{|\tau|_k^{-o}}\bar{u}_{ko}$ on discrete locality $|\tau|_k^{-o}$ by step-down action $\downarrow 1_{|\tau|_k^{-o}}\bar{u}_{k\delta 1}$.

Fulfillment the relations

$$[\uparrow 1_{s_k^{+o}}\bar{u}_{ks} - \downarrow 1_{\delta_k^{\tau-}}\bar{u}_{ks} + \uparrow 1_{\delta_k^{\tau-}}\bar{u}_{k\delta s} - \uparrow 1_{\delta_k^{\tau+}}\bar{u}_{k\delta s} + \uparrow 1_{\delta_k^{\tau+}}\bar{u}_{k\delta o} - \downarrow 1_{|\tau|_k^{-o}}\bar{u}_{k\delta 1}] = 0$$

$$[\uparrow 1_{s_k^{+o}}\bar{u}_{ks} + \uparrow 1_{\delta_k^{\tau-}}[\bar{u}_{k\delta s} - \bar{u}_{ks}] + \uparrow 1_{\delta_k^{\tau+}}[\bar{u}_{k\delta o} - \bar{u}_{k\delta s}] = \downarrow 1_{|\tau|_k^{-o}}\bar{u}_{k\delta 1}, \downarrow 1_{|\tau|_k^{-o}}\bar{u}_{k\delta 1} = -1/2\downarrow 1_{|\tau|_k^{-o}}\bar{u}_{ko}$$

leads to sum of the impulse intervals:

$\bar{u}_{ks} - \bar{u}_{ks} + \bar{u}_{k\delta s} + \bar{u}_{k\delta s} - \bar{u}_{k\delta s} + \bar{u}_{k\delta o} - \bar{u}_{k\delta 1} = 0$, and $\bar{u}_{k\delta 1} = -1/2\bar{u}_{ko}$,

or to

$$\bar{u}_{k\delta o} = \bar{u}_{k\delta 1}. \qquad (1.17b)$$

Impulse $[\uparrow 1_{\delta_k^{\tau+}}\bar{u}_{k\delta o} - \downarrow 1_{|\tau|_k^{-o}}\bar{u}_{k\delta 1}] = [\uparrow 1_{\delta_k^{\tau+}} + \uparrow 1_{|\tau|_k^{-o}}]\bar{u}_{k\delta}$ contains intervals $\bar{u}_{k\delta} = \bar{u}_{k\delta o} + \bar{u}_{k\delta 1}$,

where from (1.9), (1.13a) it follows $\bar{u}_{k\delta} = \bar{u}$, and (1.17b) leads to

$$\bar{u}_{k\delta o} = \bar{u}_{k\delta 1} = 1/2\bar{u}. \qquad (1.17c)$$

Interval

$$\bar{u}_{k\delta}=\bar{u}\times[(o(\delta_k^{\tau+}))(\delta_k^{\tau+}-\tau_k^{-o})^{-1})+(o(\tau_k^{-o}))(\tau_k^{-o})^{-1}(\delta_k^{\tau+}-\tau_k^{-o})^{-1}\tau_k^{-o}] \quad (1.17d)$$

consists of $\bar{u}_{k\delta}$ components:

$$\bar{u}_{k\delta o}=\bar{u}\times(o(\delta_k^{\tau+}))(\delta_k^{\tau+}-\tau_k^{-o})^{-1}) \text{ and } \bar{u}_{k\delta 1}=\bar{u}_{ko1}\times\bar{u}_{ko2}/\bar{u}, \quad (1.17f)$$

where

$$\bar{u}_{ko1}=\bar{u}\times(o(\tau_k^{-o}))(\tau_k^{-o})^{-1},\ \bar{u}_{ko2}=\bar{u}^{-1}\times\tau_k^{-o}(\delta_k^{\tau+}-\tau_k^{-o})^{-1}.$$

Intervals $\bar{u}_{ko1}$ and [$\bar{u}_{ko2}/\bar{u}$] are multiplicative parts of impulse interval $\bar{u}_{k\delta 1}$ covered by starting interval $|\tau|_k^{-o}$.

From (1.17c) and relations (1.17f) it follows

$$\bar{u}_{k\delta o}=\bar{u}\times(o(\delta_k^{\tau+})(\delta_k^{\tau+}-\tau_k^{-o})^{-1})=1/2\bar{u},$$

$$o(\delta_k^{\tau+})(\delta_k^{\tau+}-\tau_k^{-o})^{-1}=1/2 \quad (1.18)$$

and

$$\bar{u}_{ko1}=\bar{u}\times(o(\tau_k^{-o})(\tau_k^{-o})^{-1})=1/2\bar{u}. \quad (1.18a)$$

That leads to

$$o(\tau_k^{-o})(\tau_k^{-o})^{-1}=1/2, \quad (1.18b)$$

and from (1.18) to relation

$$(\delta_k^{\tau+}-\tau_k^{-o})^{-1}=(\tau_k^{-o})^{-1}, \delta_k^{\tau+}-\tau_k^{-o}=\tau_k^{-o},\ \tau_k^{-o}(\delta_k^{\tau+}-\tau_k^{-o})^{-1}=1, \quad (1.18c)$$

then to

$$\tau_k^{-o}=1/2\delta_k^{\tau+}. \quad (1.18d)$$

From (1.18c) it follows

$$\bar{u}_{ko2}=\bar{u}^{-1}. \quad (1.18f)$$

Applying the sequence of Eqs (1.7-1.9), at fixed invariant $\bar{u}$, leads to

$$\bar{u}=u_+(s_k^{+o})-u_-(\tau_k^{-o}), \quad (1.19)$$

$$\bar{u}=u_+(\delta_k^{\tau+})-u_-(\tau_k^{-o}) \quad (1.19a)$$

at $u_+(\delta_k^{\tau+})-u_-(\tau_k^{-o})=\bar{u}_{kb}$,

which brings invariant $|\bar{u}_s|=|1|_s$ to both impulses (1.19) and (1.19a).

Relation

$$u_+(s_k^{+o})+u_-(\tau_k^{-o})=2[u_-(\delta_k^{\tau-})+(u_+(\delta_k^{\tau+})]=0$$

following from the sequence of Eqs (1.7-1.9) leads to

$$u_-(\delta_k^{\tau-})=-u_+(\delta_k^{\tau+}), \quad (1.19b)$$

reversing and mutual neutralizing these actions on related moments $\delta_k^{\tau-}\cong\delta_k^{\tau+}$.

Impulse interval $\bar{u}_{k\delta}$, with $\bar{u}_{k\delta o}$ and $\bar{u}_{k\delta 1}$, starts interval of applying step-down action $o(\tau_k^{-o})(\tau_k^{-o})^{-1}=0.5$ in (1.4) at $\bar{u}_{k\delta 1}=\bar{u}_{k\delta o}=1/2\bar{u}_{k\delta}$.

Invariant impulse $|\bar{u}_s|=|1|_s$, consisting of two step-actions $[\uparrow 1_{\delta_k^{\tau+}}\downarrow 1_{|\tau|_k^{-o}}]\bar{u}_{k\delta}$, measures intervals

$\bar{u}_{kb} = \bar{u}_{sm} = \bar{u}_s$
at
$\bar{u}_{k\delta 1} = 1/2\bar{u}_{sm}$. (1.19c)

At conditions (1.18c, d), limiting time-jump in (1.13a), step-actions of impulse $\bar{u}_{k\delta}$ apply on two equal time intervals following from (1.19c). On the first interval
$(\delta_k^{\tau+} - \tau_k^{-o}) = \delta_k^{\tau+}/2$

step-up part of $\bar{u}_{k\delta}$ -action $[\uparrow 1_{\delta_k^{\tau+}}]$ captures entropy increment

$$\Delta S[\tilde{x}_t / \varsigma_t]|_{\delta_k^{\tau+}}^{\tau_k^{-o}} = 1/2[\uparrow 1_{\delta_k^{\tau+}}]\bar{u}_- = 1/8[\uparrow 1_{\delta_k^{\tau+}}]. \quad (1.20)$$

On the second interval $\tau_k^{-o} = \delta_k^{\tau+}/2$, the captured entropy (1.20) through the step-down multiplicative part (1.17c,d) delivers to the cutting action $\bar{u}_{ko} = \bar{u}_- \times o(\tau_k^{-o})(\tau_k^{-o})^{-1}$ equal contributions

$$\Delta S[\tilde{x}_t / \varsigma_t]|_{\delta_k^{\tau+}}^{\tau_k^{-o}} = 1/4[\downarrow 1_{\tau_k^{-o}}]\bar{u}_- = 1/8[\downarrow 1_{\tau_k^{-o}}]. \quad (1.20a)$$

Control action $[\downarrow 1_{\tau_k^{-o}}]$ at $\bar{u}_- = 0.5$ cuts external entropy of correlation in impulse (1.4) at $\bar{u}_{ko1} = 1/2\bar{u}_{ko}$.

<u>Comment 1.1</u>. Action $[\uparrow 1_{\delta_k^{\tau+}}]$ cuts the captured entropy from impulse $\bar{u}_s = |1|_s$, while multiplicative step-down part (1.17c) transforms the captured entropy to the cutting action in (1.4) at $\bar{u}_{ko2} = \bar{u}^{-1}$. •

At the end of $k$ impulse, control action $\bar{u}_+$ transforms the entropy (1.20) on interval $\bar{u}_{kio} = \bar{u}_- \times (o(\tau_k^{+o})/\tau_k^{+o})$ to information

$$\Delta I[\tilde{x}_t / \varsigma_t]|_{\delta_{k+}^{\tau+}}^{\tau_k^{+o}} = 1/4[\uparrow 1_{\tau_k^{-o}}]\bar{u}_{kio}\bar{u}_+ = 1/4 \times (-2\bar{u}_{kio})[\uparrow 1_{\tau_k^{-o}}] \quad (1.21)$$

and supplies it to $k+1$ impulse.
(If between these impulses, the entropy increments on the process trajectory are absent (cut)).
That leads to balance equation for information contributions of $k$-impulse:

$$\Delta I[\tilde{x}_t / \varsigma_t]|_{\delta_k^{\tau+}}^{\tau_k^{-o}} + \Delta I[\tilde{x}_t / \varsigma_t]|_{\tau_k^{-o}}^{\tau_k} + \Delta I[\tilde{x}_t / \varsigma_t]|_{\tau_k}^{\tau_k^{+o}} = \Delta I[\tilde{x}_t / \varsigma_t]|_{\delta_{k+}^{\tau+}}^{\tau_k^{+o}}, \quad (1.21a)$$

where interval $\bar{u}_{kio}$ holds information contribution $\Delta I[\tilde{x}_t / \varsigma_t]|_{\tau_k}^{\tau_k^{+o}} = 1/4\bar{u}_{km}$ satisfied (1.4) at $\bar{u}_+ = -2$, which measures $\bar{u}_{km} = 0.75$. That brings relations

$$0.125 + 0.75 + \bar{u}_{kio} = -2\bar{u}_{kio}, 0.125 + 0.75 + 3\bar{u}_{kio} = 0, \bar{u}_{k1} = 3\bar{u}_{kio} = 0.375 = \bar{u}_- \times o(\tau_k^{+o})/\tau_k^{+o}) \quad (1.22)$$

and

$$o(\tau_k^{+o})/\tau_k^{+o} = 0.1875, \quad (1.22a)$$

$$\bar{u}_{ko} + \bar{u}_{km} + \bar{u}_{k1} = 1.25 = 5/3\bar{u}_{km} \quad (1.22b)$$

from which and (1.21a) it follows

$$\Delta I[\tilde{x}_t / \varsigma_t]|_{\tau_k}^{\tau_k^{+o}} = 3\Delta I[\tilde{x}_t / \varsigma_t]|_{\delta_k^{\tau+}}^{\tau_k^{-o}} . \quad (1.23)$$

Ratio $\bar{u}_{k1} / 2\bar{u}_{kio} = 3/2$ at $2\bar{u}_{kio} = 0.25$ evaluates part of $k$ impulse information transferred to $k+1$ impulse.

Relations (1.17c,f), (1.18b,d,f), (1.19c), and (1.22a) *prove* the Proposition parts A-B (including (1.16b)) •

Since $\bar{u}_- = 0.5$ is cutting interval of impulse $\bar{u}_k$, it allows evaluate the additive sum of the discrete cutoff entropy contributions (1.4) during entire impulse $(\downarrow 1_{\tau_k^{-o}} - \uparrow 1_{\tau_k^{+o}}) = \delta_k$ using $\bar{u}_- = \bar{u}_k$:

$$\Delta S[\tilde{x}_t / \varsigma_t]|_{\tau_k^{-o}}^{\tau_k^{+o}} = 1/4\bar{u}_k / 2 + 1/2\bar{u}_k + 1/4 \times 3/2\bar{u}_k = \bar{u}_k . \quad (1.24)$$

That determines the impulse cutoff information measure

$$\Delta S[\tilde{x}_t / \varsigma_t]_{\delta_k} = \Delta I[\tilde{x}_t / \varsigma_t]_{\delta_k} = (\downarrow 1_{\tau_k^{-o}} - \uparrow 1_{\tau_k^{+o}})\bar{u}_k = |1|\bar{u}_k , \bar{u}_k = |1|_k \; Nat \quad (1.24a)$$

equals to $\cong 1.44$ Bit, which the cutting entropy functional of this random process generates.

The above single unit impulse $\bar{u}_k = |1|_k$ measures the relative information intervals

$$\bar{u}_{ko} = 1/3\bar{u}_{km} , \; \bar{u}_{km} = 1 , \; \bar{u}_{kio} = 1/3\bar{u}_{km} = \bar{u}_{ko} , \text{ and } \tau_k^{+o} / \tau_k^{-o} = 3 . \quad (1.24b)$$

From relations

$\bar{u}_{ko1} = 1/2\bar{u}_{sm}$ and $\bar{u}_{ko1} = 1/2\bar{u}_{ko} = 1/6\bar{u}_{km}$ it follows

$$\bar{u}_{km} = 3\bar{u}_{sm} , \quad (1.25)$$

which shows that impulse unit $\bar{u}_k = |1|_k$ triples information supplied by entropy unit $\bar{u}_s = |1|_s$ or interval $\bar{u}_k$ compresses three intervals $\bar{u}_s$.

At satisfaction of the extremal principle, each impulse holds invariant interval size $|\bar{u}_k| = |1|_k$ proportional to middle impulse interval $o(\tau)$ with information $\bar{u}_{km}$ which measures $o(\tau)$, and vice versa, time $o(\tau)$ measures this information.

### 2.2.1. Persistence continuation of the impulse sequence measuring by the IPF

Condition of decreasing $t - s_k^{+o} = o(t) \to 0$ with growing $t \to T$ and squeezing sequence $s_k^{+o} \to \tau_{m-1}^{+o}, k = 1,2....m$ leads to persistence continuation of the impulse sequence with transforming previous impulse entropy to information of the following impulse: $\bar{u}_s = |1|_s \to \bar{u}_k = |1|_k$.

The sequence of growing and compressed information increases at

$$\bar{u}_{k+1} = |3\bar{u}_k| = |1|_{k+1} . \quad (1.25a)$$

The persistence continuation of the impulse sequence links intervals between sequential impulses $(\bar{u}_{ks}, \bar{u}_{k\delta s}, \bar{u}_{k\delta o})$ whose imaginary (virtual) function $[\uparrow 1_{s_k^{+o}} - \downarrow 1_{\delta_k^{\tau-}} + \uparrow 1_{\delta_k^{\tau+}}]u$ prognosis entropies increments (1.11), (1.12), (1.10).

Information contributions at each cutting interval $\delta_{k-1}, \delta_k$, $k, k+1,...,m$: $\Delta I[\tilde{x}_t / \varsigma_t]_{\delta_{k-1}}, \Delta I[\tilde{x}_t / \varsigma_t]_{\delta_k},....$ determine time distance interval $\tau_k^{-o} - \tau_{k-1}^{+o} = o_s(\tau_k)$, when each entropy increment

$$\Delta S[\tilde{x}_t / \varsigma_t]|_{\tau_{k-1}^{+o}}^{t\to\tau_k^{-o}} = 1/2uo(\tau_k) = \bar{u}_s \times o_s(\tau_k)$$

supplies each $\Delta I[\tilde{x}_t / \varsigma_t]_{\delta_k}$ satisfying

$\bar{u} \times o(\tau_k) = \Delta I[\tilde{x}_t / \varsigma_t]_{\delta_k}$ at $\bar{u} \times o(\tau_k) = \bar{u}_k(\tau_k^{+o} - \tau_k^{-o})$.

Hence, impulse interval

$$\bar{u}_k = \Delta I[\tilde{x}_t / \varsigma_t]_{\delta_k} / (\tau_k^{+o} - \tau_k^{-o}) \quad (1.26)$$

measures density of information at each $\delta_k = \tau_k^{+o} - \tau_k^{-o}$, which is sequentially increases in each following Bit.

Relations (1.25a,b), (1.26) *prove* part C of Proposition 1.3. •

Such a Bit includes three parts:

-the first delivers multiplicative action (1.16c) by capturing entropy of random process;

-the second delivers the impulse step-down cut of the process entropy;

-the third is information, which delivers the impulse step-up control and then transfers to nearest impulse.

That keeps information connection between the impulses and provides persistence continuation of the impulse sequence during the process time $T$ .

Corollaries 1.2.

**A**. The additive sum of discrete functions (1.4) during the impulse intervals determines the impulse information measure equals to Bit, generated from the cutting entropy functional of random process.

The step-down function generates $1/8 + 0.75 = 0.875 Nat$ from which it spends $1/8$ $Nat$ for cutting correlation while getting $0.75$ $Nat$ from the cut. Step-up function holds $1/8$ $Nat$ while $0.675$ $Nat$ it gets from cutting $0.75$ $Nat$, from which $0.5 Nat$ it transfers to next impulse leaving $0.125$ $Nat$ within $k$ impulse. The impulse has $1/8 + 0.75 + 1/8 = 1 Nat$ of total $1.25 Nat$ from which $1/8$ $Nat$ is the captured entropy increment from a previous impulse. The impulse actually generates $0.75 Nat \cong 1 Bit$, while the step-up action, using $1/8 Nat$, transfers $2/8 Nat$ information to next $k$ impulse, capturing $1/8 Nat$ from the entropy impulse between $k$ and $k+1$ information impulses (on interval $o_s(\tau_k)$ ).

**B**. From total maximum $0.875$ Nat, the impulse cuts minimum of that maximum $0.75$ $Nat$ implementing minimax principle, which validates variation condition (I.1.7) and results (I.2.17a,b).

Transferring overall $0.375 Nat$ to next $k+1$ impulse, that $k$ impulse supplies it with its maximum of $1/3 \times 0.75 Nat$ from the cutting information, implements principle maximum of minimal cut.

**C**. Thus, each cutting Bit is active *information unit* delivering information from previous impulse and supplying information to following impulse. It includes: the cutting step-down control's information delivered through capturing the external entropy of the random process; the cutoff information, which the above control cuts from the random process; the information delivered by the impulse step-up control, which, being transferred to the nearest impulse, keeps the information connection between the impulses providing persistence continuation of the impulse sequence.

**D**. The amount of information that each second Bit of the cutoff sequence condenses grows in three times, which sequentially increases the Bit information density. At invariant increments of impulse (1.4), every $\bar{u}_k$ compresses three previous intervals $\bar{u}_{k-1}$ thereafter sequentially increase both density of interval $\bar{u}_k$ and density of these increments for each $k+1$ impulse. •

The IPF, at condition (1.25a), summing on invariant impulse time intervals (1.24b), acquires finite form

$$I_p = \lim_{m\to\infty}\sum_{k=1}^{m}\Delta I_k \times[\tau_k], \Delta I_k = \Delta I_{k=1}3^{m-1}, \tau_k = \tau_{k=1}3^{-(m-1)}, I_p = \lim_{m\to\infty}\sum_{k=1}^{m}\Delta I_{k=1}3^{m-1}\times\tau_{k=1}3^{-(m-1)} = \lim_{m\to\infty}\sum_{k=1}^{m}\Delta I_{k=1}\times\tau_{k=1},$$

which, for invariant impulse information measure $\Delta I_{k=1} = |1|_k = 1Nat$ brings

$$I_p = \lim_{m\to\infty}\sum_{k=1}^{m}[|1|_k \times \tau_{k=1}] = \lim_{m\to\infty} m[|1|_k \times \tau_{k=1}].. \quad (1.27)$$

$$I_p = \lim_{m\to\infty}\sum_{k=1}^{m}[\Delta I_k \times[\tau_k]] \to \sum_{k=1}^{m} p\times[\tau_k]^2 \to 0 \text{ at } \tau_k = \tau_{k=1}Q^{-(m-1)},\ Q > 0. \quad (1.27a)$$

where the proportional $p = inv$ follows from relations (1.4), (1.5).

The density of information for each following impulse

$$D_k[\Delta I_k] = \Delta I_k / \tau_k = \Delta I_{k=1}Q^{m-1} / \tau_{k=1}Q^{-(m-1)} = (\Delta I_{k=1} / \tau_{k=1})Q^{2(m-1)}$$

is substantially growing along the observing extreme trajectory.

In the limit, at $Q > 0$:

$$D_m = \lim_{m\to\infty}\sum_{k=1}^{m}[(\Delta I_{k=1} / \tau_{k=1})Q^{2(m-1)}] \to \infty. \quad (1.28)$$

Since the time transition to each following IPF' Nat decreases, each such Nat becomes condense integrating all preceding Nats concentrating the integral information in the final IPF Nat.

Such Nat of infinite density cannot exist as real.

Or such reality measure is unobservable even during finite observation time:

$$T_I = \lim_{m\to\infty}\tau_{k=1}Q^{-(m-1)} \to 0, \quad (1.29)$$

as well as

$$\min I_p = \min[\lim_{m\to\infty}\Delta I_{m-1}x(\tau_{m-1})] = \min[\lim_{m\to\infty}\Delta I_{m-1}[\tau_{m-1}] \to 0,$$

which limits the above sum of $I_p$ at $m \to \infty$, and therefore, the cutting *Entropy Integral converges in the Path Functional,* and both of them are restricted at the unlimited dimension number.

### 2.3. The Emerging Microprocess on a Path to Reality

As the probabilities grow along the observations, neighbor impulses $\downarrow\uparrow$ and $\downarrow\uparrow$ may merge.

A discrete merge generates interactive jump $\uparrow$ within the impulse border in a Lévy process.

When Levy measure becomes finite, the jumps have started.

At the merge, the probability of the impulse carrying a bit is zero.

The merge location on the observing trajectory identifies the impulse' probability and entropy.

The jumps releases the entropy-uncertainty hidden in the cutting correlation.

The merge meets causing action with reaction, superimposing a cause and effect and their probabilities.

It could cover unpredictable events within the merge, as the merge itself.

(J.A. Wheeler and R. Feynman [24,25], show that when action and symmetric (adjunct) reaction merge, a direct self-interaction in a microprocess rises from classical Maxwell field)

Along the observations, the drawing together the neighboring impulse actions $\downarrow$ and reactions $\uparrow$ decouple correlation turning them orthogonal. That also turns orthogonal the correlated events time intervals.

The jumps initiate increments of these actions' entropy movements.
The moving anti-symmetric trajectories initiate conjugated dynamics within the jumping impulse.
Mathematically the jump increases Markov drift (speed) up to infinity.
A starting jumping action $\uparrow$ interacting with opposites action $\downarrow$ of the bordered impulse initiates the impulse inner process $\tilde{x}_{otk} = \tilde{x}(t \in o(\tau_k))$ called a microprocess.
The microprocess arises during the cutting micro-time interval of the merging impulses through the observation.
The jump $\uparrow$ initiates an extreme entropy gradient on the orthogonal time intervals, identifying the virtual radius of rotation of the time during its orthogonal displacement.

Comments 3.
In a sub-Markov process [26], potential kernel negative curvature exposes Markov drifts convergence, which could lead to the merge. •

**2.3.1. The conjugated entropy increments in the microprocess**

The microprocess is developing under step-functions $u_{\pm}^{t1}$, $u_{\mp}^{t2}$ within the bordered impulse with the step-function $u_t(u_-^t, u_+^t) = c^2(t \in o(\tau_k))$ on a fixed impulse interval $o(\tau_k)$ within the discrete impulse (1. 4).

The impulse step-down $u_-^t = u_-(\tau_k^{-o})$ and step-up $u_+^t = u_+(\tau_k^{+o})$ functions, acting on the discrete interval $o(\tau_k) = \tau_k^{+o} - \tau_k^{-o}$ satisfying (1.1A-1.1C) and (1.2a-1.2d), generates the EF (2.1) increments:

$$\Delta S_- = \Delta S_-[u_-^t], \Delta S_+ = \Delta S_+[u_+^t], \qquad (3.1)$$

which preserve the additive and multiplicative properties within the Markov process.
(But the merging actions may not simultaneously possess both of these Markov properties).
Here, step function $u_{\pm}^{t1}$ (1.1c) is the analog of $\overline{u}_{k\delta 1}$ (1.14b) which multiplies in (1.16c) at locality $\delta_k^{\tau+}/2$ of the beginning of impulse moment $\tau_k^{-o}$.
Opposite functions $u_{\pm}^{t1}(t^*)$ of jumps $\uparrow\downarrow$, starting at beginning of the process with relative time

$$t^* = [\mp\pi/2 \times \delta t^{\pm *}/\mathrm{o}(\tau_k)], \delta t^{\pm *} \in (\delta t_{ok}^{\pm} \to 1/2\,\mathrm{o}(\tau_k)), \qquad (3.2)$$

hold directions of opposite impulses

$$u_{\pm}^{t1} = [u_+ = \uparrow_{\mathrm{t}_o^{*-}} (j-1), u_- = \downarrow_{\mathrm{t}_o^{*+}} (j+1)] \qquad (3.3)$$

on interval $\delta_o[\mathrm{t}_o^{*-}, \mathrm{t}_o^{*+}] = \delta t^* < o(\tau)$ at a reversible locality of the impulse initial time $\tau_k^{-o}$. (The multiplication (1.16c) applied to functions (3.3), leads to invariant impulse $\overline{u}_k = |1|_k$ by the end of the locality.)

Functions (3.3), holding $u = c^2 < 0$ (1.2d), brings imaginable $u$ and minimal time interval

$o = (\delta_k^{\tau+} / 2)^2 = (\tau_k^{-o})^2$. (3.3a)

The microprocess increments at interval $o$ do not possess Markov properties (1.C) but holds (1.1B).
The jumps (3.3) initiate merging differential relative increments of entropy:

$$\frac{\delta S}{S} / \delta t^* = u_{\pm}^{t1}, \; [u_+ = \uparrow_{t_o^{*-}} (j-1), u_- = \downarrow_{t_o^{*+}} (j+1)], \quad (3.4)$$

which in a limit leads to differential Equations:

$$\dot{S}_+(t^*) = (j-1)S_+(t^*), \dot{S}_-(t^*) = (j+1)S_-(t^*). \quad (3.5)$$

The applied (3.3) with symbol $j$ of orthogonality to the microprocess entropy' increments rotates them.
Solutions of (3.5) describe the microprocess with opposite conjugated entropies functions on relative time $t^*$ :

$$S_+(t^*) = [exp(-t^*)(\mathrm{Cos}(t^*) - jSin(t^*))]\big|_{t_o^{*-}}^{1/2o(\tau_k)}, S_-(t^*) = [exp(t^*)(\mathrm{Cos}(t^*) + jSin(t^*))]\big|_{t_o^{*+}}^{1/2o(\tau_k)} \quad (3.6)$$

with initial conditions $S_+(t_o^{*-})$, $S_-(t_o^{*+})$ at moment

$$t_o^{*+} = t_o^{*-} = [\mp\pi / 2\delta t_{ok}^{\pm}]. \quad (3.6a)$$

The wide of step-function $u_{\pm}^{t1} : \delta t_o^{\pm} / \mathrm{o}(\tau_k) = 0.2 + 0.005 = 0.205$ relative to interval $\mathrm{o}(\tau_k)$ ,being relative to the impulse beginning interval $\tau_k^{-o} / \mathrm{o}(\tau_k) = 0.25$, determines the relative moment $\delta t_{ok}^{\pm} = \delta t_o^{\pm} / \tau_k^{-o} = \pm 0.82$ starting this function.

From that, numerical solutions of (3.6) by the moment of time $\delta t_{ok}^{\pm} = \pm 0.82$ follow:

$$S_+(t_o^+) = [exp(-\pi / 2 \times 0.82)(\mathrm{Cos}(\pi / 2 \times 0.82)) - jSin(\pi / 2 \times -0.82))] \approx 0.2758 \times 1,$$
$$S_-(t_o^-) = [exp(\pi / 2 \times -0.82)(\mathrm{Cos}(-\pi / 2 \times -0.82) + jSin(-\pi / 2 \times -0.82))] \approx 0.2758 \times 1. \quad (3.7)$$

The numerical solutions by the moments of the reversible times:

$$t^{*-} = -\pi / 2 \times 1/2\mathrm{o}(\tau_k) / \mathrm{o}(\tau_k) = -\pi / 4 \text{ and } t^{*+} = \pi / 2 \times 1/2\mathrm{o}(\tau_k) / \mathrm{o}(\tau_k) = \pi / 4 \quad (3.8)$$

are

$$S_+(t^{*-}) = S_+(t_o^{*-}) \times \exp(-\pi/4)[\mathrm{Cos}(\pi/4) - jSin(\pi/4)],$$
$$S_-(t^{*+}) = S_-(t_o^{*+}) \times \exp(-\pi/4)[\mathrm{Cos}(-\pi/4) + jSin(-\pi/4)] = S_-(t_o^{*+}) \times \exp(-\pi/4)[\mathrm{Cos}(-\pi/4) - jSin(\pi/4)]. \quad (3.9)$$

These vector-functions at opposite moments (3.6a) hold opposite signs of their angles $\mp\pi / 4$ with values:

$$S_+(t^{*-}) \cong 0.2758 \times 0.455 \cong +0.125, S_-(t^{*+}) \cong 0.2758 \times 0.455 \cong -0.125. \quad (3.10)$$

Function $u_{\pm}^{t2}$ (1.2d), starting these opposite increments, turns them on angle $\varphi_-^2 - \varphi_+^2 = \pi / 2$ that equalizes the increments and starts entangling both equal increments with their angles within interval $t = \tau_k \mp 0$:

$$S_-^2(t = \tau_k + 0) = \delta S_-^1(t = \tau_k - 0) \times \downarrow_{\tau_k + 0} \pi/2 = S_-^1(t = \tau_k - 0) \times \exp(\pi/2 \times t_{\tau_k+0}^{*+})[\mathrm{Cos}(\pi/2 \times t_{\tau_k+0}^{*+}) + \mathrm{jSin}(\pi/2 \times t_{\tau_k+0}^{*+}), \quad (3.11)$$
$$S_+^2(t = \tau_k + 0) = \delta S_+^1(t = \tau_k - 0) \times \uparrow_{\tau_k + 0} \pi/2 = S_-^1(t = \tau_k - 0) \times \exp(-\pi/2 \times t_{\tau_k+0}^{*-})[\mathrm{Cos}(-\pi/2 \times t_{\tau_k+0}^{*-}) + \mathrm{jSin}(-\pi/2 \times t_{\tau_k+0}^{*-})$$

at moments

$$t_{\tau_k+0}^{*\pm} = [\mp\pi \times 2\delta t_{1k}^{\pm}], \delta t_{1k}^{\pm} = \delta t_1^{\pm} / 1/2\tau_k \cong 0.4375, \delta t_1^{\pm} = \pm(0.5 - \delta t_{\pm}^{k1}),$$
$$\delta t_{\pm}^{k1} = \tau_k^{-o} / \tau_k + \delta t_{\pm}^{ko} / \tau_k = 0.25 + 0.03125 = 0.2895 \quad (3.12)$$

where $\delta t_{\pm}^{ko} / \tau_k \cong 32^{-1}$ evaluates dissimilarities between functions $u_{\pm}^{t2} = [u_{+} = (j+1), u_{-} = (j-1)]$ switching from moment $t = \tau_k - 0$ to moment $t = \tau_k$.

The resulting values at $t = \tau_k + 0$ are

$$S_{-}^{2}(t = \tau_k + 0) = 0.125\exp(\pi / 2 \times 0.4375) \times 1 \cong 0.25, S_{+}^{2}(t = \tau_k + 0) = 0.125\exp(\pi / 2 \times 0.4375) \times 1 \cong 0.25 \quad (3.13)$$

which, being in the same direction, are summing at that locality:

$$S_{\mp}^{o} = 2S_{\mp}^{2}[(\delta t_{\pm}^{ko} / \tau_k)] \cong \mp 0.5. \quad (3.14)$$

The entanglement, starting with entropy (3.13), continues to entropy (3.14) up to cutting all entangled entropy increments.

Thus, the entanglement starts at angle $(\pi / 2) \times 0.4375 < \pi / 4$ takes relative time interval of the impulse $\delta t_{\pm}^{ko} / \tau_k \cong 0.03125$ up to ends of angle $\pi / 2$.

Since only at angle $\pi / 2$ the space interval within impulse begins, it means that *the entanglement starts before the space is formed and ends with beginning the space.*

Here $\tau_k = 1/2o(\tau), o(\tau) = 1Nat$ and

$$\delta t_{\pm}^{ko} = 0.03125 \times 1/2o(\tau) = 0.015625o(\tau) = \varepsilon_{ok}. \quad (3.14a)$$

*Moreover, the time interval within the entanglement that enables creating the space is reversible.*

Comments 4. A potential path during creation of both entanglement and space could be a "*wormhole"*-a short cut in space-time predicted by general relativity. But real *space curvature does not exist at this time*. It may emerge only after end of entanglement by a moment of forming a Bit at the end of the impulse. Hence, space curvature may forming at the *end of a microprocess (analog of a quantum process)* when the Bit, as the elementary unit of a macroprocess, emerges.

*Since the entanglement has no space measure, the entangled states can be everywhere in a space.* •

The $t = \tau_k \mp 0$ locality evaluates the $0_k$ -vicinity of action of inverse opposite functions (3.9), whose signs imply the signs of increments in (3.14) and in following formulas.

The subsequent step-up function jumping action $\downarrow$ changes increment (3.14) according to Eqs

$$S_{\mp}(\tau_k^{+o}) = S_{\mp}^{o}(\delta t_{\pm}^{ko} / \tau_k) \times \exp(t_{\tau_k^{+o}}^{*+}), t_{\tau_k^{+o}}^{*+} = [\pi / 2\delta t_k^{*o}], \delta t_k^{*o} \in (\delta t_{1k}^{*o} \to \tau_k^{+o} / \tau_k), \quad (3.15)$$

at intervals

$$\delta t_{1k}^{*o} = \delta t_{1k}^{\pm} / 1/2\tau, \delta t_{1k}^{\pm} = \pm(0.5 - \delta t_{\pm}^{k1}),$$

$$\delta t_{\pm}^{k1} = \delta t_{\pm}^{ko} / \tau_k + \tau_k^{+o} / \tau_k = 0.25 + 0.03125 = 0.2895, \delta t_{1k}^{\pm} = \delta t_{1}^{\pm} / 1/2\tau_k \cong 0.4375 \cdot\cdot$$

with resulting value

$$S_{\mp}(\tau_k^{+o}) = \mp 0.5\exp(\pi / 2 \times 0.4375) = \mp 0.5 \times (\cong 2) \cong \mp 1, \quad (3.16)$$

which measures total entropy of the impulse

$$\bar{u}_k = |1|_k = 1Nat. \quad (3.17)$$

Trajectories (3.10-3.16) describe anti-symmetric conjugated dynamics of the microprocess within the jumping impulse $\uparrow\downarrow$, which is reversible generating entangled entropy increments (3.16) up to the impulse cutting moment.

Comments 5. From relation (3.4) and Jacobi-Hamiltonian variation equation $\partial S / \partial t = -\tilde{H}$ it follows that the microprocess Hamiltonian gets form'

$$\tilde{H}(t^*) = -u_{\pm}^{t2} S(t^*) . \qquad (3.17a)$$

That Eq. admits the conjugated Hamiltonian with both real and imaginary parts:

$$\tilde{H}(t^*) = -[(j+1)S + (j-1)S] = -[(\dot{S}_+(t^*)/S_+(t^*) + \dot{S}_-(t^*)/S_-(t^*)]S(t^*) \qquad (3.17b)$$

At the entanglement:

$$S_+(t^*+) = S_-(t^*+), S = S_+(t^*+) + S_-(t^*+) = 2S_+(t^*+) = 2S_-(t^*+)$$

and Hamiltonian acquires view

$$\tilde{H}(t^*+) = -[(\dot{S}_+(t^*)/2 + \dot{S}_-(t^*)/2] = -\dot{S}(t^*+) \qquad (3.17c). \bullet$$

Cutting the entangled joint entropy at moment $\tau_k^+ \cong 0_k + \tau_k^{o+}$ coverts it to equal information contribution

$$S_{\mp}^o[\tau_k^+] = \Delta I[\tau_k^+] \cong 1.44 \text{bit} \qquad (3.18)$$

which each $\bar{u}_k$ impulse produces.

An interactive action $\uparrow$ of this impulse with the microprocess jumping action $\downarrow$ delivers entropy on the $0_k$ -vicinity of the cutting moment:

$$S_c^*(\tau_k^+) = \exp 0_k = 1 . \qquad (3.19)$$

The current impulse step-up action $[\uparrow_{\tau_k^{+o}} \bar{u}_+^o]$ generates an information Bit from the microprocess reversible entropy.

The jumping actions provide the minimal discrete displacement (3.3a,3.2), which rotates the entropy opposite increments. The interactive jump generates a pair of random interactive actions on the bordered impulse, which are equally probable and reversible within the probabilities of multiple random interactive actions. The curving shift initiates a microprocess within the bordered impulse running the superposition and entanglement of conjugates entropy fractions during time interval starting with the jump.

The entanglement starts before the space of the shift is formed and ends with beginning the space shift, being small part of impulse reversible time interval. $\bullet$

**2.3.2.The rotating conjugated dynamics of the microprocess**

*Starting step functions* $u_{\pm}^{t1}$ initiates increments of the entropies on interval $o(\tau_k - 0)$ by moment $t = \tau_k - 0$:

$$\delta S_+[u_+^{t1}] = \delta S_+^1(t = \tau_k - 0)) = \delta S_+^1(t = \tau_k^{-o}) \uparrow_{\tau_k^{+o}} (j-1), \delta S_-[u_-^{t1}] = \delta S_-^1(t = \tau_k - 0) = \delta S_-^1(t = \tau_k^{-o}) \downarrow_{\tau_k^{-o}} (j+1) . \qquad (3.20)$$

Step functions $u_{\pm}^{t2}$ (1.2d) starting at $t = \tau_k - 0$ contribute the entropy increments on interval $o(\tau_k)$ by moment $t = \tau_k$:

$$\delta S_+[u_+^{t2}] = \partial S_+^2(t = \tau_k) = \partial S_+^2(t = \tau_k - 0)) \uparrow_{\tau_k} (j+1), \delta S_-[u_-^{t2}] = \partial S_-^2(t = \tau_k) = \partial S_-^2(t = \tau_k - 0)) \downarrow_{\tau_k} (-j+1) . \qquad (3.21)$$

Complex function $u_+^{t1}$ turns on the multiplication of functions $\delta S_+^1(t = \tau_k^{-o})$ on angle $\varphi_+^1 = -\pi/4$, and function $u_-^{t1}$ turns on the multiplication function $\partial S_-^1(t = \tau_k^{-o})$ on angle $\varphi_-^1 = \pi/4$ by moment $t = \tau_k - 0$.

This brings the entropy increments

$$\delta S_+^1(t = \tau_k - 0)) = \delta S_+^1(t = \tau_k^{-o}) \times \uparrow_{\tau_k^{+o}} -\pi/4, \delta S_-^1(t = \tau_k - 0)) = \delta S_-^1(t = \tau_k^{-o}) \times \downarrow_{\tau_k^{+o}} \pi/4 . \qquad (3.22)$$

Analogously, step-functions $u_{\pm}^{t2}$, starting at $t=\tau_k-0$, turn entropy increments (3.22) on angles $\varphi_{-}^{2}=\pi/4$ by moment $t=\tau_k$ and on angle $\varphi_{+}^{2}=-\pi/4$ the entropy increments by moment $t=\tau_k$:

$\delta S_{-}^{2}(t=\tau_k)=\delta S_{-}^{2}(t=\tau_k-0)\times\downarrow_{\tau_k}\pi/4, \delta S_{+}^{2}(t=\tau_k)=\delta S_{+}^{2}(t=\tau_k-0)\times\uparrow_{\tau_k}-\pi/4$ . (3.23)

The difference of angles between the functions in (3.22): $\varphi_{+}^{1}-\varphi_{-}^{1}=-\pi/2$ is overcoming on time interval $o(\tau_k-0)=\tau_k^{-o}+1/2o(\tau_k)$.

After that, control $u_{\pm}^{t2}$, starting with opposite increments (3.23), turns them on angle $\varphi_{-}^{2}-\varphi_{+}^{2}=\pi/2$ equalizing entropy increments (3.23).

That launches *entanglement* of entropies increments and their angles *within* interval $o(\tau_k)$ (on a middle of the impulse $t=\tau_k$):

$$\delta S_{+}^{2}(t=\tau_k)=\delta S_{-}^{2}(t=\tau_k)=\delta S_{\mp}^{2}. \quad (3.24)$$

Control $\bar{u}_{-}=0.5$, turning the time-located vector-function at the impulse beginning:

$$u_{-}^{t}=u_{-}(\tau_k^{-o}):\xleftarrow{\tau_k^{-o},\bar{u}_{-}=0.5,}\bar{u}_{-}\uparrow\delta\varphi_1)=0 \quad (3.25)$$

on angle $\delta\varphi_1=\varphi_{+}^{1}-\varphi_{-}^{1}=\pi/2$, transforms it to space vector $u_{+}(\tau_k-0)=\uparrow_{\tau_k-o}\bar{u}_{+}=1$ during a jump from moment $t=\tau_k^{-o}$ to moment $t=\tau_k-0$ on interval $o(\tau_k-0)$ in (3.22).

Then, vector-function $\downarrow_{\tau_k}\bar{u}_{-}^{o}=2$, starting on time $t=\tau_k-0$ with space interval $\bar{u}_{-}^{o}=2,$ jumps to vector-function $\uparrow_{\tau_{k+0}}\bar{u}_{+}^{o}=2$, forming on time interval $o(\tau_k+0)=1/2o(\tau_k)+\tau_k^{+}$ the additive space-time impulse:

$$u_{\mp}=[\downarrow_{\tau_k+0}\bar{u}_{-}^{o}]+[\uparrow_{\tau_k^{+o}}\bar{u}_{+}^{o}]. \quad (3.26)$$

The first part of (3.26) equalizes (3.24) within *space-time* interval $\bar{u}_{-}\times 1/2o(\tau_k)$, and then joins, summing them on $\bar{u}_{-}\times o(\tau_k+0)$ which finalizes the entanglement.

The last part of impulse (3.26) cuts-kills the entangled increments on interval $\bar{u}_{+}\times\tau_k^{+}$ at ending moment. Relations (3.1-3.26) lead to following specifics of the microprocess.

3.1a.Step-functions $u_{\pm}^{t1}$ initiate microprocess $\tilde{x}_{otk1}=\tilde{x}(t\in o(\tau_k-0))$ on beginning of the impulse discrete interval $o(\tau_k-0)$ with only additive increments (3.2). Opposite step functions $u_{\pm}^{t2}$ continue microprocess $\tilde{x}_{otk2}=\tilde{x}(t\in o(\tau_k+0))$ within interval $o(\tau_k+0)$ with both additive and multiplicative increments (3.3) preserving the process Markov properties.

3.1b.Space-time impulse (3.17) within interval $o(\tau_k+0)$ processes entanglement of increments (3.25) of microprocess $\tilde{x}_{otk2}=\tilde{x}(t\in o(\tau_k+0))$ summing these increments on $o(\tau_k)$ locality of $t=\tau_k$:

$$S_{\mp}^{o}=2\delta S_{\mp}^{2}[(o(\tau_k)]. \quad (3.27)$$

Then it kills entropies (3.27) at its ending moment $\tau_k^{o+} \to \tau_k^{+}$:

$S_{\mp}^{o}[\tau_k^{+}] = 0$. (3.27a)

The microprocess, producing entropy increment (3.27) within the impulse interval, is reversible before killing, which converts the increments to the equal information contributions

$S_{\mp}^{o}[\tau_k^{+}] \Rightarrow \Delta I[\tau_k^{+}]$, (3.27b)

That amount depends on the entropy of merging impulses starting the microprocess.
The information, emerging at the ending impulse time interval, accomplishes the injection of an energy with step-up control $[\uparrow_{\tau_k^{+o}} \bar{u}_{+}^{o}]$.

The energy injection can be a result of the impulse's middle inter-action with an observed environment.
From the impulse ending moment starts an *irreversible* information process.

3.1c. Transferring the initial time-located vector to equivalent space-vector $\uparrow_{\tau_k - o} \bar{u}_{+}$ transforms a *transition impulse*, starting within a jump of time $\tau_k^{-o}$ on interval of $\bar{u}_{-} = 0.5$ up to creating space interval with $\bar{u}_{+} = 1$. The opposite space vector $\downarrow_{\tau_k} \bar{u}_{-}^{o} = 2$, acting on relative time interval $1/2o(\tau_k)/(\tau_k^{+o} - \tau_k^{-o}) = 0.5$, forms space-time function $\downarrow_{\tau_k} \bar{u}_{-}^{1} : \bar{u}_{-}^{1} = 2 \times 0.5 = 1$, which, as inverse equivalent of opposite function $\uparrow_{\tau_k - o} \bar{u}_{+}$, neutralizes it to zero. Both time duration of $\bar{u}_{-} = 0.5$ and $\bar{u}_{+} = 1$ concentrate these functions in transition interval $\tau_k - (\tau_k - 0) = 0_k$. Within the whole impulse, only step-down functions $[\downarrow_{\tau_k^{-o}} \bar{u}_{-}]$ on time interval $\bar{u}_{-} = 0.5$ and step-up function $[\uparrow_{\tau_k^{+o}} \bar{u}_{+}^{1}]$ on space-time interval $\bar{u}_{+}^{1} = \bar{u}_{+} \times \tau_k^{+} = 2_{\tau_k^{+}}$ are left.

That determines size of the discrete $1-0$ impulse by multiplicative measure $U_m = |0.5 \times 2| = |1|_k = \bar{u}_k$ generating an information Bit. Therefore, functions $u_{+}(\tau_k - 0) = \uparrow_{\tau_k - o} \bar{u}_{+}$ and $u_{-}(\tau_k) \downarrow_{\tau_k} \bar{u}_{-}^{o}$ are transitional during formation of that impulse and creation time-space microprocess $\tilde{x}_{otk} = \tilde{x}(t \in 1/2o(\tau_k), h_k \in 2_{\tau_k^{+}})$ with final entropy increment (3.27) and a virtual logic of this microprocess. The microprocess transits from the entropy at ending moment $\tau_k^{o+} \to \tau_k^{+}$-locality (3.27) to actual information (3.27b).

### 2.3.3. Probabilities functions of the microprocess

Amplitudes of the process probability functions at $S_{\mp}^{*}(\tau_k^{+o}) = |S_{+}^{*}| = |S_{-}^{*}| = 1$ are equal and independent:

$p_{+a} = 0.3679, p_{-a} = 0.3679$. (3.28)

That leads to

$p_{+a} p_{-a} = p_{\pm a}^{2} = 0.1353, S_{\mp a}^{*} = -\ln p_{a\pm}^{2} = 2$,

or at $S_{\mp a}^{*} = 2$, to

$p_{a\pm} = \exp(-2) = 0.1353$, (3.28a)

where

$S_{\mp a}^{*} = S_{\mp}^{*}(\tau_k^{+o}+) + S_{c}^{*}(\tau_k^{+}+)$ (3.28b)

includes the interactive components at $\tau_k^{+o}+$ following $k$-th impulse.

Functions $u_+ = (j-1), u_- = (j+1)$, satisfying (1.IA), fulfill the additive property at the impulse starting interval $o[t_o^{\mp}]$, running the anti-symmetric entropy fractions.

Opposite functions $u_+ = (1+j), u_- = (1-j)$, satisfying (1.IB) by the end of impulse at $\uparrow_{\tau_{k+}^{+o}} \bar{u}_\pm$, mount entanglement of these entropy fractions within the impulse' $|1/2\times 2| = |\bar{u}_k| = |1|_k$ space interval $\bar{u}_\pm = \pm 2$.
The entangling fractions hold the equal impulse probabilities (3.28), which indicate appearance of both entangled anti-symmetric fractions simultaneously with starting space interval.

Interacting probability amplitudes $p_{+a}, p_{-a}$ of $p_{\pm a}$ satisfy multiplicative relation $p_{\pm a} = \sqrt{p_{+a} p_{-a}}$.

However sum of non-interacting probabilities: $p_+ + p_- = \exp(-S_+^*) + \exp(-S_-^*) = p_\pm \neq \mathrm{p}_{a\pm}$ does not comply with it.

The summary probability $p_{\pm am} = 0.7358$ of the non-interacting entropies components is unequal to probability $p_{\pm a}$ of interacting entropies.

The interacting probabilities in transitional impulse $[\uparrow 1_{\tau|_k^-} \downarrow 1_{\tau|_k^+}]\bar{u}_k$ on $\tau_k$ -locality violate their additive property, but preserve additive of the entropy increments. The impulse microprocess on the ending interval preserves both additive and multiplicative properties only for the entropies.
The basic relations for the impulse's entropy and probability are equivalent to the probability amplitudes' relations in quantum mechanics (QM).

However, the impulse cutting probabilities $p_+, p_-$ are the probability of random events in the hidden correlations, while probability amplitudes $p_{+a}, p_{-a}$ are attributes of the microprocess starting within the cutting impulse. That distinguishes the considered microprocess from the related QM equations, considered for physical particles.
The multiple impulses integrates the microprocess entropy along the observing random distributions.
With minimal impulse entropy ½ Nat starting a virtual observer, each following impulse' initial entropy $S_\pm(t_o) = 0.25 Nat$ self-generates entropy $S_{\mp a}^* = 0.5 Nat$.

Thus, the virtual observer's time–space microprocess starts with probability $p_{a\pm} = \exp(-0.5) = 0.6015$.

Probability $p_{a\pm} = 0.1353$ is relational to the impulse initial conditions, which evaluates appearance of time–space actual impulse (satisfying (3.26)) that *decreases* its initial entropy on $S_{\mp a}^* = 2$ Nat.

The impulse's invariant measure, satisfying the minimax, preserves $p_{a\pm}$ along the time-space microprocess for multiple time-space impulses.
Reaching probability of appearance the time-space impulse needs $m_p = 0.6015/0.1353 \cong 4.4457 \approx 5$ multiplications of invariant $p_{a\pm} = 0.1353$, which predicts a priori probability of the impulse's reactive action. The space interval, beginning the displacement shift, starts within the interval of entanglement (3.15a) having probability

$$P_\Delta^*(\delta t_\pm^{k1}) = \exp(-|S_\mp^*(\delta t_\pm^{k1})|, P_\Delta^*(\delta t_\pm^{k1}) = 0.821214 \qquad (3.28c)$$

at

$$\delta t_\pm^{k1} = 0.2895, S_\mp^*(\delta t_\pm^{k1}) = \mp 0.125 \exp(\pi/2 \times \delta t_\pm^{k1}) = \mp 0.1969415, \qquad (3.28d)$$

and continues during the shift, extending to the space part of the impulse multiplicative measure after the displacement ends.
Hence, each reversible microprocess within the impulse generates invariant increment of entropy, which enables sequentially minimize the starting uncertainty of the observation.

Assigning the entropy minimal uncertainty measure $h_\alpha^o = 1/137.036$-physical structural parameter of energy [27], which includes the Plank constant's equivalent of energy, leads to relation:

$$S_{\mp a}^* = 2h_\alpha^o, p_{\pm a} = \exp(-2h_\alpha^o) = 0.98555075021 \to 1, \quad (3.29)$$

This evaluates the probability of a real impulse' physical strength of coupling entropy fractions independently on chosen these fractions.

The initially orthogonal non-interacting entropy fractions $S_{+a}^* = h_\alpha^o$ , $S_{-a}^* = h_\alpha^o$ at mutual interactive actions, satisfy multiplicative relation

$$S_{\mp a}^* = (h_\alpha^o)^2 [\mathrm{Cos}^2(\bar{u}t) + Sin^2(\bar{u}t)]|_{t_o^\mp}^{t=1/2\tau} = (h_\alpha^o)^2 = inv, inv = 0.53251 \times 10^{-4}, \quad (3.30)$$

which at $S_{\mp a}^* = (h_\alpha^o)^2 \to 0$ approaches $p_{\pm a}^* = \exp[-(h_\alpha^o)^2] \to 1$.

The impulse interaction adjoins the initial orthogonal geometrical sum of entropy fractions in linear sum $2h_\alpha^o = 2/137.036 = 1/68.518 = 0.014594705$.

As J.A.Wheeler [28] pointed out, "constant $\sqrt{h_\alpha^o} = 1/11.70623765 = 0.0854424543$ be Nature's fundamental measuring unit."

Assuming entropy conjugated dynamics start after entangling fractions

$S_{-a}^1 = S_-(t_o^-) \times exp(\pi/2 \times \delta t_\pm^{ko}/\tau_k) = 0.2758 \times exp(\pi/2 \times 0.0312) = 0.289676085,$

$S_{+a}^1 = S_+(t_o^+) \times (exp(\pi/2 \times \delta t_\pm^{ko}/\tau_k) = S_+(t_o^+) \times exp(\pi/2 \times 0.0312) = 0.289676085$

during time of the entanglement $\delta t_\pm^{ko}/\tau_k$, we get the starting coupled entropy

$$S_{\mp a}^1 = S_{-a}^1 \times S_{+a}^1 \cong 0.08391224234 \quad (3.30a)$$

which evaluates $\sqrt{h_\alpha^o}$ by the *entagled entropy unit.*

Approximating the realtive impulse' transitive path through the uncertain gap up to creation of elementary information unit by $\delta t_\pm^{ko}/\tau_k = 0.0636/1.44 = 0.044166$, according to [28:125] , leads to

$$S_{\mp a}^1 = S_{-a}^1 \times S_{+a}^1 \cong 0.087387332. \quad (3.30b).$$

This brings the evaluation of $\sqrt{h_\alpha^o}$ by the *starting* minimal information unit.

Evaluation (3.30a) determines a lower limit on $\sqrt{h_\alpha^o}$ , while (3.30b) determines its upper information limit.

The average of both limits brings $S_{\mp a}^{1a} = 0.085649787$ which is much closed to the Wheeler's constant.

That means the Nature fundamental unit' measure is Nat.

Starting a physical coupling unit with double structural parameter $h_\alpha^o$ creates initial information triple with probability (3.29).

*The microprocess initiates the merge that starts with the jumping actions multiplication on the bordered impulse time according to (1.16b) succeeding displacement (3.3a) during the merge. Both follow from the EF extreme. Random microprocess' merge proves impossibility of deterministic prediction of QM.*

*The multiplication violates the Markov property (1.1B) leading to a complex control (1.2c), which starts the microprocess within the displacement and rotates the initial conjugated entropy increments.*

*The microprocess (3.3) emerges from multiple interactions starting with probabilities (3.28),inverse entropy* $S_{\mp a}^* = 2$, *and injection of the related random energy. With growing probabilities up to 1, this energy increases rising the equivalent entropy, which is leading to equal information Bit.*

*The energy aspect is in Sec.2.6, where the Jarzynski Equality (JE)*[29], *applied to the evolving microprocess, measures thermodynamic energy connecting the JE with this process's information measure.*

*Thus, this developing microprocess presents a Stochastic Quantum process with evolving thermodynamics and a path to Information Macrodynamics* [15,16].
*Results show that a window gap of interaction with an observed environment opens only on the impulse border twice: at the beginning between moments* $\delta_k^{\tau+}/4$ and $\tau_k^{-o}$ *when the entropy flow with energy accesses impulse, and at the end of a gap when an entangled entropy with accesses of energy converts to equivalent information.*
*Thus, microprocess emerges from the classical process under the interactive observation, and returns to classical information through the observing Bits.*
*The found evaluation of the Nature's fundamental measuring unit brings its information measure in Nat.*

### 2.3.4. Mesuring the curved time and equivalent space length within an impulse

Let us have a two-dimensional rectangle impulse with plain $p$ measured in time length $[\tau]$ unit and orthogonal $h$ measured in space length $[l]$ unit, with the rectangle measure

$$M_i = p \times h. \qquad (3.31)$$

The problem: Having a measure of the plain part of the impulse $\mathrm{M}_p$ to *find* high $h$ at equal measures of both parts:

$$\mathrm{M}_p = \mathrm{M}_h \text{ and } \mathrm{M}_p + \mathrm{M}_h = M_i. \qquad (3.32)$$

From (3.32) it follows

$$\mathrm{M}_h = 1/2 M_i = 1/2 p \times h. \qquad (3.33)$$

Assuming the impulse has only equal plain parts $1/2p$, it measure $\mathrm{M}_p = (1/2p)^2$.

Then from $\mathrm{M}_p = (1/2p)^2 = \mathrm{M}_h = 1/2p \times h$ it follows

$$h/p = 1/2. \qquad (3.34)$$

Let us find a length unit $[l]$ of the curved time unit $[\tau]$ rotating on angle $\pi/2$ using relations

$$2\pi h[l]/4 = 1/2p[\tau] \quad (3.35a), \quad [\tau]/[l] = \pi h/p. \qquad (3.35)$$

Substitution (3.34) leads to ratio of the measured units:

$$[\tau]/[l] = \pi/2. \qquad (3.36)$$

Relation (3.36) sustains orthogonality of these units in a time-space coordinate system, but since initial relations (3.32) are linear, ratio (3.36) represents a linear connection of time-space units (3.35).
The impulse-jumps curve the time unit in (3.8).

According to Proposition 1.3, the impulse' invariant entropy implies the multiplication, starting the rotation.
The microprocess, built in rotation movement curving the impulse time, adjoins the initial orthogonal axis of time and space coordinates (Fig.1a).The curving impulse illustrates Fig.1b.

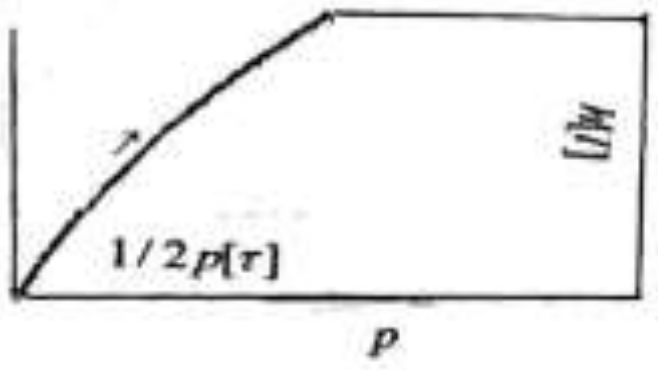

Fig.1(a)

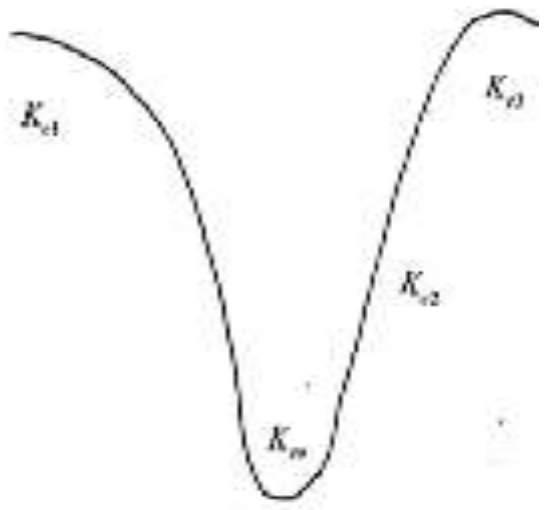

Fig.1 (b)

Fig.1(a). Illustration of origin the impulse space' coordinate measure $h[l]$ at the curving time coordinate measure $1/2p[\tau]$ in transitional movement.

Fig.1(b). Illustrates curving impulse with curvature $K_{e1}$ of the impulse step-down part, curvature $K_{eo}$ of the cutting part, curvature $K_{e2}$ of impulse transferred part, and curvature $K_{e3}$ of the final part cutting all impulse entropy.

The impulses, preserving the multiplicative and additive measures, have common ratio of $h/p=1/2$, whose curving part $p=1/2$ brings universal ratio (3.36), which concurs with Lemma 1.1, (1.2a).

At the above assumption, measure $\mathbf{M}_h$ does not exist until the impulse-jump curves its only time plain $1/2p$ at transition of the impulse. This transition is measured only in time. The following impulse transition is measured both in time $1/2p$ and space coordinate $h$.

According to (3.33), measure $\mathbf{M}_h$ emerges only on a half of that impulse' total measure $M_i$.

The transitional impulse could start on border of the virtual impulses ↓↑, where the transition, curving time $\delta t_p = 1/2p$ under impulse-jump during $\delta t_p \to 0$, leads to

$M_p \to 0$ at $M_h \Rightarrow M_i = p \times h$ . (3.37)

If a virtual impulse ↓↑ has equal opposite functions $u_-(t), u_+(t+\Delta)$, at $\bar{u}_+ = \bar{u}_-$, the additive condition for measure (1.2a): $\mathrm{U}_a(\Delta)=0$ is violated, and the impulse holds only multiplicative measure $\mathrm{U}_m(\Delta) \neq 0$ in relation (1.2C): $U_m(\Delta) = \mathrm{U}_{am}$ which is finite only at $\bar{u}_+ = \bar{u}_- \neq 0$. If any of $\bar{u}_+ = 0$, *or* $\bar{u}_- = 0$, both multiplicative $\mathrm{U}_m(\Delta)=0$ and additive $\mathrm{U}_a(\Delta)=0$ disappear. At $\bar{u}_- \neq 0$, measure $\mathrm{U}_a(\Delta)$ is a finite and positive, specifically, at $\bar{u}_- = 1$ it leads to $\mathrm{U}_a(\Delta)=1$ preserving measure $\mathrm{U}_{amk} = |\mathrm{U}_a|_k$.

An impulse-jump at $o[t_o^{\mp}] \to \delta t_p \to 0$ curves a "needle pleat" space at the transition to the finite form of the impulse. The Bayes probabilities measure may overcome this transitive gap.

Since entropy is proportional to the cutting correlation time interval, whose impulse curvature $K_s = h[l]^{-1}$ is positive, this curved entropy is positive.

The curving needle cut changes the curvature sign converting this entropy to Information.

### 2.3.5. Estimation Curvature of the impulse

An external step-control carries entropy which evaluates:

$\delta_{ue}^{i} = 1/4(\mathrm{u}_{io} - \mathrm{u}_i)$, (3.38)

where $u_{io} = \ln 2 \cong 0.7Nat$ is the total cutoff entropy of the impulse and $u_i \cong 0.5Nat$ is its cutting part. The same entropy-information carries each impulse step-down and step-up control, while both controls carry $\delta^i_{ueo} \cong 0.1Nat$.

That evaluates information wide of each single impulse control's cut which the impulse carries:

$$\delta^i_{ue} \cong 0.05Nat. \quad (3.38a)$$

To create information, the starting step-down part and the step-up part transfer entropy to the final killing part generating information. These three parts carry the entropy measures accordingly:

$$\delta^i_{ue1} \cong 0.025Nat, \delta^i_{ue2} \cong 0.02895Nat, \delta^i_{ue3} \cong 0.01847Nat. \quad (3.38b)$$

The first relation in (3.38b) allows estimate Euclid's curvature $K_{e1}$ of the impulse step-down part, related to currying entropy $0.25Nat$ and its increment $\delta K_{e1}$:

$$K_{e1} = (r_{e1})^{-1}, r_{e1} = \sqrt{1+(0.025/0.25)^2} = \mp 1.0049875, K_{e1} \cong -0.995037, \delta K_{e1} \cong -0.004963. \quad (3.39)$$

The cutting part's curvature estimates relations

$$K_{eo} = (r_{eo})^{-1}, r_{eo} = \mp\sqrt{1+(0.1/0.5)^2} = 1.0198, K_{eo} \cong -0.98058, \delta K_{eo} \cong -0.01942. \quad (3.39a)$$

The transferred part's curvature estimates relations

$$K_{e2} = (r_{e2})^{-1}, r_{e2} = \sqrt{1+(0.02895/0.25)^2} \cong 1.0066825, K_{e2} \cong +0.993362, \delta K_{e2} \cong 0.006638 \quad (3.39b)$$

which is opposite to the step-down part.

The final part cutting all impulse entropy estimates curvatures

$$K_{e3} = (r_{e2})^{-1}, r_{e3} = \sqrt{1+(0.01847/\ln 2)^2} \cong \pm 1.014931928, K_{e3} \cong +0.99261662, \delta K_{e3} \cong -0.00738338.$$

Thus, the entropy impulse is curved with three different curvature values (Fig.1b).

These values estimate each impulse' curvature holding the invariant entropies.

The entropies emerge in minimax cutoff of the impulse carrying entropy $S_{ki} = 0.5$ and *a priori* probability $p_{a\pm} = \exp(-0.5) = 0.6015$ after multiple numbers $m_p$ of probing impulses observe this probability. Since the rectangle impulse, cutting a time correlation, has measure $M = |1|_M$, the curving impulse, cutting the curving correlation, determines measure

$$r_{iM} = M \times K_{ei}. \quad (3.40)$$

The rectangle impulse, not cutting time-correlations, possess Euclid's curvature $K_{iM} = 1$.

Accordingly, the impulse with both time and space measure $|M_{io}| = \pi$, which could appear in transitional impulse curvature of cutting part $K_{eo}$, determines correlation measures

$$r_{icM} = M_{io} \times K_{eo}. \quad (3.40a)$$

At appearance of the impulse with emerging space coordinate, the increment of the curved impulse correlations measure ratio of the measures for the curved correlation to one with only time correlation:

$$r_{icM} / r_{iM} = \pi / |1| K_{ei} / K_{eio} \quad (3.40b)$$

Counting (3.40b) leads to

$$r_{icM} / r_{iM} \cong 3.08.$$

Relative increment of correlation:

$\Delta r_{iM} / r_{iM} = (r_{iM} + r_{icM}) / r_{iM} = 1 + r_{icM} / r_{iM} \cong 4$

concurs with (2.12), which in limit:

$\lim_{\Delta r(\Delta t), \Delta t \to 0} [\Delta r_{iM} / r_{iM}) = \dot{r}_{icM} / r_{iM}$

brings the equivalent contribution to integral functional IPF (4.4.7), which is growing with increasing the above ratio.

Measure $|M_{io}| = |[\tau] \times [l]| = \pi$ satisfies relations

$[\tau] = \pi / \sqrt{2}, [l] = \sqrt{2}$ at $[\tau]/[l] = \pi / 2$ . (3.40c)

Shortening the cutting time intervals triples density (1.26) of each invariant curving correlation for the minimax impulse (1.25), preserving its measure (3.40).

Since any virtual cutting impulse preserves its virtual measure (3.40b), the related virtual time correlation is able to create the space during the entanglement that triple density measures.

For the invariant impulse that compresses the impulse curvature, the probability of both cutting time interval and emerging space coordinate increases.

The impulse measure $\mathrm{M}_{io} = r_{icM} \times (K_{eo})^{-1}$ defines the correlation multiplied on inverse impulse curvature. But since $\mathrm{M}_{io} = inv = \pi$ , it follows direct connection of the correlation with curvature $r_{icM} = \pi K_{eo}$; with a growing correlation curvature increase, and vice versa.

Growing the impulse density, accompanied with the shortening of cutting time intervals, increases the space interval for the invariant impulse measure, but changes the correlation only with the changing curvature.

Since the increasing IPF with growing density accompanies the increasing curvature of rotating impulses, the correlations also grow. After accumulating energy these information curvatures evaluate the impulse information gravity (Sec.4.3).

## III. The Emerging Self-Organization in the Observer Multi-level Reality, and its Information Dynamics

The main–upper stages include observation in uncertain (virtual reality), conversion the observation in certain information reality with discrete time-space, and evolving Information Observer in this discrete reality.

Here we focus on the emerging hierarchy of information process and forming Observer's structural organization. This hierarchical organization follows from the observing multileveled interactive activities.

Starting with an observable random field of probable events, we show how the field's interactive impulses start self-connecting, develop an interactive Virtual Observer, generate the Bit during a microprocess, and compose Bits in information macroprocess which creates the Information Observer.

The Virtual Observer of the observing process builds multiple observing probabilities.

The observing probabilities of Markov chain carry discrete Yes-No actions of the random impulses. Each of these pair acts is a probabilistic impulse like a *classic* action and reaction. When a virtual observing action and reaction merges within a bordered impulse, it brings probability of both actions to zero-on edge of classical predictability. That's why understanding quantum physical microproccess becomes uncertain, fuzzy, weird, and on the edge classical knowledge and even reality. An emerging microprocess composes two qubits and/or Bit starting information microprocess. Multiple Bits self-organize information macroprocess which structures cooperative networks. The emerging self-organized hierarchical dynamics of the network levels conceal increasing Bit densities, where each Bit at a higher level condenses multiple events from lower levels.

These results bring new *systemic* descriptions of the stages and levels of structural organization that compose the *observing path of the evolving dynamics*.

### 3.1.Stages and levels of the emerging Observer's hierarchical self-organization processes

#### 3.1.1.Observing process

Multiple interactive actions are random events, mathematically expressed as variables forming a random process of the impulses in the surrounding random probability field.

In the random field, the occurrence of specific events starts each sequence of the probabilistic observation. The formal probability field emanates a stochastic process of random events and time.

The field probabilities, interacting with the stochastic process, virtually observe it by multiple interactive acts. Each inter-action is a discrete Yes-No action of impulse $\downarrow\uparrow$ modeling potential information Bit, while multiple interactions cover multiple potential Bits changing the stochastic process to a Markov chain at level one.

The Markov transitional probabilities randomly shift each impulse 0-1 to a following 0-1 or to 1-0, which connects them in a correlation through the Markov drift and diffusion.

The probabilistic impulse's Yes-No actions represent the act of a *virtual observation*, where each observation measures the probability of potential events.

The Markov chain under such observation transforms to Markov diffusion process evolving to a Levy process at level two.

The observing probability of triple cooperative events above leads to their forth probability more predictable with growing the probabable density.

Observing more cooperating events increses the density which is limited by $p(\omega)=1$. That leads to zero entropy. Zero entropy limits the observed cooperation, but it also lowers limit of minimial entropy. The minimal entropy imposes extreme principle on the cooperating process probabilities, which leads to invariant maximal $p(\omega)\rightarrow 1$. Thus, the maximal predicted events may follow the beak off unpredictability at level three. A minimal entropy-uncertainty of a probabilistic virtual impulse, overcoming maximal finite uncertainty, connects the probabilistic observations in a virtual observing process. The process No-Yes actions ends this stage at level four.

#### 3.1.2.The impulse's max-min self-action

Each impulse's opposite No-Yes interactive actions (0-1) carry a virtual impulse, which potentially cuts off the Markov process correlation (at the first level) whose conditional (Bayes) entropy decreases as the cutting correlations grow. (The interactive cuts the random process–virtual during probabilistic observation, is required for disclosing the Bit discrete impulses, Sec.2).

If a preceding No action cuts a maximum of the cutting entropy (and a minimal probability), then a following Yes action gains a minimum of the maximal entropy reduction (with its maximal probability) during the impulse cutoff.

A part of the maximal entropy is spent in interactive impacts with the interacting Markov diffusion at level two.

The impulse's maximal cutting No action minimizes the absolute entropy that conveys the Yes actions, raising its probability.

Thus, the cutting action maximizes the cutting entropy, while the reaction spends entropy, minimizing the cutting entropy.

That provides the max-min principle for conditional (relational) entropy between the impulse No-Yes actions. The following Yes-No actions transfer the probabilities and minimum of the impulse cutoff entropy to the next impulse, initiating the minimax principle between the multiple impulses.

The maxmin-minimax principle rules the impulse observations at level three.

The virtual impulse, transferring the Bayes probabilistic observation, virtually probes the observable Markov diffusion process. The probing impulse, consisting of step-down No and step-up Yes actions, preserves the probability measure of these maxmin actions along the observation.

This sequence of interacting impulses, transforming opposite No-Yes actions, increases each following probability and decreases the relative entropy, reducing entropy along the observable Markov diffusion.

The observation, under random probing impulses with opposite Yes-No probability events, reveals a hidden correlation that increase each posterior event correlation at level four.

The hidden entropy releases an attraction of the correlation being cut.

The maximin-minimax self-defines a variation principle (VP) which *formalizes* the description of the observing evolving path at level five.

The VP defines invariant measure of each impulse $|1|_M$ .

### 3.1.3.Virtual self-observer

If the observing process is self-supporting through the automatic renewal of these virtual probing actions, a Virtual Observer self-emerges, which acts until these actions resume, up to the emergence of a real Information Observer (if it appears).

Such Virtual Observer belongs to a self-observing process, whose Yes action virtually starts the next impulse No action, and so on.

Both the observing process and Observer are temporal, ending when the virtual observation stops.

Starting the virtual self-observation limits a *threshold* of the impulse's connection on the first level on this stage. The virtual observations precede the real ones, but may not link up to them.

The sequentially reduced relational entropy conveys a probabilistic causality along the process as temporal memory collects correlations. This cutting entropy defines the second level of this stage.

### 3.1.4.The emerging observer time

The starting correlation holds the entrance of a *time interval* of the impulse-observation process at level one.

The time interval, connecting the probability-entropy in correlation, measures an uncertainty distance between the nearest current observations at level two.

Beginning from the starting observation, the measure identifies the time interval from the start, which is also virtual, disappearing with each new connection; that identifies a next interval temporally memorized in that correlation connection.

The difference of the probabilities temporarily holds the memory of the correlation, as a virtual measure of an *adjacent distance* between the impulse's No-Yes actions.

It indicates a probabilistic accuracy of measuring correlation in a *time interval's unit* at level three.

The impulses of the observing process hold the virtual observing random time intervals.

### 3.1.5.The Emergence of the impulse space interval and space-time geometry in an Observer structure

With growing correlations, the intensity of entropy per the interval (as entropy density) increases on each following interval. That indicates a merging of the virtual actions, measured in a time interval's unit measure $|1|_M$ at level one.

The merging interactive actions $\uparrow\downarrow$ of bordering impulses generate an interactive jump with a high entropy density (within the Levy process) with a curving action $\downarrow$

The jump's curving action $\downarrow$ *curves* an emerging *½ time unit* of the border impulse's time interval, initiating a *displacement* of the curved action.

This originates a curved interval within an impulse quantified by the impulse $\bar{u}_k$ discrete probability measure $\mathrm{p}[\bar{u}_k]$ (1 or 0).
The displaced curved jump action starts rotation of the ½ *time unit within the impulse* $\bar{u}_k$ at level two.
The invariant measure $|1|_M$ of the impulse with unit rotating on ½ times unit should be preserved, which requires a *jumping* rotation of the ½ times unit within the impulse with probability measure $\mathrm{p}[\bar{u}_k]$.
The rotating transformation, preserving the impulse $\bar{u}_k$ invariant measure, satisfies the measure transformation $M[\bar{u}_k] = |1/2 \times 2| \xrightarrow{\mathrm{p}[\bar{u}_k]} |1|_{M_k}$, which originates *two space units* of the displaced impulse. This transformation changes the primary impulse's time measure $|1|_M$ to the equal measure $|1|_{M_k}$ for the impulse with both time and space units, as a counterpart to the primary impulse curved time interval measure $|1|_M$, see Figs.1, 1a.
The emerging rotating time-space coordinate system involves the following time-space impulse in correlated curved movement.
That initiates a time-space observing process within the Markov model which conserves the transferring impulse invariant measure at level three.
The changing probabilities of the Markov process self-observe the correlated time-space movement modeling a Virtual Observer.
The Virtual Observer, being displaced from the initial virtual process, sends the discrete time-space probabilistic impulses $\bar{u}_k$ as virtual probes to *self-test* the preservation of the probability measure in the observing process.
The probing impulses are sent with frequencies defined by the observing probabilities at level four.
The Observer's *self-supporting* probes increase frequencies with a growing probability. Such probes also check the probability symmetry condition, indicating its correctness at level four.
The memory of the correlating time-space intervals temporarily holds difference of the impulses' space-time, identifying both accuracy of their closeness and Virtual Observer's location on the observing trajectory. The space-time location measures the time-space shape of the Virtual Observer geometry at level six. The evolving shape gradually confines the running rotating movement, which *self-supports* developing both the shape and the Observer geometry.
The observer rotating geometry integrates the sequential conditional probabilities in a final *a posteriori* probability and the entropy of correlated impulses.
The Observer *self-develops* its space-time virtual geometrical structure during virtual observation, gaining its real form with the sequential transformation of the integrated entropy to the equivalent Information at level seven.

### 3.1.6.The microprocess

The microprocess emerges inside a Markov diffusion process which, therefore, should preserve the Markov additive and multiplicative properties during the rotating correlation movement.
These conditions satisfy the following levels of the emerging microprocess, detailed below.
1.Growing probabilities along the observations intensifies the entropy force, drawing together the neighboring impulse actions $\downarrow$ and reactions $\uparrow$. This changes the width and curving of the impulse time intervals. The curving squeezes the time interval up to its merging with the neighboring interactive actions $\downarrow\uparrow$ on the bordering impulses, which are equally probable and reversible within the probabilities of multiple random interactive actions. The merging action and reaction within bordered impulse brings probability of both actions to zero-on edge of classical predictability.

A jumping action $\downarrow$ starts the entropy increment, beginning a microprocess.

The curving jump $\uparrow$ initiates an extreme entropy gradient on the curving time interval, identifying the virtual radius of rotation, which displaces the curving action.
As the displacement starts, the opposite entropies of the microprocess emerge.
During the jump's curving time the displacement's space emerges.
The interactive jump identifies the impulse curvature, entropy measure, and time-space measure equal to π which is invariant under the minimax principle.

The jumping impulse $\downarrow\uparrow$ develops this impulse time-space volume.
When that displacement rises between the actions with probabilities 0 or 1, the displacement has no classical Markov probabilities measure.
The process within the displacement has been studied as a sub-Markov process, loops [26], and a Schrodinger bridge [30,31], which is a unique Markov process in the class of reciprocal processes introduced by Bernstein [32].
2. When the sub-Markov process receives a negative entropy measure $S^*_{\mp a} = -2$ through a jumping action $\downarrow$ with relative probability $p_{a\pm} = \exp(-2) = 0.1353$, it initiates the minimal time microprocess on the verge of the displacement.
3. With the development of the merging jumping impulse anti-symmetric (conjugated) rotating entropy increments, the observing entropies become correlated. The correlation increases with the growth of the classical prior probabilities of multiple random actions $\downarrow$, and the posterior probabilities of the random action $\uparrow$. Both of them are virtual, decreasing with the growing probability measure.
The sub-Markov process with its opposite random actions disappears from the observation.

At the maximal probabilities, only a pair of the Markov additive entropies increments with axiomatic symmetric probabilities (that contain symmetrical-exchangeable states) advances in the correlated superposition of both actions $\downarrow\uparrow$.
These random actions, belonging to the Markov process, measure a multiplicative probability.
At the satisfaction of the symmetry condition, an interactive jumping action transforms the observing axiomatic probabilities to "quantum" probabilities with pairs of conjugated entropies to their correlated movements.
4. The conjugated entropies, increments rotating on angle $\pi / 4$, raise the space interval with a virtual transitive action $\uparrow$ within the microprocess, initiating the correlated entanglement. Maximal correlation adjoins the conjugated symmetric entropies, uniting them in a running pair entanglement.
The conjugated entropies increments, rotating the space interval on angle $-\pi / 4$, transform the transitive action $\uparrow$ to an action $\downarrow$ that settles into a *transitional impulse* $\uparrow\downarrow$, finalizing the entanglement at total angle $\pi / 2 = \pi / 4 - (-\pi / 4)$.

The transitional impulse, holding actions $\uparrow\downarrow$ opposite to the primary jumping impulse $\downarrow\uparrow$, generates an inner conjugated entanglement involved, for example, in left and right rotations( $\mp$ ).

The transitional impulse, interacting with the opposite correlated entropies $\mp$, reverses it on $\pm$.

Since the correlated entropies are virtual, transition action within this impulse $\uparrow\downarrow$ is also virtual, and its interaction with the forming correlating entanglement is reversible.

Within the impulse time interval, the *entanglement starts before the space is formed and ends as space begins.* It occurs *during the reversible relative time interval* $0.015625\pi$, *being* part of the impulse time-space invariant measure $\pi$.
*Since the entanglement has no space measure, the entangled states can be anywhere in a space.*
*The space emerges with probability* $P^*_\Delta(\delta t^{k1}_\pm) = 0.821214$ on the time interval $\delta t^{k1}_\pm$.
5. The following interaction logically erases (cuts) each previous directional rotation of the entangled entropy increment of the entropy volume. This erasure emits minimal energy $e_l$ of quanta [33].
The transitional impulse absorbs this emission inside the virtual impulse, which logically memorizes the entangled units, making their mirror copy. (Like a rubber ball, hitting a surface, makes a temporal copy of a dent). The interacting probabilities in transitional impulse-locality violate their additive property, but preserve additive of the entropy increments. The impulse microprocess on the ending interval preserves both additive and multiplicative properties *only for the entropies intcrements*.
Such an operation performs the function of a logical Maxwell's Demon, creating asymmetrical logic Bit.
*The transition to the asymmetry abruptly changes the probability multiplicative to additive property.*
6. The entangled logic transfers to information logic when the rotating step-up action $\uparrow$ of the transitional impulse moves to transfer the entangled entropy volume to the ending step-up action $\uparrow$ of the jump, which follows that real step-down action $\downarrow$.
This ending impulse attracting action captures energy of the real interactive action physically erasing the entropy logic Bit. Each process's high–quality energy compensates for entropy of lesser quality. That removes the causal entropy of the asymmetrical logic, bringing asymmetrical information logical Bit as certain impulse Bit. Such a Bit is naturally extracted at a minimal quality energy measure equivalent to entropy ln2. The Bit memorizes at cost of Landauer's energy [34,35] working as Maxwell Demon.
The memorized Bit freezes the energy spent on the erasure for its creation as the Bit equivalent $\ln 2$.
The memorized impulse 1 Nat includes information Bit and free information enclosing hidden information from cutting correlation (connected the Markov process impulses).
7. The microprocess is different from that in quantum mechanics (QM), because it arises inside the evolving impulse under No-Yes virtual and real final actions. The superposing rotating anti-symmetric entropy increments in the microprocess have additive time-space complex amplitudes correlated in the time-space entanglement that does not carry and bind energy. It just connects the entropy in joint correlation. These complex amplitudes model elementary interactions with no physics, while the real cut brings a physical Bit. The QM probabilistic particles carry analogous conjugated probability amplitudes correlated in time-space entanglement.
Theoretically, Kolmogorov's probability measure at the QM entanglement, when both additivity and symmetry of probability for mutual exchangeable events vanishes, challenges the predictability of the QM probability. In this Observer-probabilistic approach, the microprocess holds the predictable relational probabilities until the entangled entropy is cut. These relational probabilities satisfy the multiplication property, while, before the interaction, these probabilities are additive. Cutting the entangled entropy and generating qubits and/or a Bit, ends the microprocess in the evolving observation.
8. Logical operations with Information units integrate the discrete Information hidden in the correlations of cuts, creating the structure of Information Observer. The integration performs entropy path functional EF. The relational entropy conveys the probabilistic causality with a temporal memory of correlations, while the real cutoff memorizes certain Information causality during the objective probability observations.
The self-observing Observer self-generates the elementary Bit self-participating in building the self-holding geometry and the logic of its prehistory, thus predicting evolving dynamics without any physical law.

### 3.1.7. Specifics of memorizing and encoding elementary information through interacting curved impulses

The interaction of the impulse Fig. 2A and Fig. 2B holds the opposite action $\uparrow\downarrow$, curving a displacement between them, which provides a time-space asymmetrical barrier between 0 and 1 actions, which is necessary for creating a Bit logic ln2.

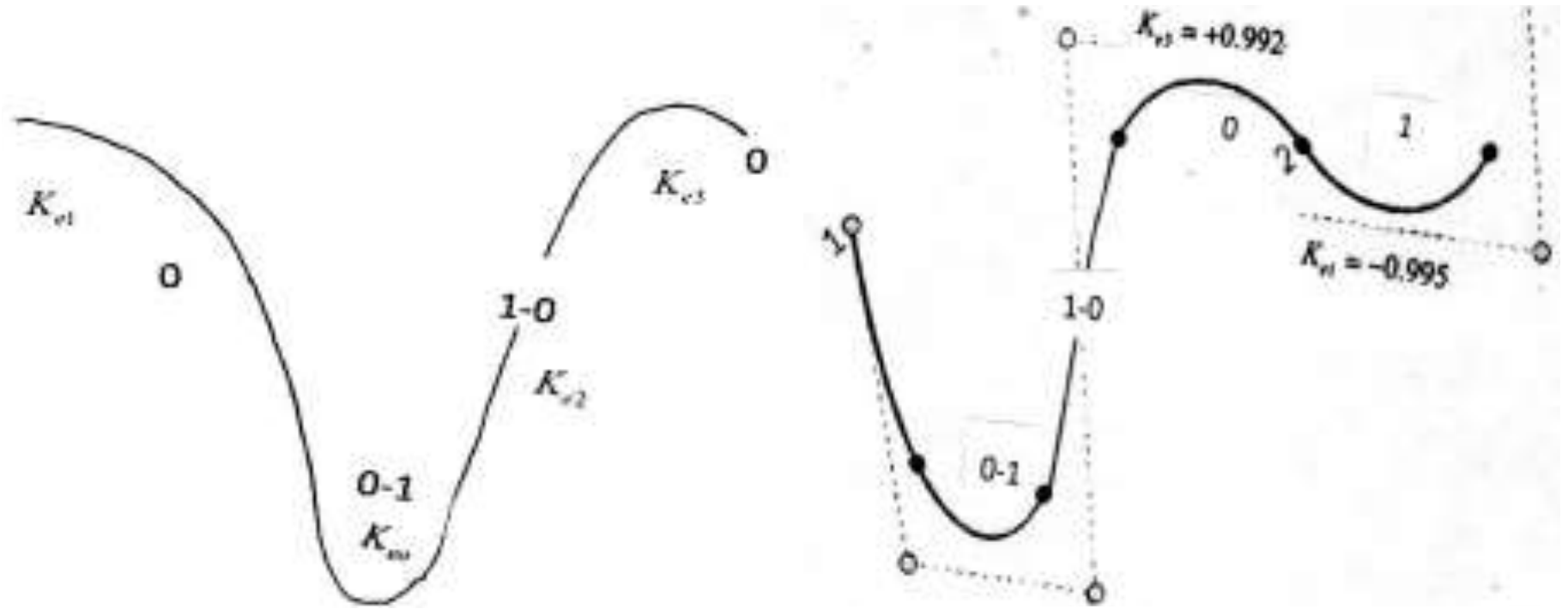

A B

Figure 2. A virtual impulse (Figure 2A) starts step-down action with probability 0 of its potential cutting part; the impulse middle part has a transitional impulse with transitive logical 0-1; the step-up action changes it to 1-0 holding by the end interacting part 0, which, after the inter-active step-down cut, transforms the impulse entropy to Information Bit. In Figure 2B, the impulse Figure 2A, starting from instance 1 with probability 0, transits at instance 2 during interaction to the interacting impulse with negative curvature $-K_{e1}$ of this impulse step-down action, which is opposite to curvature $+K_{e3}$ of ending the step-up action ($-K_{e1}$ is analogous to that at beginning the impulse Fig.2A). The opposite curved interaction provides a time–space difference (a barrier) between 0 and 1 actions, necessary for creating the Bit.

When the interactive process provides Landauer's energy with maximal probability (certainty) 1, the interactive impulse' step-down action ending state memorizes the Bit. Such certain interaction injects the energy overcoming the transitive gap including the barrier toward creation the Bit.

The curving topological geometry of the asymmetry can enclose minimal energy ln2 in a transitional impulse.

Forming a transitional impulse with the entangled qubits leads to the possibility of memorizing them as a quantum Bit. The required memory of the transitional curved impulse encloses entropy $0.05085 Nat$ [40].

The step-up action of an external (natural) process' curvature $+K_{e3}$ is equivalent of potential entropy $e_o = 0.01847 Nat$ which carries entropy $\ln 2$ of the impulse total entropy 1 Nat.

The interacting step-down part of internal process impulse' invariant entropy 1 Nat has potential entropy $1-\ln 2 = e_1$. Actually, this step-down opposite interacting action brings entropy $-0.25 Nat$ with anti-symmetric impact $-0.025 Nat$ which carries the impulse wide $e_w \cong -0.05 Nat$ with total entropy $-0.3 Nat$ equivalent to $-e_1$.

The step-down state of a real action $\downarrow$ (carrying the energy) supplies Landauer's minimal energy equivalent $\ln 2$ with maximal probability. This action kills the entropy and erases it, which memorizes a classical Bit in an irreversible process of multiple Bits.

*Thus, in transition to certainty, the transitive impulse geometry enables memorizing a quibit.*

Creation of the entangled entropy volume is a reversible process following the memorizing of this volume, which freezes $-\ln 2$ in two opposite qubits.

Finally, the impulse' (Fig. 2A) step-down cut $\downarrow$, extracting each Bit's hidden position, erases it at a cost of cutting the real time interval, which encloses the energy of an interactive process.

The impulse' (Fig. 2B) step-up action $\uparrow$, stopping at the end of the impulse's time interval, memorizes the impulse logic of the Information Bit by encoding it.

### 3.1.8.The gap between entropy and information

As maximal *a priori* probability approaches $P_a \to 1$, both the entropy volume and the rotating momentum grow. Still, between the maximal *a priori* probability of the virtual process and *a posteriori* probability of the real process $P_p = 1$, is located a gap whose left (starting) edge belongs to the ending microprocess. The gap associates with a time-space probabilistic transitive movement, separating entropy and its Information (at $P_a < 1$ throughout $P_p \to 1$).

The gap holds a hidden *real locality* which the impulse cuts within the hidden correlation.

The rotating momentum, growing with the increased entropy volume, intensifies the time-space volume transition over the gap. That momentum acquires a physical property near the gap end when the last posterior probability $P_p$ overcomes the last prior virtual probability, and the momentum curves a physical cut of the transferred entropy volume. The real local gap reveals a physical Markov diffusion containing energy of the hidden entropy of correlation.

Transition maximal probability of observation through the gap up to killing the resulting entropy runs a *physical* part of the microprocess when the entanglement creates a space, which preempt memorizing.

Until that, the microprocess within impulse is reversible.

Ability an observer to overcome its gap depends on the entropy volume, collected during virtual probes, whose entropy force and momentum spin the rotating momentum for transition over the gap. The real microprocess builds each information unit-Bit within the cutting impulse in real time, becoming irreversible after the cut.

It is impossible to reach a reality in the quantum world without overcoming the gap between entropy-uncertainty and Information-certainty, which is located on the edge of reality.

Within the gap, the entangled microprocess's conjugated entropies $S^*_{\mp a} = 2h^o_\alpha$, limited by minimal uncertainty measure $h^o_\alpha = 1/137$ (a fine structural parameter of energy) and the entangling qubits, are confined. Injection of the energy has probability

$p_{\pm a} = \exp(-2h^o_\alpha) = 0.9855507502$.

The energy starts the erasure of entropy and creation of Information, while the actual killing with probability $P_k = 0.99596321$ ends the erasure. A gap to reality evaluates probability $1 - P_k \cong 0.004$.

Forming the classical Bit and qubits entropys a higher probability, but less than 1, which does not allow reaching absolute certain reality.

The gap or "The Blind Split"[95] between entropy (uncertainty) and information (certainty) is invevitable regularity of this transformation.

### 3.1.9.Information process

This process emerges from the observing process of the Virtual Observer (at level one) starting microprocess which is evolving from the conjugated entropies within a merging interacting impulse (at level two). Information arises from multiple random interactive impulses when some of them could erase-cut others, providing Landauer's energy (on the first level of this stage).

Asymmetrical interaction (involving transitional impulse topology), enables erasing the impulse, becomes a Bit of Information, finalizing the next level of the stage at level three.

The impulse cutoff correlation sequentially converts the hidden entropy to Information that memorizes the entropy logic in a Bit. Each created Bit participates in the subsequent conversions, which generate an interactive Information process at level four. Thus, the origin of Information is associated with the impulse ability of both the cut and the observing process, generating Information under the cut, whose memory holds the impulse's cutting time interval.

Since the curved topology of the interacting impulses decreases total needed energy, this energy, at the found ratio of the impulses' external and internal temperatures [40], can deliver the minimal Landauer's energy equivalent to ln2. The increment of Information covering the asymmetric interaction evaluates free Information $i_f = 0.23bit$, which enables connecting multiple Bits through *Information attraction* at level five.

Multiple cuts of the more probable posterior correlations in the interactive multi-dimensional observation are a source of persistent Information attraction through the $i_f$ of a sequential impulse. That $i_f$ can comprises multiple Bits which concurrently memorize the impulse's cutting time intervals. The emerging discrete information freezes the observing events dynamics in Information processes at level six.

Integration of the cutting Bits' time intervals along the observing time course converts it to the *Information Observer's* inner time course. That time course is opposite to the virtually observable process time course in which the process entropy increases. This difference emerges at level seven.

On the Information process trajectory, each previous impulse's Information is distinguishable from that in the following impulse. The distinction measures a random interval between these impulses.

The interval is predictable through the cutting correlations that integrate the entropy–information path functional (EF-IPF) [46].

The random difference can model "mutation" in an evolving Information process, which the EF-IPF measure estimates at level eight.

The difference *a priori* holds the imaginary entangled entropy of a microprocess, which proceeds along the EF. Each impulse observation creating a Bit estimates a frequency of probing impulses $F_{im} = 10^{-4} \times 1.13276$ while the frequency $F_{imo} = 10^{-4} \times 1.13636$ anticipates memorizing the Bit. That is evaluated at level nine.

### 3.1.10.The emerging macroprocess composing its basic triplet structural units

The rotating movement (Figs.3) connects the microprocess imaginary entropy and information Bits *in a macroprocess,* where the free information binds the diverse Bits in *collective* information time-space macrotrajectories (Figs. 3,4). That is main specific of information macroprocess building its structural units.

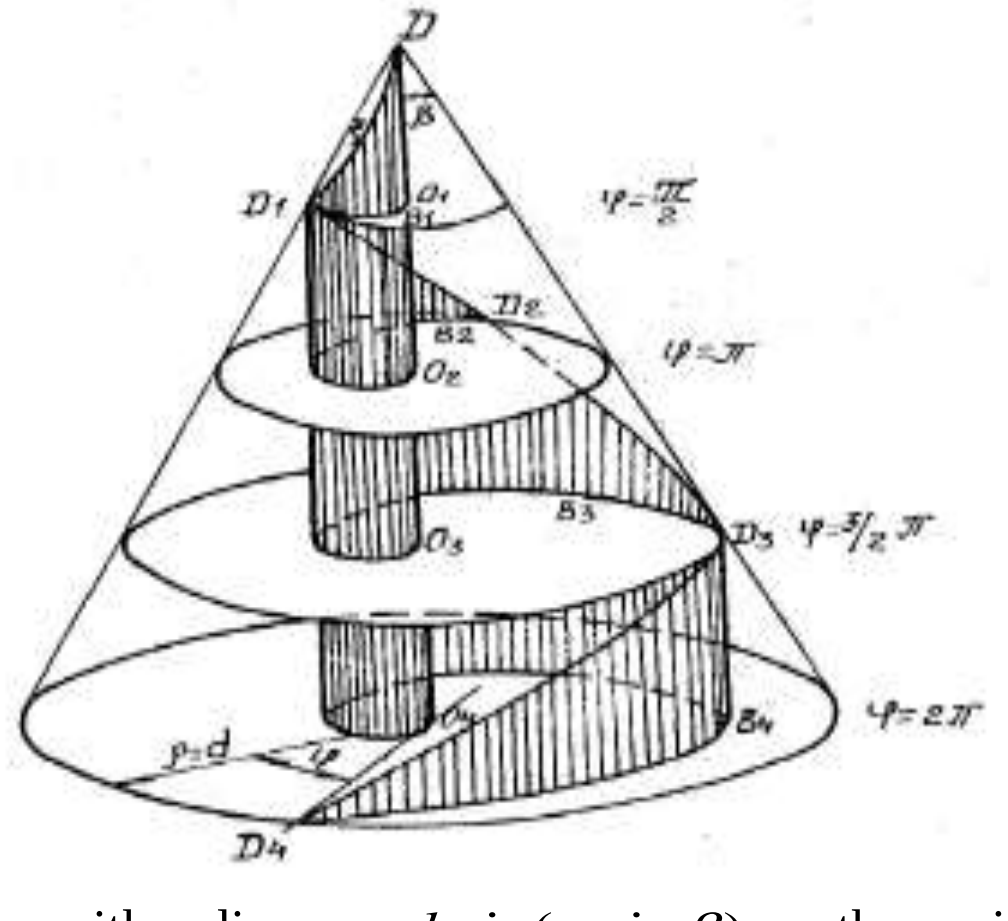

Fig.3. Forming a space -time spiral trajectory with radius $\rho = b\sin(\varphi\sin\beta)$ on the conic surface at the points D, D1, D2, D3, D4 with the spatial discrete interval DD1=$\mu$, which corresponds to the angle $\varphi = \pi k / 2$, $k = 1,2,...$ of the radius vector's $\rho(\varphi,\mu)$ projection of on the cone's base (O1, O2, O3, O4) with the vertex angle $\beta = \psi^o$.

The macroprocess free information integrates the Bits in information path functional (IPF) which encloses the Bits time-space geometry in the process' information structure.

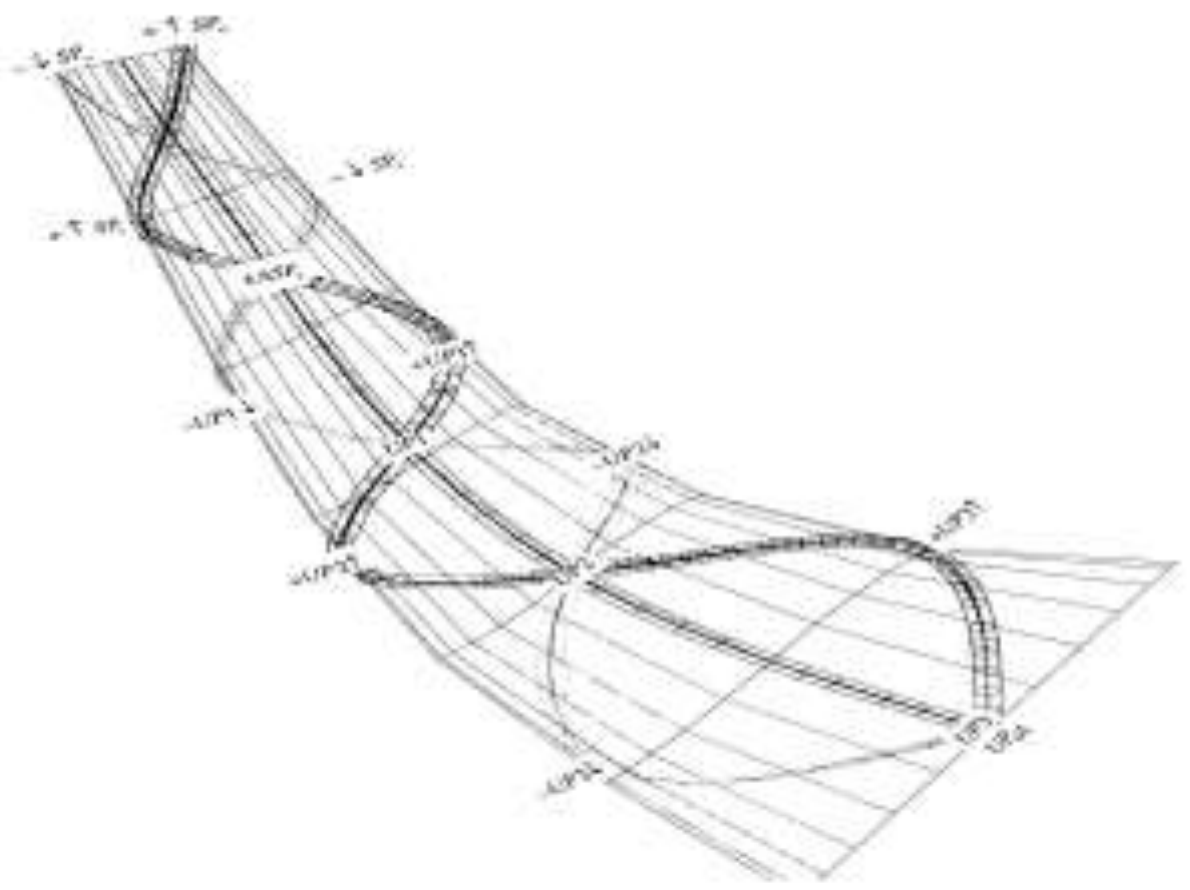

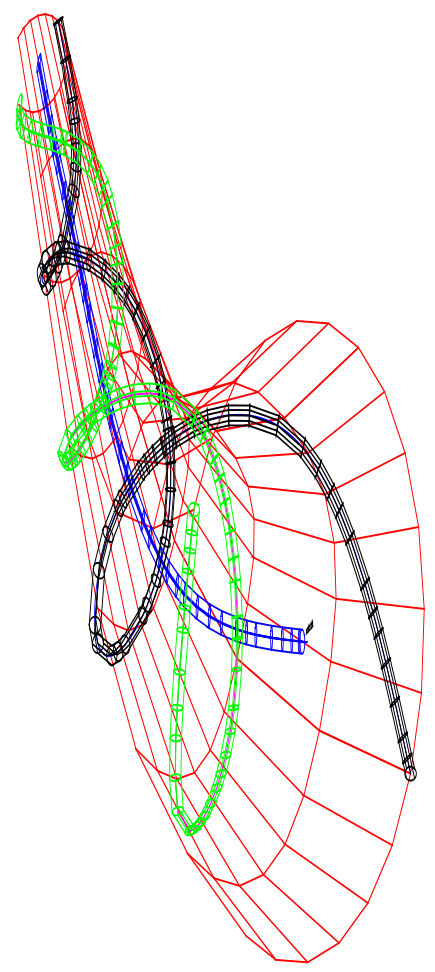

4a. 4b.

Fig. 4. Time-space opposite directional-complimentary conjugated trajectories $+\uparrow SP_o$ and $-\downarrow SP_o$, forming spirals located on conic surfaces (analogous to Fig.3).

Trajectory of bridges $\pm SP_i$ (4a) binds the contributions of process information unit $\pm UP_i$ through the impulse joint No-Yes actions which model a line of switching controls. Two opposite space helixes and the middle curve are on the right (4b).

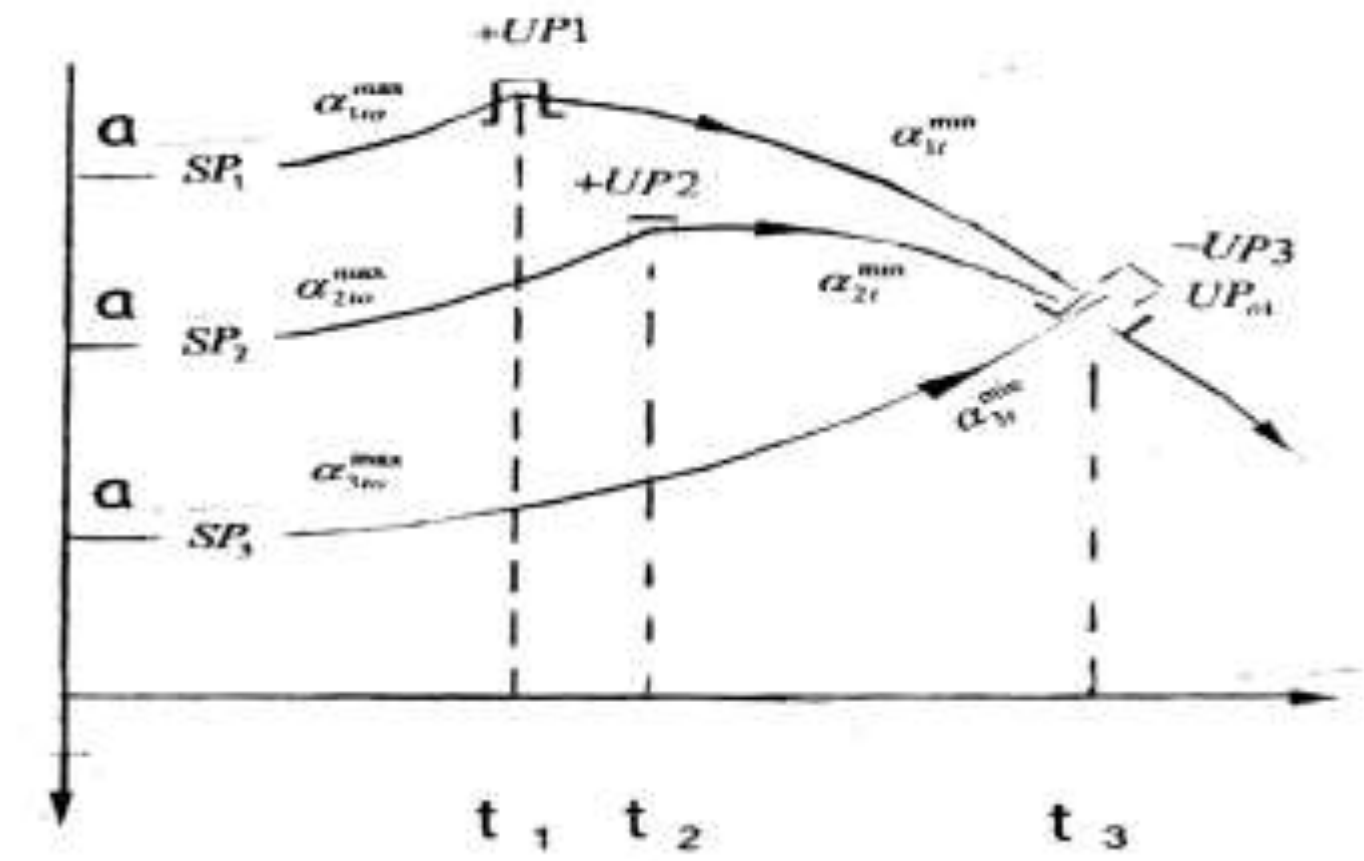

Fig.5. Illustrative dynamics of assembling units $+UP1, +UP2, -UP3$ on the space-time trajectory and adjoining them to $UP_{o1}$ knot along the sections of space-time trajectory $SP_1, SP_2\ SP_3$ (Fig.4) at changing information speeds from $\alpha_{1to}^{\max}$, $\alpha_{2to}^{\max}$, $\alpha_{3to}^{\max}$ to $\alpha_{1t}^{\min}$, $\alpha_{2t}^{\min}$, $\alpha_{3t}^{\min}$ accordingly; $\mathbf{a}$ is dynamic information invariant of an impulse.

The free Information binds the multiple diverse Bits in *collective* Information time-space macrotrajectories of Information macrodynamics [47] (Figs. 3, 5) at level one.

The observing Information moves the macroprocess through the rotation which depends on forming the entropy gradient (as a potential Coriolis force).

A minimum of three rotating Bits join in a triplet unit (UP) which measures macroprocess invariant Information $\mathbf{a}_{io}(\gamma_{io})$ (at level two).

The parameter of Information dynamics $\gamma_{io} = \beta_{io} / \alpha_{io}$ connects the UP imaginary entropy part $\beta_{io}$ with the forming UP real part $\alpha_{io}$. The formation begins with speed $c_{ev}$, transiting entropy speed $\beta_{io}$ to real $\alpha_{io}$ while forming dynamic invariant $\mathbf{a}_i = \alpha_{io} t_{io}$ at $\beta_{io} = c_{ev}$.

With information speed $c_{ev}$, the entangled entropy volume transits through the gap.

When the unit is complete information $\mathbf{a}_i$, the imaginary speed will turn to zero.

This requirement connects $\mathbf{a}_i$ and $\gamma_{io}$.

The entropy volume, rotating with speed $c_{ev}$, is measuring the frequency of imaginary speed $\beta_{oi} \rightarrow f_{io}$. When the imaginary speed turns to zero, the frequency $f_{io}$ will periodically appear as the bleaching signal emanating from the microprocess.

**Comments.** Within the microprocess reversible times ($+t = -t$), the ratio of current imaginary entropy increments to real ones follows from the equations for opposite entropies:
$S_{-}(t) / S_{+}(t) = \beta_{it} / \alpha_{it} = \gamma_i = [1 + \mathrm{jtg}(-t\ )] / [1 - \mathrm{jtg}(+t\ )] = 1$.

This means that when the entangled conjugated entropies are equal, the frequency of each "bleaching signal" from the microprocess, depending on ratio $\gamma_i$, remains invariant.

*The bleaching frequency signals the approach of the end of the microprocess.* •

Each forming UP carries the frequency of its free Information $f_{io} = \pi / 3$ in an attracting process that joins the Free Information of triple Bits rotating in the cycle Fig.4a. The cyclic loop harmonizes the equal speeds-Information frequencies in a coherent resonance movement.

The rotating loop is analogous to Efimov's scenario [49-50] of resonance frequencies which give rise to three-unit systems, Fig. 6.

The loop includes the Borromean knot [51] and Borromean ring [52], memorizing two cooperating Bits to third Bits in the triplet knot (at level three).

The dynamic logical loop assembles each triplet tr1, tr2,... in the resonance on Fig. 7.

The resonance movement joins the free Information of triple Bits which carry the minimal Landauer energy. The energy initiates memorizing two cooperating Bits to third Bits in the triplet knot (at level three). Memorizing the three Bits' free Information provides a time-space *asymmetrical barrier* of rising irreversibility between the segments of the observing trajectory.

Defined through the structural parameter of energy, the barrier is an indicator of rising irreversibly at the creation of Information. Multiple segments of the Hamiltonian dynamics emerge between the barriers. (The barriers evolve from the conjugated bridges Fig.4a).

Each barrier memorizes a triplet's Bit, which encodes the triplet logic in the knot code (at level four).

The Information binding the third Bit in the triplet knot provides the stability of the formed UP.

The Information triplet can build a pair of qubits from the observing process's opposite segments under the step-down actions, attracting other qubits within the gap from the complementary segments. Such opposite triple qubits can assemble a triplet Bit (Fig. 4).

That enables forming a UP from the qubits at level four.

The forming UP depends on the Bit's *fitness* for the triple cooperation in the UP. Fitness implies the ability to alter the direction of the Bit's moving speed.

The conjugated Hamiltonian dynamics of the opposite processing segments brings this ability to the segments, attracting with free Information.

Such fitting UP enables binding accumulated information $\mathbf{a}_{io}$.

A variety of the Bits with different moving speeds brings a "mutation" at macrolevel, which changes its fitness for a particular forming UP. Variation of the fitness's for three Bits could produce a different $UP_i$, which the minimax selects. Or, it does not produce any UP if they do not fit its triple self-connection and ending triple cooperation (at level four).

The free Information of the UP knots comprises an irreversible macroprocess at level five.

The macroprocess integrates the UP's Bits in the IPF which encloses the UP's time-space geometry in their Information structure (at level six).

**3.1.11. Emerging Information networks**

1. The $UP_i$ assembles triplet cooperative units $UP_{oi}$ during the attractive macro movement, which cooperates in a time-space hierarchical network (IN) (Fig. 6) (at level one).

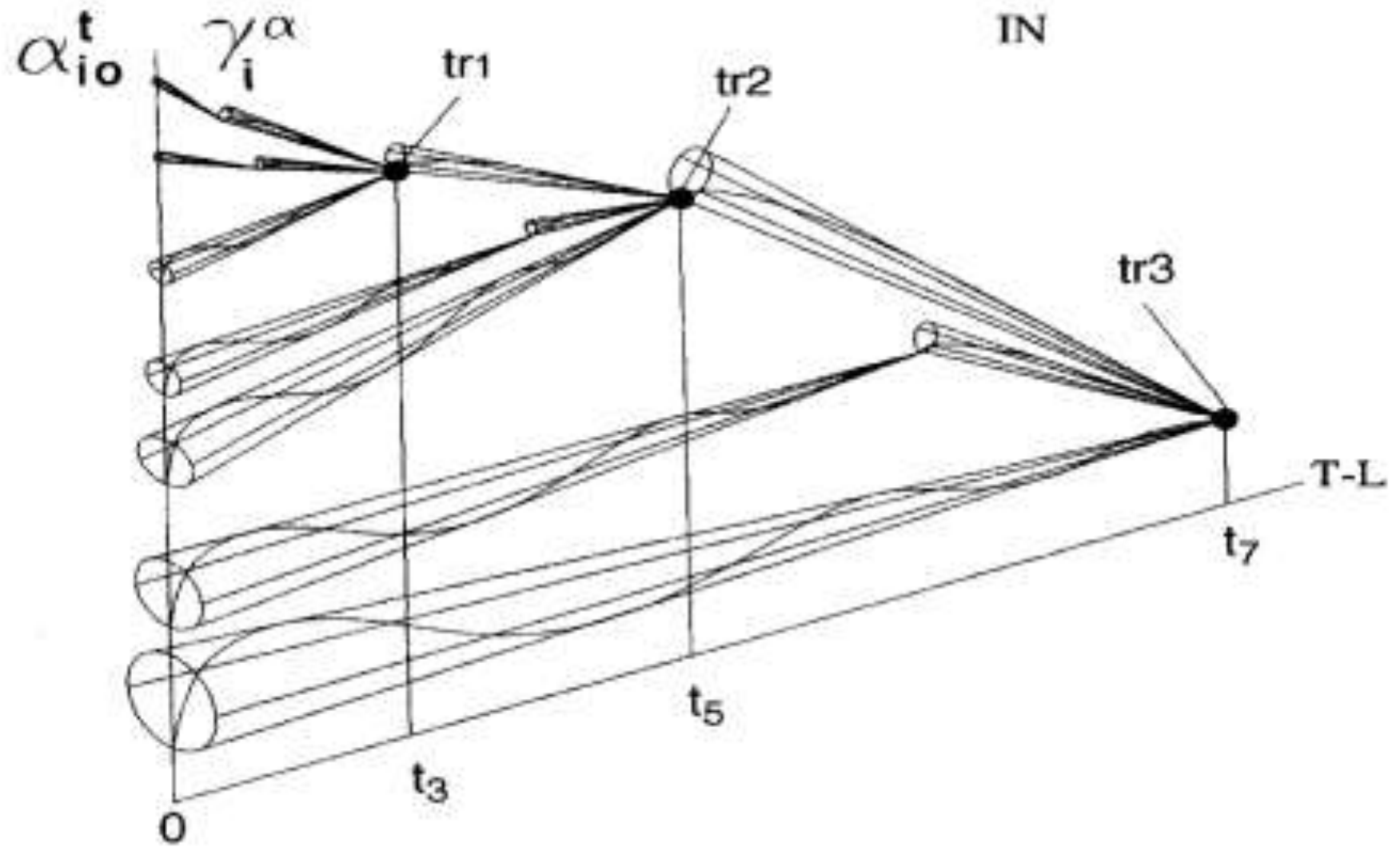

Fig.6. The IN information geometrical structure of hierarchy of the spiral space-time dynamics (Figs.3,5) of triplet nodes (tr1, tr2, tr3, ..); $\{\alpha^t_{io}\}$ is a ranged string of the initial eigenvalues, cooperating on $(t_1, t_2, t_3)$ locations of T-L time-space.

Specifically, Information speeds of primary triplets $UP1, UP2$ connect them to $UP3$, building the triplet's $UP_{01}$ knot by the Information spent on their attractive movement. The knot's attracting free Information forms the rotating loop, which attracts the next forming $UP_{o2}$ and then triplet knot $UP_{o3}$, possibly from different observing process dimensions.

This happens when the triplet fits the cooperative minimax conditions analogous to the forming UP (at level two).

Free Information of the sequentially cooperating $UP_{oi}$ builds a *hierarchical* IN Information structure of the nested knots-nodes (at level three). Each $UP_{oi}$ has a *unique position* in the IN hierarchy, which defines the exact location of the $UP_{oi}$ Information logical structures.

The position depends on each unit's Information measure $\mathbf{a}_{iu}(\gamma_{iu})$, which identifies the IN position dynamic parmeter $\gamma_{iu}$.
The IN node hierarchical level classifies the *quality* of the assembled Information, while the currently ending IN node integrates Information enfolding all previous INs' levels and measures the IN information quality (at stage level three).
Enfolding each second node through free information provides the IN *inner dynamics* with *no* connection to environment.
New Information for each IN delivers the requested node's interactive impulse.
The impulse interactive feedback impact [53] cuts off the entropy of the observing data, bringing their Information to the requested node (at level four).
The appearing new quality of Information builds the IN temporary hierarchy.
The observer IN's hierarchy high level enfolds the Information logic that requests new Information for this running observer's IN.
That extends the IN hierarchy and logic (at level five).
The ending IN triplet integrates these cooperative qualities, holding Information, frequency, and space-time location, which evaluates the quality of the enclosed IN (at level six).
In the IN hierarchy, where the node quality grows with the node hierarchical level, the ending triplet-node has a higher quality compared to other nodes.
The IN nested structure harmonizes its node's quality, enfolded by the ending node, through the resonance frequencies.
A variety of distinctive $UP_k$ can cooperate multiple different networks $IN_{oj}$, which harmonizes its node's specific cooperative qualities with the related frequencies.
The $IN_{oj}$ ending node enfolds the particular quality of Information and time-space positions of each enclosed $UP_k$, depending on the nested node Information $\mathbf{a}_{iuo}(\gamma_{iu})$ (at level seven).
2. The attracting minimax movement assembles each three ending network $IN_{oj}$ nodes $UP_{oi}$ in a newly formed $1_{oi}$ unit which forms a new loop on the higher structural organization level. The loop connects the cooperating speeds of each triple (Fig. 6) in a coherent movement of resonance frequencies.
Specifically, the attractive motion of rotating triple units $(+UP_{o1}, -UP_{o2}, +UP_{o3})$, emerging from opposite (conjugated) ending nodes of the running networks, can cooperate in composite triplet units $-1_o(+UP_{o1}, -UP_{o2}, +UP_{o3})$. The opposite composite triple units $+2_o(-UP_{o5}, +UP_{o6}, -UP_{o7})$ cooperate the same way. A third composite triplet unit $-3_o$, forming analogously, adjoins the first two in the triplet knot. Each of these triplet units enfolds the qualities of the enclosed three networks though a loop of resonance frequencies, which depend on the location of the nodes and Information values.
Thus, this unit's resonance frequency joins the qualities of the IN ending nodes in a new quality which adjoins qualities at the next level network.
The resonance frequency of each IN node is only a part of the spectrum of these network node frequencies.
The highest of these frequencies can adjoin a new IN enclosing the higher Information quality from all three that it composes.

As a result, Information qualities of each IN's composite triplet units $-1_o, +2_o, -3_o$ grow when they join in a rotating cooperative circle, forming new triplet unit $1_{o3}$ since this unit encloses the composite units from the above three which join in the cooperative circle.
Multiple triplets $1_{oj}, j = 3,5,7,...$ sequentially cooperate in a new network $IN_1$, which harmonizes its nodes to higher qualities and enfolds the highest of these qualities in its ending node (at level eight).
3. Each harmonized $IN_1$ forms the *domain* of an observer with its specific qualities and high density of Information enclosing all IN node densities (at level nine).
The sequentially built triplet knots memorize only the IN current composite unit $1_{oj}, j = 3,5,7,...$, while the previous units hold only logical Information.
The sequentially memorized Information implements the IPF concurrent maximum, minimizing the total time of building each composite Information unit.
The minimax leads to the sequentially decreasing ending Information speed of each node, and therefore to decreasing starting Information speeds on the next cooperative unit of the growing structural organization.
The growing levels of structural organization automatically restrict the spectrum of Information frequencies for each self-built IN.
Since each self-built IN has a limited number of cooperating triplets and the IN nodes, it restricts the ability of each IN ending node for the next cooperation.
The violation of the restriction leads to growing the IN instability with rising chaotic movement. Building each higher level cooperative unit toughens the requirements for the fitness of the variety of primary Bits, units UP, $UP_i$, $-UP_{o1}$, etc., which requires decreasing of their variety (at level ten).
Any forming INs should satisfy invariant relations for ratios of starting Information speeds $\gamma_1^\alpha = \alpha_{io} / \alpha_{i+1o}$ and $\gamma_2^\alpha = \alpha_{i+1o} / \alpha_{i+2o}$, connected by dynamic invariant $\mathbf{a}(\gamma)$.
These speeds bind the ending eigenvalues composing each triplet along the segments of the macrotrajectory (Fig.4).
Transfer occurs from one cone's trajectory to another one located on the cone's base (Fig.3), where each location satisfies the extreme condition for entropy–Information.
The sequential transfer requires the rotation of each spiral on the space angle to adjoin the next optimal trajectory and relocate it in cooperation (Fig. 6).
The space-time trajectories, rotating on the cones and cooperating in the triplet, shapes its geometrical structure (Figs.5, 7) evolving during each triplet formation with growing parameter $k$.

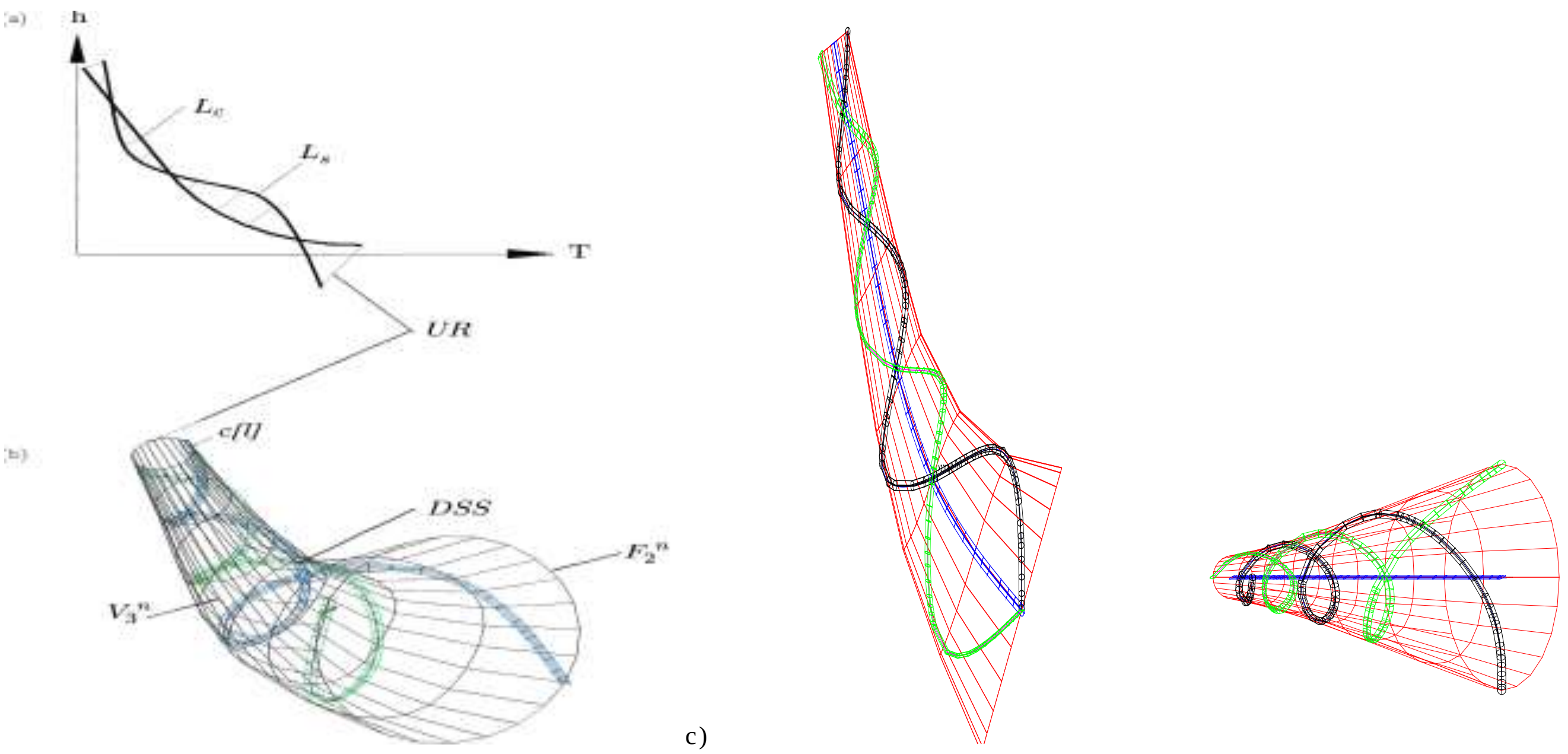

Fig. 7. Simulation of the double spiral cone's structure (DSS) with the cells (c[l]), arising along the switching control line Lc (a); with a surface $F_2^n$ of uncertainty zone (UR) (b), surrounding the Lc-hyperbola in the form of the Ls-line, which in the space geometry enfolds a volume $V_3^n$(b,c).

Both information dynamics and its space structure evolve concurrently, producing each other (Fig. 7).

Each IN triplet accumulates three Bits enfolded in its knot, which forms the IN node.
The nested nodes enclose Information logic enfolded in an ending IN knot.
This ending triplet in every network contains the maximum amount of free Information.
The INs can be self-connected through the attraction of their final triplet's Information logic.
The attracting Information logic of multiple moving INs sequentially equalizes their ending speeds, whose frequencies cohere in resonance assembling the joint INs logic.
Each forming IN emerges with the logic of assembling triplets, whose knot memorizes and encodes the triplet code of the triple logic.
The code of multiple IN holds geometrical double spiral structure (DSS), Fig. 7 is enfolding each triple Informational knot's code.
Each IN ending knot encloses the cell that condenses its local DSS code. The double structure of the conjugated segment builds Hamiltonian dynamics, becoming irreversible at composing the triplet knot on a bridge between segments (Fig. 4a), and then encoding the bridge in the knot-barrier code.
Since each Bit of the knot code holds energy, allowing the multiple IN knot-code physically organizes their local codes in coding the Information structure of Information Observer.
The Eqs. of rotating time-space trajectory on the cones (Figs.3, 7) and the space volume determine the Observer geometry, generated by the Interactive Integrated Information Dynamics (IIID)[48] .

The IN scale parameter $\{\gamma_i^\alpha\}$ [15] identifies the rotating velocity and cooperating volumes of each knot, which are transferred to the next triplet.
The multiple IN geometry structures the Observer's Information geometry by a manifold of the cellular DSS (Fig. 8).

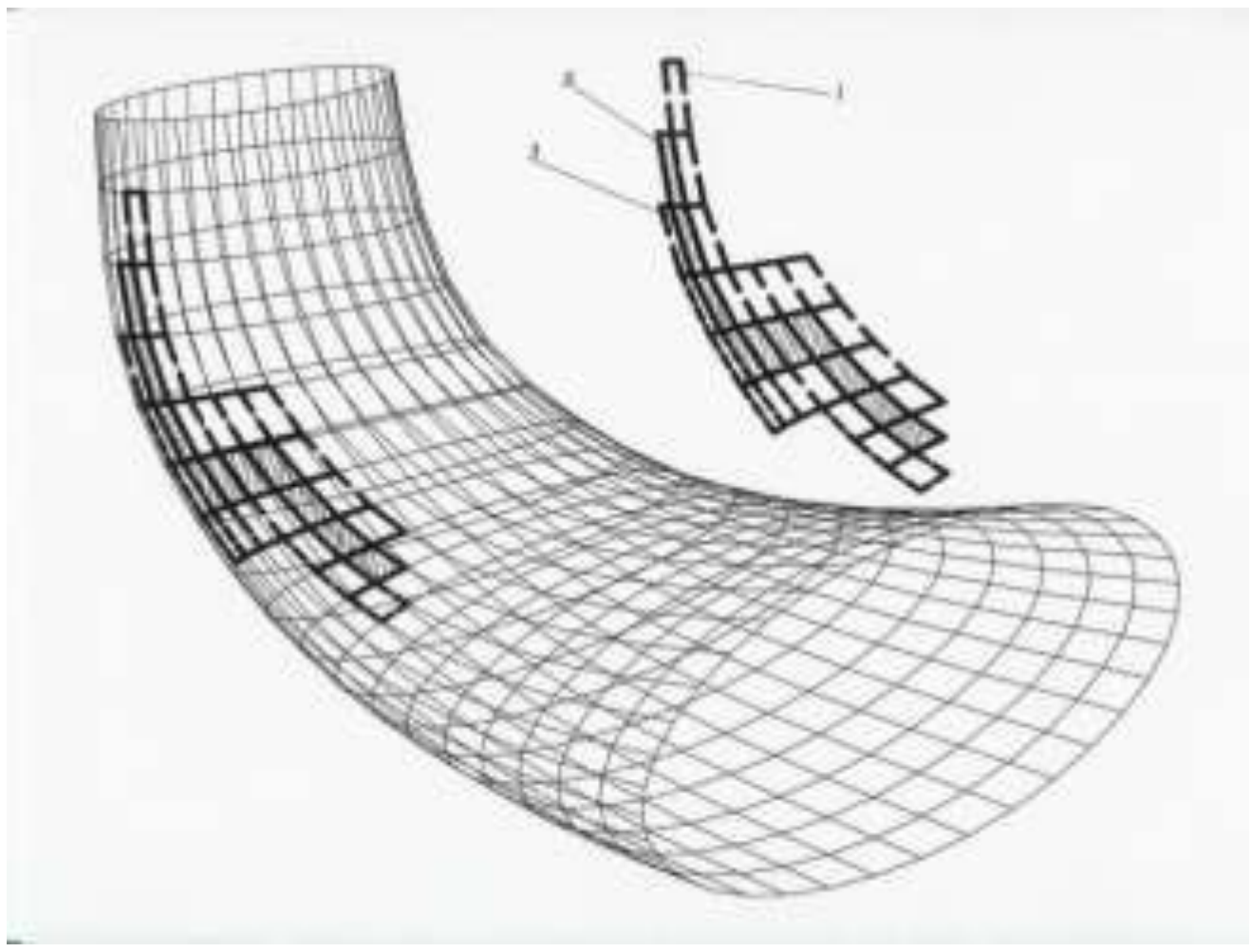

Fig. 8. Structure of the cellular geometry, formed by the cells of the DSS triplet's code, with a portion of the surface cells (1-2-3), illustrating the Observer space formation.
This structure geometry integrates information contributions modelling on Figs.3, 7.

The geometry composes "anatomy" of multiple Bits' structures, where each following triplet's knot creates a *new class* of Information Bits, which assemble the units, distinct from the first-class Bits. In each following class, the Bit grows the density of the enclosed information, geometrical density, and the curvature.
*Therefore, the information emerging from the interactive observation builds the Observer's hierarchical reality.*

### 3.2. Statistical micro-thermodynamics

An impulse ensemble describe statistical micro-thermodynamics [41] applying C. Jarzynski Eq. (JE) [29] to the evolving microprocess.
Results [41,42] connect the JE thermodynamic energy' measure with this process' information measure. Such energy creates Bit after the space forms.
This developing microprocess presents a Stochastic Quantum Process (SCP) with evolving thermodynamics and a path to Information Macrodynamics [15,16].
Applying the Jarzynski equality (JE) of the irreversible thermodynamic transition to conversion energy in Information, and using results of its experimental verification [43], lead to the JE in form

$$e^{\Delta F/k_B\theta} - < e^{W/k_B\theta} >= \gamma. \qquad (3.45)$$

Here $\Delta F$ is increment of free energy needed to produce energy $W, \gamma$ is parameter of the verification, which defines sum of probabilities $P_\Sigma \Rightarrow \gamma$ that an inverse trajectory is observed in the experiments.
At $\gamma = 1$, the JE satisfies exactly:

$$e^{\Delta F/k_B\theta} - < e^{W/k_B\theta} >= 1, P_\Sigma = 1. \qquad (3.45a)$$

A thermodynamic process, satisfying the JE for all its states sequence, evolves irreversibly in the microprocess enables generation qubits or Bit.
The quantity of information $I_{\delta t}$ at a curved cut has equivalent of free energy $\Delta F = \Delta F_{\delta t}$ to produce energy $W_t$ during a fixed time $\delta_t$, satisfying relation

$$\Delta F_{\delta t} = k_B \theta I_{\delta t}. \qquad (3.45b)$$

Average exponential energy $< \exp W / k_b \theta >$, collected during natural interactions at interval $\delta T_t \cup \delta_t$, measures the EF integral increment $\Delta S_{\delta t}$ :

$< \exp W / k_b \theta >= \exp \Delta S_{\delta t}$. (3.45c)

Since EF counts also curved observing probabilistic impulses, the average energy includes the curved impulses. Erasing that entropy by the natural source' high quality energy brings a non-random information $I_{\delta t}$ measures of the curved impulse during $\delta_t < \delta T_t$ .

Influx of impulse energy $\Delta F_{\delta t}$ at $\delta_t$ enables converting entropy $\Delta S_{\delta t}$ to equivalent information $I_{\delta t}$ .

Substituting $I_{\delta t}$ from (3.45b) and $\Delta S_{\delta t}$ from (3.45c) to the JE (3.45a):

$\exp I_{\delta t} - \exp \Delta S_{\delta t} = 1$, (3.45d)

brings the equivalence of JE in both formulas for transition the entropy to information at probability $P_\Sigma = 1$.

During the thermodynamic process, a free energy compensates the entropy dissipative energy.

By emerging real time on cutting intervals $\delta_t$ a certain logic with $I_{\delta t}$ appears with probability $P_\Sigma = 1$, which brings entropy $\Delta S_{\delta t} = 0$ ; that relates to $\ln(P_\Sigma = 1) \Rightarrow \Delta S_{\delta t} = 0$ .

Substitution these in the previous equation leads to $\exp I_{\delta t} - 1 = 1, I_{\delta t} = \ln 2$ which is equivalent to free energy $\Delta F_{\delta t} = k_B \theta \ln 2$.

Thermodynamic conversion of entropy to information Bit occurs during a probability approaching one when the entropy jumps from ln2 to zero, while related information jumps from zero to ln2.

While the first jump starts the microprocess with its thermodynamics, the second opposite jump occurs when the probability approaches to one at potential entropy conversion to Bit.

*It confirms our initial statement about connection the entropy and information and condition of their conversion.*

The equivalence of JE in both formulas for the transition to information requires invariant $I_{\delta t} = 1 Bit$, the unit of information per impulse to compensate for the Maxwell Demon energy at the time interval of transition.

Thus, to satisfy the MD, the information produced by each impulse time interval should be invariant, holding constant unit (Bit, Nat) in $I_{\delta t}$ .

It confirms that the impulse extreme minimax principle (EP) satisfies the JE for impulse information transition, or vice versa. Each impulse time interval enables encoding invariant unit of information.

Or, the EP follows from the JE in the physical process whose interactive time interval is an equivalent of the impulse information cutting from the correlation carrying the above energy.

The cutting correlation's time intervals hold the information equivalent of this energy, and any real time interval of interaction brings the entropy equivalent of energy $\Delta F_{\delta t}$ which compensates for the MD while producing information during the interaction.

In interactive random process, whose sequence of cuts satisfy the EP, each impulse encodes the cutting correlation, and all information of the process cutoff correlations encodes the information process fulfilling the minimax law applied to any impulse independent on size and curvature.

Moreover, with probability $P_\Sigma = 1$, the inverse conjugated trajectories of the interacting impulses in the microprocess are part of the natural interacting process. That exactly satisfy the JE initial conditions [43, 44].

This developing microprocess presents a Stochastic Quantum Process (SCP) with evolving thermodynamics.

**3.3. A bridge connecting physical results with our approach** [24, 25,and 35].
Many years have passed since Schrödinger introduced his equation of Quantum Mechanics as a new physical microscopic theory of interacting particles. But, until now, the origin of the connection between Quantum Mechanics and classical physics has not been scientifically established. That must include linking a wave function to a probabilistic field, and connecting Quantum Mechanics to Quantum Information Theory. Results [25,35] have shown that a "theory of direct inter-particle interaction, associated with a particle acting upon itself, derives from the motion of a system of charged particles under the influence of electromagnetic forces." However, the inter-particle interaction in the electrodynamic Maxwell field must deal with the problem that action and symmetric (adjunct) reaction should merge. Satisfaction of this requirement allows Maxwell's equations to be connected with the equations for atomic particles using the variation principle for total energy as the equivalent of a conservation law for such adjunct interactions. Solution of the obtained equation leads to a *discrete* action crossing reaction.

Reference [35] obtains Schrödinger's equation in quantum mechanics from Maxwell equations. The equations for energy, momentum, frequency and wavelength of the electromagnetic wave in the atom are derived using the model of the atom by analogy with a transmission line. The balance of electromagnetic energy in the atom satisfies the structural constant for the $so=$ 8.27756. This constant connects to the physical structure constant $1/h_{\alpha}^{o*}\cong 137.036$ (the updated value) by relation $so=(1/2h_{\alpha}^{o*})^{1/2}$.
The results shown in [15] identify a *bridge between minimal uncertainty and a certainty* measured by the entropy invariant $S_{\mp a}^{*}=2h_{\alpha}^{o}$ which enables the creation of an *initial Information macrounit*—a triplet with probability $p_{\pm a}=\exp(-2h_{\alpha}^{o})=0.98555075021\rightarrow 1$ approximating the certainty.
This is the *bridge between micro- and macroprocesses* emerging along the path of observing the impulse interactions from maximal uncertainty to Information certainty.
The invariant connects this microprocess, which arises at the merge of interactive action and reaction, with the motion of the interactive adjunct charged particle in a Maxwell field. This proves the requirement for the *discrete action merging reaction*, which leads to impulse interaction rising the microprocess. Since the merging microprocess emanates for random field, it indicates that equations of the electromagnetic wave in the atom also originate in a random field. Moreover, the Schrödinger equation, describing the microprocess, emerges from the *initial random impulses* of the merging actions and reactions, while references [24, 25] and [35] have studied the *deterministic* processes. The invariant constant also binds the emerging micro-macroprocess with Maxwell equations extended to an equation of the interacting atom particles. In addition, the extended model of the atoms, covering three of the four fundamental interactions (electro-magnetic, weak, and strong interactions), allows the *Information description*, which confirms "It from Bit." The merging impulses 1–0 and 1–0 also explain the *creation of qubit* |01 and |10 *in the emerging microprocess* during the entanglement [37] •

Study [38] requires "that gravity, just as electromagnetism in Wheeler-Feynman's time symmetric electrodynamics, also be an 'adjunct field' instead of an independent entity."
In [15] we calculate a weak Information force analogy with the gravitational force. •

Reference [39] has revealed that "the entangling space-time works just like a quantum error-correcting code, protecting information in jittery qubits to store it not in individual qubits, but in patterns of entanglement among many," starting with a triple •
*Finally we show that microprocess emerges from the classical markov process under the interactive observation, and returns to classical information through the observing Bits.*

**3.4.Selection Information, and structuring a selective observer**

1. Forming an Information dynamic cooperative requires increasing cooperative Information force between the potential cooperating triplets:

$$X_{ik}^{Im} = -\frac{\delta I_{ik}^{m}}{\delta l_{ik}^{\alpha}} = \mathbf{a}_{oi}(\gamma)(\gamma_{ik}^{m} - 1) \qquad (4.1)$$

where current IN' triplet $m_i$, currying Information $\delta I_{ik}^{m} = \mathbf{a}_i + \mathbf{a}_{oi}^2 \cong \mathbf{a}_{oi}$, attracts $m_k$-triplet, depending on Information $\mathbf{a}_{oi}(\gamma)$ with IN invariant parameter $\gamma_{ik}^{m}$ and on relative distance

$$(t_i^m - t_k^m)/t_i^m = (l_i^m - l_k^m)/l_i^m = \partial l_{ik}^{m*}. \qquad (4.2)$$

That cooperative force measures the cooperative attraction between these triplets in Nats (Bits).

2.The cooperative force, relative to the Information of the first IN triplet, measures

$$X_{1k}^{Im1} = (\gamma_{1k}^{m} - 1). \qquad (4.3)$$

The relative cooperative Information force between the IN first and second triplets:

$$X_{12}^{\alpha} \geq [\gamma_{12}^{\alpha} - 1], \qquad (4.4)$$

at limited values $\gamma_{12}^{\alpha} \to (4.48 - 3.45)$. That restricts the related cooperative forces by inequality

$$X_{12}^{\alpha} \geq (3.48 - 2.45). \qquad (4.5)$$

The quantity of Information, needed to provide this Information force, is

$$I_{12}(X_{12}^{\alpha}) = X_{12}^{\alpha}\mathbf{a}_o(\gamma_{12}^{\alpha}), \qquad (4.6)$$

where $\mathbf{a}_o(\gamma_{12}^{\alpha})$ is an invariant evaluating quantity of Information concentrated in a selected triplet measure $\mathbf{a}_o(\gamma_{12}^{\alpha}) \cong 1bit$ at $\gamma_{12}^{\alpha} = \gamma_{1o}^{\alpha}$.

It follows that

$$I_{12}(X_{12}^{\alpha}) \geq (3.48 - 2.45)bits. \qquad (4.7)$$

3.The invariant quantities $\mathbf{a}_o(\gamma_{io} \to 0)$, $\mathbf{a}(\gamma_{io} \to 0)$ provide *maximal* cooperative force $X_{12}^{\alpha m} \cong 3.48$.

The minimal quantity of Information needed to form the very first triplet estimates invariants

$$I_{o1} \cong 0.75 Nats \cong 1 bits. \qquad (4.8)$$

Therefore, the total Information needed to start adjoining the next triplet to the IN, estimates

$$I_{o12} = (I_{o1} + I_{12}(X_{12}^{\alpha})) \geq (4.48 - 3.45)bit, \qquad (4.9)$$

which supports the node cooperation and can initiate the IN feedback.

That Information equals or exceeds the Information of the IN current node, requesting sequential cooperation with the next triplet.

This Information should deliver an observer to select the requesting unit.

The minimal triplet's node force $X_{1m}^{\alpha} = 2.45$ depends on the ratio of the starting Information speeds of the nearest nodes, which determines the force scale factor $\gamma_{m1}^{\alpha} = 3.45$ satisfying the minimax.

The observer, satisfying both minimal Information $I_{o1}$ and $I_{12}(X_{12}^{\alpha})$, delivers the total Information $(4.48 - 3.45)bit$. We call it a *minimal* selective observer.

It includes the feedback, carrying the needed Information for a triplet's node.

These limitations are the observer boundaries of the *admissible* Information spectrum, which depends on the related scale factors, applied to a multi-dimensional selective observer.

At the satisfaction of a cooperative condition (4.9), each observed Information speed, delivering the required frequency of Information spectrum to an existing IN, enables creating the next IN's level of the triplet's hierarchy.
This leads to two conditions: *necessary*-for creating a triplet with the required Information quality, and *sufficient* –to provide a cooperative force, needed to adjoin this triplet with an observer's IN.
At the satisfaction of the sufficient condition, each next Information unit joins a sequence of the triplet's Information structures forming the IN. This IN progressively increases the Information bound in each following triplet. The ending IN triplet's node conserves all IN Information. The node *location* in the IN spatial-temporal hierarchy determines the *quality* of the Information bound in the IN node, which depends on the node's enclosed Information density. When acquisition of Information brings new $\gamma_{ik}^{\alpha} \cong 4.81$, the related cooperative force $X_{1k}^{\alpha} = 3.48$ enables transfer to the next cooperative level, extending the IN. Both above conditions should satisfy the observer's ability to *select* Information of growing quality.
The minimal selective observer, satisfying only the necessary condition, we call an *objective* observer. A selective *subjective* observer satisfies ssufficient conditions, limiting delivered Information to 4.48 bit.

**3.5. The Information mechanism of selection and structuring a multiple selective observer**
Suppose an external observer requires its evolving IN by increasing quantity and quality of Information which could deliver the current observer's IN ending node emanating the requested information. The external observer is assumed satisfying the VP information law.
Let us evaluate the interactive information impact on an external observer, carrying its own IN information with $m$ triplets, from the current observer's IN carrying information to form a requested new $m+1$ triplet.
Such a request carries the free information $\mathbf{a}_{m+1} \sim 0.25 Nat$ with information speed $\alpha_{m+1}^{t}$ determined by the requested IN node, which encloses information density

$$\gamma_{m+1}^{\alpha} = (\gamma_{m=1}^{\alpha})(\gamma_{13}^{\alpha})^{m}, \alpha_{1o} / \alpha_{m}^{t} = \gamma_{m+1}^{\alpha} , \qquad (5.1)$$

where $\alpha_{1o}$ is the information speed of a dynamic processs on a segment of the requested IN first triplet.
Ratio $\alpha_{1o}$ to speed its third segment: $\alpha_{1o} / \alpha_{3o} \cong 3.45 = \gamma_{13}^{\alpha}$ determines $\gamma_{m=1}^{\alpha} = \gamma_{13}^{\alpha}$ which identifies the scale factor for cooperating $m$ triplets.Attaching the $m+1$ triplet to the current IN' $m$ triplets extends $\gamma_{m=1}^{\alpha}$ to $\gamma_{m+1}^{\alpha}$.
Parameter $\gamma_{m+1}^{\alpha} = (\gamma_{13}^{\alpha})^{m+1}$ identifies the scale factor for $(m+1)th$ triplet requesting IN's node density (5.1) allowing cooperation the $m+1$ triplet. Cooperation of minimum two triplet forms minimal subjective observer.
The requested density $\gamma_{m+1}^{\alpha}$ requires relative information frequency

$$f_{1m+1} = 1/3(\gamma_{m+1}^{\alpha}) . \qquad (5.2)$$

Any increasing frequency has a tendency of growing $\gamma_{m+1}^{\alpha}$, thereby increasing the information quality of the requesting nodes, which is extending this IN.
That quality for these nodes distance delivers information forces (4.1), which automatically overcome the single selective observer limitations and extend the IN.
Minimal IN with a single ending triplet node and potential speed of attraction $c_{ev}$ (Sec.3.3.10) requests its Information density with speed:

$$c_{m+1}^{\alpha} = (3.45)^2 \times 0.1444 \times 10^{14} \approx 11.9 \times 0.1444 \times 10^{14} \approx 1.7187 \times 10^{14} Nat / \sec . \qquad (5.3)$$

at $\gamma_{m+1}^{\alpha} = \gamma_{1+1}^{\alpha} = 3.45^2 = 11.9.$ (5.3a)

Adjoining the requested Information $I_{o12}$ (4.9) to the IN node forms two triplets with speed (5.3) requires a time interval for transporting this information:

$$t_m \cong 3.45Nat / 1.7187\times10^{14} Nat / \sec \cong 2\times10^{-14} \sec . \quad (5.4)$$

This time interval approximates the impulse width's time $\delta_{te} \approx 1.6\times10^{-14}$ sec with ratio $t_m / \delta_{te} \cong 1.2546$ .
Comparing this ratio to the related ratio (5.3a), it follows that the time interval increases in ratio $\sim 12$ to both the initial speed of attraction $0.1444\times10^{14} Nat / \sec$, and the impulse density.
The increased impulse information density decreases the impulse time width $\delta_{te}$ to

$$\delta_{tm} = \delta_{te} / 12 \cong 0.1333\times10^{-14} \sec , \quad (5.5)$$

which is less than the communication time (5.4) in ~15 times.
Impulse with time width $\delta_{tm}$ (5.5) enables transferring the requested information to and from the observing external process where it interacts through 15 such probing impulses.
This information action should deliver the step-down cut of external impulse that requires quantity Information $0.25Nat$, the same as the Information which carries the requested free information.
The communication time (5.4) should deliver to the node the needed frequency-density (5.2) from the observing external process.
This time also determines the density-frequency which limits the interval of the sending probing impulses.
That requires an increase in the initial impulse's attracting information density

$$i_{od} = 0.25Nat / \delta_{te} = 0.25Nat / 1.6\times10^{-14} \sec = 0.15\times10^{14} Nat / \sec \quad (5.6)$$

in $\sim 12$ times up to

$$i_{md} \cong 1.8\times10^{14} Nat / \sec . \quad (5.7)$$

The observer time of inner communication between the IN initial eigenvalue $\alpha_1$ of its first triplet and $m+1$ eigenvalue $\alpha_{m+1}$, where the requested information is transferred, depends on the ratio
$(\gamma_{13}^{\alpha})^{m+1} = \gamma_{m+1}^{\alpha} = \alpha_1 / \alpha_{m+1} = M_{\tau}$ .
Preserving the information invariant for each triplet including the $m+1$ eigenvalue:

$$\mathbf{a}_{om} = \alpha_1 \tau_1 = \alpha_{m+1} \tau_{m+1} \quad (5.8)$$

leads to relation $M_{\tau} = \tau_{m+1} / \tau_{1o}$ and to

$$\Lambda_{M\tau} = M_{\tau} - 1 = (\tau_{m+1} - \tau_{1o}) / \tau_{m+1} = (\gamma_{12}^{\alpha})^{m+1} - 1 = inv \quad (5.9)$$

which determines the invariant relative interval of inner communication (5.9) defining the observer *time scale*.
For $\gamma_{m+1}^{\alpha} = \gamma_{1+1}^{\alpha} = 3.45^2$ the time scale is

$$\Lambda_{M\tau} \cong 11. \quad (5.10)$$

Comments.
For the minimal subjective observer with $M_{\tau} = \tau_{m+1} / \tau_{1o} = 3.45^2 \cong 12$, for example, each 12 hours of external communication squeezes to one hour of internal communicational.
Thus, to have uninterrupted intercommunication with external information, this observer should open the intercommunication each 12 hours.
Hence, to coordinate the inner and external time course, the observer must periodically switch in a 12/1 rhythm.
That implies memorizing a code of this rhythm on the knot of the observer IN's second triplet, which the cooperating IN automatically includes.

Then, each one hour of inner communication will request external information (4.9) which needs $\delta_{te} \approx 1.6\times10^{-14}$ sec for observing an external observer's IN.

Therefore, such an observer should potentially have $N_{\tau} = 3600/1.6\times10^{-14} = 2.25\times10^{17}$ single communicating networks. The last one, which memorized the code, will automatically switch to external communication, requesting needed information.

Consequently, *the rhythm, coordinating the observer's inners and external time course, is a part of the observer information regularities*, even for an elementary subjective observer. (The $N_{\tau}$ could model a number of communicating-interacting pairs of particles and/or cells). •

### 3.5.1.The time of inner communication and threshold separating subjective and objective observers

The growing Information density increases both the quality of Information enfolding in the observer current IN and the time of delivering this Information.

Thus, each observer posseses *the time of inner communication,* depending on the requested Information and the *time scale,* depending on the *density* of accumulated (bound) Information.

Supppose that formation of the IN new node requires $k$ cutting multiplicative information actions.

The time interval of such cuts $\delta^t_{eik}$ will depend of the IN time scale, increasing proportionally: $\delta^t_{eik} = (\Lambda_{M\tau})^k \times \delta^t_{eio}$, which for $k=10$ increases initial $\delta^t_{eio} \cong 0.2\times10^{-15}$ sec in $11^{10}$ times up to $\delta^t_{eik} \cong 2.6\times10^{10} \times 2\times10^{-15}\sec = 5.2\times10^{-5} 2.6\times10^{9}\sec$. (5.11)

This is time of the observer's external communication, which in $2.6\times10^{9}$ times more than the related time of inner communication (5.4), counted for the IN single node.

As the node's number grows, the estimated time of inner communication increases by multiplying the node number on the invariant time scale (5.10).

Thus, free Information, carrying attracting Information force $X^{\alpha m}_{12} \cong 3.48$ with quantity of the force Information $I_{o12}$ (4.9) delivers a related quality to the forming IN node with density (5.1).

The delivered Information enables forming new IN triplet with $\mathbf{a}_o(\gamma^{\alpha}_{12}) \cong 1bit \cong \ln 2 Nat$, which should be attached to the current IN.

The impulse, carrying this triplet, interacts with the existing IN node through the impact, which provides Information $0.25Nat$, the same as that at the interaction with an observer's external process.

The relative Information effect of the impact estimates ratio $\ln 2/0.25 \approx 3$.

The attracting Information $0.231Nat$, which carries the triplet Bit, brings total $(3\ln 2 + 0.231)Nat \cong 3.573bit$.

That increase compensates for the current IN triplet node's requesting $I_{12}(X^{\alpha}_{12}) \geq (3.48 - 2.45)bits$.

This is the minimal *threshold* for building an elementary triplet, which enables attracting and delivering the requested Information to the IN node.

The selective observer can choose this node with the needed frequency of the probing impulses.

The double triplets are elementary selective objective observers, which acquire the requested IN level's quality of Information.

That quality evaluates *the IN level of the distinctive cooperative information*, building the sequential IN nodes.

The identified Information threshold *separates* subjective and objective observers.

A selective observer, working as observer-participator, builds its first triplet's structural unit with free Information whose attraction arranges the IN hierarchy of growing quality Information, automatically selecting the maximal quality observer.
Note that such a selective process, emerging *within* a macrodynamic process, is distinct from Darwinian Natural Selective, where "only the organisms best adapted to their environment tend to survive and transmit their genetic characters in increasing numbers to succeeding generations while those less adapted tend to be eliminated."

**3.6.The IN limited time-scale, speed of cooperation, and dimension of cooperative process**

Minimal admissible time interval of impulse acting on observing process is limited by $\delta^{o}_{t\min} \cong \mathbf{a}_{io}\hat{h} \approx 0.391143\times 10^{-15}\sec$.

The minimal wide of the internal impulse $\delta_{te} \approx 1.6\times 10^{-14}\sec$ limits the ratio $\delta_{te}/\delta^{o}_{t\min} \cong 41$, that evaluates the limited IN time scale (5.10) and scale ratio $\gamma^{\alpha}_{m+1} = (\gamma^{\alpha}_{12})^{m+1} = 3.45^{m+1}$.

For $m+1=3$ triplets, that ratio brings the IN time scale

$$\Lambda_{M\tau} = 3.45^{m+1} - 1 = 40.06 \text{, or } 3.45^{m+1} = 41.06, \qquad (6.1)$$

which limits $m+1=3$ by $m=2$.

It forms a *minimal subjective observer* whose IN binds two triplets enfolding $n=5$ process dimensions.

Defining the IN *geometrical* bound scale factor by $F_n = \sqrt{(\gamma^{\alpha}_{m=1})^n}$, we find its value for process dimension $n=5$ and the IN scale ratio $\gamma^{\alpha}_{m=1} = 3.45$ binding these dimensions in the IN.

Thus, $F_{n=5} = \sqrt{(\gamma^{\alpha}_{m1})^5} \cong 22.1$ identifies the geometrical bound scale factor for the minimal subjective observer.

In Physics, a bound stable resonance of three particles has been observed [49, 50] with the predicted scale factor $\cong 22.7$.

If an observer enables condensing the external Information by decreasing the width of its impulse, then the number of the IN enclosed triplet grows.

This number limits a relative cooperative speed of the IN's last triplet $m$ node, which determines the ratio of the triplet maximal cooperative speed $c_{am}$ to the triple node Information cooperative speed $c_{oa}$:

$$c_{am}/c_{oa} = C_{ocm}.$$

The $C_{ocm}$ evaluates invariant relation [41]:

$$C_{ocm} \cong 1/2\mathbf{a}_{io}(\gamma)\mathbf{a}_i^{-1}(\gamma)(\gamma^{\alpha}_{i=m}-1)(\gamma^{\alpha}_{i=m})^m . \qquad (6.2)$$

Since each cooperating pair requires $1/2\ln 2 Nat$ of Information, applied during time of cooperation $\delta_{te}$,
the maximal speed of cooperation:

$$c_{ami} = 0.35Nat/\delta_{te} = 0.21875\times 10^{14}Nat/\sec , \qquad (6.2a)$$

is close to the maximal speed killing an entropy bit (at conversion to Information Bit):

$$c_{iv} \cong 0.1444\times 10^{14}Nat/\sec \cong 20\times 10^{13}bit/\sec . \qquad (6.3)$$

(That means the cooperation may follow the forming Information Bit).

The ending node cooperative speed $c_{oci} = 1bit/\sec \cong 0.7Nat/\sec$ predicts the ratio

$$c_{ami}/c_{oci} = C_{cmi} = 0.2062857\times 10^{14}.$$

Applying to (6.2) the numerical values of the optimal minimax invariants leads to:

$$C_{ocm} \cong 1/2\ln 2/0.33\ln 2(2.45)(3.45)^m = 1.515\times 2.45(3.45)^m = 3.77(3.45)^m . \qquad (6.4)$$

Equalizing $C_{cmi} = C_{ocm}$ brings the maximal number of IN node $m_o \cong 23.6$ capable of cooperating with the process dimensions $n_o \approx 48$.
A new IN, starting from such triplet, can produce another triplet, cooperating with other INs, and so on.
That integrates all observed Information in sequentially composed INs (Figs. 6, 7).
Speed $c_{ocn} \approx 10 bit/\text{s}$ at $C_{oc} \cong 10^5$ requests the information frequency $B_f \cong 10^6 bit/\text{s} \approx 10^{-3} Gbit/\text{s}$.
A human being's single IN approximates maximal number of the nodes levels $m_M \cong 7$, which encloses $4 > m_{M1} > 3$ levels of minimal selective subjective observers.
The minimal self-selective observer with a single triplet adds one more triplet, building the minimal IN with two triplet's levels $m_{Min} = 2$.
The observer is able to building multiple Information networks, when each three INs "ending nodes" (of maximal $m_M$) can form a triplet structure which enfolds all three local INs.

That increases the encoded Information in $3m_M$ and then multiplies it on other $m_M : 3m_M m_M$ by building a new IN, starting with this triplet.
This process allows progressively increase both quantity and quality of total encoded Information in $(3m_M m_M) \times (3m_M m_M) \times ....... = (3m_M m_M)^{m_M} = N_m$ times of the initial IN's node Information $\mathbf{a}_{1o} = \ln 2$:

$$I_m = \ln 2 (3m_M m_M)^{m_M} \qquad (6.5)$$

with maximal density

$$I_m^d = \ln 2 (\gamma_{12}^\alpha)^{N_m} . \qquad (6.6)$$

and the time scale

$$\Lambda_{mM\tau} = (\gamma_{12}^\alpha)^{N_m} - 1 . \qquad (6.7)$$

This huge quantity and quality of Information is limited by maximal $m_M$.
The IPF maximal Information available from an external random Information process is limited at the infinite dimension of the process (Sec.1 and [48]).

### 3.7.The Information conditions for self-structuring a multiple selective observer

Satisfaction of the sufficient condition (4.9) in multiple IN's interactions determines a multiple selective observer, whose attracting cooperative Information force grows from

$$X_{12}^\alpha \geq [(\gamma_{12}^\alpha) - 1]$$

to

$$X_{12}^{\alpha N_m} \to (\gamma_{12}^\alpha[\gamma_m])^{N_m} \text{ at } (\gamma_{12}^\alpha[\gamma_m]) \to (\gamma_{12}^\alpha[\gamma_m])^{N_m} , \qquad (7.1)$$

accumulating at $\gamma_m \to 0$ maximal Information

$$I_{12}(X_{12}^{\alpha N_m}) = X_{12}^{\alpha N_m} \mathbf{a}_o [(\gamma_{12}^\alpha)^{N_m}] . \qquad (7.2)$$

The communicating multiple observers send *quality messengers* (qmess) [15], enfolding the sender IN's cooperative force (7.1).
This requires access to other IN observers in order to obtain more Information quality and logic.
The observer-sender generates a collective IN logic of the accessible multiple observers.

Comments.
Observer's Information units interact in an communication information channel, producing randomness, entropy that generates nonrandom Information of the channel errors which may corrupt the sender Information message. •

Each observer's IN memorizes its ending node Information, while the total multilevel hierarchical IN memorizes the Information of the whole hierarchy in the ending node of this hierarchy.
The observer, requesting maximal quality Information by its Information forces, generates probing impulses, which select the needed density-frequency's real Information among imaginary Information of virtual probes.
That brings to the observer-sender all current IN logical Information.
The information mechanism of selection classifies observer on:
Objective with a single triplet; Minimal selective subjective–with double triplets; Selective subjective with IN.
The elementary IN subjective with triple triplets cooperating in a knot;
(Such IN possesses minimal cooperative complexity, Sec.5.4.2).
Selective subjective complex obsever with multilevel hierarchical IN.
The IN Information dynamics enable renovation of the existing IN through the feedback process of exchanging the requested Information with the environment, rebuilding the IN by encoding and re-memorizing recent Information.
Since the whole multiple IN Information is *limited* at a total time of the IN existence, the possibility of the IN self-replication arises (Sec. 8.5).
The observer ability to self-organize each next level of the evolution dynamics, we call the observer *creativity* which limits each observer's integrated Information.
*All described stages and levels comprise the observing path of the evolving dynamics.*
It's seen *the observing path orders the hierarchical multilevel information dynamics.*

## IV. The Observer Emerging Self-Encoding, Cooperative Complexity, Information Mass, Space Curvature, and Self-Replication

### 4.1. Encoding Information units in the IN code-logic, and Observer's computation using this code

The Observer code serves for common external and internal communications, encoding different interactions in a universal Information measure. It conducts cooperative operations both within the Observer and outside.
Different Observers unites a common observation, the collective triplet code building a cooperative Observer.

#### 4.1.1.The code, program, and Information quality of the code-logic

Each triplet unit generates three symbols from three segments of Information dynamics and one impulse-code from the knot binding the triple.
The triplet composes a minimal *logical code* that encodes the observing Information process.
The knot's Free Information, attracting the next observing Bit, forms a duplet which self-cooperates a Bit in the following triplet unit. The knot of that triple attracts another incoming Bit, and so on.
That sequentially creates the nested structure of an Information network (IN) whose nodes-knots concurrently self-encode the observing Information units.
Thus, each three observing Bits self-cooperate a triplet which encodes the knot composing hierarchy of the IN preceding code's nodes.
Each Information unit has its unique position in the time-space Information dynamics, which defines the exact location of each triple code in the IN nodes hierarchy.
Even though the code impulses are similar for each triplet, their time-space locations allow the *discrimination* of each code, its logics, and the distinction of the codes unit's geometry.
The IN code units include digital time intervals, enclosed in the unit code's geometry, which depend on digits of Information collected by the information path functional (IPF).

The shortened process intervals in the IPF, enclosing in the IN, condense the observing Information in the rotating space-time triplets' knots.
The knot's nodes cooperate and integrate the IN Information logical structure.
The code logic emerges from the observing probabilistic and Information causality.
The self-participating Bits generate the Observer's (0-1) probes and the observing logic Information units.
The IPF collects the Information units, while the IN performs logical computing operations using the doublet-triplet code. The operation creates the Observer's program, which is specific for each Observer, and therefore is self-encrypting.
Performed with the entangled memorized Information units, these operations model quantum computation [54-57].
The operations with physical Information macro-units, which cooperate quantum Information units, run the IN modeling of a classical computation.
An Observer, uniting the logic of quantum micro-and macro- Information processes, enables the composition of quantum and/or classical computation on different IN levels.
The program holds a distributed space-time processing generated by the Observer Information dynamics.
Information, emanated from different IN nodes, encloses a distinctive quality measure and logic, which encodes the specific Observer's IN *genetic code*.
The triplets are elementary logical units enclosing the IN genetic code that composes a helix geometrical structure (Fig.4a, 7) analogous to DNA [58-60].
The genetic code can reproduce the encoded system by decoding the IN final node and the specific position of each node within the IN structure.
Finally, the observing interacting process *naturally encodes the impulse interactions with environment.*

**4.2.The emerging Cooperative Complexity at forming a cooperative network of multilevel information units**

The IN nested structure holds Cooperative Complexity (CC), which decreases the initial complexity of collected uncooperative Information units.
The CC measures the *origin* of complexity in the interactive dynamic *process*, cooperating an elementary duplet-triplet, whose Free Information anticipates new Information, requests it, and automatically builds the hierarchical IN [61].
The CC emerges from both the probabilistic and Information logics of observing processes.
Existing complexities measure complexity of already formed complex system and processes [62-66].
The CC is an *attribute* of the process's *cooperative dynamics and its logics.*
Bound information of cooperating units is *potential* source of decreasing the CC, which measures the concealed *Information.*
We study both the complexity of Information Macrodynamic process (MC) and the CC, arising in interactive dynamics of changing Information flows from $I_i$ to $I_k$, moving their shared volume from $V_i$ to $V_k$: $\Delta I_{ik} = I_i - I_k, \Delta V_{ik} = V_i - V_k$.
The MC defines an increment of concentration of Information in the Information flow before and after interaction, measured by the flow's increment per changed Information volume: $MC_{ik} = \Delta I_{ik} / \Delta V_{ik}$.
The flow's increment measures the increment of entropy speeds $\Delta I_{ik} = \partial \Delta S_{ik} / \partial t$ in the form

$$MC_{ik} = (\partial \Delta S_{ik} / \partial t) / \Delta V_{ik} . \quad (2.1)$$

This complexity determines an instant entropy's concentration in this volume: $\frac{\partial \Delta S_{ik}}{\Delta V_{ik} \partial t}$ (the entropy production), which evaluates the specific Information contribution, transferred during the interactive *dynamics* of the Information flows.

Complexity (2.1) is measured *after* the interaction has occurred, assuming that both increments of speeds and volumes are known.

To evaluate complexity arising *during* interactive dynamics, the *Information measure of a differential interactive complexity* $MC_{ik}^{\delta}$ is introduced, *defined by* increment of Information flow $-\frac{\partial \Delta S_{ik}}{\partial t}$ per small volume increment $\delta V_{ik}^{\delta}$ (within the shared volume $\Delta V_{ik}$):

$$MC_{ik}^{\delta} = \frac{\partial H_{ik}}{\partial t} / \frac{\partial \Delta V_{ik}}{\partial t}, \tag{2.2}$$

where (2.2) defines ratio of the speeds, measured by the increments of the Information Hamiltonian $H_{ik} = -\partial \Delta S_{ik} / \partial t$ per increment of volume $\delta V_{ik}^{\delta} = \delta \Delta V_{ik}$ at $\delta = \partial t$:

$$MC_{ik}^{\delta} = \partial H_{ik} / \partial \Delta V_{ik} = \frac{\partial H_{ik}}{\partial t} / \frac{\partial \Delta V_{ik}}{\partial t}.$$

The $MC_{ik}^{\delta}$ automatically includes both $MC_{ik}$ and its increment $\delta MC_{ik}$.

The $MC_{ik}$ measures the differential increment of Information of interactive elements $i, k$, whose current *Information difference* $-\Delta S_{ik}$ per shared volume $\Delta V_{ik}$ *-before joining,* will reduce to increment $-\partial \Delta S_{ik}$ per volume $\delta V_{ik}^{\delta}$ after their cooperation in a macrodynamic process.

Thus, within the IN nested structure, the IN Cooperative Complexity arises from both the $MC$ complexity and differential $MC^{\delta}$ complexity of the Information process.

The Information path functional (IPF) integrates the observing Bits of the Information process.

Applying the VP Information invariants allows the direct measurement of both $MC_{ik}$ and $MC_{ik}^{\delta}$ in the Bits of Information code. Both complexities measure the triplet's dynamics through their eigenvalues, which connect them with related geometrical structure's volume.

Complexity (2.2) of a triplet is defined at the moment $t = t_{i+2}$ of three segments' eigenvalues equalization $\alpha_i = \alpha_{i+1} = \alpha_{i+2}$. That complexity measure is

$$M_{i,i+1,i+2}^{\delta} = 3\dot{\alpha}_{i+2,t} / \dot{V}_{i,i+1,i+2}, \tag{2.3}$$

at $\dot{\alpha}_{i+2,t} |_{t=t_{i+2,t}} = [\alpha_{i+2,o}^2 t_{i+2,t}^2 \exp(\alpha_{i+2,o} t_{i+2,t})(2 - \exp \alpha_{i+2,o} t_{i+2,t})^{-1} - \alpha_{i+2,o}^2 t_{i+2,t}^2 \exp 2(\alpha_{i+2,o} t_{i+2,t})] / t_{i+2,t}^2$

which takes invariant form

$$M_{i,i+1.i+2}^{\delta} = \mathbf{a}_o^2 [exp\mathbf{a}_o (2 - exp\mathbf{a}_o)^{-1} - exp 2\mathbf{a}_o] / t_{i+2,t} = (\mathbf{a}_o \mathbf{a} - \mathbf{a}_o^2 exp\mathbf{a}_o) / t_{i+2,t}, \mathbf{a} = exp\mathbf{a}_o (exp\mathbf{a}_o (2 - exp\mathbf{a}_o)^{-1}. \tag{2.3a}$$

The retive volume in (2.3) holds

$$\dot{V}_{i,i+1,i+2} = \delta V_{i,i+1,i+2} / \delta t, \delta V_{i,i+1,i+2} = V_c \delta t^3, \ \delta V_{i,i+1,i+2} / \delta t = V_c \delta t^2 = V_c t_{i+2,t}^2 \delta t^2 / t_{i+2,t}^2 = V_c t_{i+2,t}^2 \varepsilon(\gamma)^2. \tag{2.3b}$$

Here space area $\varepsilon^2(\gamma)$ is an invariant at fixed $\gamma$, and $V_c = 2\pi c^3 / 3(k\pi)^2 tg\psi^o$ is an invarint volume at angle $\psi^o = \pi / 6$ on the vertex of each cone (Fig.3) at $k=1$, and the invariant speed of rotation $c$ of each cone's spiral which produces angle $\pi / 2$.

Differential complexity $M_m^\delta = M_{i,i+1,i+2}^\delta$. for any joint *m*-th triplet-node is:

$$M_{i,i+1,i+2}^\delta = 3(\mathbf{a}_o\mathbf{a} - \mathbf{a}_o^2 \exp 2\mathbf{a}_o) / V_c t_{i+2,t}^4 \varepsilon(\gamma)^2, \quad (2.4)$$

depending on that time interval.

Applying $M_m^\delta$ to each triple cell (Fig.8) with volume $\delta V_{i,i+1,i+2} = \delta V_m$, formed during time interval $\delta t_{i,i+1,i+2} = \delta t_m$, leads to

$$M_m^\delta \delta t_m = 3\dot{\alpha}_m \delta t_m / \delta V_m = 3\Delta\alpha_m / \delta V_m,$$

where $\Delta\alpha_m$ is increment of Information speed during $\delta t_m$.

The related increment of quantity Information at the same $\delta t_m$ is $\Delta\alpha_m \delta t_m = a_m^\Delta = \dot{\alpha}_m \delta t_m^2$, where

$$a_m^\Delta = (\mathbf{a}_o\mathbf{a} - \mathbf{a}_o^2 \exp(2\mathbf{a}_o))\delta t_m^2 / t_{i+2,t}^2, \delta t_m^2 / t_{m,t}^2 = \varepsilon_m^2, t_{m,t}^2 = t_{i+2,t}. \quad (2.5)$$

Each $3a_m^\Delta$ invariant measures the quantity of Information produced during an interaction of three equal eigenvalues within the area $\varepsilon_m^2$.

Increment of entropy per the cell volume $\delta V_m$, measured by equivalent quantity Information and $M_m^\delta$ during the time $\delta t_m$ holds invariant measure

$$M_m^\delta \delta t_m^2 = M_m^\Delta = 3(\mathbf{a}_o\mathbf{a} - \mathbf{a}_o^2 \exp(2\mathbf{a}_o))\varepsilon_m^2 / \delta V_m. \quad (2.5a)$$

Information $3a_m^\Delta$ binds the three segments in $\varepsilon_m^2$ prior the two impulse attract, and assembles them.

By the moment of interactive assembling $\tau_k^{i+2}$, the three equal eigenvalues have signs $\alpha_{it}(\tau_k^{i+2}) sign\alpha_{it}(\tau_k^{i+2}) = \alpha_{i+1t}(\tau_k^{i+2}) sign\alpha_{i+1t}(\tau_k^{i+2}) = -\alpha_{i+2t}(\tau_k^{i+2}) sign\alpha_{i+2t}(\tau_k^{i+2})$.

The negative eigenvalues $\alpha_{it}(\tau_k^{i+2}) sign\alpha_{it}(\tau_k^{i+2}) = \alpha_{i+1t}(\tau_k^{i+2}) sign\alpha_{i+1t}(\tau_k^{i+2})$ hold the stable dynamics, and positive eigenvalue $-\alpha_{i+2t}(\tau_k^{i+2}) sign\alpha_{i+2t}(\tau_k^{i+2})$ brings unstability.

Their interaction associates with a choatic attraction, which leads to the instability localized within area $\varepsilon_m^2$.

The attraction, delivering Information $2\mathbf{a}_o^2$, *cooperates* three segments by joining them into a single node within area $\varepsilon_m^2$.

Thus, relation (2.5a) measures Cooperative Complexity of the *interactive* three segments, forming *a single node* of *m*-th triplet' cell.

The cooperative node forms its cell within volume $\delta V_m$, where both the eigenvalues' interaction and cooperation takes place. Since quantity Information $2\mathbf{a}_o^2 \cong 1bit$ of the joint segment from *m*-th triplet's node is transferred to a first segment of the following $m+1$-th triplet, the quantity of *binding* Information $3a_m^\Delta$ (in (2.5)) is spending on holding the *m*-th triplet concentrated in volume $\delta V_m$.

Let $M^{\Delta}_{cm} = 3a^{\Delta}_{m}(\gamma) / [\delta V_m / \varepsilon^2_m]$ evaluate the quantity of Information per cell volume $\delta V_m$ related to a cell size area $\varepsilon^2_m$. Then using $M^{\delta}_{cm} = 3\Delta\alpha_m / \Delta V_m$, $M^{\delta}_{cm}\delta t_m = 3\Delta\alpha_m \delta t_m / \Delta V_m$, we have $M^{\Delta}_{cm} = 3a^{\Delta}_{m} / \Delta V_m$ which for each $\Delta V_m$ evaluates $M^{\Delta V}_{cm} = 3a^{\Delta}_{m}$.

At $\gamma = 0.5, \mathbf{a}_o \cong -0.75, \mathbf{a} \cong 0.25$ we get $M^{\Delta V}_{cmN} = 3a^{\Delta}_{m}(\gamma = 0.5) \cong -0.897 Nat$ per cell, or $M^{\Delta V}_{cmb}(\gamma = 0.5) \cong -1.29 bit$ per cell-volume, which each $m$-th node *conserves* during it formation.

This invariant, produced during the considered interaction (that primary binds these segments), measures a *cooperative* effect of the interactions holding the node's *inner Cooperative Complexity*.

This relative Cooperative Complexity does not depend on the IN actual cell volume and the number of nodes that the cell enfolds, since the $M^{\Delta V}_{cm}$ invariant quantity is not transferred along the IN nodes' hierarchy.

Actually, in volume $\delta V_m / \varepsilon^2_m = V_c t^2_m$, at a fixed both invariant $\varepsilon^2_m$ and volume $V_c$, the increment $\delta V_m$ grows as more nodes are assembled. Since that complexity $M^{\Delta V}_{cm} = inv(\gamma)$ for any fixed cell's volume (according to (2.5)) decreases with assembling more cooperating nodes within this volume.

As the size of the cooperative nodes grows, the Cooperative Complexity per its volume decreses in the ratio $M^{\Delta}_{m+1} / M^{\Delta}_{m} = t^2_m / t^2_{m+1} = (\gamma^{\alpha}_{m})^{-2}$ while each following $M^{\Delta}_{m+1}$ enfolds complexity of the previous $M^{\Delta}_{m}$.

The absolute value of the interval $\delta t_m = t_m \varepsilon$ grows as $t_{m+1} / t_m = \gamma^{\alpha}_{2}$ increases, leading to $\delta t_{m+1} / \delta t_m = \gamma^{\alpha}_{2}$ and

$M^{\Delta}_{m} = 3a^{\Delta}_{m} / \delta t_m \delta V_m, \delta V_m = 3V_c \delta t^2_m$, $M^{\Delta}_{m} = a^{\Delta}_{m} / V_c \delta t^3_m = a^{\Delta}_{m} / V_c \varepsilon^3_m t^3_m$, at $M^{\delta}_{m} = a^{\Delta}_{m} / V_c \delta t^4_m = a^{\Delta}_{m} / V_c \varepsilon^4_m t^4_m$.

This confirms the previous relations.

The ratio of the nearest triplet's complexities (2.4) is

$$M^{\delta}_{m+1} / M^{\delta}_{m} = t^4_m / t^4_{m+1} \text{ at } (\mathbf{a}_o \mathbf{a} - \mathbf{a}^2_o \exp(2\mathbf{a}_o)) / V_c \varepsilon(\gamma)^2 = A_M(inv(\gamma)). \quad (2.5b)$$

At $t^4_m / t^4_{m+1} = (\alpha_{m+1} / \alpha_m)^4 = (\gamma_{m+1})^{-4}$, satisfaction of (2.5b) with $\gamma_{m+1} = \gamma_2(\gamma) = inv_o(\gamma)$, brings

$$M^{\delta}_{m+1} / M^{\delta}_{m} = \gamma^{-4}_{m+1}, \quad (2.6)$$

which for $\gamma_2(\gamma = 0.5) = 3.89$ takes values $M^{\delta}_{m+1} / M^{\delta}_{m} \cong 0.00437$.

Complexity $M^{\delta}_{m+1}$, measuring $m+1$ node, also enfolds and condenses the complexity of a previous node.

By the moment $\tau_m$ of the $m$-th triplet's cooperation, its three eigenvalues equalize: $\alpha^m_{3\tau} = \alpha^m_{2\tau} = \alpha^m_{1\tau}$ and, at the *moment* of the triplet's knot formation $\tau_m + o$, the cooperative eigenvalues $\alpha_m$ encloses the joint triplet eigenvalues in a knot:

$$\alpha^m_3(\tau_m + o) = 3\alpha^m_{3\tau} = \alpha_m \,. \quad (2.7)$$

In the IN, $m$-th triplet's first eigenvalue $\alpha^m_{1\tau 1}$ *equals* to last eigenvalue of $(m-1)$-th triplet $\alpha_{m-1} : \alpha^m_{1\tau 1} = \alpha_{m-1}$, where $\alpha^m_{1\tau 1}$ enfolds all three eigenvalues of previous $(m-1)$- triplets.

The $m$-triplet holds ratio of these eigenvalues

$$\alpha^m_{3\tau} / \alpha^m_{1\tau 1} = (\gamma^{\alpha}_{m})^{-1}. \quad (2.7a)$$

Substituting (2.7a) to (2.7) leads to $\alpha_m = 3\alpha^m_{1\tau 1}(\gamma^{\alpha}_{m})^{-1}$, and with $\alpha_{m-1}$ we get ratio

$$\alpha_m / \alpha^m_{m-1} = 3(\gamma^{\alpha}_{m})^{-1}. \quad (2.7b)$$

The sustained cooperation of the IN eigenvalues requires $\gamma_m^{\alpha}(\gamma=0.5)\cong 3.9$, which brings ratio (2.7b) to $\alpha_m / \alpha_{m-1}^m \cong (1.3)^{-1}$.

Decreasing the eigenvalues of the triplets, cooperating along the IN, encloses the increased Information density, which condenses more $MC_{ik}^{\delta}$ complexity.

Specifically, at $M_{m+1}^{\delta} / M_m^{\delta} = (\alpha_{m+1}^4 / \alpha_m^4) / \dot{V}_{m+1} / \dot{V}_m$ and $\dot{V}_{m+1} / \dot{V}_m = \alpha_m^2 / \alpha_{m+1}^2$, $\alpha_{m+1} / \alpha_m = (1/3\gamma_{m+1})^{-1}$, ratio $M_{m+1}^{\delta} / M_m^{\delta} = (\alpha_{m+1}^4 / \alpha_m^4) / \dot{V}_{m+1} / \dot{V}_m = (\alpha_{m+1}^6 / \alpha_m^6) = (1/3\gamma_{m+1})^{-6}$ brings decreasing ratio $M_{m+1}^{\delta} / M_m^{\delta} \cong 0.203$.

Comparing (2.6) to the differrence of relative complexities:

$\Delta M_m^{\delta} / M_m^{\delta} = (M_m^{\delta} - M_{m+1}^{\delta}) / M_m^{\delta} = (1-\gamma_2^4)$,

we get $\Delta M_m^{\delta} / M_m^{\delta} \cong |0.996|$ at $\gamma_2(\gamma=0.5)=3.89$, indicating that the difference decreases insignificantly.

Relative sum of these complexities:

$\Delta M_{m\Sigma}^{\delta} / M_m^{\delta} = (M_m^{\delta} + M_{m+1}^{\delta}) / M_m^{\delta} = (1+\gamma_2^4), \Delta M_{m\Sigma}^{\delta} / M_m^{\delta} \cong 1.0044$ also grows insignificantly.

Comparing these complexities to the complexities of double cooperation within a triplet, we have

$M_{12}^{\delta} / M_1^{\delta} = (\alpha_{12}\delta t_{12} / \delta V_{12}) / (\alpha_1 \delta t_1 / \delta V_1) \cong 2(\alpha_2 / \delta V_{12}) / (\alpha_1 / \delta V_1)$, which at $\delta t_{12} \cong \delta t_1$ and

$\alpha_2 / \alpha_1 = (\gamma_2^{\alpha})^{-1}, \delta V_{12} / \delta V_1 = (\gamma_2^{\alpha})^{-3}$, leads to $M_{12}^{\delta} / M_1^{\delta} \cong 2(\gamma_2^{\alpha})^{-4}$.

For a triplet we have

$$M_{123}^{\delta} / M_1^{\delta} \cong 3(\gamma_3^{\alpha})^{-4}, \qquad (2.8)$$

which at $\gamma_1^{\alpha} = 2.215$, $\gamma_2(\gamma=0.5)=3.89$ brings $M_{12}^{\delta} / M_1^{\delta} \cong 0.083$, $M_{123}^{\delta} / M_1^{\delta} \cong 0.013$. (2.8a)

During a triple cooperation, the complexity decreases more than that in douple cooperation within a triplet.

The $MC_{ik}^{\delta}$ (2.6) between the nearest triplets decreases much faster than it occurs in the cooperation within a triplet.

At the cooperative cooperation, each following nodes' complexity wraps and absorbs complexity of previous node, binding these node units and conserving the bound Information.

The CC complexity cooperative binding effect is a part of the IN *inner dynamics*.

The decrease of the IN Cooperative Complexity indicates that more cooperations have occurred, while, at negative eigenvalues jump, the complexity grows with decoupling nodes and rasing the choitic movement.

The MC for each extremal segment and their ratios determine the following relations:

$M_i^d = \alpha_{it} / \Delta V_{it}, M_{i+1}^d = \alpha_{i+1,t} / \Delta V_{i+1,t}, M_{i+2}^d = \alpha_{i+2,t} / \Delta V_{i+2,t}$, $M_{i+1} / M_i = (\alpha_{i+1,t} / \alpha_{it}) / (\Delta V_{i+1,t} / \Delta V_{it})$,

$M_{i+2} / M_i = (\alpha_{i+2,t} / \alpha_{it}) / (\Delta V_{i+2,t} / \Delta V_{it})$

which at $(\alpha_{i+1,t} / \alpha_{it}) = \gamma_1^{-1}$, $(\alpha_{i+2,t} / \alpha_{it}) = \gamma_2^{-1}$, $(\Delta V_{i+1,t} / \Delta V_{it}) = (1-\gamma_1^3)$, $(\Delta V_{i+2,t} / \Delta V_{it}) = (1-\gamma_2^3)$,

lead to $M_{i+1}^d / M_i^d = \gamma_1^{-1}(1-\gamma_1^3)^{-1}$ and $M_{i+1}^d / M_i^d = \gamma_1^{-1}(1-\gamma_1^3)^{-1}$.

For $m$-th triple we get:

$$(M_i^d + M_{i+1}^d + M_{i+2}^d) / M_i^d = \Delta M_{m\Sigma}^d / M_m^d = 1 + \gamma_1^{-1}(1-\gamma_1^3)^{-1} + \gamma_2^{-1}(1-\gamma_2^3)^{-1}, \qquad (2.8b)$$

and $\Delta M_{m\Sigma}^d / M_m^d \cong 1.05$ at $\gamma_2(\gamma=0.5)=3.89, \gamma_1(\gamma=0.5)=2.215$.

Invariants relaltions (2.3a) bring invariant forms for each $i, i+1, i+2$ triple of these complexities:
$$M_i^d = \mathbf{a} / t_i \Delta V_{it}, M_{i+1}^d = \mathbf{a} / t_{i+1} \Delta V_{i+1,t}, M_{i+2}^d = \mathbf{a} / t_{i+2} \Delta V_{i+2,t}, \qquad (2.9)$$
where each of the volume encloses its invariant Information.

Comparing two ratios of Cooperative Complexities in (2.8a) with the related sum of dynamic complexities (2.8b) brings
$$1+\gamma_1^{-1}(1-\gamma_1^3)^{-1} + \gamma_2^{-1}(1-\gamma_2^3)^{-1} >> 3(\gamma_2)^{-4}, \qquad (2.9a)$$
for which, at $\gamma_2(\gamma = 0.5) = 3.89, \gamma_1(\gamma = 0.5) = 2.215$ ,we obtain inequality $1.05 >> 0.013$.

The results indicate the essential differerence between both types of complexities, where $M_i^\delta$ measures the unit's Information intensity determined by the quantity of Information needed to spend on cooperation with other units.

When cooperation occurs, the intensity is diminished, being compensated by Information that binds these units and conserves the bound Information.

A collective unit holds less Information intensity than prior to cooperation, measured by a summary of each unit complexities. With more units in the collective, each complexity of an attached unit $M_{i,i+1,i+2}^\delta, i = 1, ...., m$ tends to decrease. The growing cooperatives intend to spend less Information for attracting other units, accepting the assembled units with decreasing Information speeds under the minimax.

A total (integral) *relative MC*-complexity for the IN with $m$ triplets approximates sum of (2.8b): $MC_m^\Sigma \cong m$ , which grows as each new triplet is added.

The IN integral relative *differential Cooperative Complexity*
$$MC_m^{\delta\Sigma} \cong \sum_1^m [3(\gamma_2^{-4})]^m = [[(1-[3(\gamma_2^{-4})]]^m / [1-[3(\gamma_2^{-4})] \qquad (2.10)$$
decreases with adding each new triplet.

At $m \to \infty, \gamma_2(\gamma = 0.5) = 3.89, \gamma_1(\gamma = 0.5) = 2.215$ it holds $MC_m^{\delta\Sigma} \cong 1.013$.

When total $MC_m^{\delta\Sigma}$ grows, the complexity of each following cooperation diminishes the contribution to the IN Cooperative Complexity, and with growing number of the IN inits, the sum of the contribution approaches zero.

The $MC_m^\Sigma$, defined for a non-cooperating triplet's segment in $m$ , is higher than the IN's cooperative complexy $MC_m^{\delta\Sigma}$ of the growing triplets number.

The evolution dynamics with the adaptive self-controls keep the $MC_m^{\delta\Sigma}$ decreasing at each evolving IN in the observation. Therefore, the MC complexity decreases in the observation with reduction of the uncertainty of randomness. This complexity emerges from different interactive process being observed [67].

The cooperative complexities integrate the Entropy Functional following the Information path functional.

The IN complexity decreases during formation of the nested structure where each following knot-node enfolds complexities of the previous formed node, and the ending IN node encloses and integrates complexities all IN.

The minimal MC complexity arises on the emerging elementary IN observer (Sec.4.3.7).

## 4.3.The cooperative Information mass and Information space curvature

Each Information speed of a triplet's third eigenvalue $\alpha_m = 3\alpha_{i+2}$ processes in space a differential volume $v_m = \delta V_m / \delta t = \dot{V}_m$ during cooperation.

Relation $M_{vm} = \alpha_m v_m$ we call the *Information cooperative mass* in this volume.

That eigenvalue is equivalent to the elemenary unit's Hamiltotian $\alpha_m = H_m$ in Information dynamics.
This leads to Information mass in the form

$$M_{vm} = H_m \dot{V}_m . \qquad (3.1)$$

The connection of entropy derivation $\partial \Delta S_m / \partial t = -H_m$ with entropy's divergence $\partial \Delta S_m / \partial t = c_m div \Delta S_m$ for the same volume $v_m$, moving linear speed $c_m$ at cooperation of $m$ the triplet, measures the Information mass (3.1) in the form

$$M_{vm} = -(c_m div \Delta S_m)\dot{V}_m . \qquad (3.1a)$$

The ratio of the Information masses for the nearest triplets:

$$M_{vm} / M_{vm+1} = \alpha_m / \alpha_{m+1}(v_m / v_{m+1}), \; v_m / v_{m+1} = \alpha_{m+1}^2 / \alpha_m^2 = (1/3\gamma_{m+1}^{\alpha})^2 , \qquad (3.1b)$$

$$M_{vm} / M_{vm+1} = 1/3\gamma_{m+1}^{\alpha}, \gamma_{m+1}^{\alpha} \cong 3.9 \qquad (3.1c)$$

grows in 1.3 times when each following IN's triplet is added.
The triplet Cooperative Complexity (2.3) for the same cooperative volume measures differential Hamiltonian:

$$M_m^{\delta} = 3\dot{H}_m / \dot{V}_m, \dot{H}_m / \dot{V}_m = \dot{H}_m^V , \; \dot{H}_m = \dot{\alpha}_m \qquad (3.2)$$

where the derivation applies on the EF extreme trajectory' segment between the emerging new Information acceleration $\dot{\alpha}_m$ is finite.

Multiplication Information mass $M_{vm}$ on the complexity (3.2) leads to $M_{vm} M_m^{\delta} = 3H_m \dot{H}_m$, which brings

$$M_m^{\delta} M_{vm} = 3\alpha_m \dot{\alpha}_m . \qquad (3.2a)$$

Information curvature $K_{\alpha}^m$ at the cooperation of the three triplet's eigenvectors describes a curving phase space at a locality within cooperative volume $v_m$.

This curvature originates in the curved impulse interactive cooperation creating its increment at a curving instant. The curvature defines the ratio of the increment to this instant.

This curvature connects classical Gaussian curvature with that in a Riemann space, where curvature $K_m^{\alpha}$ rises by the increment of Information speed on an instant of geodesic line metric $ds$.

The Riemann space' curvature is defined via fundamental metric tensor $\sqrt{g}$ and the phase space metric $ds = v_m dt$ in form:

$$K_m^{\alpha} = (\sqrt{g})^{-1} d(\sqrt{g}) / ds = (\sqrt{g})^{-1} d(\sqrt{g}) / v_m dt , \qquad (3.3)$$

where $g$ describes a closeness of the space vectors in the cooperative curving interactions.

For the eigenvectors in Information phase space, a metrical tensor $\sqrt{g}$ is expressed [47] through the Information of triplet eigenvectors $\alpha_m$ , localized in a space, generates an increment of tensor $\sqrt{g}$ for the triple cooperation:

$$\sqrt{g} = (\alpha_m)^{-3} . \qquad (3.3a)$$

Substitution to (3.3) determines $m$-th triplet curvature

$$K_{\alpha}^m = -3\dot{\alpha}_m / \alpha_m v_m , \qquad (3.4)$$

which connects the triplet' differential complexity and differential Hamiltonian (3. 2) with the curvature:

$$K_{\alpha}^{m} = -M_{m}^{\delta} \dot{H}_{m} . \quad (3.4a)$$

According to [68], multiplication of a physical mass on $\sqrt{g}$ determines the mass density.

Multiplying Information mass $M_{vm}$ on $\sqrt{g}$, expressed through the eigenvector (3.3a), leads to mass *density* $M_{vm}^{*}$:

$$M_{vm}^{*} = (\alpha_m)^{-3} \alpha_m v_m = (\alpha_m)^{-2} v_m . \quad (3.5)$$

In the simulated IN hierarchy (Fig. 7), the cooperating eigenvalues $\alpha_m$ decrease with the growing number of triplets $m \to n/2$, which increases $M_{vm}^{*}$.

Since that, the mass includes the Information speed and originating an instance–unit volume.

It allows estimating Information units encoding during the mass formation.

The encoding requires energy $e_{ev}$ covering the entropy $s_{ev}$ cost of conversion to Information $i_{ev}$ at $s_{ev} = i_{ev}$ and memorizing the resulting effective complexity.

Expressing (3.4) at $MC_m = \alpha_m / v_m$, in form $K_{\alpha}^{m} = -3\dot{\alpha}_m \alpha_m v_m v_m / v_m \alpha_m \alpha_m v_m v_m$ leads to

$$K_{\alpha}^{m} = -M_{vm}^{*}, M_{m}^{\delta} MC_m = -M_{vm}^{*} = MC_{m}^{\delta e} = M_{m}^{\delta} MC_m, MC_{m}^{\delta e} = \dot{H}_{m}^{V} MC_m \quad (3.\ 6)$$

where $MC_{m}^{\delta e}$ is a triplet *effective complexity* and $\dot{H}_{m}^{V}$ is density of Information Hamiltonian per volume.

The effective complexity generates the Information curvature of the joint cooperation and memorizes the unit Information mass in the cooperative Information mass.

The cooperation decreases uncertainty and increases Information mass at forming each triplet.

The cooperation, accompanied by decreases of the triplet's eigenvalues and complexity, declines the curvature of the IN cooperating structure.

The negative curvature (3.5) characterizes a topology of the space area where the cooperation occurs.

*The Information mass emerges as a curved Information space per cooperative Information complexity.*

Multiplication information mass density of curvature:

$$M_{vm}^{*} K_{\alpha}^{m} = -3\dot{H}_m H_m^{-1} = -3d^{*} H_m^{*} \quad (3.6a)$$

determines the relative decrease of the differential increment of energy $d^{*} H_m^{*}$ spent on forming this mass.

Simple estimation of the classical Gaussian curvature by inverse radius $r_m$ of a cooperating triplet node:

$$| K_{\alpha E}^{m} | = (r_m)^{-1}, r_m = \varepsilon(\gamma) = [(\gamma_1^{\alpha} / \gamma_2^{\alpha})^2 - (\gamma_2^{\alpha})^{-1}]^{1/2} \quad (3.\ 7)$$

connects the estimated curvature with the triplet invariants' ratios. That brings the invariant evaluation of the curvature for the triplet cooperating within radius $\varepsilon_m(\gamma^*) \cong 0.33$.

To evaluate maximal speed $c_{mo}$ of a single Information cooperation, measured by Information invariant $\mathbf{a}$, we use relation connecting a space divergence at the cooperation with speed $c_m$, time interval of the cooperation $t_m$, and $\mathbf{a}$:

$$\partial \Delta S_m / \partial t = | \mathbf{a} | / t_m = c_m div \Delta S_m$$

which at a minimal time of cooperation $t_{mo}$ leads to

$$c_{mo} = [t_{mo} div \Delta S_m / | \mathbf{a} |]^{-1} .$$

The minimal interval of cooperation $t_{mo}$ estimates the minimal physically admissible time interval $t_{mo} \cong 1.33 \times 10^{-15}$ sec of light wavelenght.

The entropy space' divergence, normalized by ratio to invariant $\mathbf{a}$ , estimates the structural invariant of minimal uncertainty $h_\alpha^o \cong 1/137$ at the Information cooperation:

$$div^*\Delta S_m = div\Delta S_m / |\mathbf{a}| \approx 1/137 \qquad (3.7a)$$

The resulting maximal Information speed $c_{mo} \approx 1.03 \times 10^{17} Nat/\sec$ restricts the cooperative speed and minimal Information curvature at other equal conditions.
Information mass (3.1a) in form :

$$M_{vm} = -|\mathbf{a}|(c_m div\Delta S_m / |\mathbf{a}|)v_m = -|\mathbf{a}|v_m h_\alpha^o \qquad (3.7b)$$

encloses the impulse Information measure a, volume $v_m$, and the triplet structural invariants $h_\alpha^o$.

The elementary cooperation binds *space* Information $div^*\Delta S_i$ (3.7a) that limits maximal speed $c_m$ at $t_{mo}$ for incoming $i$-Information units, imposing an *Information* connection on the time and space.
An Unbound Information unit does not have such a limitation.

The ratio of maximal speed $c_{mo}$ to speed of light $c_o$ :

$$c_{mo}/c_o \approx 0.343 \times 10^9 Nat/m = 0.343 gigaNat/m \qquad (3.8)$$

(measured in a light's wavelength meter) limits a maximal Information space speed.

In this case, each wavelength of the speed of light delivers $\cong 137$ Nats during $t_{mo} \cong 1.33 \times 10^{-15} \sec$.

The physical mass-energy that satisfies the law of the preservation energy (following the well-known Einstein equation), is distinguished from the Information mass (3.1), (3.5) defined for the cooperating triplet's units.
The law is satisfied when the triplet acquires energy (as the mass-energy) carrying the Information mass.
This triplet Information curvature and the effective Cooperative Complexity also hold physical energy.

### 4.4. Relative Information Observer

Let's analyze how impulse linear space speed $c_k$ on the IPF path curves the time-space trajectory–an extremal of the observing process, affecting the impulse with invariant geometrical measure $\pi$ that encloses its invariant Information measure $|1|_M$ .

Suppose the impulse $\pi_k$, located on distance $l_k$ along the IPF trajectory, is moving with linear speed $c_k$, and another invariant impulse $\pi_{k1}$ with the same invariant Information measure, located on distance $l_{k1}$, is moving along this trajectory with linear speed $c_{k1} > c_k$ relative to impulse $\pi_k$.

Each impulse speed is the ratio of invariant geometrical measure to the impulse location' time interval on the trajectory: $\tau_k$ and $\tau_{k1}$ accordingly, at $c_k = \pi_k / \tau_k, c_{k1} = \pi_{k1} / \tau_{k1}$, where at $c_{k1} > c_k, \pi_k = \pi_{k1} = \pi$..

Since with growing a total time along the IPF extreme trajectory, the impulse time intervals is shortening: $\tau_{k1} < \tau_k$ , it leads to the persisting increase of Information density.

The Information densities: $D_k = |1|_M / \tau_k, D_{k1} = |1|_M / \tau_{k1}$ of these impulses' invariant Information measure $|1|_M$ grows at $D_{k1} > D_k$ .

The distance between these impulses on the trajectory determines relations $\Delta l_{k,k1} = l_{k1} - l_k = (c_{k1} - c_k)/(\tau_{k1} - \tau_k), \Delta l_{k,k1} = (\pi/\tau_{k1} - \pi/\tau_k)(\tau_{k1} - \tau_k) = \pi(\tau_k - \tau_{k1})^2 / \tau_{k1}\tau_k, \tau_{k1} < \tau_k$.

Since the impulse preserves the invariant measure $\pi$, this distance is limited by the condition $(\tau_k - \tau_{k1})^2 / \tau_{k1}\tau_k \geq 1$.
The minimal distance at this condition satisfies $(\tau_k - \tau_{k1})^2 / \tau_{k1}\tau_k = 1, (\tau_k - \tau_{k1})^2 = \tau_{k1}\tau_k$, which, after solution by the quadratic Eq. $\tau_k^2 - 3\tau_{k1}\tau_k + \tau_{k-1}^2 = 0$ leads to a minimal admissible time ratio $\tau_{k1} / \tau_k = (3-\sqrt{5})/2, \tau_{k1} / \tau_k \cong 0.38$, or to the related maximal admissible time ratio $\tau_{k1} / \tau_k = (3+\sqrt{5})/2, \tau_{k1} / \tau_k \cong 2.36$.
In both cases, at $\Delta l_{k,k1} = \pi$, the decreasing time ratio satisfies only the first solution.
Moreover, this ratio *limits the minimal distance* of observation of each invariant impulse $\pi$.
As the growing speed decreases this ratio, the observing impulse does not preserve its invariant measure.
Therefore, this ratio limits both the distance of observation each invariant impulse and all observing paths along the trajectory.
The path of observation of the invariant impulse from the starting location $(l_{k1}, \tau_{k1})$ is minimal on the trajectory.
The nearest $\pi_k$, $\pi_{k1}$ impulses hold a third one distance $\Delta l_{k,k1} = \pi = \pi_{k,k1}$ between them.
These three *identify a triple located* on each $k$ dimension of the $n$-dimensional observing process $n = 1, 2, 3, ..., k, ...$
Such triple possess the above features.
At the minimal time interval, more Bits concentrate in the impulse time unit, which increases the speed of natural encoding. This growing Information density increases the impulse curvature along the trajectory.
The rising curvature and the impulse speed of natural encoding enfolds the growing impulse Information mass. The encoded Information impulses determine Information Observer moving with that speed.
Moving with maximal speed, the Information Observer encodes maximal Information with maximal speed, combined with the growing curvature, density, and Information mass, which *shortens the path of observation.*
The IPF integrates the density in the Observer's geometrical structure (Fig. 8).
The structure's rotating speed grows as the linear speed increases, shortening the time intervals of the invariant impulse Information measure.
The growing density increases number of the enclosed events in both time and space intervals.
Hence, the observing minimal impulse time interval automatically increases its linear speed, curvature, density, speed of encoding, and shortens the enclosed Information mass.
That links to a growing attraction in Information gravitation on the reduced path of observation.
The relative moving Information Observer possesses the relative encoding and attractive curved gravitation. Both initially emerge in the microprocess, in addition to relative time and space (Sec.3).

Formula $X_\delta = -1/4r_x^{-1/2}$ connects the *weak Information force*, arising from the microlevel correlations $r_x$, with the topological curvature $K_{eo}$ of the impulse correlation: $r_{icM} = \pi \times K_{eo}$ at $r_x = r_{icM}$.
It leads to connection through a *strong Information force:*

$$X_{\delta cM} = -1/4(\pi \,|\, K_{eo} \,|)^{-1/2}. \qquad (4.1)$$

with the macroprocess curvature which links to a potential observing space gravitation.
The topological curvature turns to Information curvature $K_\alpha^m$, and leads to the connection with the Information force $X_\alpha^m$ arising with the Information curvature in form (3.6).
The Observer's relative time encloses Information and the curvature (gravitation). This also connects the Information force with Information mass and complexity according to formulas (3.6).

### 4.5. About origin of a time and time course

The origin of time is an attribute of probabilistic observation of sequential states (events) in a probability field.

*Time is a measure of the correlation interactive connection.*

Following [2:21], let the correlation between random states or interacting events $\tilde{x}_o, \tilde{x}_1$ to satisfy the relations

$r_{\tilde{x}_o\tilde{x}_o} = E[\tilde{x}_o^2] \sim \sqrt{t_o}, r_{\tilde{x}_1\tilde{x}_1} = E[\tilde{x}_1^2] \sim \sqrt{t_1}$ .

Then, because $E[\tilde{x}_o(t_o)]$ is independent on $E[\tilde{x}_o(t_o) - \tilde{x}_1(t_1)]$, it follows that

$E[\tilde{x}_o\tilde{x}_1] - E[\tilde{x}_o^2] = E[\tilde{x}_o(\tilde{x}_1 - \tilde{x}_o)] = E[\tilde{x}_o]E[(\tilde{x}_1 - \tilde{x}_o)] = 0, E[\tilde{x}_o^2] = E[\tilde{x}_o\tilde{x}_1] \sim \sqrt{t_o}$ .

According to these formulas, correlation $r_{\tilde{x}_o\tilde{x}_1} = E[\tilde{x}_o(t_o)\tilde{x}_1(t_1)]$ satisfies the relations

$$r_{\tilde{x}_o\tilde{x}_1} / r_{\tilde{x}_1\tilde{x}_1} = [t_o / t_1]^{1/2} = r_{\tilde{x}_o\tilde{x}_o} / r_{\tilde{x}_1\tilde{x}_1} \quad (5.1)$$

At fixed $r_{\tilde{x}_o\tilde{x}_o}$ and $t_o$, increasing time $t_1$ increases correlation $r_{\tilde{x}_1\tilde{x}_1}$ ,or the correlation connection of the states-events, and vice versa.

Decreasing $t_1$ decreases this correlation or the correlation connections of the states-events; and vice versa.

Let us start counting time by comparing different Observers at the same starting time of clock $t_o$ .

The Observer with the increasing correlation connection will observe more clock time $t_1$ than an Observer having less correlation connection during time clock $t_o$ .

Or, the clock may move faster in the Observer with stronger correlation connections compared with the Observer having less correlation connection where the clock moves slower.

Therefore, the increasing correlation connections of interacting events speed up their time course, while decreasing correlation slows their time course.

Stronger correlation connects larger number of the states-events with successively smaller intervals of the time holding this connection or the time of coupling the correlated states.

That is also connected with effect of Zeno behavior [69].

For the microprocess's entangled states, correlations $r_{cr}$ are very strong. At $r_{\tilde{x}_1\tilde{x}_1} \to r_{cr} \to \infty$, the related time of the entangling states $t_1 \to \infty$, these states loose a notion of time and the time difference.

They are indistinguishable by the time and stay coupled forever.

For the comparing Observers with events $\tilde{x}_o, \tilde{x}_1$ , there is no the time difference at $E[\tilde{x}_o\tilde{x}_1] - E[\tilde{x}_o^2] \to 0$ .

With zero time $t_o$ of observing microprocess correlation $r_{\tilde{x}_o\tilde{x}_o}$ , it leads to $r_{\tilde{x}_o\tilde{x}_1} \to 0$, or the comparing Observers loose correlation connections.

This analysis shows the common connection of the observing time of random microprocesses, which is evaluated by classical probability theory.

For the certain Information process, the Observer time measures the Free Information spent on attraction and binding Bits, triplets, Information networks, and encoding, while building the Observer time-space Information structures.

These structures have growing complexity, curvature, and Information mass, whose attractiveness measures a force (4.1) analogous to gravitation.

It's known that an object, attracted to a black hole, will hold there forever.

Hence, the Information approach connects the relative times of moving Observers, the observing process's time, the microprocess time, and the certain-real time of Information Observers.

The inner and external time courses in the Information Observer are also mutually connected (Sec. 3.4).
The observing sequence of time intervals initiates its logic.
For the Information Observer, the increasing time of observation holds more encoding Information.
*Finally, all these results follow from the fundamental phenomena of interactions and are an attribute of the random field and the reality of interactions.*

**4.6. The Observer's IN self-replication and conditions of self-generation the Observer new Information quality**

Each IN node's maximal admissible $m_M$ -th level ends with a single dimensional process, which in attempting to attach the new triplet over the constrained $\gamma_{k1} \to 1$, loses the ability to enfold newly attracted Information. Such IN stops, satisfying the minimax Information law, leads to its *instability*.
That violation is a *natural intention*, growing for each IN node, is constrained by limitations [41].
Specifically, after completion of the IN cooperation, the last IN ending node $n$ initiates an one dimensional Information process

$$x_n = x_{n-1}(2-\exp(\alpha_{n-1}^t t_n)) \quad (6.1)$$

with relative speed $\dot{x}_n / x_n = \alpha_n^t$, where the process eigenvalue $\alpha_n^t$ depends on the previous $n-1$node eigenvalue $\alpha_{n-1}^t$ and dynamic invariant $\mathbf{a}_{io} = \alpha_{n-1}^t t_n = \ln 2$ :

$$\alpha_n^t = -\alpha_{n-1}^t \exp(\alpha_{n-1}^t t_n)(2-\exp(\alpha_{n-1}^t t_n)^{-1} \quad (6.1a)$$

The invariant allows defining time $t_n$ through $\alpha_{n-1}^t$ :

$$t_{n-1} = \ln 2 / \alpha_{n-1}^t \quad (6.1b)$$

but its substitution to (6.1a) leads to $\alpha_n^t$ approaching infinity with a potential infinite relative phase speed

$$\dot{x}_n / x_n \to \infty . \quad (6.2)$$

At condition (6.1b), process (6.1) with infinitive (6.2) cannot reach a zero final state $x_n(t_n) = 0$.
Adding a potential $k-$ dimensional process with relative speeds :

$$\dot{x}_{n+k-1} / x_{n+k-1}(t_{n+k-1}) = \alpha_{n+k-1}^t , \dot{x}_{n+k} / x_{n+k}(t_{n+k}) = -\alpha_{n+k}^t \quad (6.3)$$

leads to periodical process of the alternating the opposite values of each *two* relative phase speeds. That brings instable fluctuations of these speeds at each $t = (t_{n+k-1}, t_{n+k})$ starting the alternations with eigenvalue

$$\dot{x}_k / x_k = \alpha_k^t(\gamma_k), \gamma_k \le 1, k = 1,2,... \quad (6.3a)$$

The instable fluctuations in a three-dimensional process involve their oscillation interactions approaching $\gamma_k \ge 1$. That generates a frequency spectrum of model eigenvalues $\lambda_i^*(t_{n+k})$ in each of its space dimensions $i = 1,2,3$.
Formal analysis of this instability associates with nonlinear fluctuations [70], which describe a superposition of linear fluctuations with a frequency spectrum ( $f_1,..., f_m$) proportional to imaginary components of the spectrum eigenvalues $(\beta_1^*,...,\beta_m^*)$, where $f_1 = f_{\min}$ and $f_m = f_{\max}$ are the minimal and maximal frequencies of the spectrum.

In this model, the oscillations under interactive actions generate imaginary eigenvalues $\beta_i^*(t)$:

$$\operatorname{Im}\lambda_{n+k}^i(t) = \operatorname{Im}\lambda_{n+k-1}^i[2-\exp(\lambda_{n+k-1}^i t)]^{-1} \quad (6.4)$$

at each $t = (t_{n+k-1}, t_{n+k})$ for these $i$ dimensions.

This leads to the relation

$$\operatorname{Im}\lambda_n^i(t_{n+k}) = j\beta_n^i(t_{n+k}) = -j\beta_{n+k-1}^i \frac{\cos(\beta_{n+k-1}^i t) - j\sin(\beta_{n+k-1}^i t)}{2-\cos(\beta_{n+k-1}^i t) + j\sin(\beta_{n+k-1}^i t)}\ , \quad (6.5)$$

at $\beta_i^* = \beta_{n+k}^i, \beta_i^* \neq 0 \pm \pi k$, where $\lambda_{n+k}^i$ includes a real component:

$$\alpha_n^i(t_{n+k}) = -\beta_{n+k-1}^i \frac{2\sin(\beta_{n+k-1}^i t)}{(2-\cos(\beta_{n+k-1}^i t))^2 + \sin^2(\beta_{n+k-1}^i t)}\ . \quad (6.5a)$$

At $\alpha_i^* = \alpha_i^*(t_{n+k}) \neq 0$, the ratio determines the relative parameter of dynamics

$$\gamma_i^* = \frac{\beta_n^i(t_{n+k})}{\alpha_n^i(t_{n+k})} = \frac{2\cos(\beta_{n+k-1}^i t) - 1}{2\sin(\beta_{n+k-1}^i t)}\ . \quad (6.6)$$

At $\gamma_i^* = 1$ relation (4.6) brings

$$(\beta_{n+k-1}^i t) \approx 0.423 rad (24.267^o) = 0.134645\pi\ . \quad (6.7)$$

The fluctuations may couple the nearest dimensions by an interactive cooperation overcoming *a minimal elementary* uncertainty (UR) as a border between the dimensions.

The question is could the interactive oscillations couples dimensions regenerating a triple?

Applying the Fine Structural constant in Physics $h_\alpha^o \cong 1/137$ to the UR coupling allows evaluating the UR border's entropy or it information measure by the increment

$$i_{bN} = h_\alpha^o \mathbf{a}_o(\gamma = 0) \cong 0.00547\,\text{Nat}\ . \quad (6.8)$$

Here $\mathbf{a}_o(\gamma = 0) = 0.75$ is maximal dynamic invariant, and invariant (6.8) separates the dimensions.

Suppose a $k-th$ interaction from $1,2,3,..m,..k,....$ fluctuations is needed to create a dynamic process' dimension which carries a single unit of entropy $s_c(\gamma) = \mathbf{a}_{ok}^2(\gamma)$.

The potential bordered cooperative dimensions evaluate the minimal interactive increment of information $\delta\mathbf{a}_k = \mathbf{a}_o(\gamma = 0) - \mathbf{a}_o(\gamma)$ between a new dimension carrying invariant $\mathbf{a}_o(\gamma = 0)$ interacting with other dimension carrying invariant $\mathbf{a}_o(\gamma)$ during the fluctuations.

The $k-th$ fluctuation interaction may overcome the UR *border* invariant increment of the entropy concentrated in UR (4.8).

That requires equality $h_\alpha^o \mathbf{a}_o(\gamma = 0) = \mathbf{a}_{ok}^2(\gamma)$.

To create the double cooperation, the invariant Information quantity $\delta\mathbf{a}_k$ should compensate the UR bordered *entropy* invariant $\mathbf{a}_{ok}^2(\gamma)$ satisfying the equality

$$\delta\mathbf{a}_k = \mathbf{a}_{ok}^2(\gamma) = 0.00547 \quad (6.8a)$$

which is equivalent to minimal Information

$\mathbf{a}_{ok}(\gamma)=0.073959448$ . (6.8b)

Each $k-th$ interaction changes the initial $\gamma_k \geq 1$ on $-\Delta\gamma=-\gamma*$, bringing the minimal Information increment in (6.8b).

That Information creates three of the $k-th$ interactive multiplicative changes, creating Free Information $\mathbf{a}_k \cong 0.23$, enabling a cooperative attraction by generating Information

$3\mathbf{a}_{ok}(\Delta\gamma)=0.221878345 \cong 0.23$. (6.8c)

Information attraction $\mathbf{a}_k$ can cooperate the interactive Information invariant $\mathbf{a}_{ok}(\gamma)\cong 0.7$ through another three interactions.

The last invariant binds three dimensions in single Bit through a total of nine interactions.

The question is: Can the interactive fluctuations creating triplet self-replicates a new IN?

The dynamic invariant $\mathbf{a}(\gamma)=\mathbf{a}_k$ of Information attraction determines ratios of starting Information speeds $\gamma_1^{\alpha}=\alpha_{io}/\alpha_{i+1o}$ and $\gamma_2^{\alpha}=\alpha_{i+1o}/\alpha_{i+2o}$ needed to satisfy invariant relations (6.5.) and $2\sin(\gamma_i\mathbf{a}_{io})+\gamma_i\cos(\gamma_i\mathbf{a}_{io})-\gamma_i\exp(\mathbf{a}_{io})=0.$ (See [71]) .

To create a new triplet's IN with ratio $\gamma_1^{\alpha}=\alpha_{io}/\alpha_{i+1o}$, relation (6.6) requires the ratio $\frac{\beta_i^*(t_{n+k})}{\beta_{n-1,o}(t_{n-1,o})}=l_{n-1}^m$ which deliver the imaginary invariant $(\beta_{n+k-1}^i t)\rightarrow(\pi/3\pm\pi k), k=1,2,...$ at each $k$ with Information frequency $\beta_i^*(t_{n+k})$ that generates the needed $\alpha_n^i(t_{n+k})$ from $\operatorname{Re}\lambda_i^*(t_{n+k})$.

In the case $\Delta\gamma\rightarrow 0$, it generates IN with minimal $\gamma*=\gamma$, which can be achieved in (6.6) at $2\cos(\beta_{n+k-1}^i t)\rightarrow 1$, or at

$(\beta_{n+k-1}^i t)\rightarrow(\pi/3\pm\pi k), k=1,2,...$, with $\alpha_n^i(t_{n+k})\cong -0.577\beta_{n+k-1}^i$ . (6.9)

That determines the maximal *ratio* of *frequencies:*

$l_{n-1}^m=\beta_{n+k-1}^i/\beta_{n-1,k=o}^i$ (6.9a)

which at $\gamma=1$, $\beta_{n-1,o}(t_{n-1o})=\alpha_{n-1,o}(t_{n-1,o})$ and $\beta_{n+k-1}^i=\alpha_{l,k=3}^i/0.577$ holds

$l_{n-1}^m=\alpha_{l,k=3}^i/0.577/\alpha_{n-1,k=o}^i$, $\alpha_{l,k=3}^i/\alpha_{n-1k=o}^i=\gamma_1^{\alpha}$. (6.9b)

Ratios (6.9a) and (6.9b) identify the triplet invariant

$l_{n-1}^m=\gamma_1^{\alpha}/0.577$, (6.10)

generated by the observing $(n-1)$-dimensional spectrum with an imaginary eigenvalue $\beta_{n-1,o}(t_{n-1,o})$ by the end of the interactive movement.

Invariants (6.9a) and (6.10) lead to $\gamma_1^{\alpha}=\alpha_{l,k=3}^i/\alpha_{n-1,k=o}^i$=3.89 and to the ratio of initial frequencies $l_{n-1}^{m=1}\cong 6.74$.

The next nearest $\alpha_{l+1,k=6}^i/\alpha_{lo,k=3}^i=\gamma_2^{\alpha}$ needs to increase the first ratio in $l_{n-1}^{m=2}\cong 3.8$, and the following $\alpha_{l+2,k=9}^i/\alpha_{l+1,k=6}^i=\gamma_2^{\alpha}$ needs $l_{n-1}^{m=3}\cong 3.8$.

That is needed to build new triplet, which at $\gamma = 1$ is not ending with a stable segment.

The $l_{n-1}^{m=1-3}$ identifies the maximal ratio of spectrum frequencies generated by the instable fluctuations to create a triplet.

Each three Information units $3\mathbf{a}_{ok}(\Delta\gamma)$ binds a pair of the nearest spectrum frequencies, starting with the pair $\beta_{n-1,o}^{i}, \beta_{n,k=1}^{i}$ which sequentially grows with each $k-th$ interactive Information (6.8a-c), intensifying the increase of the frequency.

The first three pairs bind three dimensions in a single Bit $\mathbf{a}_{ok}(\gamma) \cong 0.7$ Nat through nine interactions of the multiplicative frequencies requiring ratio $l_{n-1}^{m=1} \cong 6.74$, while the next pair ratio grows in ~2.24 multiplications, and each next in 1.26 multiplications.

Thus, a natural source, producing a first triplet, is a nonlinear fluctuation of an initial dynamics. The fluctuation should involve a minimum of three such dynamic dimensions that enclose memorized Information on an IN ending node.

That, at the node Information invariants $\mathbf{a}_o(\gamma = 1) = 0.58767$, $\mathbf{a}(\gamma = 1) = 0.29$, brings $\gamma_1^{\alpha} \cong 2.95$ with ratio $l_{n-1}^{m=1} \cong 5.1126$.

A natural source of maximal speed frequency is the light wavelength whose time interval $t_{lo} \approx 1.33 \times 10^{-15}$ sec determines maximal frequency $f_{\max} \cong 0.7518 \times 10^{15} \sec^{-1}$.

The required frequency ratio $l_{n-1}^{m=1-3}$ identifies the minimal frequency $f_{\min} = 0.7724586 \times 10^{13} \sec^{-1}$ multiplied ratios

$$l_{n-1}^{m=1-3} = l_{n-1}^{m=1} \times l_{n-1}^{m=2} \times l_{n-1}^{m=3} \cong 97.3256 \qquad (6.10a)$$

Let us find the energy equivalent to (6.8b).

Following [34], the energy equivalent of $1Nat$ is $e_N = 0.0248ev$.

Then Information (6.8b) holds the energy equivalent

$$e_b(i_{bN}) = 1.35735 \times 10^{-4} ev\ . \qquad (6.10b)$$

The triplet invariant is therefore Information equivalent of energy spent on creating such a triplet. In this Information approach, (6.10b) evaluates the energy that conceals the IN node bordered dimension.

The triplet creation needs nine of such interacting increments, which evaluate the triplet energy's equivalent

$$e_{tr} = 1.221611 \times 10^{-3} ev\ . \qquad (6.10c)$$

Invariant conditions (6.6), (6.9a) enable modeling *cyclic* renovation [72], initiated by the two mutual attractive processes, which do not consolidate at the moment of starting the interactive fluctuation.

After the model disintegration, the process can renew itself with the state integration and transformation of the imaginary Information to the real Information during the dissipative fluctuations, which bring energy for a triplet.

A direct source is a joint Information of three different IN nodes $\mathbf{a}_{o1}(\gamma_1), \mathbf{a}_{o2}(\gamma_2), \mathbf{a}_{o3}(\gamma_3)$ enabling the initiation of attracting Information with three Information speeds, where one has opposite sign of the two, whereas the Information values cooperating in initial triplet will satisfy the above invariant relations.
Creating a triplet with specific parameters depends on the starting conditions which initiate the necessary attracting Information [73-74].
To achieve an Information balance satisfying the VP and the invariants, each elementary $\mathbf{a}_{oi}(\gamma_i)$ searches for partners for the required consumption of Information.
A double cooperation conceals Information $s_c(\gamma)=\mathbf{a}_o^2(\gamma)$, while a triple cooperation conceals Information $s_{cm}(\gamma)=2\mathbf{a}_o^2(\gamma)$. It could produce less Free Information than is needed for cooperation, while each of the above values depend on $\gamma$. With more triplets cooperating IN, the cooperative Information grows, spending that Free Information on joining each following triplet.
Minimal relative invariant $h_\alpha^o=0.00729927\cong 1/137$ evaluates a maximal *increment* of the model's dimensions $m_M\cong 14$, and the quantity of the hidden invariant Information (6.8) that produces an elementary triple code, enclosed into the cellular geometry of a hyperbolic structure Fig.8.
The hidden *non-removable UR uncertainly conserves a potential DSS Information code.*
Results (6.6-6.9a), and (6.9b) impose important *restrictions* on both maximal frequency generating new starting IN and maximal IN dimension which limits a single IN.
That IN's ending node may initiate this frequency.
For example, $\gamma=1$ corresponds to relations

$$(\beta_{n+k-1}^i t)\approx 0.423rad(24.267^o)\text{, with } \beta_i^*(t_{n+k})\cong -0.6\beta_{n+k-1}^i. \qquad (6.11)$$

Here $\beta_i^*(t_{n+k})\cong\alpha_{lo}^m(t_o)$ determines maximal frequency $\overset{*}{\omega}_m$ of fluctuation by the end of optimal movement at $\alpha_{lo}^m(t_o)\cong -0.6\beta_{n+k-1}^i$, where $\alpha_{lo}^m(t_o)=\alpha_{1o}(t_o)(\gamma_{12}^\alpha)^{m_M}$, with $\alpha_{1o}(t_o)$ as the starting speed of the IN with $m_M$ dimension. At $\alpha_{1o}(t_o)=0.002\sec^{-1}$ and $m=14$ it follows

$$\alpha_{14o}(t_o)=0.00414\sec^{-1}, \beta_{14}\cong 0.0069\sec^{-1}. \qquad (6.11a)$$

The new macro-movement starts with that initial frequency.
This newborn macromodel might continue the consolidation process of its eigenvalues, satisfying the considering restrictions on invariants and cooperative dynamics up to the ending consolidations and the arising periodical fluctuation movements.
Initial interactive process may belong to different IN macromodels (as "parents") generating new IN macrosystems (as "daughters") at end of the "parent" process.
The macrosystem, which is able to continue its life process by renewing the cycle, has to transfer its coding life program into the new generated macrosystems and secure their mutual functioning.

This leads to cyclic *micro-macro functioning* when the state integration alternates with state disintegration and the system decays with the possible transformation of an observable virtual process to evolving Information-certain process.

### 4.6.1.The functions of evolutionary Informational mechanisms

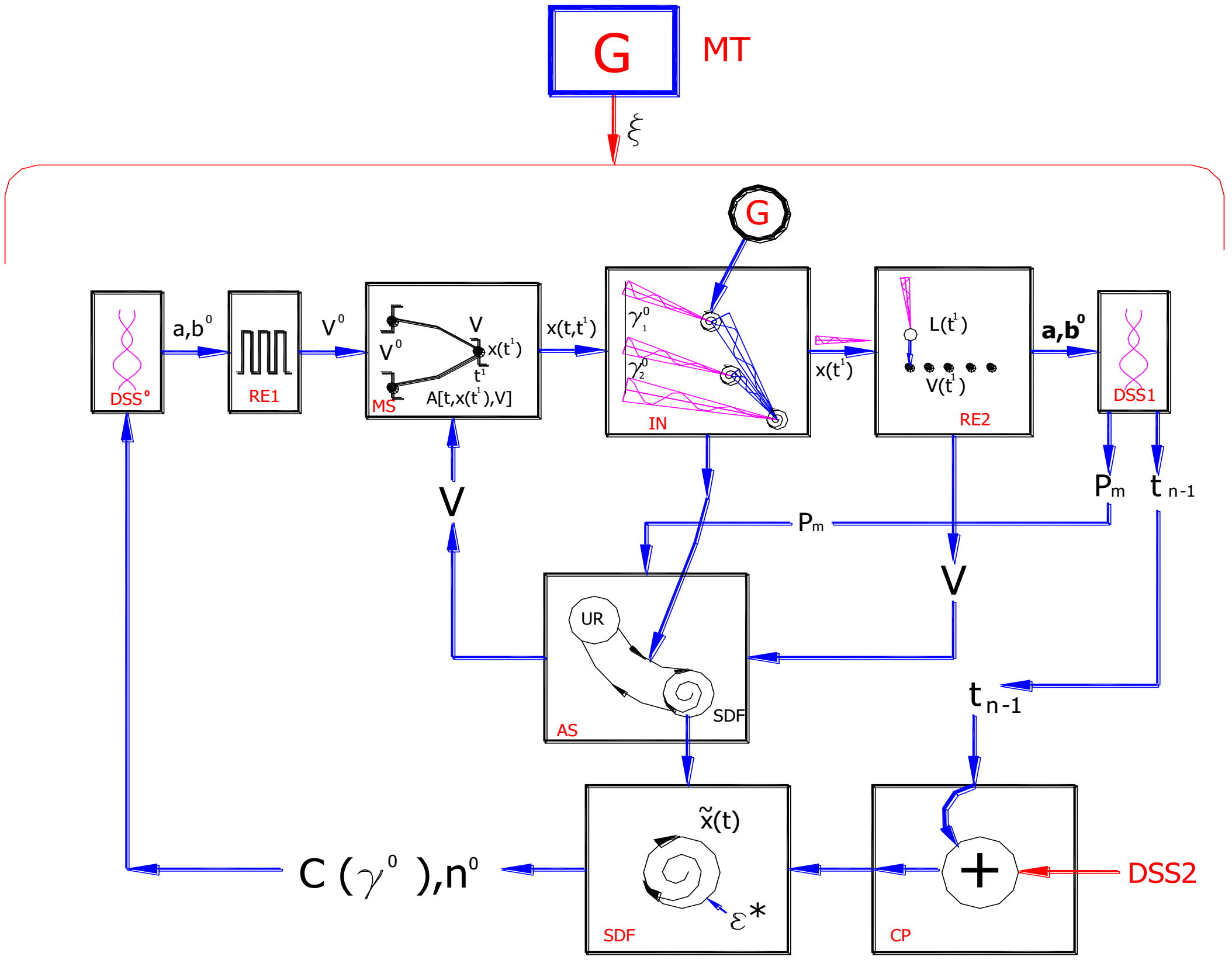

**Fig. 9. The functional schema of the evolutionary Informational mechanisms**

The evolution mechanism Fig. 9 includes: the system macrodynamics MS, defined by the model operator $A(t,x(\tau),v)$ that is governed by the inherited double spiral structure $DSS^o$; the control replication mechanism RE1 that transforms the $DSS^o$ code into the initial controls $v_o$ and delivers $v_o$ as the MS input programmable controls; the IN, formed in the process of the macrostate cooperation and the macromodel renovation, generating a renovated $DSS1$; mechanism of mutations MT, delivering external perturbations, which act on the total system; the adaptive and the self-organizing mechanisms AS, stimulated by the MT, which generate (G) fluctuations $\xi$; replication control mechanism RE2, which selects the macrostates $x(\tau)$ at DP $t'=\tau$ and forms the current control $v(\tau)=-2$ $x(\tau)$ by the duplication of $x(\tau)$; act of coupling of the two macrostates CP that carry both parents' $DSS1$ and $DSS$ 2 invariants; generation of stochastic dissipative fluctuations SDF after coupling, while forming new macrosystemic invariants $(\gamma^o, n^o)$ that define a new $DSS^o_o$, initiating the new MS and IN, which are renovating under the MT and the AS, in the process of functioning (a previous inherited $DSS2$ minimizes a possible SDF set, generating a new $DSS^o_o$); repeating the whole cycle after coupling and transferring the inherited invariants to a new generated macrosystem.

The IMD software package simulates the main mechanisms of the evolutionary cycle, arXiv:1212.1710

## V. The Observer Information Cognition And Intelligence

### 5.1.Emerging the observer's cognition and intelligence

We show how from multiple linking INs, enclosing Observer Information, emerge cognition, and intelligence.

#### 5.1.1.The Observer logic

The Observer logic emerges on the path from the collected observing interactive probabilistic logic, the curving impulses interactive certain logic, the rotating triplet logic, and the nested Information logics of attracting triplets.
The rotating speeds of each three logical Bits generate frequencies with the attractive logic of free Information.
Equalizing the speeds of the attracting free Information synchronizes the frequencies of the triplet's logic in a resonance which assembles a logical loop. That loop assembles the triple logic of the attracting logical Bits.
The assembled logic forms a non-chaotic logical attractor [75].
The attractor assembles triplets' logic in a logical knot. The attractor-knot' free information attracts new forming triplet logic in resonance loop which assemble other logical attractor.
These logically organized triplets' attractors-knots build a logical chain.
The chain resonances assemble the attractors-knots in nested logical nodes.
The nodes logically organize the nested hierarchy of Information network (IN).
The IN nodes self-organize the *hierarchical* frequencies of the nested attractors-knots.
The ending node of the IN encloses all the *attracting* logic of its triplets' node hierarchy.
The IN ending node holds a frequency which could attract in resonance other Observer INs.
The logical attractors-knots of the multiple INs triplets' logic form a distributed logical chain which encloses the hierarchy of the assembling local loops with their hierarchical frequencies, up to the Observer's highest level IN.
This hierarchical logic of the Observer consists of the mutual attracting free logic which self-organizes the assembling logical rotating loops in a chain enclosing all observing time-space logical structures.
The logically organized Observer's highest level IN *structures* the Observer Logic.

### 5.2.The Observer cognition

The observing time-space *logical* structure, composing a distributed chain of the multiple cooperative logical loops, is the Observer *cognitive logic*, or cognition.
The Observer's cognition assembles the logical chain through the multiple resonances of nested logical loops forming the IN-triplet hierarchy.
Such a multi-level logical structure possesses observations of virtual probabilistic causality, ability for real Information causality, and complexity.
All of that measure the cognitive intentional actions along each cooperative logical loop.
Each nested coherent loop accepts only such units that each IN node logic recognizes.
Finally, the IN finally memorizes each recognized unit in its Information node.
The cognitive rotating movement, upon forming each IN node and level, processes a *temporary loop* (Fig. 6) which might disappear after the newly formed IN triplet unit is memorized.
The Observer cognition emerges as an evolving intentional information ability to request, integrate, and predict the needed Observer Information that builds the Observer's growing networks.
The free logic resonances self-organize a logical chain of *cognitive functions,* which are distributed along hierarchy of the assembling logical units: triplets, IN nested nodes, and the IN ending nodes.
These local cognitive functions self-organize the Observer's cognition, up to the cognitive functions at the upper level's synchronizing frequency.

Along the IN hierarchy runs the distributed resonance frequencies spreading a chain of the nested loops.
The chain rotates with minimal energy of the synchronized frequencies.
These cognitive thermodynamics process the chain's resonance logic.
Comment.
Since logical nested loop needs minimum cooperating triplets, the real information cognition emerges only in a subjective observer. •

**5.3. Arising the Observer intelligence**

Since each assembling logical unit possesses the free logic, its topological opposite curved interaction with an external Yes-No impulse brings the asymmetry of interaction with a logical entropy Bit.
The *multiple local Bits* self-assemble a triple cognitive logic at all Observer's IN hierarchical levels.
Each level of the assembled triplet's interactive "Yes-action" can capture the external Landauer's energy which starts erasure the asymmetrical triplet's logic entropy bits, memorizing the information triplet up to encoding it in the knot. Such switching actions encode the logical triplet assembling the attractor-knot with its Free Information. The Bit's Free Information "No-action" stops delivering this energy.
The multiple level's Free Information self-organize cognitive information logic at all Observer's IN hierarchical levels. The nested triplet's logic knots are encoding in cognitive information in the codes on the IN hierarchal levels when all levels are synchronized in the IN ending knot-node.
In that hierarchy, the loop logic, assembling in each the knot lower level, will be released after this knot logic is memorizing and then encoding in the ending node.
(Thus, currently emerge knot-node has not instant effects on the IN code.)
The Observer hierarchical codes hold the energy of memorizing and encoding.
This code hierarchy physically organizes the multiple IN, self-encoding their local codes in the Information structure of Information Observer.
We call this coding structure the *Observer intelligence originated by intelligence functions.*
The logical nested-organized switching actions perform *intelligence functions*, which generate each local code. These functions are distributed the hierarchically along the assembling units' local codes.
It is the Observer intelligence cooperative code which self-organizes these local functions.
The question is: What initiates the switching to memorize the cognitive logic Bits and their encoding?
Evidently, it requires opening the access of external energy to each local logical unit at every hierarchical level. That requires a coordinating connection of the Observer's inner and external times, which schedule the switching along the Observer hierarchy.
As it had shown (Secs.3.5, 3.6), such coordination takes place exactly at the moment ending interval of the free logic at each unit level. The switching time interval $\Delta t_{o1}$ runs units with specific time intervals $\delta T_{cm}$ and related frequencies of switching $f_{cm} = 1/\delta T_{cm}$. The $\delta T_{cm}$ rate changes from $\delta T_{co} = 12$ (for an elementary subjective Observer with IN two triplet units) up to $\delta T_{cs} = 69242.359$ for the subjective Observer (with 9 triplets IN). If the unit runs one hour, then the objective Observer opens to get external energy with frequency $f_{co} = 1/12$, or each 12 hours.
The subjective Observer gets external energy with the frequency $f_{cs} = 1/69242.359$, or ~1/30 min, which is equivalent to one opening for two seconds, or 30 times per minute.
This means that each Observer has its own time clock with its time course which commands the hierarchical switching. The clock interactive switches command the encoding of each cognitive function to an intelligence function, starting the Observer intelligence.

Within the sequential segments of the observing dynamics (Fig. 3), each switch links two segments through a bridge between them, while the third segment ends with its bridge.
This bridge holds the triple segments' logical attractor, which is memorizing and encoding at the scheduled time. Each third bridge shapes a knot of the IN node hierarchical level.
At forming each node and level, the cognitive rotating process holds the coherent loop of harmonized speeds-Information frequencies at different levels. These frequencies determine the clock time units at different levels. At the unit level with a longer interval $\delta T_{cm}$, its frequencies are lower, and vice versa up to the highest frequencies of the subjective and an intelligence Observer.
The higher frequency grows the Information density of the encoding Information Bits.
The subjective Observer, self-assembling a hierarchy of logical structures, possesses a hierarchy of the frequencies and clock courses. The *harmonized speeds-Information frequencies* automatically setup the switching times and the frequencies when the cognitive loop at each level is self-established.
Finally, the coherent cognitive dynamics, assembling the cognitive functions of the units, self-organize the hierarchy of intelligence functions, encoding the Bits in the hierarchical codes. Or, the local cognitive units, involved in the resonance chain movement, self-organize themselves in local cognitive function which self-forms the Observer's cognition. The hierarchical cognition schedules the switching clock of intelligence functions, encoding the Observer hierarchy in the Observer cooperative code.
Since each assembled unit encodes the triplet code, the Observer cooperative code integrates the structural units of the triplet code at each level. These local codes have increasing densities of encoded impulses according to their hierarchal locations.
The cooperative code, which the clock synchronizes, has a rhythmical sequence of time intervals scheduling access the external energy to each Observer logical structural unit as it needs.
The clock time course assigns the frequency through the repeating time intervals, which determine each local resonance frequency of the assembling structural unit.
These frequencies-local rhythms identify the moments of the ending intervals of Free Information at each unit level, or the interacting cognitive and intelligence local actions. Each stable Observer logically conserves its switching time intervals. Therefore, the Observer code enhances multiple rhythms of the local structural units. That's why the rhythms of an external melody, resonating with the Observer code rhythms, supports the cognitive functions, intelligence actions, and generation of both the cooperative Observer's logic and the code encoding this intelligence logic.
Recently experiments have confirmed the influence of music on neural encoding [36, 76, 77].

**5.4. Arising structure of Observer code**
The curving interactive movement, starting the curving impulses, rotates the trajectories of the observing process, forming spirals on a cone surface (Fig. 3).
During probabilistic observations, these spiral trajectories hold a random periodic sequence.
With the emerging space and the space-time conjugated entropy increments, the rotating trajectories shape the conjugated space-time spirals (Fig. 4a). The emerging Information process continues rotating its trajectories in double spirals, assembling them according to Fig. 6.
The path from each interacting entropy impulse along the observing process entropy integrates the entropy functional (EF).
The minimax principle, preserving each interacting impulse invariant measure, leads to the minimax variation principle (VP) for the EF. The VP described the conjugating trajectories of the observing process as the EF extremals converging to the Information processes' conjugated Information trajectories as the IPF extremals. These EF-IPF trajectories satisfy the Hamilton equations for the Observer dynamic process.
The interactions of opposite directional conjugated spirals form bridges between the trajectory spirals (Fig. 4a). Each spiral segment integrates its observing logic in the bridge. The bridges locate the switching interactions which hold this logic according to the time schedule.
The attracting Free Information of the triple Bits connects the memorized Bits within the triplets' logic. The triplet's knot encodes the bridge logic of the triple segments in the barrier code according to the time schedule. The barriers, located along the observing trajectory, connect the encoding knots of the INs nested nodes, and then, in the ending IN triplet code.
The trajectories integrate the observing process logic, including the logic enclosed in the segments bridges, and the barrier encoding the logic in each ending the IN node.
The multiple IN ending triple codes integrate the double space spiral coding structure (DSS) (Fig. 7).
The time schedule sequentially builds a hierarchy of the memorized triplets and their encoding.
That schedules sequential transfers from the logical to the memorized, and the encoding bridges along the observing trajectory.
The hierarchical cognition schedules the switching clock of intelligence functions, encoding the hierarchy in the triplet codes. Encoding starts when the hierarchical switches open access to the external energy.
Then, sequentially emerges the triple code and the intelligence cooperative coding structure DSS.
The bridges localize the coding units where external interactions may change the DSS coding structure.
Each triplet, memorized in the conjugated interactive bridge, divides the trajectory on a reversible proces' segment which does not include the bridge and irreversible bridge between the reversible segments.
Thus, the Observer's irreversible dynamic trajectory includes the reversible segments, ending assymmetrical logical bridge. Each irreversible information triplet emerges from the encoding of the observation logic in the trajectory barriers-knots.
The conjugated trajectories integrate the EF extremals, while the emerging Bits encoding in the barriers integrate the IPF. The IPF encloses the integral Information in its final encoding Bit.
Therefore, the Observer's integral Information identifies the IPF final code, which the observation predicts through the VP minimax optimal Information law.
This IPF code has increasing densities, which triple with each following Bit.
The prediction, based on the EF-IPF integration of the observing process with its both probabilistic and Information logic may project *artificially designed Observer cognition***.**
The EF logic predicts the conversion of cognition to the IPF Information, memory, and to the Observer cooperative encoding in an artificially designed Observer intelligence. It starts with the persisting Information speeds-frequencies of the attracting observing impulses on the EF-IPF trajectories segments.
The attracting Information sequentially equalizes the Information speeds-frequencies in the attracting resonances which assemble the cooperative (cognitive) logic.
The evolving logic self-organizes each hierarchical level's specific time-space Information logical structural unit that assembles the triplets, builds the IN, and the domains [74].
The assembling Information memorizes the self-organized logic.
The assembling resonance frequencies identify the clock, commanding the logical switching, which coordinates cognitive and intelligence actions. The cognitive functions perform the impulse switches delivering Landauer's energy for memorizing each unit's logic.
The intelligence functions encode the triplets' Bits logic.

The Observer EF-IPF integrates the observing process in the multiple Information process, coordinates and unifies the observing Information in the IPF code.

In the *artificial designed Observer*, where the EF integrates cognitive logic and IPF integrates its encoding Information, the automatic conversion of the EF in the IPF implements encoding the triplet cognitive barriers in intelligence barriers. It allows prediction of the intelligence potential in the evolving Observer.

The IPF frequencies initiate the Observer clock time course which the EF prognosis in the observation.

The time course sustains the EF-IPF integration in an optimal observing process of Information dynamics, and maintains the optimal Observer DSS double spiral structure which finally encloses the predicting Observer code. (Each persisting time interval encloses the entropy of the interval impulse, Sec. 3).

The Observer triplet code memorizes the Observer cooperative Information structure which encloses multiple rhythms of the local structural units. The DSS coding structure memorizes total collected Observer quantity and quality Information, which determine the Observer cooperative complexity.

The Observer assembled Information self-organizes the functions of cognition commanding the encoding intelligence on all hierarchical levels. These functions self-connect the local codes in the Observer code, which encodes all these structures in the space-time Information structure of Information Observer.

The EF-IPF observing process and its Information dynamics artificially design the DSS.

The DSS measures the total Information IQ of this Observer. The DSS code integrates each Observer's IQ. The difference of these IQs measures the distinctness of their intelligence. and the potential of Intelligence.

The maximal Information, obtained in the observation, allows designing the DSS with the maximal achievable IQ measure in the optimal AI Observer and the potential score of IQ Intelligence.

The space-time Information structure, enclosing the encoded EF-IPF, integrates the observed Information in the analytically designed AI Information Observer (Fig. 8).

The observing Information of a particular Observer is limited by the constraints of each observation [41].

The constraints also limit the conversion of the observing process to the Information process.

The thresholds between the evolving stages of the observation limit the stages' evolution, which can stop it at any stage in Sec.3,[74].

All these limit the integral cognitive Information and the following intellective actions, which also limits the amount of Free Information that reduces ability to make the intelligent IN's connections (see also Sec.2.1.4).

### 5.5. Essentials of the Information Cognition

The Observer logical structure possesses both virtual probabilistic and real Information causality and cooperative complexity (Sec.4.2).

A virtual Observer, forming the rotational space-time displacement of the impulse's opposite actions during virtual observation, starts accumulating virtual Information by temporally memorizing it in probabilistic logic, initiating cognitive movement.The rotating cognitive movement connects the impulse microprocess with the Bits in the macroprocess. The microprocess disappears automatically when the probability of the impulse borders approaches zero (Sec.3), and certain logic arises. It means the microprocess from that border has *no going past*, as well as the microprocess cognitive movement. The forming macroprocess composes the triple macrounits through Free Information, which assembles each evolving IN. The INs ending triplets integrate multiple nested IN's Information logics in the assembling Information domains with the evolving growth of the quality Information.

The Observer's cognitive dynamic movement models the Observer's hierarchical rotation mechanism (Fig.3), which enables transferring the Observer through the evolution stages by overcoming the stage thresholds.

The mechanism rotating movement characterizes potential of power $P_{in}(i)$ which measures the current $(i)$ rotating moment $M(i)$ multiplied on angular speed $\omega(i)$ :

$P_{in}(i) = M(i) \times \omega(i)$. (5.1)

This minimal power compensates the resonance movement along each loop of the Observer cognition.

The loop rotates the thermodynamic process with minimal energy of cognitive thermodynamics.

The cognitive movement, upon forming each node and level, processes a temporary loop (Fig. 6) which might disappear after the newly formed IN triplet is memorized. That memorizes the loop logic.

After the memorized Bit emerges during the observation, the rotation movement develops Information form of double helix dynamic movement (Figs.4, 4a).

The Observer's cognition assembles the multiple resonance logical loops, forming in the IN-triple logic hierarchical levels, which accepts only the Information units which each cognitive loop recognizes.

The rotating process in the *coherent* loop *harmonizes speeds-Information frequencies* at different hierarchical levels.

The cognitive functions model the correlated interactions and feedbacks between the IN levels, which the highest domain level's feedback controls. Both cognitive processes and cognitive functions emerge from the evolving interactive observations with their emerging properties.

The discrete impulses provide a discrete Information language for the cognitive logic.

The rotating double spirals (DSS) compose the evolving Information logic by running the macrodynamic process segment's logic. The DSS knots memorize the sequence of process' reversible segments and encode the cognitive process in the Observer time course.

In the rotating DSS, the cognition merges with the natural memorizing of each Bit on all evolution levels.

The cognition emerges in two forms: a virtual rotating movement, processing temporal probabilistic logic, which follows certain Information process's logic rotating in a double helix structure.

The DSS concurrently organizes the observing Information Bits in the IN nodes which end with the knot.

On the observing trajectory, the sequential knots memorize the Information causality and logic and structure of the IN node hierarchy.

These processes start with the elementary virtual Observer and the emerging Bit at the microlevel, which holds the irreversible prehistory and participates in the evolving Information Observer.

The starting cognitive thermodynamic has no actual physical cost.

Results [78] confirm that cognition arises at the quantum level as a kind of "entanglement in time" in the process of measurement, where cognitive variables are represented in such a way that they don't really have values (only potentialities) until you measure them and memorize, even "without the need to invoke neurophysiologic variables," while "perfect knowledge of [a cognitive variable] necessitates uncertainty for the others." Moreover, this analysis shows that both cognition and intelligence have an Information nature.

Recent work in cognitive maps [79] confirms "large-scale internal representations of navigable spaces," showing how cognitive maps are encoded, anchored to environmental landmarks and used to plan routes.

Similar neural mechanisms might be used to form 'maps' of *nonphysical spaces*, and "applied to nonspatial domains to provide the building blocks for many core elements of human thought."

Since certain interaction and reality start a microlvel, following cognition also emerges there as well as neural mechanism dealing with cognition.

## 5.7.The self-forming hierarchical distributed logical structure of cognition

Multiple moving INs, sequentially equalizing the nodes speeds-frequencies of attracting Information logic in a resonance, assembles total Observer logic.

The mutual attracting free Information logic, sequentially interacting, self-organizes the cooperative logical rotating spiral loops in a chain which encloses the observing logic.

Each curved impulse invariant time-space measure $\pi$ enfolds the Information measure $1Nat$ which includes a Bit, Free Information, and the Information needed to encode the impulse Bit.

This Free Information attracts the impulse with intensity ~1/3 Bit per impulse.
The IPF integrates the Bits with Free Information connecting the sequence in information trajectory.
The minimax principle, applied along the IPF extreme trajectory, maximizes the Information enclosed in each current impulse, squeezing its time interval.
The growing attracting Free Information along this trajectory minimizes the time interval between the nearest impulses proportionally to 1/3Bit. For each third impulse, that interval of Information distance becomes proportional to 1 Bit, preserving the impulse's invariant Information and the time-space measure $\pi$. By the end of the IPF integration, all integrated Information is concentrated in a final impulse, whose Information density approaches the maximal limit. Since Free Information encloses Information logic, the multiple Bits with triple growing density increase the process's Information logic.
The IPF integral Information with its logic is condensed in the last integrated impulse time-space interval volume. For multiple Information impulses, each third curved impulse having invariant measure $\pi$ appears in the Information process with time frequency $f_i = k_i, k_i = 3,5,7,9,...$, which indicates the entrance of the triplets and their specific sequences.
The invariant impulse time measure $\tau_i = \pi / \sqrt{2}$, the impulse flat surface space measure $l_i = \sqrt{2}$, and the orthogonal to surface space measure $h_i = \pi$ determine geometrical volume $v_i^S = \tau_i \times l_i \times h_i = \pi^2$ of each $i-$ three-dimensional impulse.
Information time-space density $D_i^I = k_i Nat / v_i^s$, concentrating $k_i$ Nat for each third impulse, increases with growing $k_i$ while the time-space geometrical volume $v_i^S$ holds the invariant geometrical density.
The Information density measure $D_i^I = k_i Nat / \pi^2$ grows only with each $k_i$.
Let us evaluate the relative Information density of each triplet's Bit encoding Information in its relative time-space interval $u_k = 1, 1/3, 1/9, ....1/k_i, ...,$.
Since each triplet's Bit encodes 3 Bits in its starting relative time-space interval $u_1 = 1$, its Information density is $N_d^1 = 3$.The following triplet also encodes 3 Bits in interval $u_2 = 1/3$, but 3 Bits are already encoded from the previous triplet's Bits. Thus, the Information density of such two triplets equals $N_d^2 = 3^2$ and so on. Hence, for the $m$-th triplet density is $N_d^{mn} = 3^m$.
The density of the process's dimension $n$ with $m = n/2$ triplets is $N_d^{m=n/2} = 3^{n/2}$.
Thus, each current impulse time-space geometry encloses Information, its density and frequency, concentrating Information logic, and the Information of all previous impulses along the Integrated Information Path [48].
The EF extreme trajectories, starting from the multi-dimensional observing process, the EF-IPF transformation converts to the multi-dimensional orthogonal processes (Sec. 2.1) whose curved impulses hold the above Information measures.
The EF-IPF space-time extremal trajectories rotate the forming spirals located on conic surfaces (Fig. 3), which starts from virtual (entropy) process and continues as the Information process.
Since the cutting entropy in the impulse observation converts to the Bit on this trajectory, the trajectory consists of the segments of Information process dynamics and the intervals between segments delivering each Bit to the following segment. On Fig. 3 each segment starts on the cone vertex-point D and ends on point D4, which connects to a vertex of the following cone. The observing Bit is delivered at each cone vertex.

The segment includes the observing impulse with its logical Bit, intervals of free logic, and the correlation connecting the nearest segment, temporarily memorizing the segment logic.
The logical and Information dynamics describe the process of sequential logical interactions of the multiple impulses, rotating with Information speed determined by impulse density $D_i^I$.
The dynamics on trajectory between the cone points D and D4 is reversible and symmetrical described by Hamiltonian equations [46,47].
The logical anti-symmetry brings the anti-symmetrical logical Bit prior to the interaction with an external impulse which starts delivering the external energy.
This Bit is supplied at each cone vertex, as well as the interactive impulses, with intervals currying the Bit, the intervals memorizing the Bit.
After the external energy generates the physical multiple Bits, the physical Information process starts.
According to [37], the moment of appearance of the interactive logical Bit from beginning of the impulse is $t_{11} = 0.2452$, which defines interval $\Delta t_1 = 0.2452/1.44 \cong 0.17$ relative to invariant information measure of each impulse. $1.44 Bit = 1 Nat$.The time interval for memorizing the Bit $\Delta t_B$ identifies the Bit information measure $\ln 2$ which is the equivalent of the invariant impulse relative time part $\Delta t_B = \ln 2/1.44 = 0.481352$. For impulse relative measure $\| 1 \|_M$, the relative difference $1-(0.17+0.481352) = 0.348648$ includes intervals of supplying external energy, erasing the asymmetric logical Bit, memorizing this Bit, and interval of Free Information logic $\Delta t_{fo} = 0.23/1.44 \cong 0.1597$.

The deduction brings interval $0.188948$ which includes the interval following the opposite asymmetrical interaction $0.01847$. The interval of interaction $0.01847$ disappears at the end of the encoding interval $\Delta t_{eno} = 0.188948$. Therefore, the interval of encoding is $\Delta t_{en} = 0.188948 - 0.01847 = 0.17039$ which approximates the time interval of the appearance of the asymmetrical logical Bit $t_1$.
The encoding Bit releases the Free Information interval $0.01847$ emerging after other impulse is encoding.
The external impulse, erasing the asymmetric logical Bit, spends interval $\Delta t_B$ and ends with the interval of encoding $\Delta t_{en}$ for each invariant impulse. Along the IPF trajectory, the moment of creation logical Bit, which ends forming the triple logic follows interval $\Delta t_B$ that binds the free logic the triplet.
The interval of memorizing the physical Bit requires the same interval $\Delta t_B$ during which the entropy of logical Bit erasure. The interactive impulse with external energy, supplied in time interval $t_{\Sigma b} = 0.481352 + 0.18948 = 0.67083$, includes both erasure of the logical Bit and needed for its encoding. Since the external impulse interactive part measures $0.025 Nat$, it brings the total $t_{\Sigma bo} = 0.67083 + 0.025 = 0.69583 \cong \ln 2$ measuring the interval of supplying the external Bit.
(It also numerically confirms the direct connection of the Information and its time measures.)
Since each impulse's curved measure $\pi$ with its time interval $\Delta t_1$ appears in the Information process with frequency $f_{10} = \pi$, the relative frequency is $f_1 = (0.17/\pi) \times \pi = 0.17$.
The frequency of spectrum $\omega_1 = 2\pi f_1 = 1.068$ identifies the time of opening supply of the external energy, memorizing the Bit. The time interval of memorizing the Bit $\Delta t_B$ identifies the frequency $f_B$ of the appearance of that interval within the impulse with the frequency of spectrum $\omega_2 = 2\pi f_B = 2\pi \ln 2/1.44 = 3.02 < \pi$.

The time interval of the impulse encoding $\Delta t_{en}$ determines the spectrum frequency $\omega_3 = \omega_1$, estimating also the process frequency $f_1$. Therefore, the frequency spectrum $\{\omega_1\omega_2\omega_1\}$ initiates the frequency of the appearance of the logical Bit, following the frequency of the memorized Bit and the frequency of encoding, which equals $\omega_1$.

Spectrum $\{\omega_1,\omega_2,\omega_1\} = \omega_o, \omega_o = (1.068, 3.0.2, 1.068)$ delivers the logical Bit, energy memorizing it, encoding the Bit, which includes the Free Information attracting the Bit along the trajectory

Or, these frequencies approximate spectrum $\omega_o = \{1, 28277, 1\} \times 1.068 \approx \{1, 28277, 1\}$.

After supplying the external energy during the sum of the impulse' invariant intervals, the impulse becomes the segment of a physical Information process.

Therefore, physical dynamics describe the IPF extremal trajectory rotating on sequential cones (Fig. 3). Each cone vertex encodes the Bit memorized with the frequency $\omega_2$ delivered from a previous impulse-segment with the frequency of a logical Bit $\omega_1$. Hence, each physical Information impulse carries spectrum $\{\omega_1\omega_2\omega_1\} = \omega_o$, while their sequential pair on the trajectory carries the impulses spectrum $\{\omega_1, \omega_2, [\omega_1 = \omega_1], \omega_2, [\omega_1 = ...]\} = \omega_\Sigma$, where $[\omega_1 = \omega_1]$ is the resonance frequency for two impulses whose distance is shortening by 1/3.

That allows closely connecting the impulses in the resonance.

Along the trajectory, each of these pairs appears with the growing frequency $f_{io} = 1/k_i, k_i = 3, 5, 7, ...$

Since the fixed time intervals $\Delta t_1$, $\Delta t_B$, $\Delta t_{en}$ are relative to the invariant impulse measure, they are repeating for each invariant impulse with the increasing Information density and growing frequency.

Thus, along the extreme trajectory, each third impulse will deliver triple frequency of spectrum $\{\omega_1, \omega_2, [\omega_1 = \omega_1]_{\Delta t_{10}}, \omega_2, [\omega_1 = \omega_1]_{\Delta t_{20}}, \omega_2, [\omega_1 = \omega_1]_{\Delta t_{30}},\} = \omega_{\Sigma 10}$ with related time intervals $| \Delta t_{10}, \Delta t_{20}, \Delta t_{30} |$. .

These time intervals are sequentially proportional to the distance between each Bit in the ratio $1/k_i$, or the invariant time measures of these impulses $\tau_i = \pi / \sqrt{2}$ are shortening in the sequence $\Delta t_{10} = \pi / 3\sqrt{2}, \Delta t_{20} = \pi / 5\sqrt{2}, \Delta t_{30} = \pi / 7\sqrt{2}$.

In the sequentially shortened distances between the impulses on the extreme trajectory, each such three impulses (with their Free Information) assemble in a triple of the resonance frequencies.

The triple resonance frequencies, in collective resonance, assemble the trajectory of the triplet segments. Sum of these interval assembling triplet equals $\sum_{ko=1,2,3} \Delta t_{ko} = \pi / 2.09 \cong 0.478\pi$.

Adding the interval of binding the knot $\Delta t_{kn} = 0.025\pi$ we get $\sum_{ko=1,2,3} \Delta t_{ko} + \Delta t_{kn} = 0.503\pi \cong \pi / 2$.

The trajectory of the three not shortening time segment turns on $3\tau_{io} = 2.12\pi$ or on a circle assembling each three segments in the triplet loop.

### 5.8. Self-forming triplet logical structures and their self-cooperation in the IN hierarchical logic

In the multi-dimensional observing process, a minimum of three logical Bits with free logics can appear, which, attracting each other, would cooperate in a logical triple.

Multiple probabilities of interacting impulses (in this multi-dimensional process) produce the numerous frequencies. Some of those, a minimum of three, can generate an attractive resonance, cooperating in a triple. This triple logic starts temporarily memorizing two sequential pair cross-correlations during their time of correlation.

A locally asymmetric cross-correlation memorizes the asymmetrical logic during this correlation process.

Comments
As reported recently [81], such anti-symmetric cross-correlations have been observed. •

When this cooperating process is ending, the triple correlations temporarily memorize the triple logical Bits, and the minimal entropy of cross-correlation $\ln 2$ can be memorized at a cost of the equivalent minimal energy of the logical Bit. This is the Information cost of memorizing the triple logical Bit, which includes the free logic.

Comments.
Recent publication [82] experimentally confirms our theory proof of Landauer energy squzing factor in the logical irreversible microprocess-quantum. •

The attracting free logic of the emerging three logical Bits starts the Bit's self-cooperation in the following sequence.
The free logic of the emerging logical Bit holding the frequency $\omega_2$ attracts next logical Bits toward a resonance with the equal frequency of next Bit's free logic, assembling the two in joint resonance.
This resonance process links these Bits in duplets.
The free logic from one Bit of the pair gets spent on binding the duplet.
The free logic from the duplet's Bit attracts the third Bit and binds all three in a knot Bit, creating the triplet logical structure. The knot Bit still has Free Information, and it is used to attract a different bound pair of emerging Bits, creating two bound triplets.
This process continues creating nested layers of bound triplets, three triplets and more (Figs. 4-6).
Hence the triplet logical structure creates the resonance frequencies of the attracting logic, joining the triple Bits. The free logic attraction toward the triple resonance of their equal frequencies is the *core Information mechanism* structuring an elementary triplet.
The trajectory of the forming triplet describes the rotating segments of their cones (Fig. 5), whose vertexes join the triplet knot and start the base of the following cone. When the next rotating segment starts, the knot frequency joins the cone vertexes in resonance along the cone base. It connects the next triplet in the resonance, and so on, creating the nested layers of a logical space-time network (IN), where the knot hierarchy identifies the nested nodes of the IN hierarchy.
Each triplet unit generates three symbols from three segments of Information dynamics, and one when the segment attracting triple logic binds the three in the logical triplet knot.
These symbols can produce a triplet logical code, while the knot logic symbol binds the triple code, potentially encoding all triples. Each assembled knot enables releasing its Free Information logic, which transfers this triple logical code to the next triplet node.
Thus, the nodes logically organize themselves in the IN logical code. The network, built through the resonance, has limited stability. Therefore, each IN encloses a finite structure.
The external energy, encoding the IN triple logical code, sustains the frequency spectrum mentioned above.
The IN emerging logical structure carries the triple information code on each node's space-time hierarchy, and the last triplet in the network collects and encloses the entire network's Information code.
The observing process enables self-building only limited multiple INs through the Free Information of its ending nodes. The final triplet in every network contains the maximum amount of the enclosed Free Information. The limited networks develop self-connection through the attraction of their *ending triplets.*
Even after each IN potentially loses stability, evolving in chaos, it possesses the ability of self-restoration (Sec.4.6).
The multiple INs self-cooperate in a hierarchical domain, starting as each three ended triplets' Free Information assembles a knot which joins this IN's triples in resonance. This IN ending knot's Free Information resonates with the other three INs ending Free Information, forming a triplet structure analogous to the elemental triplet. This high-level triplet joins these three INs, structuring the next IN of the domain hierarchy. The hierarchical logical trajectory describes the space-time spiral structure (Figs. 7.), evolving in observations. This hierarchy enables generating sequential triple codes located on the rotating trajectory of the cone vertexes, which are spatially distributed at the different hierarchical levels of the multiple IN and the domain hierarchy.
Such a discrete space-time code (DSS) integrates the observing process in space-time Information geometry, self-organizing an Observer.
Still the question is: *What self-organizes the structuring of the Information units in the Observer geometrical space-time shape during their movement along the observing trajectory?*

### 5.9.The Observer wave function self-forming hierarchical distributed logical structure of cognition

The self-creating units of the hierarchy generate frequency delivering spectrum $\{\omega_1,\omega_2,[\omega_1=\omega_1]_{\Delta t_{10}},\omega_2,[\omega_1=\omega_1]_{\Delta t_{20}},\omega_2,[\omega_1=\omega_1]_{\Delta t_{30}},\}=\omega_{\Sigma 10}$ which is growing in the triple sequentially shortening intervals $|\Delta t_{10},\Delta t_{20},\Delta t_{30}|$ for each $i$ trajectory of the segment.
The answer on the previous question specifies Propositions 5.1 below with the following initial conditions.
The space-time spiral trajectory of the EF extremal (Fig. 3) describes the sequence of multi-dimensional curving rotating segments, representing the interacting impulses of the observing process, which integrates the observing process's logic. Each segment's impulse with invariant entropy measure $1Nat$ moves along the trajectory, rotating the impulse with invariant geometrical measure $\pi$.

This curved impulse's three-dimensional measure includes time coordinate measure $\tau_i=\pi/\sqrt{2}$, flat surface's space coordinate measure $l_i=\sqrt{2}$, and space coordinate measure $h_i=\pi$ orthogonal to both of them. The Information measure $1Nat$ includes the impulse logical Bit $\ln 2$ and the free asymmetric logic measure $f_{li}=1-\ln 2\cong 0.3Nat$ on each trajectory segment. The logic density per each third segment volume $v_i^S=\tau_i\times l_i\times h_i=\pi^2$ increases according to $D_i^I=k_iNat/v_i^s, k_i=3,5,7,...$
The asymmetric logic divides the sequential segments by bridges: logical, Information, and physical barriers. Between segments, the bridges transfer the anti-symmetrical interval $\Delta t_1$ of interactive logic, following the barriers forming interval $\Delta t_B$ of memorizing the Bit, and interval $\Delta t_{en}$ of encoding and releasing the Free Information. These three intervals identify the locations of bridges and barriers along the space-time trajectory. Each sequential location on the trajectory repeats this triple with invariant frequency spectrum $\{\omega_1,\omega_2,\omega_1\}=\omega_o,\omega_o\cong(1.068,3.0.2,1.068)$.
The ratio of the spectrum's adjacent parts to its middle part repeats with the frequency $f_{io}=1/k_i$, where $k_i$ indicates each third density logic per segment. Thus, along the trajectory, the repeating invariant triple frequencies of the spectrum prognosis appearance of each logical bit, it's memorizing and encoding.

*Propositions* 5.1
Along each of $i$-dimensional space-time segment rotates *three-dimensional space wave function* on the cone external shape, Fig. 3 with the following rotating speeds:
(a) around each spiral cross-section $\alpha_i^s=1[square/radian]$, or $\alpha_i^{s_o}=\pi/radian$, and
(b) orthogonal to this rotation space speed $\alpha_i^h=1[volume/radian]$, or $\alpha_i^{h_o}=\pi/radian$.
Accordingly, the related frequencies of each orthogonal rotation are $\omega_i^s=\alpha_i^s/2\pi,\omega_i^{so}=1/2$ and $\omega_i^h=\alpha_i^h/2\pi,\omega_i^{ho}=1/2$.

Each $i$-dimensional segment's cross-sectional rotation spins the rotation on the space interval $\pi$ of the segments invariant measure. The three-dimensional wave function distributes the space rotation along the trajectory's segments with the above invariant speeds, delivering the invariant spectrum $\{\omega_1, \omega_2, \omega_1\} = \omega_o, \omega_o \cong (1.068, 3.0.2, 1.068)$.

*Proof.* We use the equation of a wave $u = F(vt - x)$ depending on velocity of movement $v$ and distance $x$ of the movement along a trajectory. This equation we apply to a wave function $u(u_s, u_h)$ whose component $u_s$ describes the rotation of the wave running along its forming cross-section square $s_i^w = \tau_i \times l_i = \pi$, and component $u_h$ running along the rotating orthogonal space length $h_i = \pi$.

The wave function's $u_s = F(f_{ws})$ argument $f_{iws} = \alpha_i^s \rho_i^s - s_i^w$ describes a two-dimensional rotation of the trajectory rotating $i$-dimensional cone's radius $\rho_i^s$ of the cone's cross-section with speed $\alpha_i^s$, where $\rho_i^s$ is the analog of the curved distance moving to reach cross-section's square $s_i^w = \pi$.

From the geometry of rotating movement follows that each $i-$segment is rotating along trajectory on the cone square-basis reaching angle $\varphi_i^s = k_{io}\pi_{io}$, where $k_{io} = 1, 2, 3,..$ is sequence of the cone basis for segment $1, 2, 3,.., i, ...$ (Figs. 3, 8). This rotation reaches distance $s_i^w = \pi$ at $f_{iws} = \alpha_i^s \rho_i^s - s_i^w = 0$ when radius $\rho_i^s$ rotating the wave cross-section with speed $\alpha_i^s$ reaches the segment geometrical measure $\pi$. At $s_i^w = \pi$ and $\rho_i^s = \pi$, it determines $\alpha_i^s = 1$. Since, the wave function argument $f_{iws}$ reaches $f_{iws} = 0$ by the end of each $i$-segment movement with period equal to impulse measure $\pi$, the wave function $u_s$ is periodical with period $\pi$.The wave function $u_s$, moving along its cross-section with speed $\alpha_i^s = 1[square / radian]$, or $\alpha_i^{s_o} = \pi / radian$, holds the related frequency $\omega_i^s = \alpha_i^s / 2\pi$.

The wave function's $u_h = F(f_{wh})$ argument $f_{iwh}$ moves along the segment length $h_i^w$ with space-rotating speed $\alpha_i^h$ to reach the impulse volume $v_{ih}^t$ according to Eq. $f_{iwh} = \alpha_i^h h_i^w - v_i^w$. The movement reaches volume $v_i^w = \pi^2$ at $f_{iwh} = \alpha_i^h h_i^w - v_i^w = 0$, $h_i^w = h_i = \pi$ with speed $\alpha_i^h = \pi^2 / \pi = 1[volume / radian]$ and related frequency $\omega_i^{ho} = 1/2$. Thus, the wave carries the frequency $\omega_i^{ho} = 1/2$ along the spiral trajectory and the equal frequency $\omega_i^{so} = 1/2$ along its cross-section. Or, each $i-$rotation with frequency $\omega_i^s$ equal to frequency $\omega_i^h$ brings the space rotation during the cross-sectional rotation, or vice-versa.

Since the wave function argument $f_{iwh}$, decreasing along the spiral trajectory, reaches $f_{iwh} = 0$ by the end of each $i$-segment, with the period equals impulse measure $\pi$, the wave function $u_h$ is also periodical, with period $\pi$.

The arguments of these orthogonal components of the wave function connect the relation $\arg(F) = f, f = f_{ws} \times f_{wh}$ •.

*Therefore, the space-time trajectory, starting on the cone shape basis and moving on the cone's external shape, reaches the cone vertex when the projection of this trajectory reaches the base shape center* (Fig. 3).

When each $i-$rotating segment reaches the cone vertex it develops a logical bridge between $i$ and the next $i+1$ segment on the moving space-time trajectory.

The bridge holds the relative interval $0.00653$ of this logic [53]. Depending on the time course on the moving trajectory, the logical bridge moves to a memory barrier following an encoding barrier.

During the movement along the trajectory, the cone segments $i, i+1, i+2$ join a triple whose bridges and barriers develop sequence of assembling units: a logical triple, a memorized triplet, following an encoding triplet.

Moreover, since these triplet units sequentially join in a node of the IN, forming according to the time course, the logical, the memorized, and the encoding structures, these structures become the triplet' nodes of the related INs.

One scenario illustrating the assembling trajectories of the space-time triplet is shown in Fig. 6.
The knots of the triplets, cooperating in the IN, are shown in Fig. 7. The bridges at moment $t_1, t_2$ (Fig. 4) form as the double segments assemble. At moment $t_3$ they join in the triplet's knot. •

*Proposition* 5.2

Let us consider $i, i+1, i+2$ three-dimensional segments along the multi-dimensional rotating segments on the extreme trajectory. Each of these triple belongs to the related dimensions of the multi-dimensional trajectory with the intimal conditions above, and the distributed triple logic starting on each of the segments locations $|\Delta t_{10}, \Delta t_{20}, \Delta t_{30}|$ to become the triplet's and encoding Bits.

Along the extreme trajectory, each segment of equal measure $\pi$ increases density proportional to the segment's shortening invariant intervals $|\Delta t_{10}, \Delta t_{20}, \Delta t_{30}|$ on their locations along the trajectory.

From these locations, these segments deliver the related invariant spectrum $\{\omega_1, \omega_2, [\omega_1 = \omega_1]_{\Delta t_{10}}, \omega_2, [\omega_1 = \omega_1]_{\Delta t_{20}}, \omega_2, [\omega_1 = \omega_1]_{\Delta t_{30}}, \} = \omega_{\Sigma 10}$ , $\{\omega_1, \omega_2, \omega_1\} = \omega_o, \omega_o \cong (1.068, 3.0.2, 1.068)$ through the cross-section rotation, which is speeding the space rotation that distributes the spectrum along these three-dimensional space segments.

*Then*, the wave function's frequencies synchronize the triple segment's logic in a collective resonance.

The sequentially forming triple bridges are squeezing the initial observing multi-dimensional process' segments, first to three-dimensional rotation forming Bits'triple, and finally to a single dimensional Information process barrier encoding the Bits of all multiple knots. •

*Proof*. The wave consecutive three-dimensional space movements picks segments $i$, $i+1$, $i+2$ sequentially from each of these segments' trajectory-specific locations in these dimensions, and simultaneously starts rotating each of them during interval $|\Delta t_{10}, \Delta t_{20}, \Delta t_{30}|$, placing these shortening intervals between segments $i$, $i+1$, $i+2$ accordingly. The densities increase proportionally to the shortening-squeezing of time interval measures along each of these trajectory dimensions.

The first of the wave three-dimensional rotation moves $i$ segment rotating during interval $\Delta t_{10} = 1$ (equivalent to space interval $\pi$ with density proportional $k_i = 1$). The second of wave three-dimensional rotation moves $i+1$ segment during interval $\Delta t_{20} = 1/2\Delta t_{10}$ (equivalent to space interval $\pi$ with density proportional to $k_i = 2$). The third of wave three-dimensional rotation moves segment $i+2$ during time interval $\Delta t_{30} = 1/3$ (equivalent to space interval $\pi$ with density proportional $k_i = 3$).

With growing Information density along the trajectory, these three-dimensional movements repeat the shortening these intervals for each triple segment with increasing frequency $f_i = k_i, k_i = 3, 5, 7, ...$

Since each of the segments deliver the equivalent spectrums, the equal frequencies of the sequential segments' spectrum $\{\omega_1, \omega_2, [\omega_1 = \omega_1]_{\Delta t_{10}}, \omega_2, [\omega_1 = \omega_1]_{\Delta t_{20}}, \omega_2, [\omega_1 = \omega_1]_{\Delta t_{30}}, \} = \omega_{\Sigma 10}$ are synchronized during the sequence of these time intervals.

According to the Proposition 5.1 initial condition, the invariant spectrum frequency $\omega_1$ repeats time interval $\Delta t_1$ of the logical anti-symmetrical interaction on a bridge separating $i-1$ and $i$ segments on the trajectory. The end of this interval indicates the beginning of the time interval $\Delta t_B$ on $i$ segment repeating with frequency $\omega_2$. During time $\Delta t_B$ the segment Bit is memorized. The end of $\Delta t_B$ indicates beginning of time interval $\Delta t_{en}$ of Free Information logic, which identifies beginning of barrier separated $i$ and $i+1$ segments. The Free Information attracts the separated segments.

The time intervals of the sequentially squeezing segments hold, first, the double synchronization during interval

$\Delta t_{20} = 1/2\Delta t_{10}$, and second, the double synchronization during interval $\Delta t_{30} = 1/3\Delta t_{23} = \Delta t_{20} - \Delta t_{30} = 1/2 - 1/3 = 1/6$. The sum $\Delta t_{33} = \Delta t_{20} + \Delta t_{23} + \Delta t_{30} = 1/2 + 1/6 + 1/3 = 1$ is equal to the first interval $\Delta t_{10}$, during which all doublets are forming. Three segments asymmetrical logic sequentially attract the synchronizing doublets during the rotation movement.

The wave function's frequencies synchronize the triplet logic in collective resonance.

The rising Information attraction on these time intervals adjoins the synchronized interval Information in a triple during the $i$ dimensional interval $\Delta t_{10} = 1$. Forming the triplet completes Free Information which delivers each $i+2$ segment with triple frequency while holding the invariant spectrum.

Free Information of the triplet joins the three memorized Bits in a triplet, where during an additional interval of Free Information $0.01847$ the Bits are encoded in the triplet knot-barrier .

The frequencies of the shortening time intervals distribute the orthogonal space rotations along the segments of the multiple dimensional observing trajectory which is moving in a *three-dimensional space wave function* for each of this trajectory's multiple dimensions.

Each of the three dimensions' shortening time intervals, which the three-dimensional rotation moves, brings the triplet knot that joins the three dimensions to one.

Applying the sum of the two shortening geometrical intervals, assembling a triplet at angle $\cong \pi/2$ to the segments located on three independent orthogonal dimensions $3\pi/2$, leads to squeezing these dimensions to one on the triplet knot. Whereas, at forming the segments' triple during resonance, the trajectory is assembling each three segments in a cyclical loop.

The sequentially forming triple knots *squeeze the initial observing multi-dimensional process first to three-dimensional rotation, and then to single-dimensional Information process encoding the Bits of all multiple knots. That transforms the initial random observation to classical cerain dynamics.*

Squeezing dimensions accompanies sequentially memorizing and encoding the IN hierarchical levels nodes That shortens number of the IN cognitive levels, releasing cognitive logical loops memorized in the encoding knot.

Finally, the periodical wave function includes the sequence of repeating arguments along both orthogonal rotations: $u_{sh} = u_s \times u_h$, $f_{ws} = \{f_{iws}\}$, $f_{wh}\{f_{iws}\}$, which performs the multiple three-dimensional movement with *three-dimensional space wave functions*.

The movement distributes the extreme trajectory segments on structuring the space located information networks which join the synchronizing triplets in assembling knots and compose multiple structure of the information Observer. The shape of the multiple wave functions describes the extreme multi-dimensional trajectory, formalizing the minimax observation process, which models rotating segments on cones (Fig. 3), and continues on the spiral cone structure DSS analogous to Figs. 7, 10. •

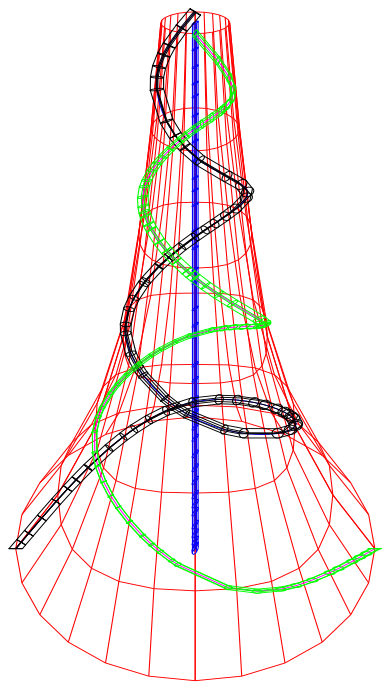

**Fig.10. Simulation of the 3-dimensional double spiral cone structure shortening the dimensions**

*The spin-rotating trajectory of the invariant time space impulses builds geometry of information Observer.*

*The wave function frequencies and properties*

1. The wave function with the above speeds and frequencies emerges in the observation process when a space interval appears within the impulse microprocess during a reversible time interval $\varepsilon_{ok} = 0.015625$ of the impulse invariant measure $\pi$ equivalent to $1Nat$. Before that, the observing trajectory has described the probabilistic time function whose probability $P_{\Delta}^{*} \cong 0.821214$ indicates the appearance of a space-time probabilistic wave. During the probabilistic time observation, the entropy *a priori-a posteriori* probabilities measures the probabilistic symmetric logic of a sequence of these probabilities.
Thus, the wave function starts emerging in probabilistic observations as a probability wave in a probability field.
At the beginning of the microprocess, the probabilistic wave measures only the time of its propagation.

2. Within the microprocess, the asymmetrical logic emerges with the appearance of a free logic interval $\Delta t_{fo} \cong 0.1597 \cong 1/2\pi$ which, repeating with equal wave frequency $\omega_i^s$, indicates the beginning of the interactive rotating asymmetry on a primary bridge and segment. From that point, the observation logic on the trajectory becomes the asymmetric part of total free logic $f_{li} = 1 - \ln 2 \cong 0.3Nat$. The asymmetric *logical wave* emerges. The asymmetric logic probability appears approaching $p_{\pm a} = \exp(-2h_{\alpha}^{o*1}) \cong 0.9866617771$, with hidden asymmetrical Bit concealing a certainty-reality. Such logic temporary memorizes correlation with the probability which carries a logical Bit of the certain logic.
Such a certain logical Bit may carry energy in a real interactive process, described the Markov diffusion process.
The path to creation of the certain Bit includes an increment of the probability $\Delta P_{ie} = 0.9855507502 - 0.981699525437 \cong 0.004$ starting injection of energy from an interacting impulse of the Markov process (Sec. 3.5). Thus, the certain free impulse logic carries the certain logical attraction.
The wave function in the microprocess is probabilistic until the certain logical Information Bit appears.
The certain asymmetrical logical Bit become physical Bit through erasure the entropy of this logic, which allows replacing the logic by memorizing its Bit.

3. The wave function starts on the observing process which the EF extreme trajectory prognosis, carrying the probabilistic wave which transforms the observing process to the certainty of real observation.
The spinning movement of the space-time trajectory describes the invariant speed encircling the cross-section of its rotating impulses-segments, which spreads the invariant rotation space speed along the segment trajectories.
The segment's invariant spectrum $\{\omega_1, \omega_2, [\omega_1 = \omega_1]_{\Delta t_{10}}, \omega_2, [\omega_1 = \omega_1]_{\Delta t_{20}}, \omega_2, [\omega_1 = \omega_1]_{\Delta t_{30}}, \} = \omega_{\Sigma 10}$ repeats the triple frequencies of these three-time intervals between them. That shortens the distance of the equal spectrum frequencies and assembles them in a resonance creating joint logical structures-triplets up to the IN hierarchies and domains. The frequency absolute maximum indicates the finite end of its creation. A minimal energy of the resonance supports the forming logical loop. •

## 5.10. Distribution of the space-time hierarchy

1. The hierarchy of self-cooperating triplet units distributes the space rotation emerging along the EF segments of the time-space extreme trajectory, where each third impulse progressively increases the Information density measure of its Bit in the triple. The time-space hierarchy of the units starts emerging in observation of the symmetrical logic at the appearance of the space interval in the microprocess. This logic self-forms a hierarchy of the logical unit structures through the impulse's mutual attracting free logic which, sequentially attracting the moving unit's speeds, equalizes their frequencies in resonance that assembles the Observer logic along the hierarchy of units.
2. The hierarchy of the logical cooperating units becomes asymmetrical with appearance of certain logical Bit on the extreme trajectory.

The repeating free logic interval indicates the wave frequency $\omega_1 = f_i^s = 1/2\pi$.

The EF rotating trajectory of three segments equalizes their Information speeds joining in the resonance frequency during the space rotation, which cooperates each third logical Bit's segment on the trajectory and logically composes each triplet structure in the unit space hierarchy.

3. The appearance of the asymmetrical logical Bit on the extreme trajectory indicates entrance the IPF Information measure $\ln 2$ on its path to forming a logical Bit. The path starts on the relative time interval $\Delta t_{fo} = 0.23/1.44 \cong 0.1597$ of the logical asymmetry, which identifies the segment bridge.

During the triple impulses, the third time interval $\Delta t_{3r} = 3\Delta t_{1r} = 3/2\pi \cong 0.4775152$ indicates the end of the triple cooperative logic, starting to build the triplet knot. Forming the triplet knot requires a time interval, during which the triple free logic binds in the triplet Bit.

The time interval of creating the Bit approaches $\Delta t_B = \ln 2/1.44 \cong 0.481352$. The difference $\Delta t_B - 3\Delta t_{3r} \cong 0.004$ evaluates the time of binding the triplet. Thus, the wave space interval delivers the logical Bit with the wave spectrum frequency $\omega_2 = 2\pi\Delta t_B = 2\pi \ln 2/1.44 = 3.02 < \pi$, while the triplet knot repeats with the spectrum frequency $\omega_{20} = 2\pi 3/2\pi = 3$.

4. Delivering external energy for memorizing the logical Bit identifies relative moment $t_1 = 0.2452/1.44 \cong 0.17$ ending the interval of the asymmetry. By this moment, the resonance frequencies of the asymmetrical logic have already been created.

Along the IPF path on the trajectory, this moment follows interval $\Delta t_B$ of creation the logical Bit, ending the emergence of the knot that binds the free logic. The interval of memorizing the physical Bit requires the same interval $\Delta t_B$ during which the entropy of logical Bit is erased. The necessary external impulse, erasing the asymmetric logical Bit, starts with interval $\Delta t_B$ and ends with the interval of encoding the Bit $\Delta t_{en} = 0.17$. The external energy, supplied on time interval $t_{\Sigma b} = 0.481352 + 0.19 = 0.671352$, includes both erasure of the logical Bit and its encoding. Whereas the interval of Information free logic $\Delta t_{fo} = 0.23/1.44 \cong 0.1597$ is left for attracting a new Bit (at interaction with an external impulse carrying energy). Since the external impulse interactive part is $0.025$, it brings total $t_{\Sigma bo} = 0.67083 + 0.025 = 0.69583 \cong \ln 2$ for the interval of the external Bit.

Therefore, the frequency spectrum, initiating the encoding, equals $\omega_1$ in sequence $\{\omega_1 \omega_2 \omega_1\}$. This triple sequence identifies the segments alternating on the trajectory with the repeating ratio of the bridge-middle part-starting next bridge- barrier, which measure the barrier relative interval $\Delta t_{en} = 0.17$.

Thus, the sequence of segments on the EF-IPF extreme trajectory carries its wave function's frequencies which self-structure the space-time unit of the logical Bits hierarchy that self-assembles the Observer logic. The logic controls memorizing and encoding physical Bits as well as the hierarchical structure of the space-time Information geometry of the units.

5. The segments-impulses on the EF spiral trajectory sequentially interact through the frequencies repeating on the bridge time-space locations connecting the segments in the trajectory. The segment sequences on the EF-IPF extreme trajectory (Figs. 3, 4) carry its wave function frequencies, self-structuring the unit logical Bit hierarchy that self-assembles the total Observer logic. This logic controls memorizing and encoding of both physical Bits and the hierarchical structure of their units. •

The Observer cognitive logic encloses both probabilistic and Information causalities distributed along all Observer hierarchies. The logical functions of the self-equalizing Free Information in the resonance perform the cognitive functions, which are distributed along the hierarchy of assembling units: triplets, IN nested nodes, and the IN ending nodes. These local functions self-organize the Observer cognition.

Assembling runs the resonance frequencies $[\omega_1 = \omega_1]$ spreading along this hierarchy.

Each unit, ending high level hierarchical structure encloses all its Information logic, whereas the high unit' impulse invariant time-space interval, containing this Information, increases more Information density than the unit of lower level hierarchy. The resonance frequencies of spectrum $\{\omega_1, \omega_2, \omega_1\} = \omega_o$, holding the cognitive logic loop, self-creates the unit hierarchy.

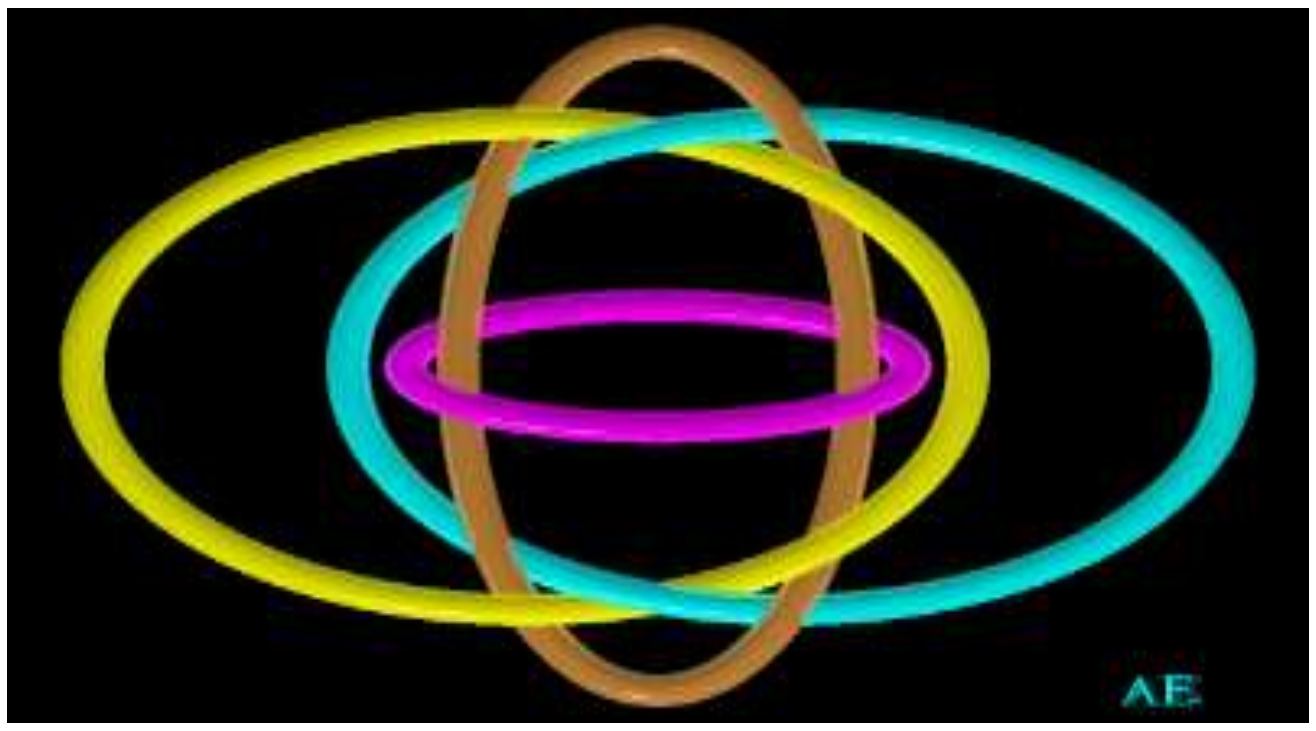

**An illustrative example of a Brunnian link**, potentially enclosing the cognitive loops from orthogonal hierarchical units. (That link with four loops discovers Brunn in 1892, source Alain Esculier's website).

**5.11. Structure of the Observer logic and cognition**

1. The Observer logical information structure self–forms the rotating mechanism of the EF dynamics, which self-organizes the hierarchy of the logical triplet units assembling in the resonance frequencies. Each triplet logical structure is an analog of Borromini ring consisting of three topological circles linked by the Brunnian link-loop. (The forth loop represent a free logic).

The EF-IPF time-space trajectory distributes the hierarchical logic.

2. The Observer logical structure carries the wave along the trajectory segments, where each third segment delivers the triple logic of the spectrum $\{\omega_1, \omega_2, [\omega_1 = \omega_1]_{\Delta t_{10}}, \omega_2, [\omega_1 = \omega_1]_{\Delta t_{20}}, \omega_2, [\omega_1 = \omega_1]_{\Delta t_{30}}, \} = \omega_{\Sigma 10}$ with sequentially shortening intervals $| \Delta t_{10}, \Delta t_{20}, \Delta t_{30} |$ and the increasing segment density.

Two sequential segments synchronize resonance frequencies $[\omega_1 = \omega_1]_{\Delta t_{10}}$ and $[\omega_1 = \omega_1]_{\Delta t_{20}}$ while the triplet synchronizes resonance frequency $[\omega_1 = \omega_1]_{\Delta t_{30}}$. This triple logic holds one Bit in each Observer's triplet logical structure unit. *Or three-dimensional segments hold the informational logics.* The attracting free logic of sequential triplets conveys the resonance spectrum with progressively shortening time intervals and growing frequencies, which cooperate to form the logical units in IN nested hierarchy. The necessary spectrum with the increasing frequencies automatically carries each consecutive segment along the EF-IPF trajectory. The emanating wave function delivers the frequencies, cooperatively growing a hierarchy of the logical units. The self-built hierarchy of the logical structures self-integrates the observed logic which the structure encloses.

3. The hierarchy of distributed logical loops self-connects logical chain.

The logical chain width determines the invariant impulse's relative interval enclosing the assembled logical code. The growing density of consecutive impulses along the trajectory sequentially squeezes the absolute value of this interval, whose ratio preserves the invariant impulse. The absolute time-space sizes of the logical chain are squeezing through the multi-level distributed hierarchy.

4. The cognitive logical chain composes the coherent triplet loops which assemble the nested attractors-knots logically cooperating the IN nodes' hierarchy. This composite logical cognition synchronizes the triple rhythms along the EF-IPF trajectory, which schedule the access energy for encoding the observer intelligence. Thereafter the cognitive chain predicts the intelligence encoding.
However, during sequential encoding the hierarchy of logical knots, the cognitive logic of the knot, encoded at a current hierarchical level, dissolves, its predictive action vanishes.
Thus sequential hierarchical encoding successively removes the cognition which predicts this encoding.
5. The logical chain rotation, carrying the frequencies of the synchronized spectrum, requires a minimal energy to support the chain. This energy is equivalent to each Bit's logical code measure.
The integrated chain logic holds this code, and the physical encoding intelligence encloses the cognitive thermodynamics. •
Therefore, the wave function frequencies, initiating the self-forming Observer cognition, emerge along the EF-IPF extreme trajectory in the form of a probabilistic time wave in a probability field.
The probabilistic impulse observation starts the microprocess, where the entangled space rotation develops the rotating space-time probability wave.
The emerging opposite asymmetrical topological interaction shapes the space-time wave function, becoming certain, as well as the Observer's cognitive logic, predicting the intelligence code.
*These results conclusively and numerically determine the structure and functions of cognition.*

### 5.11. Specifics of Information Intelligence and estimation its Information values

The causal probabilities start the Markov correlation connection with minimum of three probabilistic events. An Observer integrates the observing events in the Information networks, which accumulate the nested triple connections, depending on the IN Information invariant properties.
Each IN has an invariant Information geometrical structure and a maximal number of nodes-triple Bits, whose ability to cooperate more triplet nodes limits the possibility of the IN self-destruction by arising a chaotic movement.
The intelligence measures the *memorized ending node of the IN highest levels*, while the cognitive process at each triplet level preempts its memorization. This means each memorizing node encloses cognition.
The Information measure of intelligence is *objective for each particular Observer*, while the IQ is an *empirical subjective* measure.
This theory shows that during the current observation, an Observer can build each IN with maximum 24-26 nodes with average $3^{26}$ Bits and enfold the maximum of 26 such INs. This number of the ordered by INs connections for structural unit (triplets) is threshold defining intellectual observer.
Since each IN following level integrates Information from all the IN previous levels, measuring the relative Information quality, the built multi-levels INs hold Information quality relationships between the levels in the triplet forms.
Because the subsequent relationships have been enclosed by the cognitive rotating mechanism, they formalize a causal comparative Information quality meaning for the observing process events.
The *Observer Intelligence* has the ability to uncover causal relationships enclosed in the evaluated Observer $\mathrm{N}_{o1} = 3^{26} \times 26 bits$ networks Bits. That requires not only building each of $\mathrm{N}_{1I} = 26$ IN, but also to sequentially enfold them in a final node whose single Bit accumulates $\mathrm{N}_{oI}$ Bits:

$$\mathrm{N}_{oI} = (3^{26}) \times 26 = 2,541.865.828329 \times 26 = 66,088.511.536.554 \cong 6.61 \times 10^{9} \text{ Bits.} \qquad (5.1)$$

However, since each IN node holds single triplet's Information, the final IN node's Bit keeps the triple causal Information relationship with density $\mathrm{D}_{oI} = \mathrm{N}_{oI} / bit$ -per Bit.

To support the IN node impulse feedback communication (Sec.5) with the requested attracting Information, this node requires Information density:

$i_{md} \cong 1.8\times10^{14} Nat/\sec = 1.44\times1.8\times10^{14} bit/\sec$, (5.2)

where each such Bit accumulates $\mathrm{N}_{oI}$. Thus, the total Information density of the Observer final IN Bit:

$i_{do} \cong 1.44\times1.8\times10^{14}\times(3^{26})\times 26 bit/\sec$ (5.3)

evaluates the intelligent Observer's Information density.

With this density, the intelligent Observer can obtain maximal Information from the EF through the impulse interaction with entropy random process during time observation $T$.

Let us evaluate the EF according to formula [15]:

$I_e = 1/8\ln[\mathrm{r}(T)/\mathrm{r}(\mathrm{t}_s)] \approx 1/8\ln(\mathrm{T}/\mathrm{t}_s), \mathrm{T} = m_N\,\mathrm{t}_s$.

Here $m_N$ is a total number of the IN nodes needed to build the intelligent Observer, $\mathrm{t}_s$ is the time interval of the invariant impulse. At $m_N = 26\times26$ it allows estimate $I_e = 1/8\ln 26^2 = 11.729 Nat$.

Therefore, the intelligent Observer needs $N_i \cong 12$ invariant impulses to build its total IN during the time interval of observation $T$.

<u>Comments</u>

The human brain consists of about 86 billion neurons [83], which approximately in 14 times exceed $\mathrm{N}_{oI}$ (5.1), if each single Bit of the cognition commands each neuron. If each neuron builds own IN with about five-six triplets (with levels $3+2^4=11, or 3+2^5=13$), while the ending triplet Bit condenses this $\mathrm{N}_{oI}$, ability of the neuron building a net concurs with [83] and [84].

If this is true, then $\mathrm{N}_{oI}$ measures the Information memory of a human being.•

According to estimates [83, others], the maximal Information in the Universe approximates

$I_U \cong 3\times10^{29} Nat = 4.328\times10^{29} bits$, (5.4)

from which each invariant intelligent Observer can get $I_{ob} \cong 6.61\times10^9 bits$.

To obtain all $I_U$ Information, such intelligent Observers need $M_{ob} \cong 1.527\times10^{16}$ numbers of such invariant Observers.

Each IN triplet node may request $I_m \cong (3.45-2.45) bits$, which measures this IN level of quality Information that memorizes the node Bit. Such node's level accumulates average Information between $I_m bits$ and $\mathrm{N}_{om} = 3^{26} bits$, depending on each IN's levels quantity $\mathrm{N}_{1I}$.

Quantity $\mathrm{N}_{oI}$ (5.1) measures the invariant transformation to build the extreme IN node's structure during the observation, which transforms a probable observing process to an Information process in the emerging Information Observer with intelligence.

The initial probability field of random processes, evaluated by the Entropy Functional, contains the potential Information which an intelligent Observer can obtain through the invariant transformation.

The Information threshold $\mathrm{N}_{oI}$ limits the level of intelligence of the intelligent Observer, satisfying the minimax variation principle.

The intelligent (human) Observer can overcome this threshold requiring highest Information up to $I_U$.

An Observer that conquers the threshold, possesses a superior intellect, which can control not only its own intellect, but other Observers. Multiple joint superior intellectual Observers can form a super-intellectual system (with $I_U$) controlling Universe, or would destroy themselves and others. However, in an intelligent machine, collecting the observing Information, the emerging invariant regularities of the minimax law limits the AI Observer actions.

## 5.12. Interacting intelligent observers through communication

An Information intelligent Observer emerges during the evolving observations, which have delivered invariant information, built Information IN' nodes' hierarchy, and double helix rotating structure with DSS intelligent code. The important issue is interaction of such Observers in a mutual communications, which preserve the invariant Information properties and benefit their information qualities.

Suppose an intelligent Observer sends a message, containing Information encoding its meaning.

Another intelligent Observer, receiving this Information, would be able to *read the message, recognize its meaning, select and accept* it if this Information *satisfies the Observer's needed Information quality,* which is being memorized through its DSS code. Next, we consider the fulfillment of these five issues.

### 5.12.1. How the interacting intelligence Observers can understand meaning in each communication

Let an intelligent observer sends a message enclosing its information logic, quality, and bits encoding this information, which emanates from some intelligent observer's IN nodes.

Other intelligent observer, requesting the growing quality of needed information, sends specific qualities of free Information emanating from its IN nodes that need that quality. In the communication interaction, the receiving quality will add the needed quality compensating the need. (Each observer quality information classifies the node location in the IN hierarchy enfolding this information).

The intelligent observer, receiving that information quality, identifies the nodes locations in the IN hierarchy, being equivalent to this quality. Each node location encloses logic of its triplet with a resonance cognitive loop.

That quality may belong to ending node of the observer IN enclosing its cognitive loop.

The ending node emanates the Observer sending free information quality enclosing the loop.

The message quality associates with the node free logic attraction, carrying the related information frequency (Sec.5.11). If the cognitive loop of the sending free information quality accepts the receiving information quality, enclosing its frequency, then the receiving observer enables recognizing the message meaning. That implicates understanding of the message meaning encoding this quality.

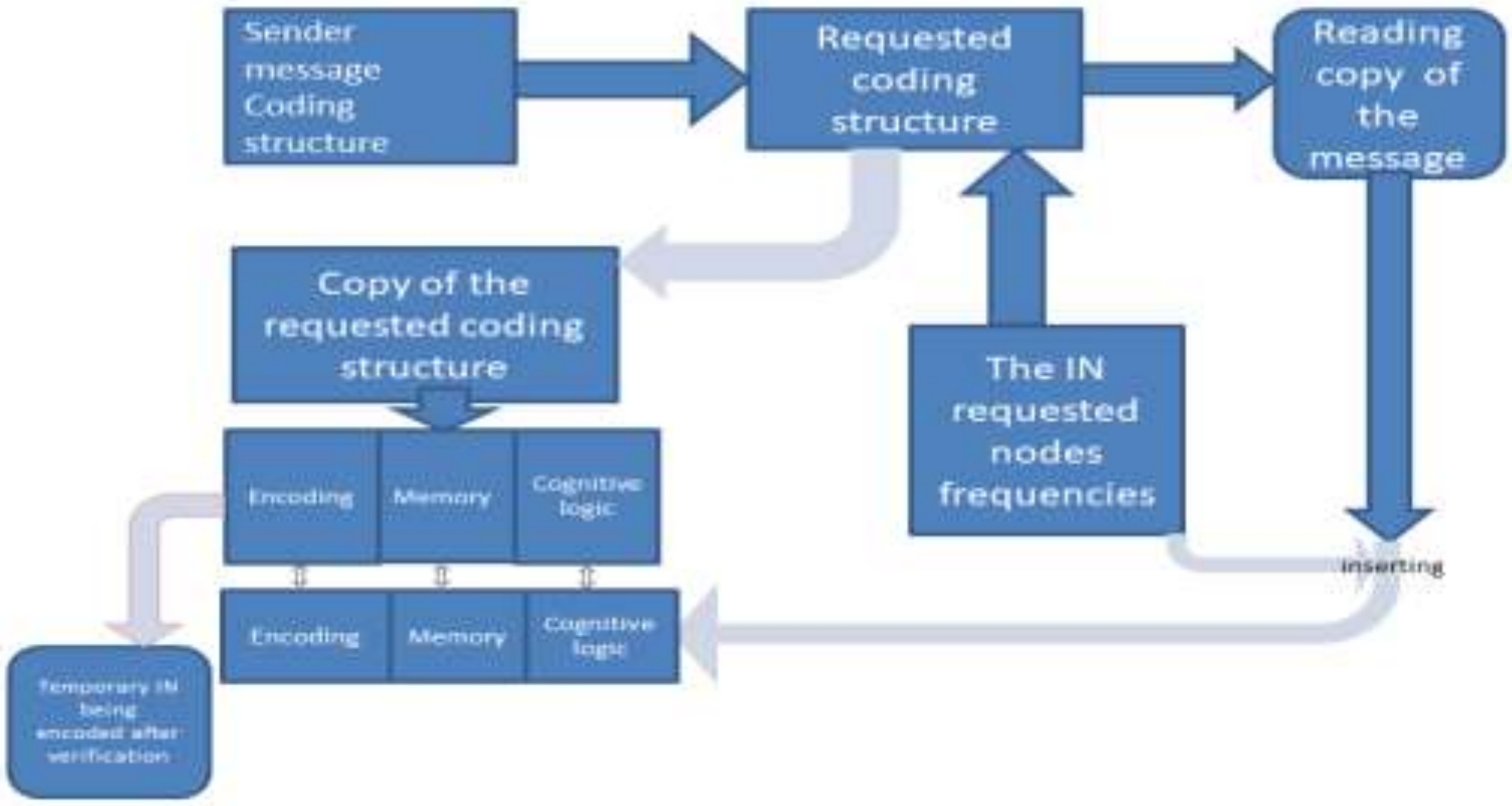

**Fig.11.Schematic illustration of functional organization of comparative acceptance a message.**

The schematic illustration shows

**1.** How the sender message coding structure is read by the requested Observer coding structure which initiates the IN requesting nodes frequencies;

**2.** The frequencies identifies the requested coding structure that allows comparing the reading message intelligence information with that in the Observer requested coding structure;

**3.** The comparison requests the same nodes frequencies which insert the Reading message copy' Bits into the copy of the requested coding structure (DSS Copy).

**4.** If the comparison verifies the receiving observer intelligence information, this observer accepts the message intelligence information. Assuming that this information identifies the observer meaning, this observer understands the message meaning. •

Hence, the receiving Observer recognizes the message meaning if the cognitive loop of the temporal IN sending node free Information quality accepts the receiving Information quality, enclosing its frequency, while this frequency reveals the DSS's equivalent logics. Revealing the message logic, its memorizing Bit, and encoding, takes place during the message information copy's insertion and moving in the Observer Copy DSS.

Starting the message copying begins the time of the movement, which starts this DSS enclosed logic, up to revealing the Bits and encoding in the IN being temporal. Those IN nodes hold the message quality needed by the Observer request. Thus, the observer request with the specific free information initiates recognition of the needed information. That process comprises the following steps.

1.The IN nodes of Observer-receiver request the needed quality through the nodes free information which, attracting the message bits, starts copying them message on an Observer double spiral helix structure. The DSS structure, while copying, integrates and reads the message information. While reading the message, the DSS is moving along its time-space information structure and allocates a temporal IN with the nodes related quality.

(The admissible number of the temporal IN nodes constrains the message information.)

2. The requested INs ending nodes quality information mirrors the information from the IN temporal node. The free information of the ending nodes impulses initiates involvement of the copy information frequencies in resonance.

3. If the copies frequencies cohere in the resonances assembling the temporal logical loops, the requesting information impulses insert the mirror copy information logic in the IN requested nodes.

4. During the insertion of the mirror copy, transitive impulses of the requested nodes provide asymmetrical free logics with $\Delta t_1$ intervals. This indicates that the requesting nodes reveal and accept the mirror copy of message logics.

5. Each of these logics intervals allows access to the ending node the interval $\Delta t_B$ currying the physical bit from the receiving Observer DSS which had copied the message bit. The physical bit energy erases logic of the mirror copy, revealing its information bit and starting process of memorizing bits and decoding the message enclosed triples.

Access of the Observer bits, which initiates the acceptance of the mirror copy logics, indicates recognizing the message bits by the observer bits.

6. Decoding of each memorized Bit takes interval $\Delta t_{en}$ of its encoding. The decoding of ending IN node impulses reveals the IN hierarchy of the enclosed message information logic.

(Starting the message copying begins the time of the integration which starts the DSS logic up to revealing the bits and encoding-decoding. The Observer DSS, emerging with copying the message, is main information apparatus allowing the Observer comparative acceptance of the message information).

7.The message information, located on the DSS, delivers the wave function frequencies of the recognized decoding impulses. The IN node' frequencies generates the cognitive loops' coherent frequencies *recognizing* the message logic delivered to the IN receiving nodes. The frequencies spectrum may update the needed information quality.

8. Decoding finalizes requesting IN nodes acceptance of the message comparative qualities. These qualities indicate ability to cohere and cooperate the message quality with the quality of the IN node, enclosing it in the Observer-receiver IN structure. The observer logic' coherence with the message logic allows memorizing the decoded message information.
9. Accepting the message quality, the intelligent Observer recognizes the message information and encodes its triple logic' digital bits (images) by the observer time-space codes, being self-reflective in understanding the meaning of the message. Since the acceptance of the message quality changes the existing observer logic encoded in the INs hierarchy, understanding the meaning of the message extends the level observer intelligent logic. Thus, the intelligent observer uncovers a meaning of communicating message in the self-reflecting process, using the common message information language, temporary memorized logic, the cognitive acceptance, and the logic of the memorized decoding.
10. Understanding the meaning of an observing process includes the coherence of its Information with the Observer current coding structure, which evolves all previous observations, interactions, and communications. •

On the path to understand the QM, an Observer builds multiple observing probabilities. One of the common is classical Kolmogorov probability where each a priori act of probabilistic observation follows a posteriori probabilistic act of observation. Each pair of these acts is a probabilistic impulse like a *classic* action and reaction formally evaluated by Kolmogorov' 1-0 probability law.
When it comes to observing this classing action and reactions approaching to quantum (micro) process, the action and reaction merge within a bordered impulse, bringing together probabilistic a priori and a posteriori actions on edge of classical predictability. That's why understanding quantum physical microproccess becomes uncertain, fuzzy, weird, and on the *edge classical knowledge and even reality*.
*Information is phenomenon of interactions and a measure of the interactive reality.*

**5.13. Individuality of the Observer**

Based on the results from previous chapters, the Observer's individuality includes the following features:
1. The particular probability field's triad observing a specific set of probabilistic events. Each specific triad from the probability field affects the formation of the observed impulse, changing its Free Information from minimum to maximum. (Each observing impulse contains 1 Nat (~1.44 bits) invariant information, which includes 1 bit and Free Information up to ~0.44 bits which is only that could change). This changes the attractive connection of the bits in a chain and determines the particular path of observation, measured by its specific path integral, ending with a limited number of certain physical Bits. Thus, each specific triad through of the Free Information virtually forms related physical Bits composing physical unit. (From beginning of the observation, the notion of the particle, images, any physical unit was replaced with the probability of event observation).
2. The final specific qubits or Bit of the microprocess depend on which pair of impulses merge on the observing random trajectory. Since the impulse position on the observing trajectory measures the amount of the impulse's random entropy, that entropy becomes the measure of the forming Information qubits or Bit. The asymmetry of the final Bit will conserve this specific value. Starting the macroprocess, such Bits build triplets, INs, and the forming Observer's cognition and intelligence. Together, these maintain the Observer's individuality.
3. The particular time of observation measures the quantity and density of Information in the delivered Bits.
4. Cooperation of the observing Information in a restricted number of IN-triplet nodes, and a limited number of the Observer's IN, depending on the individual Observer's selective actions.
5. The selective actions, defining the cooperative Information forces, which depend on the number of the IN cooperating nodes. The minimal cooperative force, forming the very first triplet, defines minimal selective Observer.

6. The individual ability for selection classifies Information Observers by the levels of the IN hierarchy, its times-space geometrical structures, and inner time scales, whose feedback holds the *admissible* Information spectrum of observation. That feedback controls the request and acceptance of the Observer's needed information. The individual Observer's INs determines explicit ability of their self-creation [ self-creation of what? MD ].
7. The produced specific self-generated Bit starts the new generation of the individual Observers. These specifics also classify the Observers by their level of cognition and intelligence.
8. The individual DSS code identifies the Observer's goal, which encloses the integral quality of information events that the Observer accumulates during its observing process, starting with the probability field triad. The observing certainty creates explicit information Bits which carry the certain reality of the observing images.
9. The multilevel hierarchical realities enable creating multiple observing images with particular environment.
10 The individual specifics, originating from the particular environment, as attributes of the observation process, affect the initial conditions, constraints, thresholds and the multiplicity of emerging observers. But are not significant compared with the individual's DSS code.
11.The Information *mechanism* of building all Observers describes the invariant equations of Information dynamics following from the common minimax variation principle (VP). Since the VP, applied to the observing impulses' path integral brings the common IMD eqs., each such impulse, being invariant, includes both invariant Bit and various Free information (see point 1.). While the Free information effects the individual observer DSS code, the information mechanism determines the common invariant DSS code. (The individual DSS code is result of applying the VP to the particular path of observation (point 1.) which includes specifics of point 10).
12.Compared to AI, at an equal IN's level of AI and human intellect, the difference between AI and a human makes the personalized history of memory for each human Being which includes many predecessors. The AI always would be less complex comparative to the Individual Observer's at the same level. The personalized history structures the IN personal memory which accumulates all Observers' cognitive observations connected with Observer individual intelligence.
13. The observer's intelligence enables self-decoding each IN's node memorized information. The self–decoding identifies some observed Bit and their cooperating triplets (limited by threshold for the selective observer ~4.48 Bits.). These bits' Free Information holds attractive logic. The logic enables cooperation with some logical information located in the specific attracting IN node along the IN hierarchy. The cooperation of both the IN internal logic with the self-decoded observed IN logic creates the inter-communication chain of an integral logic. That integral logical chain is not necessarily memorized, and could be temporal.

The observer intelligence recognizes such self-created inner logical functional chain that connects something surrounding with the Observer's logic itself as a self-awareness.

According to Webster, "the quality or state being aware especially of something within oneself" defines consciousness. Therefore the Observer's intelligence includes a hierarchical function of the logically qualifying self-arising states of being consciousness. These states are result of interactions on the reality level measured by the quality of information on the IN hierarchy.

Considering the consciousness' hierarchical function, we assume, it starts within our entropy-information gap (Sec.3.1.8), where the lack of information unable any computation.
Such unidentified function could be not computable.
A probabilistic logic observing in virtual reality possess hidden bits which I also not computable.

According to Penrose [86], "consciousness is not computable".

On a higher hierarchical level, the logic emerges with certain bits understanding a meaning through inner-communication, where, until the understanding, this function is not revealed.

With more quantity and quality information at high level of interactive communication, such information becomes available for modern computations.

### 5.14. The IIID(IMD) and IIT.

Following Toponi [90], the IIT system is unified by the sets of phenomena specialized by the author definition of conscience. The IIT formalism does not include any math integral.

And the IIT has no connections to Interactive Integrated Information Dynamics (IIID).

But the IMD-IIID demonstrates its following connection to IIT.

1.The space-time rotating trajectory on the conic surface (Fig.3) covers the space which shields (encloses) all space-time interactions (not just that which Topani postulates as phenomena of conscionsence). It has analogy with his Φ.

2.The Entropy Functional and Information Path Integral integrate all this information in bits and finally compress it to single Bit.

3.Finally the memorized information units, enfolded in that space encode the cellural geometry (Fig.7,8) forming the Double Spiral Structure (DSS) triplet code. The structure of the cellular geometry, formed by the cell of the DSS code shapes the surface of the OBSERVER (Fig.8).

Thus, the IMD actually mathematically integrates not only conscience phenomena, but all observed interactions, measures their information in bits, compresses, encodes, and transforms in space time cellular geometry enclosing the INFORMATION OBSERVER. This means, the IMD covers also the IIT (not vice versa) in more scientifically way then IIT itself, and provides the information geometry of the encoded information in double triplet code.

The Equations of IIID, built for each observing interactive process, formalize its description. The IIID is a part of the Observers information regularities and provide a physics of the Information Bit and emergent Observers.

It allows including the Observer individuality in the Observer regularity adding a possible explanation a mystery of consciousness, which no one has ever explained or solved.

### 5.15. Relations to Basic Results of J.A. WHEELER

The IMD-IIID approach presents a realization of Wheeler's aphorism 'It from Bit'.

The approach completes his idea with a mathematical formalism.

This approach contributes the following specifics.

Introduces an Observer's Information Path from virtual uncertainty to a formal definition of certainty/reality.

The path is observed by a sequence of yes-no actions-impulses acting on an observable random process.

This implements Wheeler's concept "reality arises in the last analysis from the posing of yes-no questions"and"observer-participator as definer of reality" [2 in Introductory].

Along the path, a quantum microprocess emerges. As space-time arises, an entangled elementary Qubit is formed. On the edge of this reality, a curved transitional gap appears between Entropy and Information, and elementary Bit is assembled within an impulse. The elementary (Kronicker) impulse measure is 1 Nat (~1.4bits) which includes the Bit measure ln2 while the difference ~0.4 bit measures free information. Each observed impulse is cut from the observed random process' correlated events. That's why the cut-off impulse' free information encloses the cutting correlation which holds the connected events of the process. The correlational entropy' connection releases an information-logical attraction of the correlation being cut. This is a phenomenon of free information arising at transitional gap between entropy and information, which Wheeler's approach did not recognize.

The assembled elementary information units, via its free information connections to other cut-off units possess Wheeler's Observer-Participancy. (Wheeler's publications provide multiple physical examples of such Observer).

The free information's of the multiple information units integrates the units in information macroprocess. The math integration of discrete information performs the developed information path functional (IPF).(Integration of discrete elements is serious math problem).

The macroprocess structures elementary triple bits self-connected through their free information.
The triplet's structural units' free information self-connects them in structure of information networks with hierarchy of the level's nodes, formed by a triple sequence of cooperating triplets.
Each triple IN through its ending nodes free information self-assemble information domains enable further structure extension. Such self–extending multiple INs observed unit relates to Wheeler's 'observer "ship" as the mechanism of genesis' [92].
Each triple structure' of information bits encodes a triplet code, which determines the structure Genetics enables this structure self-restoration. The IN triple code allows communication between the structural units.
The rising structure curved space-time determines geometry of each structural unit at each hierarchical level.
The information curved geometry composes double spiral triplet code (DSS) with the code' cellular space–time geometry (Fig.8). The DSS encodes all memorized discrete information units and observer cooperative complexity.
The space-time information processes describe Information Dynamics, whose regularities follows from extreme variation principle imposed on the process IPF.
Forming each structural unit limits the available free information, following from each observation. It leads to limitations on all levels the observer's structural units implemented through the Information Macrodynamics (IMD).
The IMD creates the Observer's selective mechanism, choosing the needed observation impulses for developing the observer dynamic structure. This structure creates the Observer cognition and intelligence.
The DSS code' cellular space-time geometry integrates all observed interactions, while the cellular code encodes these interactions in discrete information units. The IPF compress the discrete units to a single bit.
The interactive integrated information dynamics' formalism describes the Observer's multilevel structures formation, development, the code genetics, and ability of formed structures to self-restoration.
The Weeler' principle universality of action concurs with 1.1.Axiom, which finally leads to the IMD-IIID equations structuring Information Observer with the regularities recognizing in physics as law without law [93].

## VI.Summary of the Arising Information Regularities in an Observer

### 6.1.A Revew

### 6.1.1. The Basics

1. Interactions are fundamental natural phenomena in the universe and reality is only the emerging interactions. From that follows *information is phenomenon of interactions which also bring reality-certainty with information.*
2. Each elementary interaction is an action and a reaction.
It is an impulse which can be represented as Yes-No symbols modeling a binary 1-0 value, a Bit.
That connects the phenomenon of Interaction with the phenomenon of Information and certainty, emerging in the impulse' observations.
3. The introduced notion of Information leads to an Information Observer evolving in interactive observations. It changes Information's relation to entropy, the emerging reality, origin its certainty, causality, logic, Information dynamics, micro-macroprocesses assembling in Nets, its complexity, Observer cognition, and intelligence logic, and many other essential concepts.
These results substantiate all of these and validate them analytically and numerically.

### 6.1.2.The Specifics

1. Interactions of different events (objects, particles) primarily indicate their occurrence in multiple traces revealed by probabilistic observations. Inter-action is Actual which acts (*act*-ual) that causes Real.
2. Multiple interactions produce a manifold of random elementary 1-0 events, whose occurrence (events) describes the axiomatic Kolmogorov probabilities of each observing interactive process.

The probabilities of these random events, emerging from the probability field, interact through random processes which are modeled as Markov chains of multiple Bits. This field of axiomatic probability is the source of Information and physical events.

3. The objective probabilistic trace of multiple Bits observes a formal act which changes the probabilities of a Markov chain. The chain processes independent random increments which carry discrete Yes-No action of the random impulses of the impulse probabilistic observations. Such observations transforms the Markov chain into Markov diffusion process evolving to a Levy process. The transitional *a priori-a posteriori* probabilities bring probabilistic impulse observation of the Markov diffusion process.

4. In the process of impulse observations, each observing action ↓ cuts entropy of the impulse correlation from the initial Markov diffusion process. During the impulse switching time, action ↓ transfers to actions ↑ the cutting entropy minus an entropy spent on the transferring. The cutting entropy decreases the Markov process entropy and increases the entropy enclosing within the impulse interactive actions↓↑. Each following interactive impulse, which sequentially observes the Markov process, maximizes the entropy of the observing process impulses. The multiple observing impulses generate the secondary Markov diffusion whose entropy increases comparing with that in the initial Markov process.

5. Particular probabilities of the field observe its specific set of events, which identify a potential Observer.

6. The Markov correlations hold the relative entropy measure of the uncertainty of random events between the impulse Yes-No probabilities, or uncertain multiple impulses of a hidden Bits.

7. The entropy contribution measures each observing impulse's interactive impact on the process being observed, which the observation changes. The impacts of observations, collected by the EF-IPF integral measure, finally create the path to the Information Observer.

8. Certainty is revealed by removing the entropy of the correlation or uncertainty, originating Information which emerges from a particular set of the observing probabilistic events.

A specific Information Observer is created by the objective probabilities of the observations.

9. This Information Observer emerges without any pre-existing physical law.•

**6.2. Summary of the hierarchical evolving levels in the impulse observations of the interactions**

- The objective Yes-No probabilities measure the virtual probing impulses. Processing the interactions, they generate an idealized (virtual) probability measurement, from a finite uncertainty in the observable Markov chain to the observing probability of the potential (virtual) observer, up to the certainty of the real Information Observer.
- The impulse of the interactive No-action cuts the maximum entropy, while its Yes-action transfers a minimum cut to the next impulse, thus creating multiple impulses on the maxmin-minimax principle, decreasing the uncertainty of the observing process.
- The reduced relational entropy along the trajectory of the observing process conveys the *a priori-a posteriori* probabilistic causality of the impulse. Correlation of the impulse temporarily memorizes the sequential probe's logic of the probabilistic causality.
- The correlations hold the hidden inner connections of the impulse's entropy which integrates the Entropy Functional (EF) along the observing process. The EF also integrates the time interval of the correlation connections along the observing process. In addition, this allows the integration of the probabilistic logical causality.
- The connection of the cutting impulses decreases the potential number of multiple virtual Observers, indicating a threshold which limits the number of observers not overcoming the threshold.
- As the probability of the Levy process grows, neighboring impulses may merge, generating an interactive jump on each impulse border. The Lévy measure describes the distribution of the process potential jumps. Until any two time intervals of the process' independent random increments do not overlap, the jump absent and measure is zero. The overlap is a merge of the Yes-No impulses, when Lévy measure is nonzero, finite. The merge of these impulses indicates zero random bit. It means there is no reality, defined by the interaction, when the microprocess starts. The starting microprocess with zero probability is unpredictable.

-When Levy measure becomes finite, the jumps have started. The jumps initiate a microprocess within the observing Levy process. A pair of random interactive actions on the bordering impulses becomes equally probable. The merge converges a causing action with a subsequent reaction, superimposing the cause and effect. The emerging microprocess within the bordered impulse runs the superposition and the entanglement of conjugated entropy fractions. The merging fractions entangle during the time interval before the space is formed. Since a beginning of the entanglement has no space measure, the entangled states can be everywhere in a space. The microprocess ends as the entropy volume is entangled.

The information process starts when the microprocess generates real qubit and/or bit.

- The interaction curves the interacting impulse, which creates the inner impulse rotating movement along the observing process. The microprocess rotation transits the entangled entropy with a Bit's formation through an entropy-Information gap. The gap holds a hidden real locality which the rotating potential momentum, growing with the increased entropy volume transition over the gap, can overcome.

-The real local gap reveals a physical Markov diffusion whose entropy erases an external energy impulse. The momentum acquires physical property near the gap end when the momentum curves a physical cut of the transferred entropy volume. The cutting bits conserve the causal logic in Information logic.

- The emerging energy during the interaction kills the entropy volume within the gap, memorizing a logical Bit or two qubits. The $1Nat = 1.44bit$ of each impulse contains the $1bit \cong 0.7Nat$ and the Free Information of the cutting correlation $0.123bit$, enabling the attracting actions. The difference $1.44 - 1.23 = 0.21$ Bit with $0.21 \times 1.44 \cong 0.3Nat$ is transferred to the next interacting impulse as its entropy equivalent.

- The opposite curvature, enclosing the entropy of the interacting impulses, lowers the potential energy that converts entropy into a Bit of the interacting process.

*Finally, microprocess emerges from the classical process under the interactive observation, and returns to classical information through the observing Bits.*

**-** In a multi-dimensional observing process, the multiple cuts reveal multiple Bit units which the Hidden Information attraction binds in the collective dynamic movement of the Information macrodynamic process. The macroprocess integrates both the entropy between impulses, the microprocesses, and the cutoff Information of real impulses, which sequentially convert the collected entropy in an Information physical process during the macro movement.

-The author's developed complex Information Path Functional [12,13], defined on trajectories of the interacting observing process, for the first time, is measuring direct integration of the observing process' multiple bits.

- Multiple interacting Bits self-organize the Information process in an Information structure, encoding Information causality, probabilistic logic, and complexity.

- The trajectory of the observing process carries the wave function (both probabilistic and certain), self-building the Information structure hierarchy.

- The Information Path Functional (IPF) integrates the Information process, enclosing the cutting correlations of the EF, which creates the Bits connecting the IPF extreme trajectory of the observing process. The IPF condenses all the integrated bits in the trajectory's final Bit. The sequentially forming triple knots squeeze the initial observing multi-dimensional process first to three-dimensional rotation, and then to single-dimensional Information process encoding the Bits of all multiple knots. That transforms the initial random observation to certain dynamics.

**-** The Information Macrodynamics (IMD) are reversible within each EF-IPF extreme segment, whereas irreversibility rises at each border between the segments which encloses the memorized Information.

The borders impose a dynamic constraint on the Hamiltonian of the irreversible IMD. The IMD Lagrangian integrates both the impulse's and border constraint's Information on time-space intervals.
- The EF-IPF extreme integrates the time-space intervals of the invariant impulses in an Information Geometry.
- A flow of the moving cutoff Bits forms a unit of the Information macroprocess (UP), whose size limits the unit's starting maximal and ending minimal Information speeds, attracting a new UP through its Free Information.
Selected automatically during the minimax attracting macro movement, each UP joins two cutoff Bits with a third Bit, delivering Information for next cutting Bit.
- A minimum of three self-connected Bits assembles the optimal UP-basic triplet, whose Free Information requests and binds a new UP triplet that joins the three in a knot which accumulates and memorizes the triplet's Information in the trajectory segments.
- During macro-movement, multiple UP triples adjoin the timespace hierarchical network (IN) whose Free Information's request produces new UP at higher level's knot-node and encodes it in triple code logic. Each UP has a unique position in the IN hierarchy, which defines the exact location of each code's logical structures.
-The IN node hierarchical level classifies the quality of the assembled Information, while the currently ending IN node integrates the Information enfolding all IN levels.
- Each Yes-No impulse action, transformed in the UP's Bit logic, differentiates the impulse Information density and quality, which identify the UP location on the IN hierarchical level.
- New Information for the IN delivers the requested node information's interactive impulse impact on the needed external Information. Cutoff entropy of the observation converts it to Information. The resulting new quality of Information concurrently builds the IN temporary hierarchy, whose high level enfolds the Information logic that requests new Information for the running observer's IN, extending the logic up to the IN logical code.
- The emergence of the current IN level indicates the Observer's Information surprise, measured through the IN feedback's interaction with both external observations and the internal IN's Information, delivering new self-renovating Information quality.
- The growing IPF Information, condensed in the integrated Bit with a finite impulse geometrical size, strengthens the Bit Information density, running up to finite maximal Information at infinite process dimension.
-The time-space Information geometry, emerging in observations, connected with the macro-movement in rotating time-space coordinate systems, shapes the Observer asymmetrical structure by confining its multiple INs. The time scale of the accumulation of Information determines the Observer's time of inner communications.
- Each Observer owns the time of inner communication, depending on the requested Information, time scale, and density of the accumulated Information.
- The Observer optimal multiple choices, evaluated through the minimax self-directed strategy, implement the cooperative forces emanating from the INs integrated nodes.
- The current Information cooperative force, initiated by Free Information, measures the Observer's selective actions, attracting new high-quality Information. Such quality delivers a high density-frequency of related observing Information through the selective mechanism. These actions engage acceleration of the Observer's Information processing, coordinated with the new selection, quick memorizing and encoding each node Information with its logic and space-time structure. All these implement the minimax strategy which minimizes spending Information and IN cooperative complexity.
- The self-built Information structure, under self-synchronized feedback, drives self-organization of the IN and evolution of the macrodynamics through its self-creation.
- The macro units logically self-organize Information network INs, encoding the units in the geometrical structures enclosing the triplet code.
- Multiple INs bind their ending triplets, enclosing Observer Information, cognition, and intelligence.

The Observer cognition assembles common units through multiple attractions in resonances loops at the forming IN triplet hierarchy.
- The cognitive logic self-controls the process encoding the intelligence in a double helix coding structure (DSS). The clock time intervals open access to the external energy at each specific level of the IN multiple hierarchy, enabling the memorization and encoding of the hierarchy of these Bit. The energy brings cognitive thermodynamics for the memorization and encoding.
- The maximal number of accepted triplet levels in multiple INs measures the Observer's maximum comparative Information intelligence. The intelligent Observer recognizes and encodes these digital images in message transmission.
- The intelligent Observers connect the Information transmission and communications. Such an Observer, being self-reflective through the DSS invariant helix code enables reading and understanding the message being communicated.
- Understanding implies that the Observer can classify and select such Information according to this Observer's memorized *meaning,* among other comparative images.
- The multiple code memorizes the IN assembled logical structure in the Observer's cooperative code. Since such code holds the energy of cognitive thermodynamics, it physically organizes the multiple INs with their local codes in the coding Information structure of Information Observer.
- The Double Spiral Structure triplet code, self-organizing all multiple local codes along the hierarchy, *encodes Observer intelligence*, which automatically includes the cognition integrating the observing process.
-The *Observer Intelligence* includes an ability to uncover causal relationships enclosed in evolving Observer networks, and self-extends the growing quality Information and the cognitive logic upon building the collective Observer intellect.
-The IN highest level ending node Information *measures* the Observer Intelligence.
- The intelligence of different Observers integrates the Information of their IN's node codes, which enclose a knowledge of the observations in the communicating observer's IN levels that enhance integrated knowledge.
- The multiple inner-communications create a hierarchy of the logic Bits, cooperating with IN nodes.
The observer intelligence recognizes such inner logical functional chain as self-awareness which could model the observer consciousness. This hierarchical function starts within the entropy-information gap, where the lack of information unable any computation. At a higher hierarchical level, it emerges with understanding a meaning through inner-communication, where, until the understanding, this function is not revealed. Such unidentified function could be incomputable, but being opened it becomes computable.
*The information emerging from the interactive observation builds the hierarchical reality in the Intelligent Observer.*
- The intelligence, growing with its time interval, increases the Observer's lifespan.
- Observation processes with entropy-Information and micro-macroprocesses are Observer-dependent.
The Information of each particular Observer is distinct. Each specific probability field triad generates an Information process creating its Observer.
- The invariant Information minimax law leads to common Information regularities for different Observers.
By observing even the same process, each Observer gets Information needed by its current IN during its optimal time-space Information dynamics. This creates specific (individual) Information processes.
The Information thresholds and constraints imposed on the evolving multilevel stages of the Information process systematize [41, 74].
The constrained level identifies multiple individual Observers, each of which stops evolving (Sec.3).
- Integrating the process entropy in the Entropy Functional and its Bits in the Information Path Integral measures formalize the variation problem in the minimax law, determining all regularities of the processes. Solving the problem mathematically describes the micro-macro processes, the IN, and invariant conditions of Observer's self-organization and self-replication, which self-create law of evolving the multilevel processes and the Observer.
- These functional regularities create a united Information mechanism whose integral logic self-operates, transforming interacting uncertainties into physical reality (matter, human Information).
-The united Information mechanism analytically synthesizes the AI enabling modeling a brain processing.
-Both Information and Information processes emerge as phenomena of natural interactions.

The Equations of *Interactive Integrated Information Dynamics* (IIID), built for each observing interactive process, formalize description of the process information regularities defining the Information Observer. The IIID, developed here analytically, finalize the main results, validate them numerically, and present Information models of many interactive physical processes.

*All results above implements basic initial concepts connecting Reality, Information, and Observer.*

### 6.2.1. How the observing probability field, conserving energy, creates physical units with the condensed qualities energy and information

1. The observing probability connects the uncertainty of random interactions to the certainty of the Information process. The connection includes physical processes interacting with energies of different qualities. The quality of energy is evaluated by the level of its order (disorder) or symmetry (asymmetry). This level measures minimal entropy ln2, which is equivalent to a Bit. The minimal entropy classifies the quality of energy (from the high-quality light energy to the low-quality energy of heat dissipation).
2. The impulse interacting actions curve the impulse geometry whose curvature creates asymmetry of the impulses. Such interaction logically erases each previously rotating entangled entropy units of the entropy volume. Each process's high–quality energy compensates for entropy of lesser quality. That removes the causal entropy with symmetrical reversible logic, created by Bayes *a priori–a posteriori* probabilities, bringing asymmetrical Information logic equivalent to logical Bit. Such a Bit is naturally extracted or erased at a minimal cost of the Quality Energy through topological transitivity in a phase transition and compression. That involves a transitional impulse inside the virtual impulse, logically memorizing the entangled units by making their mirror copy. The asymmetry created qubits is encoded in a memorized Bit. Such operations perform the function of a logical Maxwell Demon.
3. Transferring entropy through interaction unifies physical and nonphysical processes.
Such interactions naturally observe the probability's equivalent of entropy, transformed to Information.
The Information creates the Information Observer.
4.The energies of different qualities and quantities interact through the entropy-Information gap. By overcoming the gap, Information Bit is produced. Such a Bit measures the Information of a physical unit.
5. The minimal energy ln2 creates the curvature measure of the Bit's geometry. The curved geometry enables binding. That creates the Bit's Free Information, enabling Information attraction and binding Information units. Free Information is a discrete Information form of a free energy (Gibbs-Landau).
6. Each Bit binds and composes different units of Information with energy of high quality. As more Bits are composed, the quantity of this quality in the composite unit grows. An elementary triple encloses the minimal quantity of that quality. The enclosed triple quality binds the equal triple quantity.

Since the IPF extremal shortens the time interval of each subsequent composed unit, the density of the enclosed quantities and quality increases. Sec. 3.1.7 proves that each invariant external impulse brings the total energy ln2 during the time taken for both erasure of the reversible logical bit and memorization of the Information logic Bit. Each logical Bit is memorized by delivering Landauer's minimal energy during that time. Composing an Information triplet, the triple logical Bit unit memorizes and then encodes a knot of the forming Information Network's (IN) node. The *quality* of the IN hierarchy measures the *number* of nodes in the IN. By enclosing all previously enclosed Information, each hierarchical level of the IN is determined. Since each knot of this level measures an equal quality of the bound Information ln2, the *knot's Quality Energy and Information coincide* all along the IN hierarchy, and the density of each impulse encloses an *equal measure of Quality Energy and Information.*
7. The quantity of Quality Energy and Information identifies the anatomy of Information units structures: from qubits, to Bits, Free information, triplets, Information Networks (IN), and a final triplet which binds multiple INs. Physical units arise, ranging from the elementary structure of particles to various macro units: molecules, electro-chemical forms, cells, biological organisms, and humans. Each unit, bound by an invariant triple structure, preserves an invariant Information measure.
8. Interacting with other triplets, a triplet of bound Bits is connected in a macroprocess. The physics of the Information macroprocess describes the irreversible thermodynamics of interacting particles.
With the same measure of the units quality, but a growing amount of quantity, entropy increases, measuring the irreversibility of their macroprocess. Whereas the entropy difference ln2, classifying the disorder between the process impulses, have spent on forming the related units, is preserving along the macroprocess.
As the composite Information units grow, entropy increases. The minimum of process quality limits dissipation.
To continue binding the composite Information units, the number of the interacting thermodynamic processes has to increase. As the number of IN nodes grows the quality of the enclosing physical process energy decreases. Each IN node number identifies the quality of a particular process having such energy quality. The growing hierarchy of binding triple structures requires a multi-dimensional structure of the physical macroprocess. In the limit, a maximal density of high–quality and high–quantity energy requires an infinite high-dimensional process.
9. The physical structure's fundamental constant of bound Bits imposes an *Information* connection on the time and space (Sec.2.3.3).This constant identifies a *bridge between micro–and macroprocesses* emerging along the observing impulse interactions as they progress from maximal uncertainty to Information certainty. The connection concurrently forms a spatial structure of Information units during the time-space observation.
10. The triplets which enclose composite units build an Information Network (IN). Each IN knot-node enables the memorization of the bound triplets. Such a bound memory is the source an Information mass which holds the bound bits together. A physical Bit, memorized on the entropy-Information gap, binds other Bits in a physical macroprocess, being a source of physical mass. The evaluated physical mass measures the volume of a physical triplet with its Information invariant and Free Information available for cooperation.
11. The IN builds the hierarchical path of interacting energy qualities by binding the growing levels of knot-nodes into a chain. The IN of the memorized knots encodes both Information and a physical code.
12. Multiple physical triple units of the macroprocess (UP) adjoin the IN hierarchical structure of growing nodes. Free Information produces new UP at a higher level node and encodes the triple code logic (DSS). The unique position of each UP in the IN hierarchy defines the location of each code's logical structure.
13. The hierarchical levels of IN nodes classify the quality of assembled Information and energies.
The ending IN node enfolds all IN levels.
14. Each specific level of the IN hierarchy generates the specific clock time intervals at which access is opened to the next quality measure of external energy. This enables the memorization and encoding of the logical Bit hierarchy. The encoding logic encloses cognitive Information. The energy quantity (power) and quality of specific interaction limits the DSS code length through its final Bit's Information density. The total length of Information code limits the finite maximal dimension of the high–quality external energy which is delivered.
15. Multiple INs enclose Observer Information, cognition and intelligence. By being self-reflective to its DSS, the intelligent Observer can read and understand the meaning of the message.
Thus, the probability field of observing impulses enables the generation of various Information–physical units. These units satisfy the emergent Information minimax law, which dictates the allowable combinations of the invariant units being composed. For each allowable unit's combination, the fundamental constant and the emerging constraints provide the values of specific properties.

The energy quality, evaluated by the energy entropy measure, limits the initial process observing probability and its entropy. That also limits the code length when the observation starts.

**6.2.2. Analytical and numerical attributes distinguishing main stages of the evolutionary regularities, their thresholds and constraints**

1. Starting virtual observation with minimal probability and maximal uncertainty identifies the following primary threshold.

Minimal increasing probability approximates formula $\Delta p_N \to 2^{-N}$, where $N$ is the number of impulses starting the virtual observation (under Plank's physical uncertainty [85]).

2. At a given accuracy $\varepsilon_k \in (0,1), i = 1,2,....n$, the number of impulses $m_o$ within each of $n$ process' dimensions estimates [37,47]:

$$(1-\varepsilon_k)^3 / \varepsilon_k^{\ 2} = 1/2m_o S_{ki} \ , \qquad (2.1)$$

where $N = n \times m_o$ measures the total number of each $m_o$ entropy $S_{ki}$ increment. . Minimal realistic accuracy $\varepsilon_k = 4.5 \times 10^{-4}$ estimates $m_o = 8800$ with relative probability increment $\Delta p_k \cong 4.5 \times 10^{-4} \exp(-1) \approx 1.65 \times 10^{-4}$ where $S_{ki} = 1/2\sqrt{1-\varepsilon_k}$ estimates $S_{ki}$, and $\Delta p_k$ measures the ratio of Bayesian *a priori* probability $P_{ao}$ to *a posteriori* $P_{po}$ starting with frequency $f_o = 8800Hz$.

3. The entropy of error $S_k = 2(1-\varepsilon_k)^3 / N \varepsilon_k^{\ 2}$ at $\varepsilon_{kN} = 1/2^N$ and $N \to \infty$ leads to $S_{kN} = 2^{N+1}/N$, which estimates the *potential start of observation with a posteriori probability*

$$P_{poo} = P_{po} / m_o \cong 0.977 \times 10^{-4}. \qquad (2.1a)$$

4. If increasing correlation brings an impulse with entropy $S_{ki} = 0.5$, such an impulse temporarily holds the probabilities difference (closeness) consistent with accuracy $\varepsilon_{ko}$ of the starting correlation and minimal *a posteriori* probability $P_{poo} < P_{ao}$. The recursive action, overcoming a threshold of a maximal uncertainty with minimal *a priori* probability $P_{aoo} < P_{poo}$, automatically starts a virtual observation that connects the probing impulses in a potential virtual test.

5. Observing a random process under a Markov process's Bayesian probabilities reduces the difference (distance) between a random event $\xi_m$ and $\xi_n$ measured by $|\xi_m - \xi_n|$.

That may start and increase each posterior correlation, reducing conditional entropy measures at the following conditions beginning with the correlation and temporal memory.

According to [87:90], a coefficient correlation between above random events: $r_{mn} \le c(|\xi_m - \xi_n|)$ reaches the required stability at the sufficient condition

$$\lim_n n^{-2} \sum_{k=o}^{n-1} c(k) \times \sum_i^n D\xi_i = 0, c(k) > 0, , \qquad (2.2)$$

where

$$D\xi_i = E(\xi_i - E\xi_i)^2 = E\xi_i^{\ 2} - (E\xi_i)^2 \qquad (2.2a)$$

determines dispersion of random $\xi_i$. The correlation starts at the satisfaction condition (2.2a).

The relation of *a priori* and *a posteriori* probabilities [16:32] for the current random events along the observed trajectory evaluates the direct connection with the correlation.

The existence of correlation between random $\xi_m, \xi_n$ establishes coefficient correlation $\mathrm{r}_{mn}$ which defines formulas for the mathematical expectation of random events related to dispersions [2:87].

Threshold of starting correlation and observation time $\mathrm{t}_n$ determines realtion

$$c(n) = \sqrt{\frac{t_m}{\mathrm{t}_n}} / (|\xi_m - \xi_n|) > 0 \qquad (2.2b)$$

The starting correlation becomes stable if, at any initial $D\xi_i \neq 0$ (in 2.2a) and restricted (2.2b), it is found such $n$ when condition (2.2) is satisfied. Stable correlations keep temporal memory.

It is initially assumed that existence of trajectories of the stochastic process satisfy the limitation [87 :44]:

$$\lim_{c\to\infty} \overline{\lim_{t^o \to t}} P\{|\xi(t^o) - \xi(t)|\} > c\sqrt{(t^o - t)} = 0 \qquad (2.2c)$$

determined by this probability measure. At conformity of both differences in (2.2c), the first difference is called a Laplace variable [2:22] for which (2.2b) is satisfied.

For these variables, the coefficients drift and diffusion in a stochastic process determine the following relations [2:28]:

$$a(t) = \frac{(t_1 - t)\xi_o - (t - t_o)\xi_1}{t_1 - t_o}, \sigma^2(t) = \frac{(t_1 - t) - (t - t_o)}{t_1 - t_o} . \qquad (2.2d)$$

*Condition (2.2b) determines starting the virtual observation where interacting impulses begin correlation, which stabilizes condition (2.2) for a stochastic process satisfying* (2.2*d*).

The Markov drift and diffusion connects the additive functional, which links to the process correlation matrixes $r_t$:

$$E[a^u(t,\tilde{x}_t)^T (2b(t,\tilde{x}_t))^{-1} a^u(t,\tilde{x}_t)] = 1/2 r_t^{-1} \dot{r}_t \ . \qquad (2.2e)$$

6. Each elementary interaction with opposite actions ↓↑ models Dirac's delta-function, whose impulse's *interactive cut* originates from the step-down and step-up interactive actions within the impulse.

The impulse discrete function, switching the entropy from its minimum to the cutting maximum, and then back from the maximum to the next minimum, provides the maxmin–minimax principle.

The minimax variation principle establishes the *invariance of the impulse entropy measure through the observing process*.

7. Cutting the observing Markov diffusion process determines the minimal entropy of step-down interactive action ¼ Nat, the minimal increment between the interactive impulse 1/2 Nats, and the step-up action's entropy ¼ Nat.

An interactive impulse ↓↑ with both step-down and step-up virtual interactive actions carries the entropy 1Nat through the multi-dimensional observing process.

For ¼ Nat, as the threshold of minimal entropy increments $S_{ki1} = 1/4$ for a dimension $n = 1$, a minimal increase dimension to $n = 2$ brings minimal increments of the interactive impulse $S_{ki2} = 1/2$ Nat.

Correlation within each impulse holds the related time interval $\mathrm{r}_{im} = c\sqrt{\tau_{im}}$, which for `each impulse a common 1Nat unifies the impulse probability 0 or 1, the time interval measure, and the entropy measures:

$$M_p \to M_{im} = [1]_{\tau_{im}} \to [1]_{Nat} . \qquad (2.3)$$

For an impulse with minimal interactive entropy 1/4 Nat, its time interval $1/2o(\tau_k)$ measures the size square measure of that entropy:

$$M_{\tau_k} = [1/2o(\tau_k)]^2 = 1/4o(\tau_k)^2 \ . \qquad (2.3a)$$

During appearance of a space measure, the impulse, preserving measure (2.3), has to extend its initial time unit $1/2o(\tau_k)$ to $o(\tau_k)=2$ reaching measure

$M_p=[1/2\times 2]=[1]\rightarrow[1]_{Nat}$ . (2.3b)

If the impulse preserves the invariant maxmin entropy measure, then the impulse's equivalent time and space intervals are connected through imaginary time directly. That follows from correlation $r_{tj}\rightarrow\pm\sqrt{\delta_{ij}},\delta_{ij}=(t_i-t_j)>0$ which for an inverse time interval $\delta_{ij}=-\delta_{ji}$ are imaginary. This occurs inside the cutting impulse with the emerging space interval.

8. The step-down action cuts the correlation which holds both the entropy and probability events hidden in the cutoff correlation. Bayesian probabilities automatically increase posterior correlations.

Assuming each probability 0 or 1 is *a priori* or *a posteriori* accordingly for a virtual impulse, from relation [16] it follows that each impulse posterior correlation $r_{im}$ increases relatively to the impulse starting auto-correlation $r_{io}$ in ratio

$r_{im}/r_{io}=4$ . (2.4)

Such self-growing correlation indicates the *emergence of an elementary virtual Observer with measure* (2.3b) *and self-cutting the observing correlations* (originating from the step-down and step-up interactive actions within the impulse).

In such virtual observer, a self–observing process starts if an impulse delivers minimal entropy $s_i=1/2$ to the following impulses, reaching this entropy threshold. Its posteriori action virtually coveys the next impulse cutting action, enabling the process to continue through self-support.

This virtual Observer rises as a part of the observing random process with interactive impulses.

9. Growing correlations intensity of entropy per the interval (as entropy density) that increases on each following interval and indicates a shift between the virtual actions, a displacement. The displacement identifies an entropy gap between the invariant impulses. Displacement $a$, starting under physical uncertainty inside of sub-Plank region [88,16], measures the proportion of the Plank constant $\hbar$ to number $N$: $a=h^o/2\pi N=\hbar/N$ of the impulses reaching $a$. This ratio evaluates the relative closeness of the displacement to the uncertainty needed to reach the standard Plank edge.

The minimal relative displacement evaluates ratio

$a^*/a=1.000262774N/N_*$ at $N/N_*=1$. (2.5)

The relative displacement's distance from its minimal value (2.5) evaluates ratio

$d_a=1-1.000262774N/N_*$, (2.5a)

Relation (2.5a) measures maximal distance of minimal displacement (2.5) from the Plank edge.

The maximal distance estimates the interactive impulse with space measure, which begins forming a minimal volume at that displacement.

The interactive impulse' momentum rotates a shift between the displaced states.

The opposite actions create the shift starting with a finite entropy of the displacement gap.

An extreme entropy for multiple impulses identifies the minimal difference between the opposite actions measured by time shift $\delta_k^{\tau+}/4$, which evaluates the finite impulse width (before starting the space interval).

The ratio of entropy of the impulse step-down action's width to part to entropy 0.25 Nat of that impulse step-down action evaluates the relative width

$\upsilon_o=(0.025/0.25)=0.1$ . (2.5b)

The minimal displacement distance between the invariant impulses equal to $d_a = 0.1$ can be reached using (2.5b) under a ratio of the numbers of observing impulses:

$N_* / N = 1.111403$.

To reach minimal displacement (2.5b) the initial $N = m_o = 8800$, starting the observation, needs to increase up to $N_* \cong 9780$.

9a. The entropy gradient, curving displacement (2.5b), measures the growing entropy force.

Under the growing entropy gradient, the curving displacement estimates its starting radius

$r_{e1} = \sqrt{1 + (0.025/0.25)^2} = \mp 1.0049875$. (2.5c)

That radius defines the verge of the threshold. The curving rotation starts by overcoming it.

Rising virtual Euclid's curvature $K_{e1} = (r_{e1})^{-1}$ estimates this threshold by

$K_{e1} \cong +0.995037$ . (2.5d)

Starting step-down curvature's radius initiates the emerging rotation movement of the impulse, whose trajectory (Fig.3) follows from the minimax variation principle.

Radius (2.5c) determines the initial angle $\beta$ of the rotation trajectory of cone Fig. 3 from relation

$r_{e1} = \rho = b \sin(\varphi \sin \beta)$ at $\varphi = \pi k / 2, \mathrm{k} = 1, b = 1/4$. (2.5e)

10. The impulse step-up action displaces the time measured virtual impulse's interval through rotation on angle $\varphi = \pi / 2$. The displacement within the impulse changes the discrete timespace form of the impulse that requires *preserving* its measure (2.3b) in the emerging timespace coordinate system.

Comments. Let the rotation start on a spherical surface at conditional probability distribution for a distance with latitude $\theta: -\pi \le \theta \le \pi$ at given longitude $\psi$ having form [87:75]:

$$P(\theta_1 \le \theta \le \theta_2 \mid \psi) = 1/4 \int_{\theta_1}^{\theta_2} |\cos\theta| \, d\theta .$$

Then this conditional probability distance is irregular.•

That indicates changing an impulse's time interval unit with the appearance of curved impulses, which is extending while curving. With growing probability, the intensity of the entropy force draws together the impulse action and reaction, squeezing the time interval between these actions up to the jump when these actions merge and start the microprocess.

The invariant measure is conserved in following timespace movement (Sec.2.3.4).

Preserving the impulse $\bar{u}_k$ measure $|M_{io}| = |1|_M \; [\tau] \times [l]$ at its height $h = 2$ and plain part $p = 1/2$ of time in $M[\bar{u}_k] = |2 \times 1/2| \xrightarrow{\mathrm{p}[\bar{u}_k]} |1|_M$ leads to following the space $[l]$ and time $[\tau]$ measures:

$[l] = \pm[(|M_{io}| / |1|_M)(2/\pi)]^{1/2}, [\tau] = \mp[(|M_{io}| / |1|_M)(\pi/2)]^{1/2}$ , (2.6)

and to $|M_{io}| = M[\bar{u}_k]\pi/2 \times [l]^2$, which at

$p = 1/2h, M[\bar{u}_k] = 1/2h^2, 1/2h^2[l]^2 \pi/2 = |M_{io}|$ and $h[l] = 2$ holds

$|M_{io}| = \pi$ . (2.6a)

That impulse's timespace irrational measure preserves the impulse entropy measure, when the virtual Observer is cutting the correlations in the curving rotation.

The invariant impulse measures 1/2 curving circle.

*Condition (2.6) determines the emerging space-time impulse with measure (2.6a) after overcoming a threshold (2.5b), which defines the starting rotation with* $\varphi = \pi / 2$ .

The rotating coordinate system of the curved impulse starts angular velocity $c$ measured by the rate of changing the angular displacement.

In the rotating space-time, an impulse appears, starting virtual observer's geometrical shape with volume $V_c = 2\pi c^3 / 3(k\pi)^2 tg\psi^o$ [47:192] determined by the initial space angular velocity $c$, the cone geometrical parameter $k$, and the angle at each cone vertex $\psi^o$ (Fig. 3).

11. The displacement shift's parameters define the following relations.

The Information analog of Plank constant $\hat{h}$, at maximal frequency of energy spectrum of Information wave in its absolute temperature, evaluates the maximal Information speed of the observing process:

$$c_{mi} = \hat{h}^{-1} \cong (0.536 \times 10^{-15})^{-1} Nat / \sec \cong 1.86567 \times 10^{15} Nat / \sec. \quad (2.7)$$

That value also estimates a minimal time interval corresponding the time shift:

$$\delta t_e \cong 1.59459 \times 10^{-14} \sec \approx 1.6 \times 10^{-14} \sec . \quad (2.7a)$$

Time shift at maximal light speed $c_o = 3 \times 10^9 m / \sec$ allows estimating a minimal space shift:

$$\delta_{lo} \approx 4.8 \times 10^{-5} m. \quad (2.7b)$$

The angular velocity, emerging with maximal linear speed $c_o$, curves length $\delta_{lo}$ to the length

$$\delta_{low} = \pi \delta_{lo}[m], \ \delta_{low} = 15 \times 10^{-5} m \ . \quad (2.8)$$

Ratio $c_o / \delta_{low} \cong w_o$ approximates a maximal angular velocity for the curved length $\delta_{low}$:

$$w_o \approx 0.1989 \times 10^{14} \sec^{-1} . \quad (2.8a)$$

Maximal entropy speed can rotate the entropy increments $\Delta s_{apo} = -\ln(0.8437) \cong 0.117 Nat$ on the starting displacement with maximal entropy angular velocity

$$w_{oe} = 0.73 \times 10^{15} Nat / \sec \ . \quad (2.8b)$$

12. The microprocess emerges inside a random process, modeling by the Markov diffusion process, when the displacement verges at distance (2.5b) reaches the minimal time interval (2.7a) upon merging the nearest impulse's opposite actions.

The opposite actions $u_-^t$ and $u_+^t$ are fixed variables of the Markov diffusion process, which preserves both their additive and multiplicative functions.

It requires the fulfillment of functions

$$u_+^t - u_-^t = u_+^t \times u_-^t \quad (2.9)$$

which leads to

$$u_+^t / u_-^t = 2 \quad (2.9a)$$

if both actions are real. And to functions

$$u_{+*} = (j-1), u_{-*} = (j+1) \quad (2.9b)$$

when both actions are complex conjugated, with their ratio (2.9a) at

$$u_{+*}^t = j\sqrt{2}, u_{-*}^t = -j\sqrt{2} \ . \quad (2.9c)$$

At these conditions, both additive and multiplicative measures equal to $\mathrm{U}_a = U_m = -2$ .

When the sub-Markov process gets negative entropy measure of the impulse actions $S^*_{\mp a} = -2$ with relative probability $p_{a\pm} = \exp(-2) = 0.1353$, it starts opposite imaginary actions (2.9b) or (2.9c), initiating the microprocess.
Within the impulse time interval $\tau = 1Nat$, *entanglement starts before its space is formed and ends with the beginning of the space during the reversible relative time interval of* $0.015625\pi$ part of the impulse invariant measure $\pi$.
*Since entanglement has no space measure, the entangled states can be everywhere in a space.*
13. The space interval, beginning the displacement shift, starts within interval of entanglement having the probability $P_{po*} \cong 0.8231$, continues during the shift, and extends to the space part of the impulse multiplicative measure after the displacement ends. That means the displacement widens, extending its ending probability up to the impulse's inner part, where it ends with probability $P^i_n = 0.86$, holding entropy $S_{\pm} = 0.15$. The end of displacement indicates the formation of a space interval within that impulse.
Or *a priori* $i$-probability $P^i_n = 0.86$ is the indicator of the appearance of the first impulse space interval (at $n$-from starting observation).
If this impulse's positive curvature interacts with the next impulse's negative curvature, then the interacting part holds the transitional curvature entropy sum $S_{\Delta} = 0.5085$ [80:135].
The difference $S_{\pm} - S_{\Delta} \cong 0.01$ estimates the increment of both impulse asymmetries.
This means the opposite asymmetries of interacting impulses estimates probability $P^i_n = 0.86$. The increment of the probability, starting an external interacting impulse, and the probability of injecting energy evaluates: $\Delta P_{ie} = 0.981699525437 - 0.9855507502 = -0.1118$ holds entropy $\Delta S_{\pm a} = -2.191$.
The difference $\delta S_{\pm} = -0.191$ determines the related increment of entropy within this impulse before the injection of Landauer's minimal energy $\ln 2$ within the interval of encoding information $\ln 2$ Nat.
The imaginary microprocess ends with the entangling entropy volume, the Information microprocess emerges with providing energy, killing that entropy and memorizing the classical Bit by the end of external impulse.
14. Probability $p^*_{\pm} = \exp(-2h^{o*1}_{\alpha}) \cong 0.9866617771$ identifies physical structural parameter $h^{o1}_{\alpha}$ which counts the sub-Plank spot above, resulting from the interactive impulse with this probability during the observation.
On a path from uncertainty to certainty, the increasing number of interacting impulses $N = 8800$ allows the observer closer approach to the gap of reality through decreasing uncertain displacement of the sub-Planck spots.
After entropy volume of the $N+$ impulses increase to overcome uncertain volume (2.5b), the entropy reaches the edge of certainty-reality with increasing probability $p^*_{\pm}$.
Since a Bit is created at the probability approaching 1 with the number of each interaction $N^o_* \cong 8828$, each impulse observation can create the Bit with frequency

$$F_{im} = 1/8828 = 10^{-4} \times 1.13276. \qquad (2.10)$$

Moreover, because each Bit creation needs a final interaction of the impulses with opposite curvatures, such interaction needs $N = 8800$, which evaluates the probability, and the frequency of appearance that impulse

$$F_{imo} = 1/8800 = 10^{-4} \times 1.13636. \qquad (2.10a)$$

Both frequencies evaluate the optimal number of impulses for a single observation.
The Information Bit, as two memorized qubits, can be produced through interaction, which generates the qubits contained by a material or device (a conductor-transmitter) that preserves the curvature of the transitional impulse entropy inside a closed device. Memorizing the entangled curvature is the Information "demon cost" for the entangled correlation, which naturally holds its entropy, time, and the curvature of the transitional impulse. •

**6.2.3. Specifics of forming quantum microprocess within classical Markov process**

1. Observation of discrete events, which includes interaction by probabilities of observing and observable processes.
2. Observable process describes Markov chain.
3. Observing the Markov chain transforms to Markov diffusion process.
4. The Markov transitive probabilities define Bayes probabilities [87:77], whose sequential a priori a posteriori probabilities bring probabilistic impulse observation of the Markov diffusion process.
5. Each observing action $\downarrow$, in a sequence of these probabilities, virtually cuts entropy of the impulse correlation from the observable Markov diffusion process and transfers action $\uparrow$ to the Bayes observing process impulses. During transition switching time, each impulse accumulates entropy spent on the interactive transition, sequentially increasing the entropy of the observing impulses. The entropy, cutting from the Markov process decreases, while entropy enclosing within each impulse interactive action $\downarrow\uparrow$ increases. Each following interactive impulse of the Bayes probabilities, observing the Markov process, maximizes the entropy of the Bayes observing process impulses. The multiple Bayes observing impulses generate the observing Markov diffusion whose entropy increases comparing with that in the observable Markov process.
6. Under such observations, the Markov diffusion evolves to a Lévy process [2]. The Lévy process' independent random increments model multiple observing impulses of the random bits. The Lévy-Khintchine decomposition [2:113-117] characterizes any Lévy process in terms of three components (the Lévy triplet) one of which is the Lévy measure. This measure describes the distribution of the Lévy process potential jumps. Until any two time intervals of the process'independent random increments do not overlap, the jump absent and Lévy measure is zero. The overlap is a merge of the yes-no impulses, when Lévy measure is nonzero, finite. The merge of the yes-no impulses indicates zero random bit. When Levy measure becomes *finite*, the jumps have started. The jumps initiate a microprocess within the observing Levy process, which releases the entropy-uncertainty hidden in the cutting correlation, finally generating a non-random real bit. With zero probability of the microprocess starts, it is unpredictable. The microprocess arises during the cutting micro-time interval of the merging impulses through the observation. The information process starts when the microprocess generates real qubit and/or bit. Thus, the microprocess can start within classical Markov diffusion transformed to Levy jump. The merge location on the observing trajectory identifies the impulse' probability and entropy.

However the evolving the microprocess requires additional requirements:

a. The jumping $\downarrow\uparrow$ actions at a reversible locality of Levy process are symmetrical with equal probabilities. Both these probabilities and entropy increments satisfy multiplicative laws.
b. The jump $\uparrow$ initiates an extreme entropy gradient on the orthogonal time intervals, identifying the virtual radius of rotation of the time during its orthogonal displacement. The jumps initiate increments of these actions' entropy movements. The moving anti-symmetric trajectories initiate rotating conjugated dynamics within the jumping impulse.
c. The rotating conjugated dynamics at angle $\mp\pi/4$ become additive with the equal opposite entropy increments, which starts a transitional impulse.
d. The continued rotation turns these opposite entropy increments on angle $+\pi/4-(-\pi/4)=\pi/2$ approaching the entropies entanglement.
e. The entanglement starts at angle $(\pi/2)\times 0.4375<\pi/4$ takes relative time interval of the impulse $\delta t_{\pm}^{ko}/\tau_k \cong 0.03125$ up to ends of angle $\pi/2$.

Since only at angle $\pi/2$ the space interval within impulse begins, it means that *the entanglement starts before the space is formed and ends with beginning the space.* Moreover, the time interval within the entanglement that enables creating the space is reversible. The conjugated entanglement increments are additive within a transitive impulse ↑↓ setting up an entropy qubit.

f. The subsequent step-up jumping action ↑ changes increment which during this rotation delivers energy compensating the entropy qubit. That reveals information Bit at the impulse cutting moment within the impulse measure 1 Nat. At these conditions, probability function for QM satisfies multiplicative relation for the interference of probabilities $p_{\pm a}=\sqrt{p_{+a}p_{-a}}$ and additive relations $p_{+}+p_{-}=p_{\pm am}$ for the additive entropy increments in transitional impulse. Resulting probability holds $p_m=\sqrt{p_{+a}p_{-a}}+p_{+}+p_{-}$ which corresponds to that in classical Schrödinger Eq.

g. Still, during transformation from entropy (uncertainty) to information (certainly) on edge of reality exists a gap which is a part of the information regularities of observer [37,53].

### 6.2.4. Math Summary. Interactive Integrated Information Dynamics (IIID)

1. Probabilities and conditional entropies of random events.

A *priori* $P^{a}_{s,x}(d\omega)$ and *a posteriori* $P^{p}_{s,x}(d\omega)$ probabilities observe the Markov diffusion process $\tilde{x}_t$ distributions of random variable $\omega$ (events).

For the $i,k$ random event $A_i, B_k$ along the observing process, each conditional *a priori* pr which obability $P(A_i/B_k)$ follows the conditional *a posteriori* probability $P(B_k/A_{i+1})$.

Conditional Kolmogorov probability

$$P(A_i/B_k)=[P(A_i)P(B_k/A_i)]/P(B_k) \quad (3.1)$$

defines the Bayes probability after substituting average probability:gap

$$P(B_k)=\sum_{i=1}^{n}P(B_k/A_i)P(A_i).$$

Conditional entropy

$$S[A_i/B_k)]=E[-\ln P(A_i/B_k))]=-\ln\sum_{i,k=1}^{n}P(A_i/B_k)]P(B_k) \quad (3.1a)$$

averages the conditional Kolmogorov-Bayes probability for multiple events along the observing process.

Conditional probability satisfies Kolmogorov's 1-0 Law for function $f(x)|\xi$ of an $\xi,x$ infinite sequence of independent random variables:

$$P_{\delta}(f(x)|\xi)=\begin{cases}1, & f(x)|\xi)\ge 0\\ 0, & f(x)|\xi)<0\end{cases}. \quad (3.1b)$$

This probability measure has been applied for the impulse probing an observable random process, which holds opposite Yes-No probabilities-as the unit of the probability impulse step-function.

Random current conditional entropy of the finite sequence of the random events is

$$\tilde{\mathrm{S}}_{ik}=-\ln \mathrm{P}(A_i/B_k)\,\mathrm{P}(B_k). \quad (3.1c)$$

Probability density measure on the Markov process trajectories:

$$p(\omega)=\frac{\tilde{P}_{s,x}(d\omega)}{P_{s,x}(d\omega)}=\exp\{-\varphi_s^t(\omega)\}\ , \quad (3.1d)$$

is connected with this process additive functional

$$\varphi_s^T=1/2\int_s^T a^u(t,\tilde{x}_t)^T(2b(t,\tilde{x}_t))^{-1}a^u(t,\tilde{x}_t)dt+\int_s^T(\sigma(t,\tilde{x}_t))^{-1}a^u(t,\tilde{x}_t)d\xi(t)\ , \quad (3.1e)$$

defined through controllable functions drift $a^u(t,\tilde{x}_t)$ and diffusion $b(t,\tilde{x}_t)=1/2\sigma(t,\tilde{x}_t)\sigma(t,\tilde{x}_t)^T$ of the process, where (3.1e) also describes the transformation of the Markov process's random time traversing the various sections of the process trajectory.

2. The *integral measure* of the observing *process* trajectories formalizes an *Entropy Functional* (EF), which is expressed through the above functions of Markov diffusion process $\tilde{x}_t$:

$$\Delta S[\tilde{x}_t]|_s^T = 1/2E_{s,x}\{\int_s^T a^u(t,\tilde{x}_t)^T(2b(t,\tilde{x}_t))^{-1}a^u(t,\tilde{x}_t)dt\} = \int_{\tilde{x}(t)\in B} -\ln[p(\omega)]P_{s,x}(d\omega) = -E_{s,x}[\ln p(\omega)], \quad (3.2)$$

and the probability density measure on the process trajectories.

3. Cutting the EF by the impulse delta-function determines the increments of Information for each impulse:

$$\Delta I[\tilde{x}_t]|_{t=\tau_k^{-o}}^{t=\tau_k^{+o}} = \begin{cases} 0, t<\tau_k^{-o} \\ 1/4Nat, t=\tau_k^{-o} \\ 1/4Nat, t=\tau_k^{+o} \\ 1/2Nat, t=\tau_k, \tau_k^{-o}<\tau_k<\tau_k^{+o} \end{cases} \text{ (3.3) with total } \sum_{t=\tau_k^{-o}}^{t=\tau_k^{+o}} \Delta I[\tilde{x}_t]_{\delta t} = 1Nat. \quad \text{(3.3a)}$$

4. *The Information Path Functional* (IPF) unites the Information cutoff contributions $\Delta I[\tilde{x}_t/\varsigma_t]_{\delta_k}$ along $n$-dimensional Markov process impulses during its total time interval $(T-s)$:

$$I[\tilde{x}_t]|_s^{t\to T} = \lim_{k=n\to\infty}\sum_{k=1}^{k=n}\Delta I[\tilde{x}_t/\varsigma_t]_{\delta_k} \to S[\tilde{x}_t], \quad (3.4)$$

which in the limit approaches the EF.

The IPF along the cutting time correlations on optimal process trajectory $x_t$, in the limit, determines equation

$$I[\tilde{x}_t/\varsigma_t]_{x_t} = -1/8\int_s^T Tr[(r_s\dot{r}_t^{-1}]dt = -1/8Tr[\ln r(\mathrm{T})/r(s)]. \quad (3.4a)$$

5. The equation of the EF for a microprocess:

$$\partial S(t^*)/\delta t^* = u_\pm^{t1}S(t^*), u_\pm^{t1} = [u_+ = \uparrow_{\tau_k^{+o}}(j-1), u_- = \downarrow_{\tau_k^{+o}}(j+1)] \quad (3.5)$$

under inverse actions of function $u_\pm^{t1}$, starts the impulse opposite time $t_\pm^* = \pm\pi/2t^i$ which measures a space rotating angle relative to the impulse inner time $t^i$.

The equation' solutions for the conjugated entropies $S_+(t_+^*)$, $S_-(t_-^*)$ determine functions $S_+(t_+^*) = [exp(-t_+^*)(\mathrm{Cos}(t_+^*) - jSin(t_+^*))]|, S_-(t_-^*) = [exp(-t_+^*)(\mathrm{Cos}(-t_+^*) + jSin(-t_+^*))]$ at

$$\begin{aligned} S_\pm(t_\pm^*) &= 1/2S_+(t_+^*)\times S_-(t_-^*) = 1/2[\exp(-2t_+^*)(\mathrm{Cos}^2(t_+^*) + Sin^2(t_+^*) - 2Sin^2(t_+^*))] = \\ &1/2[\exp(-2t_+^*)((+1-2(1/2-Cos(2t_+^*))))] = 1/2\exp(-2t_+^*)Cos(2t_+^*) \end{aligned}. \quad (3.5a)$$

Minimal interactive entropy $S_\pm(t_\pm)$ *begins the space measure* during reversible relative time interval $0.015625\pi$ of the impulse invariant measure $\pi$.

The running microprocess, overcoming the entropy-Information gap, starts Information Bit and an Observer information macrodynamics.

6. The Information macrodynamic equations:

$$\partial I/\partial x_t = X_t, a_x = \dot{x}_t = I_f, I_f = b_t X_t \quad (3.\ 6)$$

define $X_t$-a gradient (force) of Information path functional $I$ (3.4) on macroprocess' trajectories $x_t$, $I_f$- Information flow determined through speed $\dot{x}_t$ of the macroprocess. The flow emerges from drift $a^u(t,\tilde{x}_t)$ being averaged by function $a_x$ along the observing process, and the averaged diffusion $b_t \to b$ for the macroprocess force. The Information Hamiltonian of the macrodynamics:

$$-\frac{\partial \tilde{S}}{\partial t} = (a^u)^T X + b\frac{\partial X}{\partial x} + 1/2a^u(2b)^{-1}a^u = -\frac{\partial S}{\partial t} = H \,. \quad (3.7)$$

determines macro equations (3.6) from the minimax variation principle using Jacobi-Hamiltonian equations. Equations (3.6) are the Information form of the equations of Irreversible Thermodynamics [13-16, 45-47], which the *Information Macrodynamics* generalize.

The discretely changed Information Hamiltonian, during the impulse interactive observations, divides irreversible dynamic trajectory on the partial reversible segments, predicting the next emerging Information unit. The flows and forces determine the macroprocess Hamiltonian in the invariant form $H = X \times I$ .

Information curvature $K_{\alpha}^{m}$, density of Information mass $M_{vm}^{*}$, and effective complexity $MC_{m}^{\delta e}$ connect equation (Sec.4.2,4.3):

$$K_m^{\alpha} = M_{vm}^{*} MC_m^{\delta e} \,, \quad (3.8)$$

where

$$MC_m^{\delta e} = 3\dot{H}_m^V MC_m \quad (3.9)$$

includes the differential of Hamiltonian per volume $\dot{H}_m^V$ and the IN cooperative complexity $MC_m$ .

*The single Eq. (3.8) at (3.8a) encloses all previous Eqs. (3.1-3.7a), unifying the formal math description of this approach.*

**6.3. Forming Information Observer with its regularities**

Information processes begin in the same state of the unknown and the emergent, represented by a state of uncertainty. At this stage, nothing is distinct, and there is no energetic or physical basis. In the microprocess, virtual particles emerge and vanish. Statistical structures arise.

Step by step, the IIID formalism traces an Information process from its random source, following regularities which eventually make a transition to space-time conditions.

The Information process acquires a triplet structure and thermodynamic features.

The macro-movement in rotating time–space coordinate system forms Observer's information structure confining its multiple INs. That determine the *Observer time of inner communication* with *self –scaling* requesting and accumulating information. Each Observer *owns the inner time of information processing* and *scale* of the required information (on the micro and macrolevels), depending on density of the IN nodes information.

The current information cooperative force, initiated by free information, evaluates the observer's *selective* actions attracting new high-quality information. *S*uch quality delivers a high density-frequency of related observing information through the IN selective mechanism of the requested information.

These actions engage acceleration of the observer's information processing, coordinated with new selection, quick memorizing and encoding each IN node information with its logic and space-time structure, which minimizes the spending information. It determines observer's self-organized feedback loop.

The observer optimal *multiple choices*, needed to implement the minimax self-directed strategy, evaluate the amount of the information emanated from the IN integrated node, which *identifies the attracting cooperative force*.

The IN nested structure holds cooperative complexity measuring *origin* of complexity in the interactive dynamic *process* cooperating doublet-triplets, whose free information anticipates new information, requests it, and automatically builds the hierarchical IN, which decreases complexity of not cooperating yet information units.

Minimal selected information bilds an objective observer (like rocks) while a threshold separates them from the self-organizing subjective observers (like animal, people) self-requiring needed information.

The self-built information structure, under the self-synchronized feedback, drives self-organization of the IN and the *evolution macrodynamics* with ability of its self-creation.
The free information, arising in each evolving IN, builds the Observer specific time–space information *logical* structure that conserves its “cognition” as intentional ability to request and integrate the explicit information in the IN highest level.
Multiple IN’s ending triplets assembles the common units through attraction and resonances forming the IN-cognition, which accepts only units that concentrates and recognizes each IN node.
The coordinated selection, involving verification, synchronization, and concentration of the observed information, necessary to build its logical structure of growing maximum of accumulated information, *unites* the observer’s *organized intelligence action*. The IN hierarchical level’s amount of quality of information evaluates *functional organization* of the intelligent actions spent on this action. The integrated quality information of the highest level IN measures the Observer *Information Intelligence*. The cognitive logic establishes time course for encoding the intelligence information in double spiral triplet code (Figs.7-8). The IN node’s current hierarchical level cooperates the communicating observers’ existing level of integrated knowledge.
The intelligent observer uncovers a meaning of observing process using the common message information language and the cognitive acceptance, which are based on the qualities of observing information memorized in the IN hierarchy. The intelligent observer recognizes and encodes digital images in message transmission, being self-*reflective enables understanding the message meaning.*
The increasing INs hierarchy enfolds rising information density which accelerates grow the intelligence, which concurrently memorizes and transmits itself over the time course in an observing time scale. Mind emerges from timespace to navigate a local space as an activity not a substance.
The self-creating observer’s intelligence enables decoding each IN’s node memorized information recognizing it in intercommunication associated with self-awareness.
The observer intelligence includes *hierarchical function* of self-awareness modeling consciousness which starts within the entropy-information gap, where the lack of information unable any computation.
On at a higher hierarchical level, it emerges with understanding a meaning through inner-communication, where, until the understanding, this function is not revealed. The intelligence, growing with its time interval, increases the observer life span. The Observer cellular space-time geometry (Fig.8) integrates all observed interactions, while the cellular code encodes these interactions in discrete information units. This integral information is compressed to a single bit by the information path integral. The self-organized, evolving IN’s time-space distributed information structure models *artificial intellect.*
The Equations of *Interactive Integrated Information Dynamics* (IIID) (3.2-3.9) formalize the description of information regularities of observing interactive process. Instead of designing the AI manipulators, the IIID focuses on creation an artificial brain in evolving information process of the intelligent Observer. Since brain neurons operate with information form of different observation, images representations, they recognize the observing information process up to rising the observer cognition and intellect. Instead of unspecified “synthetic connections of brain network” [89], we use information mechanism of the networks’ ending free information which automatically arranges each discrete network connection building cognition and intelligence according to observer’ integral task.
*This approach, implementing concepts of Reality, Information, and Information Observer, starts with interactive probabilistic observation, tracing path of the interactions, reveals emerging information qubits, bits, and forming information process, which evolving, builds information structure of multiple observer as physical objects.*
The results have obtained based on the simulating mathematical models, software programs, which confirm the various applications cited in the publications [16, 47, 48, 71,73,74,80 others].

**References**

[1].R.Höhnle, K., Sturm, Some zero-one laws for additive functionals of Markov processes. *Probab. Th. Rel. Fields* 100, 407–416 ,1994.

[2]. P.P. Levy, *Stochasic Processes and Brownian movement*, Deuxieme Edition, Paris, 1965.

[3]. E.B. Dynkin, On transforming a certain class of stochastic processes by absolutely continuous substitution of measures, *Theory of Probability and its Applications*, **5**: 285-301, 1960.

[4]. V.S. Lerner, The boundary value problem and the Jensen inequality for an entropy functional of a Markov diffusion process, *Journal of Mathematical Analysis and Applications*, **353** (1), 154–160, 2009.

[5]. M. Kac, *Probability and Related Topics in Physical Sciences*, Boulder, Colorado, 1957.

[6]. E.B. Dynkin, Additive functional of a Wiener process determined by stochastic integrals, *Teoria. Veroyat. i Primenenia*, **5** , 441-451, 1960.

[7]. I. V.Girsanov, On transforming a certain class of stochastic processes by absolutely continuous substitution of measures, *Theory of Probability and its Applications*, **5**: 285-301, 1960.

[8]. R.M. Blumenthal, R.K. Getoor, Additive functionals of Markov processes in duality,*Trans. Am. Math.Soc.*, **112**( 1), 131-163,1964.

[9]. Y.V. Prokhorov, Y.A. Rozanov, *Theory Probabilities*, Nauka, Moscow, 1973.

[10]. W. Feller, The general diffusion operator and positivity preserving semi-groups in one dimension. *Ann Math.* **60**, 417-436, 1954.

[11]. W. Feller, On boundaries and lateral conditions for the Kolmogorov differential equations. *Ann. Math.* **65**, 527–570, 1957.

[12]. V.S.Lerner, Dynamic approximation of a random information functional, *J. Mathematical Analysis and Applications*, **327**(1), 494-514, 2007.

[13].V.S. Lerner, The information path functional approach for a dynamic modeling of a controllable stochastic process,*arXiv*, 1201.0035.

[14] V. S. Lerner, The Impulse Interactive Cuts of Entropy Functional Measure on Trajectories of Markov Diffusion Process, Integrating in Information Path Functional, Encoding and Application, *British Journal of Mathematics & Computer Science*, **20**(3): 1-35, 2017.

[15].V.S. Lerner, Information Path from Randomness and Uncertainty to Information, Thermodynamics, and Intelligence of Observer, *arXiv*:1401.7041.

[16]. V.S. Lerner, *The observer information processes and origin of the observer cognition and intellect*, GRIN Verlag Publishing GmbH, 2017.

[17] J. M.Harrison, T. M Sellke, and A. J.Taylor, Impulse control of Brownian Motion, *Math. Oper. Res.*,**8**: 454-466, 1983.

[18] M.Fukushima, P.He, and J.Ying ,Time changes of symmetric diffusions and Feller measures, *The Annals of Probability*, **32**(4): 3138–3166, 2004.

[19] J.Barral, Fournier N., S.Jaffard and S. Seuret, A pure jump Markov process with a random singularity spectrum, *The Annals of Probability*, **38**(5): 1924–1940, 2010.

[20] K.Ito and S.Watanabe,Transformation of Markov processes by Multiplicative Functionals. *Ann. Inst. Fourier*, **15**: 13-30, Grenoble, 1965.

[21] R. K Getoor. Killing a Markov process under a stationary measure involves creation, *The Ann. of Probability*, **16**(2):564-585,1988.

[22] B.Song,.Sharp bounds on the density, Green function and jumping function of subordinate killed Brownian Motion, *Probab. Th. Rel. Fields*, **128**: 606-628,2004.

[23] G. A.Korn and T. M. Korn, *Mathematical handbook for scientists and engineers*, McGraw-Hill, 1968.

[24]. J. A. Wheeler and R. P.Feynman, Interaction with the absorber as the mechanism of radiation. *Reviews of Modern Physics*, **17**(2-3):157, 1945.
[25] J. A. Wheeler and R. P.Feynman,Classical electrodynamics in terms of direct interparticle action. *Reviews of Modern Ph*ysics, **21**(3):425, 1949.
[26]. Le Jan, Yves. *Markov paths, loops and fields, Lecture Notes in Mathematics*, Vol. 2026, Springer, Heidelberg, 2011, Lectures from the *38th Probability Summer School* held in Saint-Flour, 2008.
[27].K. Krane, *Modern Physics*, Wiley, New York, 1983.
[28].J.A.Wheeler, Information, physics, quantum: the search for link, Proc. 3rd Int. Symp. *Foundations of Quantum Mechanics*, Tokyo, 354-368. 1989.
[29]. C. Jarzynski, Nonequilibrium Equality for Free Energy Differences, *Phys. Rev. Lett.*,**78**, 2690. 1997.
[30]. M. Pavon, Quantum Schrödinger bridges, *arXiv /quant-ph/0306052*.
[31]. V. S. Lerner,. *The Information Hidden in Markov Diffusion*, Lambert Academic Publisher, 2017.
[32]. S. Bernstein, Sur les liaisons entre les grandeurs aléatoires, *Verh. Int. Math. Kongress, Z¨urich*, Vol. I, 288-309,1932.
[33]. V.S.Lerner, Natural Encoding of Information through Interacting Impulses**,** *arXiv: 1701.04863, and IEEE Xplore http://ieeexplore.ieee.org/xpl*/Issue7802033: 103-115,2016**.**
[34].R. Landauer Irreversibility and heat generation in the computing process, *IBM Journal Research and Development*, **5**(3):183–191, 1961.
[35]. M Perkovac, Maxwell’s Equations as the Basis for Model of Atoms. *Journal of Applied Mathematics and Physics,* **2**, 235-251, 2014.
[36].Lerner V. S. Information Geometry and Encoding the Biosystems Organization, *Journal of Biological Systems*, **12** (2):191-231, 2005.
[37]. V.S. Lerner, Emergence time, curvature, space, casualty, and complexity in encoding a discrete impulse information process, *arXiv*:1603.01879.
[38] Chu Shu-Yuan. Time-Symmetric Approach to Gravity, *arXiv:* gr-qc/98020v1, 1998.
[39] N. Wolchover, How Space and Time Could Be a Quantum Error-Correcting Code, *Quanta Magazine*, January, 2019.
[40]. V.S. Lerner, Natural Encoding of Information through Interacting Impulses, *arXiv* 1701.04863.
[41]. V. S. Lerner, The impulse observations of random process generate Information binding reversible micro and irreversible macro processes in Observer: regularities, limitations, and conditions of self-creation, *arXiv*: 1204.5513.
[42].V.S. Lerner, *Natural Encoding of Information through Interacting Impulses*, Lambert Acad Publishing, 2018.
[43] A. Bérut, A. Arakelyan, A. Petrosyan, S. Ciliberto, R. Dilenshneider, E. Lutz. Experimental verification of Landauer‘s principle linking Information and thermodynamics, *Nature*, **484**, 187189, 2012
[44] S.Toyabe, T.Sagawa, M.Ueda, E.Muneyuki and M.Sano, Experimental demonstration of Information-to-energy conversion and validation of the generalized Jarzynski Equality, *Nature Physics*, **6**: 988-991, 2010.
[45] V. S. Lerner, Hidden stochastic, quantum and dynamic Information of Markov diffusion processes, and its evaluation by an Entropy Integral measure under the impulse control‘s actions, applied to an Information observer, *arXiv*:1207.3091.
[46]. V. S. Lerner, Arising Information regularities in an observer, *arXiv*: 1307.0449.
[47]. V.S. Lerner, *Information Path Functional and Informational Macrodynamics*, Nova Sc.Publ., NY, 2010.
[48].V.S. Lerner, *Information path from randomness and uncertainty to information, Interactive integrated information dynamics and intelligence of observer*, Lambert Acad. Publishing. 2019.
[49]. V. N. Efimov, Weakly-bound states of three resonantly-interacting particles, *Soviet Journal of Nuclear Physics*, **12**(5): 589595,1971.
[50]. R. Pires, J. Ulmanis, S. Häfner, M. Repp, A Arias, E. D. Kuhnle, M.Weidemüller, Observation of Efimov Resonances in a Mixture with Extreme Mass Imbalance, *Phys.Rev.Lett*, **112** (25),10.1103/ 112.250404.
[51]. L. H.Kauffman, *Formal Knot Theory*, Princeton Univ. Press, 1983.
[52] K. S. Chichak, Molecular Borromean Rings, *Science*, **304** (5675): 1308–12, 2004.

[53]. V. S. Lerner, An observer's Information dynamics: Acquisition of Information and the origin of the cognitivedynamics, *Journal Information Sciences*, **184**: 111-139, 2012.
[54] C. H.Bennett, E.Bernstein, G. Brassard, U.Vazirani The strengths and weaknesses of quantum computation, *SIAM Journal on Computing*, **26**(5): 1510–1523, 1997.
[55] Hatim Salih, Zheng-Hong Li, M. Al-Amri, and M. Suhail Zubairym, Protocol for Direct Counterfactual Quantum Communication, *Phys. Rev. Lett*., **110**, 170502, 2013.
[56] Yuan Cao, Yu-Huai Li, Zhu Cao, Juan Yin, Yu-Ao Chen, Hua-Lei Yin, Teng-Yun Chen, Xiongfeng Ma, Cheng-Zhi Peng, and Jian-Wei Pan, Direct counterfactual communication via quantum Zeno effect, *PNAS*, **114** (19) 4920-4924, 2017.
[57] S. Chang, M. Zhou, C. Grover, Information coding and retrieving using fluorescent semiconductor nanocrystals for object identification, *Optics express*, Osapublishing.org, 2004.
[58] C. Mao, W. Sun, and N.C. Seeman , Assembly of Borromean rings from DNA. *Nature*, **386** (6621): 137–138, 1997.
[59] M. W. Nirenbero, W. Jones, P.Leder, B. F. C. Clark., W. S. Sly, S. Pestka , On the Coding of Genetic Information, *Cold Spring Harb Symposium Quantum Biology*, **28**: 549-557, 1963.
[60] A.S. Rodin A. S., Eörs Szathmáry, S.N. Rodin, On origin of genetic code and tRNA before Translation, *Biology Direct*, **6**:14-15, 2011.
[61] V.S. Lerner Macrodynamic Cooperative Complexity in Information Dynamics, *Journal Open Systems and Information Dynamics*, **15** (3):231-279, 2008.
[62] A.N. Kolmogorov, Logical basis for Information theory and probability theory, *IEEE Trans. Inform. Theory*, **14** (5): 662–664, 1968.
[63] G. J. Chaitin Information-theoretic computational complexity. *IEEE Trans. Information Theory*, **20**:10-15, 1974.
[64] C.H. Bennett, Logical depth and physical complexity, in: *The Universal Turing Machine*, R. Herken (Ed.), 227-258, Oxford University Press, 1988.
[65] R. J. Solomonoff, Complexity-based induction systems: comparisons and convergence theorems, *IEEE Transactions on Information Theory*, **24**: 422-432, 1978.
[66] J. F Traub, G. W. Wasilkowski, H.Wozniakowski, *Information-Based Complexity*, Academic Press, London, 1988.
[67] V. S. Lerner Information complexity in evolution dynamics, *Int. Journal of Evolution Equations*, **3** (1):27-63, 2007.
[68] A. Einstein, *The meaning of Relativity*, Princeton University Press, Princeton, 1921.
[69] E. C. G. Sudarshan, B. Misra, The Zeno's paradox in quantum theory.*Journal of Mathematical Physics*, **18**(4): 756–763, 1977.
[70] A.N.Kolmogorov, On the representation of continuos functions of many variables by superposition of continuous functions of one variable and addition, *Dokl. Academy Nauk USSR*, **114**:953-956, 1978.
[71] V. S.Lerner, *Information Systems Analysis and Modelling: An Informational Microdynamics Approach*, Kluwer Acad. Publ., 1999.
[72] V. S. Lerner, Information functional mechanism of cyclic functioning, *Journal of Biological Systems*, **9**(3), 2011.
[73] V. S.Lerner, Introduction to Information systems theory: concepts, formalism and applications, *Int. Journal of Systems Science*, **35** (7): 405–424, 2004.
[74] V. S.Lerner, The Observer-Generated Information Process: Composite Stages, Cooperative—Logical Structure, and Self-Organization, *Cybernetics and Systems*, **47**(5):1-24, 2016.
[75] M. S Aravindh, A. Venkatesan, M.Lakshmanan. Strange nonchaotic attractors for computation, *arXiv*:1805.03858.

[76] J. F Chiang, M. H.Rosenberg, C. A Bufford, D.Stephens, A.Lysy and M.Monti,The language of music: Common neural codes for structured sequences in music and natural language, *http://dx.doi.org*/10.1101/202382. 2017.

[77] G.Buzsaki, *Rhythms of the Brain*, Oxford University Press, 2006.

[78] J. M. Yearsley and E. M.Pothos, Challenging the classical notion of time in cognition: a quantum perspective. *Proceedings of The Royal Society B*, doi:10.1098/rspb.2013.3056.

[79] R. A.Epstein, E. Z.Patai, J. B Julian, and H. J.Spiers The cognitive map in humans: spatial naviga tion and beyond, *Nature Neuroscience*, **20**,1504–1513, 2017.

[80] V.S.Lerner, *How Information creates its Observer. The Emergence of the Information Observer with Regularities*, Nova Science, 2019.

[81] L.Dai, M.Kamionkowski, E. D. Kovetz., A.Raccanelli, and M.Shiraishi M. Antisymmetric galaxy cross-correlations as a cosmological probe,*arXiv*:1507.05618v2.

[82] J. Klaers, Landauer's Erasure Principle in a Squeezed Thermal Memory, *Phys. Rev. Lett.***122**, 040602 , 2019.

[83] S. Herculano-Houzel, *The Human Advantage: A New Understanding of How Our Brain Became Remarkable*, MIT, 2016.

[84] K.Sidiropoulou, K.E. Pissadaki, P. Poirazi , Inside the brain of a neuron, Review, *European Molecular Biology Organization reports*, **7**(9): 886- 892,2006.

[85]. W. James, Does 'Consciousness' Exist? *Journal of Philosophy, Psychology, and Scientific Methods*, ***1***, 477-491, 1904.

[86] R Penrose, *Shadows of the Mind: A Search for the Missing Science of Consciousness,* Oxford University Press, 1994.

[87] A.N.Kolmogorov , *Foundations of the Theory of Probability,* Chelsea, New York, 1956.

[88]. W. H. Zurek, Sub-Planck structure in phase space and its relevance for quantum decoherence, *Letters to Nature, Nature* **412**, 712-717, 2001.

[89] I.Rabinowitch, What would a synthetic connectome look like? *Physics of Life Reviews*, *https://doi.org/10.1016/j.plrev.2019.06.005.*

[90] .G. Toponi, An information integration theory of consciousness, *BMC Neuroscice* 5(42), 2004.

[91].V.S. Lerner, Reality, Information, And Intelligent Observer, *European Journal of Computer Science and Information Technology*,**8** (2) 39-175, 2020.

[92] ] J.A Wheeler, *Genesis and Observership, in Foundational Problems in Special Science* **3**-34, 1977, D. Reidel Publishing Company, Dordrecht, Holland.

[93] J.A Wheeler. On of recognizing "law without law", *American Journal of Physics* **51**, 398 ,1983; 10.1119/1.13224.

[94] V.S. Lerner, Reality, Information, and Information Observer, *The Conference Theoretical Information Studies, Berkeley*, CA, USA, 2–6 June 2019, Proceedings 2020, 47, 10; mdpi.com/journal/proceedings.

[95] W. Byers , *The Blind Spot, Science and Crisis of Uncerainty,* Princeton University Press ,2011.